\documentclass[11pt]{amsbook}
\allowdisplaybreaks

\numberwithin{section}{chapter}
\numberwithin{equation}{chapter}
\usepackage[english]{babel}
\usepackage[pdftex,textwidth=400pt,marginratio=1:1]{geometry}
\usepackage{amsfonts}
\usepackage{graphics}
\usepackage[colorinlistoftodos]{todonotes}
\usepackage{amsmath}
\usepackage{amsthm}
\usepackage{amssymb}
\usepackage{bbm}
\usepackage{cancel}
\usepackage{color}
\usepackage{xcolor}
\usepackage[curve]{xypic}
\usepackage[all,cmtip]{xy}
\usepackage{graphicx}
\usepackage{mathabx}
\usepackage{tikz-cd}
\usepackage{epigraph} 
\usepackage{hyperref}
\usepackage{stmaryrd}
\usepackage{enumitem}
\usepackage[mathscr]{eucal}
\usepackage{soul}
\usepackage{xfrac}
\usepackage{slashed}
\newtheorem{theorem}{Theorem}[section]
\newtheorem{corollary}[theorem]{Corollary}
\newtheorem{proposition}[theorem]{Proposition}

\newtheorem{conjecture}[theorem]{Conjecture}

\theoremstyle{definition}
\newtheorem{definition}[theorem]{Definition}
\newtheorem{example}[theorem]{Example}
\newtheorem{exercise}[theorem]{Exercise}
\newtheorem{remark}[theorem]{Remark}

\DeclareFontFamily{OT1}{rsfs}{}
\DeclareFontShape{OT1}{rsfs}{n}{it}{<-> rsfs10}{}
\DeclareMathAlphabet{\curly}{OT1}{rsfs}{n}{it}

\newcommand\I{\mathcal I}

\newcommand\LL{\mathbb L}

\renewcommand\O{\mathcal O}
\newcommand\PP{\mathbb P}

\newcommand\D{\mathcal D}
\newcommand\cE{\mathcal E}
\newcommand\EE{\mathbb E}
\newcommand\cA{\mathcal A}

\newcommand\E{\mathbb E}
\newcommand\C{\mathbb C}
\newcommand\cC{\mathcal C}

\newcommand\cF{\mathcal F}

\newcommand\sfZ{\mathsf Z}

\newcommand\cO{\mathcal O}
\newcommand\Q{\mathbb Q}

\newcommand\R{\mathbb R}
\newcommand\ccR{\mathcal R}

\newcommand\Z{\mathbb Z}
\newcommand\cZ{\mathcal Z}

\newcommand\Res{\mathrm{Res}}

\newcommand\bfmu{{\boldsymbol \mu}}
\newcommand{\bfk}{{\boldsymbol k}}

\newcommand\s{\mathfrak s}
\renewcommand\t{\mathfrak t}

\newcommand\Quot{\mathrm{Quot}}

\newcommand\SU{\mathrm{SU}}

\newcommand\vd{\mathrm{vd}}
\newcommand\pt{\mathrm{pt}}
\newcommand\vir{\mathrm{vir}}

\newcommand\SW{\mathrm{SW}}
\newcommand\VW{\mathrm{VW}}
\newcommand\td{\mathrm{td}}
\newcommand\rk{\operatorname{rk}}

\newcommand\tr{\operatorname{tr}}

\newcommand\im{\operatorname{im}}
\newcommand\ch{\operatorname{ch}}

\newcommand\Hom{\operatorname{Hom}}
\renewcommand\hom{\mathcal{H}{\it{om}}}
\newcommand\Ext{\operatorname{Ext}}

\newcommand\Spec{\operatorname{Spec}\,}
\newcommand\Hilb{\operatorname{Hilb}}

\newcommand\Sym{\operatorname{Sym}}

\newcommand\INTO{\ar@{^{(}->}[r]}

\DeclareRobustCommand{\SkipTocEntry}[4]{}

\def\cM{\mathcal{M}}
\def\cN{\mathcal{N}}
\def\cL{\mathcal{L}}

\def\cB{\mathcal{B}}

\def\Quot{\mathrm{Quot}}

\def\Pic{\mathrm{Pic}}

\def\PGL{\mathrm{PGL}}
\def\SL{\mathrm{SL}}

\def\HH{\mathbb{H}}

\def\cH{\mathcal{H}}

\def\and{\quad\mathrm{and}\quad}

\def\bfn{\boldsymbol{n}}

\title{Quantum field theory and four-manifolds: \\ Perspectives from physics and geometry}
\author{
Niklas Garner \\ 
\vspace{-0.2cm} {\small Mathematical Institute, University of Oxford} \\ \vspace{.2cm}
Martijn Kool \\ \vspace{-0.2cm} {\small Mathematical Institute, Utrecht University} \\ \vspace{.2cm} Jan Manschot \\ \vspace{-0.2cm} {\small School of Mathematics, Trinity College Dublin}\\ \vspace{.2cm} Yuuji Tanaka \\ 
\vspace{-0.2cm} {\small Beijing Institute of Mathematical Sciences and Applications}\\ \vspace{.2cm} }

\date{\today}

\begin{document}

\maketitle

\begin{abstract}
    These lecture notes are based on lecture series by the authors for the Summer School 2025 on “Physical mathematics of quantum field theory” at the University of Massachusetts Amherst organized by Owen Gwilliam, Chris Elliott, and Matteo Lotito. The lectures centred around algebraic structures arising from twists of supersymmetric quantum field theories, correlation functions of topologically twisted supersymmetric gauge theories, and their mathematical counterparts as intersection-theoretic invariants associated to moduli spaces of instantons, semistable sheaves, or Higgs sheaves on four-manifolds or complex projective surfaces.
\end{abstract}

\tableofcontents

\chapter{Introduction}
\label{Chap:intro}

These lecture notes are based on lecture series by the authors for the Summer School on ``Physical Mathematics of Quantum Field Theory" in 2025 at the University of Massachusetts Amherst organized by Owen Gwilliam, Chris Elliott, and Matteo Lotito. The school consisted of four lecture series of four lectures each:
\begin{itemize}
\item \emph{Algebraic aspects of twisted super Yang--Mills theories}, by Niklas Garner;
\item \emph{Topological correlators of \texorpdfstring{$\cN=2$}{N=2} supersymmetric Yang--Mills theory and four-manifold invariants}, by Jan Manschot;
\item \emph{From the Donaldson invariants to the Vafa--Witten theory}, by Yuuji Tanaka;
\item \emph{Algebraic geometry and Vafa--Witten theory}, by Martijn Kool.
\end{itemize}

The connecting theme of the lectures was the interaction between four-dimensional supersymmetric gauge theory, low-dimensional topology, algebraic geometry, and enumerative geometry. In particular, the lectures centred around correlation functions of topologically twisted supersymmetric gauge theories and their mathematical counterparts as intersection-theoretic invariants associated to moduli spaces of instantons, semistable sheaves, or Higgs sheaves on four-manifolds or complex projective surfaces.
\begin{figure}[h]
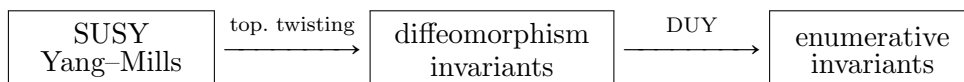

\begin{equation*}
\fbox{\parbox[c][0.8cm][c]{2.5cm}{\centering \shortstack{SUSY \\ Yang--Mills}}} \stackrel{\textrm{top.~twisting}}{\xrightarrow{\hspace*{1.6cm}}}  \fbox{\parbox[c][0.8cm][c]{3cm}{\centering \shortstack{diffeomorphism \\ invariants}}} \stackrel{\textrm{DUY}}{\xrightarrow{\hspace*{1.6cm}}} \fbox{\parbox[c][0.8cm][c]{2.5cm}{\centering \shortstack{enumerative \\ invariants}}}
\end{equation*}
\caption{The first arrow is provided by the procedure of topological twisting and the second by the Donaldson--Uhlenbeck--Yau theorem.}
\end{figure}

These ideas have their roots in several closely intertwined developments. In the 1950s, Yang and Mills formulated gauge theories for a general compact Lie group, which in the following decennia developed to the theory describing the fundamental interactions between matter particles in the Standard Model.\footnote{More precise references for the developments discussed in this introduction are provided in subsequent chapters.} The deep links of these theories with differential geometry were recognized in the 1960s and 1970s. This was manifested by the realization of non-perturbative solutions in Yang--Mills theory, including monopole solutions by Wu--Yang and 't Hooft--Polyakov, and self-dual instantons by Belavin--Polyakov--Schwarz--Tyupkin. Around the same time, the notion of supersymmetry was introduced relating bosonic and fermionic degrees of freedom, which was first combined with Yang--Mills theory by Wess and Zumino. Over time, such theories have proven to provide a particularly fruitful and inspiring connection to differential geometry and mathematics more broadly. 

In a parallel development beginning in the 1960s, the Atiyah--Singer index theorem provided a fundamental tool, nowadays used to compute the expected dimensions of the moduli spaces. In the 1970s, the geometry and topology of moduli spaces of semistable vector bundles on Riemann surfaces were studied in algebraic geometry, notably through the work of Harder and Narasimhan, which made use of the Weil conjectures in the study of these moduli spaces. In the 1980s, Atiyah and Bott brought gauge theory into this picture, using the Yang--Mills functional and equivariant Morse theory to study the topology of moduli spaces of bundles over Riemann surfaces. In particular, their gauge-theoretic approach related the Morse stratification of the Yang--Mills functional to the Harder--Narasimhan stratification. This circle of ideas was further developed by Kirwan and others in the study of the topology of symplectic and geometric invariant theory quotients, revealing increasingly rich connections between gauge theory, the geometry and topology of moduli spaces, and algebraic geometry.

Turning back to developments in the late 1970s and early 1980s, the mathematical foundations of the instanton moduli spaces on four-manifolds were also developed rapidly. Atiyah, Hitchin and Singer studied the deformation theory of self-dual connections, giving a local description of instanton moduli spaces in terms of an elliptic deformation complex. At the same time, Atiyah, Drinfeld, Hitchin and Manin gave their celebrated algebraic construction of instantons, now known as the ADHM construction, providing an explicit description of framed instantons on $\mathbb{R}^4$ in terms of finite-dimensional linear-algebraic data. The hard analytic work of Uhlenbeck and Taubes, particularly on the compactness of instanton moduli spaces and the gluing construction of instantons, was then decisive in making the compactification of instanton moduli spaces on four-manifolds possible.

Building on these analytic and geometric foundations, Donaldson successfully studied moduli spaces of anti-self-dual instantons in the early 1980s and used them to uncover exotic phenomena of smooth four-manifolds, as well as to construct powerful diffeomorphism invariants. During the same decade, Hitchin's work on self-duality equations and Higgs bundles opened another major direction in the study of gauge-theoretic moduli spaces and their links with algebraic geometry and integrable systems.

A further decisive development came from quantum field theory. In the late 1980s, Witten interpreted Donaldson theory in terms of a modification of $\mathcal{N}=2$ supersymmetric Yang--Mills theory known as a ``topological twist". Notably the invariants of four-manifolds arise as correlation functions of the topologically twisted supersymmetric Yang--Mills theory. On the mathematical side, Kronheimer and Mrowka subsequently established remarkable structure theorems for Donaldson invariants. These developments formed part of the background to the breakthrough of Seiberg and Witten in the mid-1990s: their solution of four-dimensional $\mathcal{N}=2$ supersymmetric gauge theory led to the discovery of a new and strikingly effective set of invariants of smooth four-manifolds. Around the same time, Vafa and Witten applied a topological twist to $\mathcal{N}=4$ supersymmetric Yang--Mills theory, bringing modularity and the geometry of moduli spaces of instantons into the picture. Subsequently, the approach of low energy effective field theory by Moore--Witten and Losev--Nekrasov--Shatashvili  made the connection between Donaldson invariants and the Seiberg--Witten curves and their modular properties more explicit, with modular forms entering directly into the calculation of four-manifold invariants through the $u$-plane integral.

In the 2000s, Nekrasov's instanton partition function and subsequent developments further deepened the connections between supersymmetric gauge theories, instanton counting, and enumerative geometry. The work of G\"ottsche, Yoshioka, Nakajima, Mochizuki, and many others on moduli spaces of sheaves and their invariants established increasingly precise links between four-manifold invariants, algebraic geometry, and representation theory.

One of the main tools in developing the physical perspective on the enumerative invariants appearing in Donaldson and Vafa--Witten theories is Witten's notion of topological descent, whereby extended observables naturally constructed from local ones.
Beyond the four-dimensional setting relevant to these notes, this process uncovered many remarkable algebraic structures possessed by local observables themselves; these include the string-theoretic Gerstenhaber and homotopy-algebraic structures uncovered by Lian--Zuckermann, Penkava--Schwarz, Witten--Zwiebach and others during the early '90s.

Besides the topological twists described above, an important development is the recognition that less-topological twists are also permissible and offer yet richer algebraic structures; these theories retain a dependence on holomorphic coordinates of space-time and lead to a rich algebraic structure with, for example, connections to vertex algebras.
The first example of this is Witten's half-twisted sigma model, which was connected to the chiral de Rham complex by Kapustin, but four-dimensional examples were first described in work of Johansen and Losev--Moore--Nekrasov--Shatashvili in the mid-to-late '90s, which were reformulated by Costello in a more mathematical language in the early 2010s, and by Kapustin in the mid 2000's.

Altogether, these developments continue to provide a remarkably fertile meeting point of quantum field theory, differential geometry, low-dimensional topology, algebraic geometry, enumerative geometry, and representation theory.

A recurring principle throughout these notes is that the same invariant can often be approached from several quite different directions. On the physics side, supersymmetry and topological twisting turn suitable path integrals into quantities which are formally independent of the Riemannian metric and which localise to moduli spaces of solutions of gauge-theoretic equations. At low energy, duality and the geometry of the Coulomb branch lead naturally to modular and mock modular forms. On the differential-geometric side, elliptic systems such as the anti-self-duality, Seiberg--Witten, and Vafa--Witten equations give rise to gauge-theoretic moduli spaces, whose geometry and analysis encode diffeomorphism invariants of smooth four-manifolds.

On a complex projective surface, the Donaldson--Uhlenbeck--Yau correspondence translates gauge-theoretic moduli problems into moduli spaces of semistable sheaves and, in its Higgs-bundle counterpart, semistable Higgs sheaves. Algebraic geometry then provides natural compactifications, perfect obstruction theories, virtual cycles, and localisation techniques with which the resulting invariants can be defined and, in many cases, computed explicitly. 
\vspace{.5cm}\\
{\bf Content and prerequisites.}
The present manuscript covers the lecture series mentioned above. Although they are written from different points of view, they are intended to be read as a connected account of the passage from supersymmetric gauge theory, and their topological twisting, through four-manifold invariants to, ultimately, an algebro-geometric formulation of Vafa--Witten theory. We realize that the topic of quantum field theories and their relation to four-manifold geometry is a vast subject with an extremely rich history. As a result, we do not attempt to give a comprehensive treatment and rather focus on the topics covered in the lectures (which necessarily reflect the personal tastes and expertise of the lecturers). The reader is encouraged to consult various references provided throughout the chapters for further exploration.

For Chapter \ref{chap:algTwist}, we assume prior knowledge of quantum field theory and the basics of supersymmetry; textbook treatments of this material can be found in e.g.~Weinberg \cite{Weinberg:2000cr} and Wess--Bagger \cite{Wess:1992cp}. We also make use of some notions from algebraic and differential topological, which can be found e.g. in the textbook of Bott--Tu \cite{BottTu}. A complementary discussion of many of the notions in this lecture series appeared in a collection of notes by N.G. with Natalie Paquette \cite{GaPa}, albeit with a very different focus.

For Chapter \ref{chap:ySYM}, we assume some prior knowledge of quantum field theory and supersymmetry. Examples of textbooks are Peskin--Schroeder \cite{Peskin:1995}, Tachikawa \cite{Tachikawa:2013kta}, Weinberg \cite{Weinberg:2000cr} and Wess--Bagger \cite{Wess:1992cp}. Some prior knowledge of manifold theory and characteristic classes will be useful for Lecture \ref{sec:JMLect2}. This is for example discussed in Nakahara \cite{Nakahara:2003nw}. A general reference is the textbook by Labastida and Marino \cite{Labastida:2005zz}.

For Chapter \ref{chap:fromDtoVW}, we assume some familiarity with the basics of differential geometry and manifold theory, in particular connections and curvature on vector and principal bundles, as well as characteristic classes. Some background in complex and algebraic geometry, including holomorphic vector bundles, coherent sheaves, stability, and basic deformation theory, is useful for Lectures 3 and 4 there. Standard references for the gauge-theoretic background are Freed--Uhlenbeck \cite{FrUh} and Donaldson--Kronheimer \cite{DoKr}; for Hermitian--Einstein connections, stability, and the Donaldson--Uhlenbeck--Yau theorem, we refer to Kobayashi \cite{Koba} and L\"ubke–Teleman \cite{LuTe}. 

For Chapter \ref{ch:MK}, we require familiarity with basic algebraic geometry (e.g.~such as \cite{Har}, including Appendix A). The focus of this chapter is on the algebro-geometric techniques used for computing Seiberg--Witten, Donaldson, and Vafa--Witten invariants. The relevant moduli spaces are introduced rather quickly and for a thorough treatment of their construction and basic properties references are provided to Huybrechts--Lehn's book \cite{HL}, Lehn's lecture notes \cite{Leh1}, and Nakajima's lecture notes \cite{Nak}. 
\vspace{.5cm}\\
{\bf Acknowledgments.}
We thank all the participants of the school. Their questions, comments and discussion helped both the lectures as well as the writing of these lecture notes. We are especially grateful to the organisers, Owen Gwilliam, Chris Elliott, and Matteo Lotito, for organising the school and for giving us the opportunity to present these lectures.

We would also like to thank our collaborators and colleagues, including Noah Arbesfeld, Johannes Aspman, Dirk van Bree, Christopher Beem, Matthias Dennemann, Tudor Dimofte, Paul Feehan, Elias Furrer, Amin Gholampour, Lothar G\"ottsche, Yunfeng Jiang, Dominic Joyce, Heeyeon Kim, Georgios Korpas, Ties Laarakker, Oliver Leigh, Woonam Lim, Henry Liu, Tom Manopulo, Gregory Moore, Nick Kuhn, Jerry Obour, Natalie Paquette, Surya Raghavedran, Andrea Ricolfi, Samson Shatashvili, Runkai Tao, Richard Thomas, Weisheng Wang, Brian Williams, Edward Witten, Ethan Yang, and Xinyu Zhang, for helpful discussions and comments.

The summer school was funded by DMS Grant 2042052 from the U.S. National Science Foundation.
N.G. was previously supported by ERC Consolidator Grant \#864828 “Algebraic Foundations of Supersymmetric Quantum Field Theory” (SCFTAlg) and is currently supported by the Simons Collaboration on Celestial Holography CH-00001550-1.
M.K. was supported by NWO Grant VI.Vidi.192.012 and ERC Consolidator Grant FourSurf 101087365. Y.T. was supported by JSPS KAKENHI Grant Numbers JP21H00973 and JP21K03246, a Startup Grant at BIMSA, and the Beijing Natural Science Foundation Beijing International Scientist Project IS25031. Y.T. would also like to thank Kyoto University and the Okinawa Institute of Science and Technology (OIST) for their support and hospitality during his visits, where part of these lecture notes were written. \\

\noindent \textbf{AI disclosure.}
We have not used an AI in the writing of these lecture notes, except to improve the fluency and readability of the text in Chapters \ref{Chap:intro} and \ref{chap:fromDtoVW}. 

\chapter[Algebraic aspects of twists]{Algebraic aspects of twisted super Yang-Mills theories}
\label{chap:algTwist}
\vspace{-.4cm}
\begin{center}
\it{Niklas Garner}
\end{center}
\vspace{.4cm}

The aim of these lectures is to describe in general terms the algebraic structures that may be extracted from supersymmetric quantum field theories and play a crucial role in the four-manifold invariants studied in greater detail in the subsequent chapter.

\section[Lecture 1]{Lecture 1: The four-dimensional super Poincar\'{e} algebra and twisting}
\label{sec:NGLect1}
In this first lecture we provide a brief introduction to the four-dimensional super Poincar\'{e} algebra and the notion of twisting a supersymmetric quantum field theory.
The notion of twisting will allow us to extract from the underlying supersymmetric quantum field theory a simpler object, e.g. a topological quantum field theory, which may then be amenable to precise mathematics.
In the context of the four-dimensional $\cN \geq 2$ theories and (some of) the topological field theories that come from them, this is precisely the aim of Chapter \ref{chap:fromDtoVW} and \ref{ch:MK}; the complementary physical analysis described in Chapter \ref{chap:ySYM} of the same objects can further shed light on this mathematics, e.g. by offering explicit predictions for the resulting topological invariants and structural properties they should possess.

We will work entirely in Euclidean signature.

\subsection{4d super-Poincar\'e algebras}
\label{sec:superPoincare}
The main object underlying this series of lectures is four-dimensional supersymmetry.
The four-dimensional Euclidean super-Poincar\'e algebra with $\cN$ supersymmetries is a super Lie algebra extending the Lie algebra of the spin cover of the four-dimensional Euclidean group
\begin{equation}
	\textrm{Spin}(4)\ltimes\R^4 ~\cong~ \big(\SU(2)_+\times\SU(2)_-\big)\ltimes \big(S_+\otimes S_-\big),
\end{equation}
where $S_+$ is the positive (right-handed) and $S_-$ the negative (left-handed) spinor representation of $\textrm{Spin}(4)$.
Of course, $\textrm{Spin}(4)$ is the (spin cover of the) subgroup of rotations and $\R^4$ is subgroup of translations; together they realize the group of isometries of four-dimensional Euclidean space.

The extension is by Grassmann odd/fermionic generators, called ``supercharges'' and generating ``super translations,'' that transform in the representation
\begin{equation}
	W\otimes S_+ \oplus W^*\otimes S_-,
\end{equation}
where $W$ is a complex vector space of dimension $\dim W = \cN$ and $W^*$ its linear dual.
These supercharges commute with translations, whereas the action of $\textrm{Spin}(4)$ on these supercharges is as dictated by the representation they live in; the Lie bracket of two supercharges gives a (complex) linear combination of translation generators.
We choose a basis $P_\mu \in \R^4$, $\mu = 1, \dots 4$, for translations and $Q^I_\alpha \in W \otimes S_+$ and $\widetilde{Q}_{I \dot{\alpha}} \in W^* \otimes S_+$ for the supercharges, where $I = 1, \dots, \cN$, $\alpha \in \pm$, and $\dot{\alpha} \in \dot{\pm}$.
In this basis, the non-trivial Lie brackets between 
\begin{equation}
\label{eq:QQt}
	\big[Q^I_\alpha, \widetilde{Q}_{J\dot\alpha}\big] = 2\,\delta^I_J\,\sigma^\mu_{\alpha\dot\alpha}P_\mu = 2\,\delta^I_J \begin{pmatrix}
		P_3 + i P_4 & P_1 - i P_2\\
    	P_1 + i P_2 & -P_3 + i P_4
  \end{pmatrix},
\end{equation}
where $\sigma^\mu_{\alpha \dot{\alpha}}$ denotes the four-vector of Pauli matrices that identifies the vector representation of $\mathrm{Spin}(4)$ with the tensor product of left- and right-handed spinors; the rows and columns of the matrix are labelled by $(\alpha\dot\alpha)=(+\dot{+}),\,(+\dot{-})$ in the first row and $(-\dot{+}),\,(-\dot{-})$ in the second.

The above super Lie algebra admits interesting central extensions by $\bigwedge^2W \oplus \bigwedge^2 W^*$; these correspond to additional generators $Z^{IJ} = -Z^{JI}$ and $\widetilde{Z}_{IJ} = - \widetilde{Z}_{JI}$ called central charges and appear in the Lie brackets of two $Q$'s or two $\widetilde{Q}$'s:
\begin{equation}
  \big[Q^I_\alpha, Q^J_\beta\big] = \varepsilon_{\alpha\beta} Z^{IJ}~, \qquad \big[\widetilde{Q}_{I\dot\alpha}, \widetilde{Q}_{J\dot\beta}\big] = \varepsilon_{\dot\alpha\dot\beta} \widetilde{Z}_{IJ}~.
\end{equation}
Although we will mostly focus our discussion on the case of vanishing central charges $Z = \widetilde{Z} = 0$, they play an important role in Chapter \ref{chap:ySYM}.
We also note that the (Euclidean) super Poincar\'{e} algebra admits a continuous outer automorphism group $\mathrm{U}(W) = \mathrm{U}(\cN)$, referred to as the group $R$-symmetries.
It may \emph{not} be realized in a given physical system; e.g. it may be explicitly broken or anomalous.

\subsection{Twisting}
\label{sec:twisting}
One of of the most important operations that supersymmetry allows us to perform is the process of \emph{twisting}.
In fact, this notion was first introduce by Witten in order to understand the physical origin of Donaldson theory, as described in much more detail in Chapter \ref{chap:ySYM}.

To a first approximation, twisting amounts to choosing a nilpotent supercharge $\mathbb{Q}$, i.e. one satisfying
\begin{equation}
	\big[\mathbb{Q}, \mathbb{Q} \big] = 0,
\end{equation}
and then restricting attention to $\mathbb{Q}$-invariant observables.
If the $\mathbb{Q}$-symmetry is not spontaneously broken, i.e. we work in a state that is preserved by $\mathbb{Q}$, then correlation functions of $\mathbb{Q}$-closed operators vanish if one such observable is itself a $\mathbb{Q}$-variation: if $\cO_i$ are $\mathbb{Q}$-invariant observables then
\begin{equation}
	\langle (\mathbb{Q}\widetilde{\cO}) \cO_1 \cdots \cO_n \big\rangle = \langle \mathbb{Q}(\widetilde{\cO} \cO_1 \cdots \cO_n) \big\rangle = 0~.
\end{equation}
Because $\mathbb{Q}$ squares to zero, we can view it as a differential on the vector space of all observables; the above observation then can be recast by saying correlation functions of $\mathbb{Q}$-invariant observables only depend on their classes in $\mathbb{Q}$-cohomology.
We refer to $\mathbb{Q}$-invariant observables/states as $\mathbb{Q}$-closed and say that $\mathbb{Q}$-variants are $\mathbb{Q}$-exact.
Putting this together, twisting in its simplest form amounts to choosing a nilpotent supercharge $\mathbb{Q}$ and then considering the $\mathbb{Q}$-cohomology of observables and their correlations.

\begin{remark}
	A few remarks are in order.
	\begin{enumerate}
		\item[1)] If there is a supercharge $Q_\mu$ with $[\mathbb{Q},Q_\mu]=P_\mu$, then
		$\mathbb{Q}\cO=0$ implies
		\[
			\partial_\mu\cO = i P_\mu\cO = \mathbb{Q} \big(i Q_\mu\cO\big),
		\]
		so correlation functions of $\mathbb{Q}$-closed observables are independent of $x^\mu$.
		The behavior of the twist is characterized by $\im[\mathbb{Q},-] \subset \C^4$, which is at least two-dimensional.
		If the dimension of the image is maximal $\dim \im[\mathbb{Q},-] = 4$ then all translations are trivial in the twist, and $\mathbb{Q}$ is referred to as a ``topological'' supercharge; the $\mathbb{Q}$-twisted theory behaves topologically.
		If the dimension of the image is minimal $\dim \im[\mathbb{Q},-] = 2$, e.g. if it is spanned by $P_1+iP_2$ and $P_3+iP_4$, then this subspace of (complexified) translations defines a complex structure on spacetime; such a supercharge is referred to as a ``minimal'' or ``holomorphic'' supercharge and $\mathbb{Q}$-twisted theory behaves holomorphically: i.e. correlation functions of local observables are holomorphic functions of their insertion points.
		The remaining case $\dim \im[\mathbb{Q},-] = 3$, e.g. if it is spanned by $P_1+iP_2$, $P_3$, and $P_4$, is somewhere in between: the subspace of (complexified) translations equips spacetime with a transverse holomorphic foliation; such a supercharge is referred to as ``holomorphic-topological'' and correlation functions in the $\mathbb{Q}$-twisted theory are constant along the foliation.
		
		\item[2)] $\mathbb{Q}$-invariance is \emph{not} preserved by rotations: $\mathbb{Q}$ is a spinor!
		That said, one can often choose a ``twisted spin''
		\[
			G \hookrightarrow \textrm{Spin}(4) \times G_R
		\]
		(or $\textrm{Spin}(4)~\times~G_R~\times$ flavor symmetries) under which $\mathbb{Q}$ is invariant.
		If $\pi: \textrm{Spin}(4) \times G_R \to \textrm{Spin}(4)$ denotes the first projection, the $\mathbb{Q}$-twisted theory is thus compatible with the subgroup $\pi(G)$ of rotations acting via the inclusion $G \hookrightarrow \textrm{Spin}(4) \times G_R$.
		This additional data is referred to as a twisting homomorphism and is a requirement if one wishes to put the twisted theory on curved spacetimes, as will be necessary for the later discussions on four-manifold invariants.
        At the level of the underlying supersymmetric field theory, this amounts to turning on a background $G_R$ (or $G_R \times$ flavor symmetry) bundle tied with the spin bundle of spacetime; turning on these background fields is often what is meant by twisting in much of the literature on supersymmetric localization, see e.g. \cite{Moore:2024vsd} and references therein for a recent discussion of these ideas as it pertains to Donaldson invariants.
		
		\item[3)] If there is a $\mathrm{U}(1) \hookrightarrow G_R$ (or $G_R~\times$ flavor symmetries) that commutes with $G$ and acting non-trivially on $\mathbb{Q}$, we can view its weights as a ``cohomological'' grading of the twisted theory.
		It is common to require that $\mathbb{Q}$ has weight $1$ with respect to this $\mathrm{U}(1)$ and that the Grassmann parity of observables agrees with $\mathrm{U}(1)$-weight modulo $2 \Z$, thereby providing a $\Z$ lift of Grassmann parity.
	\end{enumerate}
\end{remark}

\subsection{Classifying twists of 4d SUSY}
\label{sec:classifyTwists}
For a given super Poincar\'{e} algebra, we can consider the space of all possible twists.
To a first approximation, this can be identified with the space of all nilpotent elements in the super Poincar\'{e} algebra: if we expand $\mathbb{Q}$ in the above basis
\begin{equation}
	\mathbb{Q} = q^\alpha_I~Q^I_\alpha + \widetilde{q}^{I \dot{\alpha}}~\widetilde{Q}_{I \dot{\alpha}}
\end{equation}
then the defining equation $\mathbb{Q}^2=0$ reads
\begin{equation}
	q^\alpha_I~\sigma^\mu_{\alpha\dot\alpha}~\widetilde{q}^{I\dot\alpha}=0~.
\end{equation}
If we remove the zero supercharge, whose twist is simply the original theory, and realize that the cohomology is independent of overall scale $\mathbb{Q} \sim \lambda \mathbb{Q}$, we are lead to the \emph{nilpotence variety}
\begin{equation}
	\Big\{[q^\alpha_I,\widetilde{q}^{I \dot{\alpha}}] \in \C\mathbb{P}^{4 \cN - 1} \Big| ~ q^\alpha_I~\sigma^\mu_{\alpha \dot{\alpha}} ~ \widetilde{q}^{I \dot{\alpha}}=0 \Big\}
\end{equation}
introduced in \cite{EaSaWa}; see also \cite{ES}.

In addition to the scaling we mentioned above, which does not change the cohomology, the rotational and $R$-symmetries of the underlying QFT naturally act on the nilpotence variety and a general element thereof acts nontrivially on a given twisting supercharge $\mathbb{Q}$ without changing the resulting physics.
In fact, the nilpotence variety is naturally acted upon by its complexification $\mathrm{SO}(4, \C) \times \mathrm{GL}(\cN,\C)$, although they do not naturally act on the underlying QFT.
The natural ``moduli space of twists'' is then most naturally the space of $\mathrm{Spin}(4) \times G_R$ orbits of the nilpotence variety;  if the supersymmetric theories of interest are parity invariant then one can include those in the action, exchanging the two chiralities of supercharges.
We also note that there is a larger moduli space that could be of interest: the space of nilpotent supercharges together with compatible twisting homomorphism and cohomological grading, up to the action of $\textrm{Spin}(4) \times G_R$.
This larger moduli space naturally fibers over the aforementioned moduli space; the fiber of such a point accounts for the different possible twisting data for a given twist.

Representative twisting supercharges for $\cN = 1$, $2$, and $4$ are collected in Table~\ref{tab:twists}; we do not account for the additional twist data.
Here $\mathrm{H}^2$ denotes a holomorphic twist, $\mathrm{HT}^2$ a holomorphic-topological twist, and $\mathrm{T}^4$ a topological twist; ``$\times$'' means no such twist exists.
These families of twists were first cataloged in \cite{ElSaWi}.%
\footnote{We note that there are supersymmetric field theories with $\cN=3$, see \cite{GER} for the first recorded example, supersymmetry and that the nilpotence variety of this supersymmetry algebra was studied in \cite{EaSaWa}.
We do not include representatives for the twists in this case as our interests are primarily in supersymmetric gauge theories and a famous result of \cite{GIKOS} asserts that a Lagrangian gauge theory with at least $\cN=3$ supersymmetry necessarily has at least $\cN=4$; see \cite{Weinberg:2000cr} for a textbook treatment or \cite{AGF} for a related statement for sigma models.} %

\begin{table}[ht]
\centering
	\begin{tabular}{c|c|c|c}
	& $\cN=1$ & $\cN=2$ & $\cN=4$\\
	\hline
	\hline
	& & &\\
	$\mathrm{H}^2$ & $\widetilde{Q}_{\dot{+}}$ & $\widetilde{Q}_{1 \dot{+}}$ & $\widetilde{Q}_{1 \dot{+}}$\\
	& & &\\
	\hline
	& & &\\
	$\mathrm{HT}^2$ & $\times$ & $\widetilde{Q}_{1 \dot{+}} + Q^2_+$ (K) & $\widetilde{Q}_{1 \dot{+}} + Q^2_+$ (K)\\
	& & &\\
	\hline
	& & &\\
	$\mathrm{T}^4$ & $\times$ & $\widetilde{Q}_{1 \dot{+}}+\widetilde{Q}_{2 \dot{-}}$ (DW) & $\begin{array}{c}
		\widetilde{Q}_{1 \dot{+}} + \widetilde{Q}_{2 \dot{-}} + \lambda Q^3_+~\text{(EY)}\\ u\big(\widetilde{Q}_{1 \dot{+}} + \widetilde{Q}_{2 \dot{-}}\big) + v \big(Q^3_+ + Q^4_-\big) ~ \text{(KW)}\\ \text{($v=0$: DW/VW)}
	\end{array}$\\
	\end{tabular}
	\caption{Representative supercharges for the twists of four-dimensional QFTs with $\cN$ supersymmetries. The parameter $\lambda$ is a arbitrary, nonzero complex number and $u, v$ are a pair of complex numbers, up to overall scale.
	K refers to the twist first described in \cite{Kapustin:2006hi} and known as the Kapustin or holomorphic-topological twist.
	DW refers to the twist first described in \cite{Witt88} and known as the Donaldson-Witten twist and studied in greater detail in Chapters \ref{chap:fromDtoVW}.
	KW refers to the family of twists first described in \cite{KaWi} and known as the Kapustin-Witten or geometric Langlands twists.
	VW refers to the twist first described in \cite{VW}; the supercharge belongs to the DW and KW families but uses a yet different twisting homomorphism and will be studied in greater detail in Chapters \ref{chap:fromDtoVW} and \ref{ch:MK}.
	EY refers to the family of twists first described in \cite{EY} and deforms the Donaldson-Witten twisting supercharge but does not admit a twisting homomorphism compatible with full rotational symmetry; they are not known by another name.}
\label{tab:twists}
\end{table}

\begin{remark}
	When the underlying supersymmetric quantum field theory is actually a super\emph{conformal} field theory, there are more twists that one can consider.
	The two most famous examples of this mechanism are in three, four, and six dimensions: in \cite{BPR} the authors extract a 1-dimensional topological field theory from a three-dimensional $\cN = 4$ superconformal field theory and in \cite{BLLPRvR, BRvR} the authors extract a two-dimensional holomorphic field theory from four-dimensional $\cN = 2$ and six-dimensional $\cN = (2,0)$ superconformal field theory.
	See \cite{ElGwLo} for the first steps in categorizing the allowed twists of superconformal field theories in low dimensions and the assortment of field theories that result form these twists.
\end{remark}

\section[Lecture 2]{Lecture 2: Crash course on (twisted) topological quantum field theory}
\label{sec:NGLect2}

With the basics of twisting supersymmetric field theories under our belts, the aim of this next lecture is to better understand some of the algebraic structures that arise from such a construction in the simplest case: topological twists.
Many aspects of this discussion apply to more general field theories that are not topological on the nose but instead are define cohomologically and are topological in cohomology; to remain in contact with the perspective offered by twisting, we continue to denote the differential $\mathbb{Q}$.
The physical mechanism underlying the algebraic structures we are describe first appeared in the physical study of Donaldson theory \cite{Witt88} that will be the central focus of the later lecture series.
The algebraic structures themselves have developed from many observations over the years, see e.g. \cite{BBZBDN, ES, GaPa} and references therein for a taste of this history; we mention some of these developments below.

\subsection{The algebra of local observables}

We start this discussion with local observables in a topological field theory, i.e. observables whose spacetime support is in an arbitrarily small neighborhood of a point.
We will be fairly general about our discussion, so many of the structures we will describe appear in other dimensions; our only really assumption will be that the dimension of spacetime is larger than 1.

Suppose we are given $\mathbb{Q}$-closed local observables $\cO_1$, $\cO_2$, $\dots$, $\cO_n$ and place them at points $x_1$, $x_2$, $\dots$, $x_n$ in spacetime.
The resulting correlation function
\begin{equation}
	\langle \cO_1(x_1) \cO_2(x_2) \dots \cO_n(x_n) \rangle
\end{equation}
is generally a function of the insertion points.
Of course, as this theory is topological, the dependence on the insertion points must be trivial.
To see that this is the case, we can take a derivative with respect to, say, $x_1^\mu$:
\begin{align}
	\partial_{x^\mu_1} \langle \cO_1(x_1) \cO_2(x_2) \dots \cO_n(x_n) \rangle = \langle (\partial_\mu \cO_1)(x_1) \cO_2(x_2) \dots \cO_n(x_n) \rangle~.
\end{align}
We then note that this infinitesimal translation is implemented by the symmetry generator $P_\mu \sim -i \partial_\mu$, which is $Q$-exact:
\begin{align}
	i\langle (P_\mu \cO_1)(x_1) \cO_2(x_2) \dots \cO_n(x_n) \rangle = i\langle (\mathbb{Q} (Q_\mu \cO_1))(x_1) \cO_2(x_2) \dots \cO_n(x_n) \rangle~.
\end{align}
As we saw above, correlation functions of $\mathbb{Q}$-closed observables necessarily vanish if (at least) one is $\mathbb{Q}$-exact and so we conclude that the above correlation function is independent of $x_1^\mu$.
An identical computation shows that the correlation function must be independent of all of the insertion points, as desired.

Instead of talking about a given correlation function, we can ask what structures are present in any correlation function.
One such structure is known as the operator product expansion (OPE): it asserts that given two local observables $\cO_1$ and $\cO_2$ there are other local observables $\cO^k$ and functions $f_k(x)$ such that
\begin{equation}
	\cO_1(x_1) \cO_2(x_2) = \sum_k f_k(x_1 - x_2) \cO^k(x_2)
\end{equation}
in all correlation functions, so long as $x_1$ and $x_2$ are sufficiently close.%
\footnote{Strictly speaking, the right-hand side should be viewed as a formal expression and has vanishing radius of convergence outside of the nicest examples, e.g. conformal or topological field theories.} %
We saw above that correlation functions are independent of insertion points, so the only function that appears on the right-hand side of the OPE in a topological quantum field theory is constant.
Namely, we see that for every two observables we get a third local observable by colliding them, i.e. bringing them close together and performing the OPE.
Collision/OPE thus endows the vector space of local observables with a bilinear operation that we refer to as the product and denote the product of $\cO_1$ and $\cO_2$ by $\cO_1 \star \cO_2$.

\begin{remark}
Several remarks are in order.
\begin{enumerate}
	\item[1)] The product $- \star -$ is readily seen to be associative; given local observables $\cO_1$, $\cO_2$, and $\cO_3$ we can collide $\cO_1$ and $\cO_2$ with one another and then collide $\cO_1 \star \cO_2$ with $\cO_3$ or instead collide $\cO_2$ with $\cO_3$ and then $\cO_1$ with $\cO_2 \star \cO_3$ and the result will be the same (in $\mathbb{Q}$-cohomology).
	Thus, collision equips local observables with the structure of an associative algebra (in $\mathbb{Q}$-cohomology).
	\item[2)] So long as we are working in spacetimes of dimension larger than $1$, the products $\cO_1 \star \cO_2$ and $\cO_2 \star \cO_1$ are essentially the same (in $\mathbb{Q}$-cohomology) as we can rotate the observables around one another and to relate collision in the opposite order.
	The only possibility is that they differ by a sign correlated with their Grassmann statistics.
	Thus, in spacetime dimensions bigger than $1$, local observables furnish a graded-commutative algebra (in $\mathbb{Q}$-cohomology).
	\item[3)] There is always the trivial local observable $1$ corresponding to inserting no local observable; it is thus a unit for the algebra of local observables.
\end{enumerate}
\end{remark}

\subsection{Descent of local observables}
\label{sec:descentlob}
We have just seen that collision equips the space of local observables with the structure of a unital, graded-commutative algebra.
That said, we belabored the point that this was a statement at the level $\mathbb{Q}$-cohomology.
This structure can be enriched by keeping track of \emph{how} we related various products to one another.
These requires quite a bit of extra \emph{data}: we need to supply the observables/homotopies that make each of the properties hold in $\mathbb{Q}$-cohomology.
As we shall see, this data is neatly encapsulated in the additional supercharges $Q_\mu$ that trivialize infinitesimal translations in $\mathbb{Q}$-cohomology by means of Witten's descent formalism \cite{Witt88}.
We will assume in the following that the symmetries these homotopies generate commute with one another $[Q_\mu, Q_\nu] = 0$ for all $\mu$, $\nu$; this is certainly the case in topologically twisted quantum field theory.

For a local observable $\cO$, not necessarily $\mathbb{Q}$-closed, we consider the observable-valued 1-form 
\begin{equation}
	\cO^{[1]} := \textrm{d} x^\mu ~ (-i Q_\mu \cO)~.
\end{equation}
We refer to $\cO^{[1]}$ as the (first) descendant of $\cO$.
This descendant satisfies the defining equation
\begin{equation}
	\mathbb{Q} \cO^{[1]} = -\textrm{d} x^\mu \big(-i P_\mu\cO - Q_\mu(\mathbb{Q} \cO)\big) = \textrm{d}\cO + (\mathbb{Q}\cO)^{[1]}~.
\end{equation}
which we refer to as the descent equation.
If $\cO$ is $\mathbb{Q}$-closed and we choose a path $\gamma$ from $x$ to $x'$ then we can integrate $\cO^{[1]}$ along $\gamma$ to define
\begin{equation}
	\cO[\gamma] := \int_\gamma \cO^{[1]}
\end{equation}
which satisfies
\begin{equation}
	\mathbb{Q} \cO[\gamma] = \int_\gamma \textrm{d} \cO = \cO(x') - \cO(x)~,
\end{equation}
where we used the descent equation in the first equality and Stokes' theorem in the second.
Just as $Q_\mu \cO$ is a homotopy for infinitesimal translations of $\cO$, we thus see that $\cO[\gamma]$ defines a homotopy for \emph{finite} translations of $\cO$.

One might worry that this could depend on the path $\gamma$ in some delicate way.
If we have two such paths $\gamma_1$, $\gamma_2$ that are themselves homotopic then we can descend further and integrate the resulting observable-valued two-form over the homotopy between them to see that $\cO[\gamma_1]$ and $\cO[\gamma_2]$ are themselves equivalent in $\mathbb{Q}$-cohomology.
More generally, in a $d$-dimensional field theory we can use the $Q_\mu$'s to define an observable-valued $n$-form $\cO^{[n]}$ called the $n$th descendant of $\cO$ for $n = 0$, $\dots$, $d$, where $\cO^{[0]} = \cO$, that satisfies the equation 
\begin{equation}
	\mathbb{Q} \cO^{[n]} = \textrm{d} \cO^{[n-1]} + (\mathbb{Q} \cO)^{[n]}~.
\end{equation}
We can thus integrate $\cO^{[n]}$ over $n$-dimensional submanifolds of spacetime; these will provide the homotopies for homotopies and so on.
We will find it useful to collect these descendants to an observable-valued (inhomogeneous) differential form
\begin{equation}
	\cO^\bullet = \cO^{[0]} + \dots + \cO^{[d]}
\end{equation}
that satisfies the equation
\begin{equation}
	\mathbb{Q} \cO^\bullet = \textrm{d} \cO^\bullet + (\mathbb{Q} \cO)^\bullet~.
\end{equation}

We motivated the notation of descent as a way to implement various identities satisfies by local observables within arbitrary correlation functions, e.g. the associativity and graded-commutativity of the collision product $\star$, but they have much more to offer.
For example, we can use them to construct genuinely $\mathbb{Q}$-closed \emph{extended} observables.
If $\Sigma$ is an $n$-dimensional submanifold of spacetime and $\omega$ is a $k$-form on spacetime (pulled back to $\Sigma$) then we define
\begin{equation}
  \cO[\Sigma,\omega] := \int_\Sigma \omega~\cO^\bullet = \int_\Sigma \omega~\cO^{[n-k]}~.
\end{equation}
which is an observable supported on $\Sigma$.
Applying the descent equation and Stokes' theorem we find that it satisfies
\begin{equation}
	\mathbb{Q} \cO[\Sigma, \omega] = \pm \cO[\partial \Sigma, \omega] \pm \cO[\Sigma, \textrm{d} \omega] \pm (\mathbb{Q} \cO)[\Sigma, \omega]~.
\end{equation}
Therefore, $\cO[\Sigma,\omega]$ is
\begin{itemize}
  \item $\mathbb{Q}$-closed if $\partial\Sigma=0$, $\textrm{d}\omega=0$, and $\mathbb{Q}\cO=0$;
  \item $\mathbb{Q}$-exact if $\Sigma$ is a boundary, $\omega$ is a coboundary, or
  $\cO$ is exact.
\end{itemize}
In particular, correlation functions including this observable only depends on the classes $[\Sigma]_\partial$, $[\omega]_\textrm{d}$, $[\cO]_\mathbb{Q}$.
As we will see in Chapter \ref{chap:ySYM}, these extended operators obtained from descent play an important role in Witten's physical interpretation of Donaldson theory \cite{Witt88} and, related, in the construction of Vafa--Witten invariants \cite{VW}.

\subsection{Secondary products and $E_d$ algebras}

We now return to local observables and ask what additional algebraic structure(s) descent equips local observables with.
The key observation is that we can define an action of local observables upon themselves by integrating them around one another, cf. \cite{BBZBDN}.
We place a $\mathbb{Q}$-closed local observable $\cO_2$ at $0$ and choose a differential form $\omega \in \Omega^\bullet(\R^d - 0)$; for any other $\mathbb{Q}$-closed local observable $\cO_1$ we can consider
\begin{equation}
	(\cO_1 \star_{r, \omega} \cO_2)(0) := \int_{S^{d-1}_{r,0}} \omega \cO_1^\bullet(x) \cO_2(0)
\end{equation}
where $S^{d-1}_{r,0}$ is a sphere centered at $0$ with radius $r$.
Using the above formula, we see that this observable is $\mathbb{Q}$-closed if $\textrm{d} \omega = 0$.
Moreover, its class in $\mathbb{Q}$-cohomology is independent of $r$ so we can choose $r$ small enough to make the observable supported in an arbitrarily small neighborhood of $0$ and hence can itself be viewed as as a local observable.
The $\mathbb{Q}$-cohomology class of this local observable only cares about the $\mathbb{Q}$-cohomology classes of $\cO_1$ and $\cO_2$ and the de Rham cohomology class of $\omega$.

As $\R^d - 0$ contracts to $S^{d-1}$, we see that $\mathbb{Q}$-cohomology is equipped with two independent bilinear operators: we can take $\omega = 1$ the constant function or $\omega = \Omega$ the volume form on $S^{d-1}$ (normalized to give it unit volume).
If we take $\omega$ to be the volume form $\Omega$ we recover the collision product we described described above:
\begin{equation}
	([\cO_1] \star [\cO_2])(0) := [(\cO_1 \star_{r, \Omega} \cO_2)(0)] \equiv \bigg[\int_{S^{d-1}_{r,0}} \Omega~\cO_1(x) \cO_2(0)\bigg]~.
\end{equation}
This can be seen by taking OPE of $\cO_1$ and $\cO_2$; the only term that is not $\mathbb{Q}$-exact is the coefficient of the constant function.
The bilinear operation we get from $\omega = 1$ is new and referred to as a secondary product in \cite{BBZBDN}, whereas the former is the primary product.
We denote it by
\begin{equation}
	\big\{ [\cO_1], [\cO_2] \}(0) := [(\cO_1 \star_{r,1} \cO_2)(0)] = \bigg[\int_{S^{d-1}_{r,0}} \cO_1^{[d-1]}(x) \cO_2(0)\bigg] ~.
\end{equation}
The secondary product the OPE of $\cO_1^{[d-1]}$ and $\cO_2$, which is not prohibited having interesting functional dependence as $\cO_1^{[d-1]}$ is not $\mathbb{Q}$-closed.

\begin{figure}
	\begin{tikzpicture}[scale=0.8]
		\node (b1) at (0,2) {$\bullet$};
		\node (b1) at (0,2.25) {$x_1$};
		\node (b2) at (0,1) {$\bullet$};
		\node (b2) at (0,1.25) {$x_2$};
		\node (b2) at (0,0) {$\bullet$};
		\node (b2) at (0.5,0) {$0$};
		
		\draw (0,0) circle[radius = 1];
		\draw (-1,0) arc (180:360:1 and 0.333);
		\draw[dashed] (1,0) arc (0:180:1 and 0.333);
		
		\draw (0,0) circle[radius = 2];
		\draw (-2,0) arc (180:360:2 and 0.667);
		\draw[dashed] (2,0) arc (0:180:2 and 0.667);
	\end{tikzpicture} \quad \raisebox{1.6cm}{$=$} \quad \begin{tikzpicture}[scale=0.8]
	\node (b1) at (0,2.5) {$\bullet$};
	\node (b1) at (0,2.75) {$x_1$};
	\node (b2) at (0,2) {$\bullet$};
	\node (b2) at (0,2.25) {$x_2$};
	\node (b2) at (0,0) {$\bullet$};
	\node (b2) at (0.5,0) {$0$};
	
	\draw (0,0) circle[radius = 2];
	\draw (-2,0) arc (180:360:2 and 0.5);
	\draw[dashed] (2,0) arc (0:180:2 and 0.5);
	
	\draw (0,2) circle[radius = 0.5];
	\draw (-0.5,2) arc (180:360:0.5 and 0.166);
	\draw[dashed] (0.5,2) arc (0:180:0.5 and 0.166);
	\end{tikzpicture} \quad \raisebox{1.6cm}{$+$} \quad \begin{tikzpicture}[scale=0.8]
	\node (b1) at (0,2) {$\bullet$};
	\node (b1) at (0,2.25) {$x_2$};
	\node (b2) at (0,1) {$\bullet$};
	\node (b2) at (0,1.25) {$x_1$};
	\node (b2) at (0,0) {$\bullet$};
	\node (b2) at (0.5,0) {$0$};
	
	\draw (0,0) circle[radius = 1];
	\draw (-1,0) arc (180:360:1 and 0.333);
	\draw[dashed] (1,0) arc (0:180:1 and 0.333);
	
	\draw (0,0) circle[radius = 2];
	\draw (-2,0) arc (180:360:2 and 0.667);
	\draw[dashed] (2,0) arc (0:180:2 and 0.667);
	\end{tikzpicture}
	\caption{Illustration of the cycles appearing in the composition of primary and secondary products and the relation they satisfy that implies the derivation property of the secondary product.}\label{fig:deriv}
\end{figure}
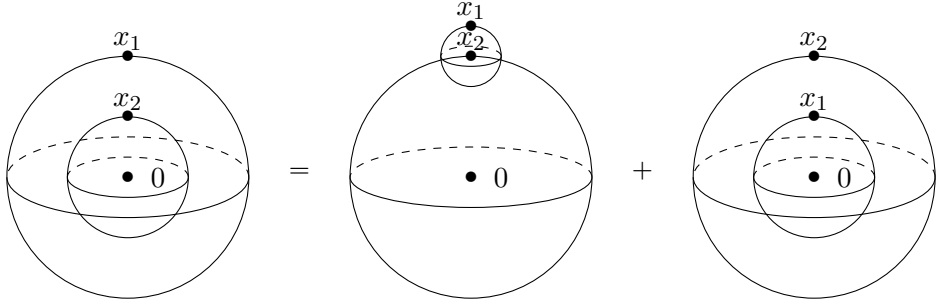

The various relations satisfied by the primary and secondary products can be understood by carefully examining and manipulating the cycles underlying these products.
For example, it is fairly straight-forward to see that the secondary product defines a (graded) derivation of the primary product.
This follows from the coincidence of homology classes depicted in Figure \ref{fig:deriv} which should be familiar to those who have seen the derivation of Borcherds' identity.
Continuing in this fashion, one finds that the secondary product is (graded) skew-symmetric and satisfies a (graded) Jacobi identity; see \cite[Sec.~3.2]{BBZBDN} for more details.
Putting these together, one deduces that local operators in a topological quantum field theory have the structure of a Poisson algebra when $d$ is odd (and bigger than 1) and a Gerstenhaber algebra when $d$ is even; this observation generalizes the appearance of Gerstenhaber structures in the context of topological conformal field theory and string field theory \cite{LZ, PS, WZ}.
When the topological field theory admits a cohomological grading then so too does this algebraic structure and can be succinctly captured in the notion of a $(d-1)$-shift Poisson algebra.
Together these form a \emph{$(d-1)$-shifted Poisson algebra}.

In verifying the fact that the secondary product defines a derivation of the primary product, as well as the fact that it satisfies the Jacobi identity, we needed to consider cycles and chains defined not just on spacetime but on the configuration space of points on spacetime.
Namely, we made use of a slightly more general structure: if we consider an insertion of $n$ local observables 
\begin{equation}
	\cO^\bullet_1(x_1)~\cO^\bullet_2 (x_2)~\dots~\cO^\bullet(x_n)
\end{equation}
then we can pull this observable-valued differential form to the configuration space of $n$ points on spacetime and then integrate it over a chain on this configuration space, perhaps weighted by some preferred cochains thereon.
So long as the chains are bounded (to ensure we can apply a homotopy to ensure it is can be enclosed in a sufficiently small neighborhood of a point) this will give rise to an $n$-ary operation on the space of local observables.
This collection of $n$-ary operations labeled by chains and cochains on the configuration space of points on spacetime can be encapsulated in the notion of an algebra over the operad of little disks (or, rather, the Fulton-MacPherson operad) in $d$-dimensional space, or simply an $E_d$-algebra.

One might wonder whether this equips local observables with any additional algebraic structures at the level of $\mathbb{Q}$-cohomology.
The short answer is no: all $n$-ary operations defined on $\mathbb{Q}$-closed local observables are $\mathbb{Q}$-cohomologous to iterated primary and secondary products.
This is a result largely due to Kontsevich \cite{Kont}, details of which were filled in by Lambrechts and Volic \cite{LV} and a different proof is due to Tamarkin \cite{Tam}.

\begin{remark}
	When working on a spacetime with symmetries that the topological quantum field theory is compatible with, e.g. rotations of $\R^d$, it is possible to turn on equivariance with respect to them.
	This amounts to the deforming the differential
	\begin{equation}
	  \mathbb{Q} \rightsquigarrow \mathbb{Q}_\varepsilon
	\end{equation}
	that no longer squares to zero, instead satisfying
	\begin{equation}
	  \mathbb{Q}_\varepsilon^2 = \varepsilon_1 R_1+\dots+\varepsilon_n R_n,
	  \qquad n\le\lfloor d/2\rfloor,
	\end{equation}
	where $R_1,\dots,R_n$ are the generators of the above rotations on observables, which are required to commute with one another.
	There is then an equivariant version of descent; the equivariant descent equation takes the form
	\begin{equation}
	  \mathbb{Q}_\varepsilon~\cO^\bullet = \textrm{d}_\varepsilon \cO^\bullet + (\mathbb{Q}_\varepsilon \cO)^\bullet,
	\end{equation}
	where $\textrm{d}_\varepsilon$ is the equivariant Cartan differential.
	This leads to extended observables labeled by the equivariant (co)homology of configuration spaces and equips local observables with the structure of an $E_{d-2n}$-algebra structure; see e.g. \cite[Sec.~6]{BBZBDN} and references therein for more details.
	Physically, this process can be viewed as twisting a supersymmetric field theory in the presence of the Nekrasov-Shatashvili $\Omega$-background \cite{Nekr03, NS}.
\end{remark}

\section[Lecture 3]{Lecture 3: Observables in holomorphic-topological QFT}
\label{sec:NGLect3}

Many of the ideas appearing in the study in the topological setting above can be extended to the holomorphic and holomorphic-topological setting, which is still an active area of research.

\subsection{Holomorphic-topological descent}
We will suppose that the there are $t$ topological directions and $h$ holomorphic directions; such a theory can be placed on a spacetime looks locally like $\R^t \times \C^h$ with coordinates $x^\mu$, $z^a$, and $\overline{z}^a$.
In addition to the twisting supercharge $\mathbb{Q}$ there supercharges $Q_\mu$ and $Q_a$, where $\mu = 1, \dots t$ and $a = 1, \dots h$, satisfying
\begin{equation}
	[\mathbb{Q}, Q_\mu] = P_\mu \sim -i \partial_{x^\mu} \qquad [\mathbb{Q}, Q_a] = P_a \sim -i \partial_{\overline{z}^a}~.
\end{equation}
In much the same way as in the topological setting, correlation functions of $\mathbb{Q}$-closed observables will only depend ``holomorphically'' on their support.

The global geometry that this algebraic structure is compatible with is that of a \emph{transverse holomorphic foliation}, see e.g. \cite{DK, Rawnsley, Asuke} for more details.
In brief, this is a foliation $\cF$ of even codimension whose space of leaves can be equipped with a complex structure.
If $T_\cF \subset T_M$ is the subbundle/sheaf of the tangent bundle/sheaf consisting of tangent vectors along the foliation, then this means that the complexified quotient has a natural decomposition into holomorphic and anti-holomorphic tangent vectors:
\begin{equation}
	Q:= T_M / T_\cF \rightsquigarrow Q \otimes \C = Q^{1,0} \oplus Q^{0,1}~.
\end{equation}
In the above local coordinates, $Q^{1,0}$ is spanned by $\partial_{z^a}$ and $Q^{0,1}$ by $\partial_{\overline{z}^a}$.
We then consider the subbundle/sheaf of the complexified tangent bundle defined by
\begin{equation}
	V_\cF := \ker(T_M \otimes \C \to Q^{1,0})
\end{equation}
which is locally spanned by $\partial_{x^\mu}$ and $\partial_{\overline{z}^a}$.
Holomorphic functions are then defined by those smooth functions that are annihilated by the vector fields in $V_\cF$
\begin{equation}
	\cO_\cF := \{f | L_X f = 0~\forall X \in V_\cF\}~.
\end{equation}
This is the global way of saying that holomorphic functions only depend on $z^a$.
The configuration space of points on spacetime, which is where correlation functions live, inherits a transverse holomorphic folation from spacetime; the correlation functions are therefore holomorphic functions on this configuration space.

There is an analogue of differential forms compatible with a transverse holomorphic foliation using the fact that forms are dual to vector fields.
Explicitly, we consider
\begin{equation}
	\Omega^{(p);q} = \bigwedge\!{}^p (Q^{1,0})^\vee \otimes \bigwedge\!{}^q (V_\cF)^\vee
\end{equation}
which locally given by sums of elements of the form
\begin{equation}
	f(x,z,\overline{z}) (\textrm{d} z^{a_1} \wedge \dots \wedge \textrm{d} z^{a_p}) \otimes (\textrm{d} \overline{z}^{b_1} \wedge \dots \wedge \textrm{d} \overline{z}^{b_m} \wedge \textrm{d} x^{\mu_1} \wedge \dots \wedge \textrm{d} x^{\mu_n})
\end{equation}
where $n + m = q$.
The exterior derivative on forms induces a map $\textrm{d}' : \Omega^{(p);q} \to \Omega^{(p);q+1}$ that is locally given by
\begin{equation}
	\textrm{d}' = \textrm{d} \overline{z}^a \partial_{\overline{z}^a} + \textrm{d} x^\mu \partial_{x^\mu}~.
\end{equation}
which squares to zero; we denote the resulting complex $\Omega^{(p),\bullet}$.

With this notation in hand, we are in a position to describe holomorphic-topological descent.
For a (not necessarily $\mathbb{Q}$-closed) local observable $\cO$ we define the first (holomorphic-topological) descendant by
\begin{equation}
	\cO^{[1]} = \textrm{d} \overline{z}^a (-i Q_a \cO) + \textrm{d} x^\mu (-i Q_\mu \cO)
\end{equation}
which satisfies the (holomorphic-topological) descent equation
\begin{equation}
	\mathbb{Q} \cO^{[1]} = \textrm{d}' \cO + (\mathbb{Q} \cO)^{[1]}~.
\end{equation}
As before, we can continue this process until we reach the $(h+t)$th descendant; we collect them together into the observable-valued element of $\Omega^{(0), \bullet}$ and denote it by $\cO^\bullet$.
This collection of descendants satisfies
\begin{equation}
	\mathbb{Q} \cO^{\bullet} = \textrm{d}' \cO + (\mathbb{Q} \cO)^{\bullet}~.
\end{equation}

\subsection{Higher operators in holomorphic-topological field theory}

Just as with the topological theories described in the previous lecture, observables field theories that are cohomologically holomorphic-topological admit rich algebraic structures coming from descent.
The basic ideas underlying this construction are essentially unchanged when going to the holomorphic-topological setting.
The key difference will be that there is, at present, no minimal model for this algebraic structure.
That said, there has been a great deal of work on building computationally-tractable models in the purely holomorphic setting ($t = 0$) \cite{FGY, GuWaWi} and certain aspects of the general theory \cite{GKW}.

The simplest piece of this algebraic structure is involves two insertions and is the analogue of the binary operations $-\star-$ and $\{-,-\}$ described in the topological setting.
Explicitly, we consider an insertion of two operators $\cO_1$ and $\cO_2$ and integrate descendants of $\cO_1$ over the sphere linking $\cO_2$; we can weight this integral by a choice of form $\omega \in \Omega^{(h), \bullet}(\R^t \otimes \C^h - 0)$, leading to the expression
\begin{equation}
	\mu_2[S^{2h+t-1}, \omega](\cO_1, \cO_2) :=  \int_{S^{2h+t-1}_0} \omega~\cO_1^\bullet(z, \overline{z}, x) \cO_2(0)~.
\end{equation}
Acting on this expression with $\mathbb{Q}$, and using the fact that $\omega$ has as many $\textrm{d}z$'s as possible, we find
\begin{equation}
\begin{aligned}
	\mathbb{Q} \mu_2[S^{2h+t-1}, \omega](\cO_1, \cO_2) & = \pm \mu_2[S^{2h+t-1}, \textrm{d}' \omega](\cO_1, \cO_2)\\
	& \qquad \pm \mu_2[S^{2h+t-1}, \omega](\mathbb{Q}\cO_1, \cO_2) \pm \mu_2[S^{2h+t-1}, \omega](\cO_1, \mathbb{Q}\cO_2)
\end{aligned}
\end{equation}
so that if $\textrm{d}' \omega = 0$, $\mathbb{Q} \cO_1 = 0$, and $\mathbb{Q} \cO_2 = 0$ then this defined a $\mathbb{Q}$-closed observable.
Moreover, in this case we find that it only depends on the the homology class this $S^{2h+t-1}$, so that its radius is inconsequential, the $\textrm{d}'$ cohomology class of $\omega$, and the $\mathbb{Q}$-cohomology classes of $\cO_1$ and $\cO_2$.
In particular, this operation will define a new $\mathbb{Q}$-closed \emph{local} observable.
There are two distinguished families of these $\textrm{d}'$ closed forms:
\begin{equation}
	\omega = \textrm{d}^h z \partial^{k_1}_{z^1} \dots \partial^{k_h}_{z^h} \omega_{\textrm{BM}} \qquad \textrm{d}^h z (z^1)^{k_1} \dots (z^h)^{k_h}
\end{equation}
where $\omega_{\textrm{BM}}$ is a generalized Bochner-Martinelli kernel, see e.g. \cite{WaWi}; it is a kernel for the differential operator $\textrm{d}'$, i.e. satisfies
\begin{equation}
	\textrm{d}' \omega_{\textrm{BM}} = \textrm{d}^h \overline{z} \textrm{d}^t x~\delta(z, \overline{z}, x)
\end{equation}
and can be viewed as an analogue of $z^{-1}$ in the case $h = 1$, $t = 0$ and the volume form on the sphere when $h = 0$.
The first family of binary operations simultaneously generalize the product $-\star-$ in the topological setting and the normally-ordered product familiar from two-dimensional chiral conformal field theory; it defines a regularized product of $\partial^{k_1}_{z^1} \dots \partial^{k_h}_{z^h} \cO_1$ and $\cO_2$.
The second family of binary operations generalizes the bracket $\{-,-\}$ in the topological setting and the coefficients of singular terms in the OPE in two-dimensional chiral conformal field theory.
It is often convenient to collect these brackets into a generating function depending on a vector of complex parameters $\lambda \in \C^h$, called the $\lambda$-bracket:
\begin{equation}
	\{\cO_1 {}_\lambda \cO_2\} := \int_{S^{2h+t-1}_0} \textrm{d}^h z e^{\lambda \cdot z} ~\cO_1^\bullet(z, \overline{z}, x) \cO_2(0) = \int_{S^{2h+t-1}_0} \textrm{d}^h z e^{\lambda \cdot z} ~\cO_1^{[h+t-1]}(z, \overline{z}, x) \cO_2(0)~.
\end{equation}
In the context of two-dimensional chiral conformal field theory, the algebraic structure resulting from the $\lambda$-bracket goes under the name of Lie conformal algebra.
Together, these operations give local observables the structure of a $t$-shifted, $h$-dimensional Poisson vertex algebra; see \cite{OY1} where the case where $h = 1$ was first described and \cite{GaWi} for another mathematical formulation of the case $h = t = 1$ similar to the structure of a vertex algebra.

We can similarly define $n$-ary operations by considering an insertion of $n$ local observables $\cO_1^\bullet$, $\dots$, $\cO_n^\bullet$, a bounded chain $\Sigma$ on the configuration space of $n$ points on spacetime, and a form $\omega$ to weight the integral; this will define a $\mathbb{Q}$-closed observable so long $\partial \Sigma = 0$, $\textrm{d}' \omega = 0$, $\mathbb{Q} \cO_1 = 0$, and $\mathbb{Q} \cO_2 = 0$.
That said, it is important to require $\omega$ have as many $\textrm{d}z$'s (typically $(n-1)h$) as possible so that the result will only depend on the homology class of the $\Sigma$ and hence will define a local observable.
Unlike the purely topological setting, when $h \neq 0$ these higher arity operations need not be iterated combinations of the $\lambda$-bracket and the regulated product.
Although it is not at present known what is the minimal collection of $n$-ary operations survive cohomology, a portion of it is known: there is an $n$-ary analogue of the $\lambda$-bracket.
Together with the binary $\lambda$-bracket, this gives local observables the structure of a $t$-shifted $\mathcal{L}_\infty$ conformal algebra.
See \cite{GKW} for a detailed description of this algebraic structure; this reference provides an explicit method for computing these bracket in the context of perturbative quantum field theory.

\begin{remark}
\begin{enumerate}
	\item We note that the paper \cite{GKW} conjectures that there is an analogue of Kontsevich formality when $t \geq 2$.
	Concretely, they observe the vanishing of many loop-integrals contributing to the higher operations described in the perturbative quantum field theories they study with $t \geq 2$ and conjecture that it is true for all loop integrals; see \cite{BG} for a proof of this conjecture.
	Of course, this is far from proving some operadic formality statement; in the topological cases ($h = 0$) this absence of no loop corrections is a \emph{consequence} of formality when restricted to observables in a field theory defined by pertubing around a free theory.
	A fully-fledged form of formality for these holomorphic-topological field theories would require both an better understanding of what algebraic structures are present at the level of $\mathbb{Q}$-cohomology as well as a handle on the dynamics of these field theories beyond perturbation theory.
	
	\item Just as in the topological setting, it is possible to turn on equivariance with respect to symmetries of the holomorphic-topological quantum field theory.
	When turning on equivariance with respect to a rotational symmetry of a topological plane, this will lead to a lower-dimensional holomorphic-topological field theory.
	See e.g. \cite{OY2, Jeong} for an example of this construction in four dimensions whereby a holomorphic-topological field theory (with $h = 1$, $t = 2$) is reduced to a two-dimensional holomorphic field theory, thereby providing complementary construction of the vertex algebra of \cite{BLLPRvR}.
\end{enumerate}
\end{remark}

\section[Lecture 4]{Lecture 4: The holomorphic twist of four-dimensional supersymmetric gauge theories}
\label{sec:NGLect4}

In this final lecture we return our attention to four-dimensional supersymmetric gauge theories and, in particular, their holomorphic twists.
These examples will provide the reader with some specific field theories that showcase some of the above structures in the context of four-dimensional quantum field theory; the resulting holomorphic field theories first appeared in \cite{Johansen, LMNS1, LMNS2} and later described mathematically in \cite{Costello1, Costello2}.
Many aspects of the above structures are studied in detail in \cite{BGKWWY, BoWu, BGWW}.

By examining the holomorphic, or minimal, twist of these field theories we will see that a remarkable amount of information can be extracted from its study.
For example, inspired by \cite{EY}, we will see that the holomorphic twist, which only requires $\cN = 1$ supersymmetry, of a theory with extended supersymmetry will actually admit further deformations to more topological twists; this rests on the observation that the representative supercharges for many of the twists in Table \ref{tab:twists} can themselves be seen as a deformation of the minimal supercharge.%
\footnote{Of course, the holomorphic twist will not be able to access every further twist, e.g. the $u = 0$ Kapustin-Witten twist. Nor will these deformations necessarily manifest the topological nature of the theory. We will comment on this below.} %

\subsection{Lightning intro/review of SUSY gauge theory}

We will start with a lightning review of the field theories of interest.
As our interest will eventually be in the holomorphic twits, we will emphasize the description of each of these theories as a special case of $\cN = 1$ supersymmetry.
In each of these cases, we simply record the classical field content and comment on various anomalies thereof that are relevant for our later discussions; we will not attempt to provide a description of their dynamics.
See e.g. \cite{Weinberg:2000cr, Wess:1992cp} for textbook treatments of these theories.

We start with a very general class of $\cN = 1$ Lagrangian gauge theories; they can be viewed as four-dimensional analogues of gauged Landau-Ginzburg models.
There is a very elegant, off-shell description of these field theories using the language of superspace, see e.g. for more details.
The basic data of this theory includes a choice of \emph{gauge group} $G$, a compact Lie group, a (finite-dimensional) unitary representation $V$ of $G$, and a $G$-invariant holomorphic function $W: V \to \C$ called the \emph{superpotential}; one must also choose coupling constants $g$ and $\theta$-angles for each simple or abelian factors in $\mathfrak{g} = \textrm{Lie}(G)$ as well as an Fayet-Illiopolis (FI) parameter for each abelian factor.%
\footnote{Strictly speaking, there are additional (often discrete) choices one must make in defining the full quantum field theory.
These include, for example, specifying a collection of allowed line operators, also called discrete $\theta$-angles \cite{ASY}.
We will not attempt to catalog all of the possibilities.} %
We denote by $V^\vee$ the dual representation of $V$.
For this theory to exhibit a $U(1)_R$ (at least classically) then there must be a $U(1)_R$ action on $V$ commuting with $G$, so that we can write it as
\begin{equation}
	V = \bigoplus_{r \in \Z} V_r
\end{equation}
such that the superpotential $W$ is quasi-homogeneous of weight $2$.
The fields of the theory include a collection of ``gauge fields''
\begin{itemize}
	\item $A_\mu$ a connection on a principal $G$ bundle $P$~,
	\item $\lambda_\alpha$, $\bar{\lambda}_{\dot \alpha}$ a Dirac spinor field valued in $\textrm{ad}(P)$~,
\end{itemize}
transforming as a \emph{vector multiplet}, as well as a collection of ``matter fields''
\begin{itemize}
	\item $X$, $\overline X$ a complex scalar field valued in the associated bundles $P \times_G V$ and $P \times_G V^\vee$~,
	\item $\psi_\alpha$, $\overline{\psi}_{\dot\alpha}$ a pair of Weyl spinor fields valued the associated bundles $P \times_G V$ and $P \times_G V^\vee$~,
\end{itemize}
where $(X,\psi_\alpha)$ transform as a \emph{chiral multiplet} and $(\overline{X},\overline{\psi}_{\dot\alpha})$ as an \emph{anti-chiral multiplet} with respect to supersymmetry.

This classical field theory may be quantum-mechanically ill-defined: if $\{T_a\}$ denote a basis of generators for the Lie algebra $\mathfrak{g}$ then we must require
\begin{equation}
	\tr_V(T_a(T_b T_c + T_c T_b)) = 0
\end{equation}
and 
\begin{equation}
	\sum_{r \in \Z} q_{r,n} = 0
\end{equation}
for each abelian factor $U(1)_n$, where $q_{r,n}$ is the weight of $V_r$ under $U(1)_n$, to ensure there are no perturbative gauge anomalies.
There are additional \emph{global}, non-perturbative gauge anomalies that one needs to account for, commonly referred to as Witten anomalies \cite{Witten:1982fp}; we will not belabor them here, but they will appear later in Section \ref{chap:ySYM}.
We also note that the $U(1)_R$ symmetry is often broken by an ABJ anomaly; the anomaly-free conditions requires that
\begin{equation}\label{eq:Ranom1}
	T_{G_i}(\textrm{ad}_i) + \sum_{r \in \Z} (r-1) T_{G_i}(V_r) = 0
\end{equation}
for each simple and abelian factor $G_i$ in $G$, where $T_H(R) = 0$ denotes the quadratic index of the $H$-representation $R$, and additionally that
\begin{equation}\label{eq:Ranom2}
	\sum_{r \in \Z} (r-1)^2 q_{r,n} = 0
\end{equation}
for each abelian factor $U(1)_n$, where $q_{r,n}$ is the weight of $V_r$ under $U(1)_n$.

An $\cN = 2$ Lagrangian gauge theory arises as the special case where $V = \mathfrak{g}_\C \oplus \mathcal{R}$, where $\mathfrak{g}_\C$ is the complexified Lie algebra and $\mathcal{R}$ is a pseudoreal representation of $G$, and the superpotential is (up to an overall numerical factor) $W = \langle \mu, \phi\rangle$ pairs the $\mathfrak{g}_\C$-valued field with the moment map for the $G$-action on $\mathcal{R}$.
The pseudoreal representation $\mathcal{R}$ is often taken to be of the form $\mathcal{R} = U \oplus U^\vee$ for $U$ a unitary representation of $G$.
The $\cN = 2$ supersymmetry transforms the $\cN = 1$ vector multiplet and adjoint-valued chiral and anti-chiral multiplets into one another to form an $\cN = 2$ vector multiplet: it rotates the $\mathcal{R}$-valued chiral and anti-chiral multiplets are rotated into one another to form a \emph{half-hypermultiplet}; if $\mathcal{R} = U \oplus U^\vee$ then one says it is a $U$-valued \emph{hypermultiplet}.
The $SU(2)_R \times U(1)_r$ $R$-symmetry is always present classically and $SU(2)_R$ always survives quantization, but $U(1)_r$ may suffer from an ABJ anomaly: it rotates the adjoint chiral with charge $2$ and does not rotate the $\mathcal{R}$-valued chirals, so that the vanishing of this perturbative anomaly requires
\begin{equation}
	2 T_{G_i}(\textrm{ad}_i) - T_{G_i}(\mathcal{R}) = 0
\end{equation}
for each simple and abelian factor.
In fact, this condition is precisely the condition that the gauge couplings have vanishing $\beta$-function and so the resulting $\cN = 2$ theory is actually \emph{superconformal}.
Note that this condition is never satisfied if $G$ has any abelian factors unless they act trivially on the matter fields.

Finally, an $\cN = 4$ Lagrangian gauge theory arises in the special case where we take $\mathcal{R} = \mathfrak{g} \otimes \C^2$ or, equivalently, $V = \mathfrak{g} \otimes \C^3$.
In this case, the $\cN=1$ superpotential takes the form $W = \tr(\phi_1 [\phi_2, \phi_3])$, where we recognize $[\phi_2, \phi_3]$ as the moment map for the $G$-action on $\mathcal{R}$.
When $G$ is non-abelian, this theory admits an $SU(4)_R$ $R$-symmetry that is never anomalous and exhibits $\cN = 4$ superconformal symmetry regardless of choice of gauge group $G$.

\subsection{A field theory for the holomorphic twist}

Given one of the $\cN \geq 1$ Lagrangian field theories described above, we can consider its holomorphic twist.
When this field theory is formulated in the Batalin-Vikovisky (BV) formalism, the resulting twisted theory admits a (perturbatively) equivalent description as a holomorphic $BF$ theory \cite{Costello1, Costello2, ElSaWi, SaWi1}.%
\footnote{The theory presented below should be thought of as a simplification of the holomorphic twist that arises from genuinely performing the holomorphic twist of the underlying supersymmetric field theory and then removing fields that cancel each other in cohomology.} %
The resulting theory can be formulated on an arbitrary Calabi-Yau two-fold for any choice of the above data; when the $U(1)_R$ symmetry is nonanomalous, i.e. Eqs. \eqref{eq:Ranom1} and \eqref{eq:Ranom2} are satisfied, then the theory can be formulated on an arbitrary complex surface.
We will denote this spacetime by $\cM$ and its holomorphic volume form $\omega$.

This bolomorphic BF theory has the following field content; we work around the trivial holomorphic $G$ bundle with complex structure $\overline{\partial}$.
The vector multiplet fields give rise to a pair of fermionic and bosonic ``superfields''
\begin{equation}
	\boldsymbol{A} = c + A + B^* \in \Omega^{0, \bullet}(\cM, \mathfrak{g}_\C)[1] \qquad \boldsymbol{B} = B + A^* + c^* \in \Omega^{0, \bullet}(\cM, \mathfrak{g}_\C^\vee)~,
\end{equation}
where the decompositions are into fields of homogeneous Dolbeault form degree; the symbol $[1]$ denotes a shift in cohomological degree, so that $c$ is in degree $1$, the components of $A$ are in degree $0$, and the components of $B^*$ are in degree $-1$.
We place the differential forms $\textrm{d}\overline{z}$ are given cohomological degree $1$ so that both $\boldsymbol{A}$ and $\boldsymbol{B}$ have homogeneous cohomological degree.
To compare with the fields described in the previous subsection, the component $A$ is the twisted remnant of the physical gauge field $A$, $c$ is the usual BRST ghost but its derivatives $\partial_a c$ are cohomologous to the spinorial fields $\overline{\lambda}_{\dot \alpha}$, and the component $B$ is identified the component $F_{++}$ of the self-dual part of the curvature.
Similarly, the chiral multiplet fields give rise to a pair of fermionic and bosonic ``superfields''
\begin{equation}
\begin{aligned}
	\boldsymbol{\gamma} & = \gamma + \Pi^* + \beta^* \in \bigoplus_{r \in \Z} \Omega^{0,\bullet}(\cM, V_r^\vee)[r]\\
	\boldsymbol{\beta} & = \beta + \Pi + \gamma^* \in \bigoplus_{r \in \Z}\Omega^{0,\bullet}(\cM, V_r)[1-r]
\end{aligned}
\end{equation}
The component field $\gamma$ is the twisted incarnation of the anti-chiral multiplet scalar field $\overline{X}$; $\psi$ comes from the component $\psi_+$ of spinorial field in the chiral multiplet; and $\Pi$ is the anti-holomorphic canonical momentum of $\overline{X}$ and is cohomologous to the holomorphic derivatives $\partial_a X$.
The dynamics of these fields are governed by an action functional of the form
\begin{equation}
	\int_\cM \omega \left[ \boldsymbol{B} \left(\overline{\partial} \boldsymbol{A} + \tfrac{1}{2}[\boldsymbol{A}, \boldsymbol{A}]\right) + \boldsymbol{\beta} \left(\overline{\partial} \boldsymbol{\gamma} + \boldsymbol{A} \boldsymbol{\gamma} \right) + W(\boldsymbol{\gamma})\right] + \tau \tr(\partial \boldsymbol{A} \partial \boldsymbol{A})
\end{equation}
where $W(\boldsymbol{\gamma})$ is the pullback of the (complex conjugate of the) superpotential to spacetime and $\tau = \frac{\theta}{2\pi} + \frac{4 \pi i}{g^2}$ is the complexified coupling constant.
As noted in \cite{BGKWWY} this last term comes from the $F$-term encoding the kinetic term for the gauge fields; it the holomorphic avatar of the second Chern class and does not alter the equations of motion or perturbative dynamics of the twisted theory -- it alters the contribution of non-trivial gauge bundles to the path integral.
This theory can be thought of as a four-dimensional version of the gauge $\beta\gamma$ system familiar to two-dimensional conformal field theories and superstring theorists.

In the BV formalism, the equations of motion and their redundancies are encoded in a differential $\mathcal{Q}$, just as how the redundancies of a gauge theory are encoded in a differential in BRST quantization.
This differential acts on the above superfields as follows:
\begin{equation}
\begin{aligned}
	\mathcal{Q} \boldsymbol{A} & = \overline{\partial} \boldsymbol{A} + \tfrac{1}{2}[\boldsymbol{A}, \boldsymbol{A}] & \quad \mathcal{Q} \boldsymbol{B} & = \overline{\partial}_{\boldsymbol{A}} \boldsymbol{B} + \boldsymbol{\nu}\\
	\mathcal{Q} \boldsymbol{\gamma} & = \overline{\partial}_{\boldsymbol{A}} \boldsymbol{\gamma} & \quad \mathcal{Q} \boldsymbol{\beta} & = \overline{\partial}_{\boldsymbol{A}} \boldsymbol{\beta} + \frac{\partial W(\boldsymbol{\gamma})}{\partial \boldsymbol{\gamma}}\\
\end{aligned}
\end{equation}
where $\boldsymbol{\nu}$ is the moment map for the $G$ action on the matter fields; $\langle\boldsymbol{\nu}, T_a\rangle = \boldsymbol{\beta} T_a \boldsymbol{\gamma}$.
In particular, it asserts that the (1-form part of) $\overline{\partial}_{\boldsymbol{A}} = \overline{\partial} + \boldsymbol{A}$ defines a holomorphic $G$ bundle on $\cM$ and that $\boldsymbol{\gamma}$ is a holomorphic section of an associated $V$-bundle.
Additionally, the equations of motion for $\boldsymbol{\beta}$ asserts that $\boldsymbol{\gamma}$ should reside in the critical local of the superpotential.

\begin{remark}
	As mentioned above, when the underlying supersymmetric gauge theory admits an action of $U(1)_R$ that survives quantization we do not have to restrict attention to complex surfaces with trivial canonical bundle.
	If we denote by $K$ the canonical bundle of $\cM$ and choose a square-root $K^{1/2}$, then we can modify the places where the fields take values:
	\begin{equation}
	\begin{aligned}
		\boldsymbol{A} & \in \Omega^{0,\bullet}(\cM, \mathfrak{g}_\C)[1] & \qquad \boldsymbol{B} & \in \Omega^{0,\bullet}(\cM, K \otimes \mathfrak{g}^\vee_\C)\\
		\boldsymbol{\gamma} & \in \bigoplus_r \Omega^{0,\bullet}(\cM, K^{r/2} \otimes V_r^\vee)[r] & \qquad \boldsymbol{\beta} & \in \bigoplus_r \Omega^{0,\bullet}(\cM, K^{1-r/2} \otimes V_r)[1-r]\\
	\end{aligned}
	\end{equation}
	We can then take the action to be as given above but without $\omega$; each term is necessarily a section of $K$ and hence can be safely integrated.
\end{remark}

\subsection{Infinite-dimensional symmetries in the minimal twist}

One of the remarkable properties of the holomorphic twist is that it admits myriad infinite-dimensional symmetries in much the same way as two-dimensional conformal field theories; see e.g. \cite{FHK, GwWi, SaWi2} and references therein.
An advantage of the above presentation is that it makes manifest these symmetries.

The first such symmetry enhances spacetime translations, rotations, and conformal transformations, cf. \cite{SaWi2,BoWu,BGWW}: when the quantum theory admits a $U(1)_R$ symmetry, the above twisted theory admits an action of all holomorphic vector-fields on $\cM$, which allows is what it to be defined on arbitrary complex surfaces.
This action of holomorphic vector fields enhances the \emph{twisted} superconformal algebra.
Namely, one can view the twisting supercharge as a differential on the $\cN = 1$ superconformal algebra and the its cohomology naturally acts on the twisted theory: the closed symmetries are those that survive the twist and the exact symmetries are those whose action is trivial.
In this case, the (complexified) $\cN = 1$ superconformal algebra can be identified with $\mathfrak{sl}(4|1)$ and its twist $\mathfrak{sl}(3)$; these correspond to the holomorphic vector fields $\partial_{z^a}$ (holomorphic translations), $z^{a} \partial_{z^b}$ (holomorphic rotations and dilations), and $z^a (z^b \partial_{z^b})$ (holomorphic special conformal transformations).
The action of this symmetry on local observables is realized by a pair of holomorphic conserved current $\boldsymbol{S}_a$ coming from the physical theory's supercurrent multiplet \cite{BoWu}; the action is encoded by $\lambda$-brackets with this conserved current.
For example, the $\lambda$-bracket of these currents with themselves take the form
\begin{equation}
	\{\boldsymbol{S}_a {}_\lambda \boldsymbol{S}_b\} = \partial_{z^a} \boldsymbol{S}_b + \lambda_a \boldsymbol{S}_b + \lambda_b \boldsymbol{S}_a
\end{equation}
which says that it transforms as a section of the holomorphic cotangent bundle of $\cM$ and, moreover, captures the fact that the symmetry they generate is the algebra of holomorphic vector fields on $\C^2$.
In fact, the regulated products with $\boldsymbol{S}_a$ extend this action of holomorphic vector fields on $\C^2$ to an action of $H^{0,\bullet}(\C^2 - 0, T^{1,0})$; this is the two complex dimensional analogue of the Witt algebra.%
\footnote{Unlike $\C - 0$, the higher-dimensional punctured affine spaces are not longer affine.
Consequently, the sheaf of holomorphic differential operators on $\C^h - 0$ has higher cohomology -- it is concentrated in degree $h-1$.} %
This is entirely analogous to how the OPE of the stress tensor $T$ of a two-dimensional holomorphic field theory  captures the Virasoro algebra generated by its modes and how their action on other observables encode their transformation properties under infinitesimal coordinate transformations.
When there is no $U(1)_R$, this symmetry is broken to an action of holomorphic, divergence-free (with respect to $\omega$) vector fields, commensurate with the fact that the theory can only be placed on Calabi-Yau two-folds.
In much the same way, if the underlying supersymmetric field theory admits a flavor symmetry $H$, e.g. a group $H$ acting on $V$ that preserves $W$ and commutes with the action of $G$ (and with $U(1)_R$ when applicable), then the holomorphically twisted theory admits an action of holomorphic $H$ transformations generated by a conserved current $\boldsymbol{J}$; if $t_A$ are a basis of the Lie algebra of $H$ and $[t_A, t_B] = f^{C}_{AB} t_C$ then the $\lambda$ bracket of these currents takes the form
\begin{equation}
	\{\boldsymbol{J}_A {}_\lambda \boldsymbol{J}_B\} = f^C_{AB} \boldsymbol{J}_C~,
\end{equation}
which captures the Lie brackets of $H^{0,\bullet}(\C^2 - 0, \textrm{Lie}(H))$.
These currents are used to couple the twisted theory to an auxiliary holomorphic $H$ bundle.
See \cite{BGKWWY, BGWW} for explicit expressions for these currents in the holomorphic field theories described above.

Each of these symmetries typically have 't Hooft anomalies capturing the dependence of the underlying supersymmetric field theory's dependence on a given choice of background, e.g. the choice of spacetime $\cM$ or background flavor symmetry bundle.
In two-dimensional holomorphic field theories, these are captured in central terms in the OPE algebra, e.g. central charges appearing in the OPE of a stress tensor and itself or levels of affine currents.
In higher dimensions, these anomalies still appear as central extensions, but no longer as central extensions of these symmetries as Lie algebras: they are central extensions \emph{as $\mathcal{L}_\infty$-algebras} \cite{FHK, GwWi}.
Namely, these anomalies are neatly captured in the holomorphic twist in terms of the higher arity $\lambda$-brackets of the corresponding conserved currents \cite{BoWu, BGWW}.
For example, when the theory has an action of a $U(1)_R$ then the twisted theory recovers the conformal Euler $a$ and Weyl $c$ anomalies from the ternary brackets of the $\boldsymbol{S}_a$'s.
Similarly the 't Hooft anomalies for the $H$ flavor symmetry are captured by the ternary bracket of three of the $\boldsymbol{J}$'s and the ternary brackets of two $\boldsymbol{S}$'s and $\boldsymbol{J}$ or one $\boldsymbol{S}$ and two $\boldsymbol{J}$'s captures mixed flavor/gravitational anomalies.
For example, the ternary bracket of three $\boldsymbol{J}$'s takes the form
\begin{equation}
	\{\boldsymbol{J}_A {}_{\lambda_1} \boldsymbol{J}_B {}_{\lambda_2} \boldsymbol{J}_C\} = \frac{k_{HHH}}{2\pi^2} d_{ABC}~\varepsilon^{ab}\lambda_{1a} \lambda_{2b}
\end{equation}
where $k_{HHH} \in H^4(BH,\Z)$ is the $H$-symmetry's 't Hooft anomaly coefficient, $d_{ABC}$ is the totally symmetric totally symmetric invariant tensor on the Lie algebra of $H$, and $\varepsilon^{ab}$ is the totally anti-symmetric two-index tensor with $\varepsilon^{12} = 1$.

Finally, we note that when considering the holomorphic twist of a theory with extended supersymmetry the action of holomorphic vector fields is itself extended.
These enhancements have a clean geometric interpretation in terms of superspaces: the algebra of holomorphic vector fields on $\C^{2} - 0$ gets replaced by the algebra of holomorphic vector fields on a punctured superspace \cite{SaWi2}.
When we consider an $\cN = 2$ theory with an unbroken $U(1)_r$, the minimal twist admits an action of the algebra of holomorphic vector fields on $\C^{2|1} - 0$.%
\footnote{While this statement is technically true, it is better to think of this symmetry algebra as being identified with the algebra of holomorphic vector fields on the total space of the super vector bundle
\[
	\Pi K^{1/3} \to \C^2 - 0~,
\]
where $\Pi$ denotes a shift in Grassmann parity.
This difference is tied to a choice of twisting homomorphism: this presentation arises when taking the twisting homomorphism to use the superconformal $R$-charge when viewed as an $\cN=1$ theory.
The theory, with this choice of twisting homomorphism is compatible with complex surfaces equipped with a choice third-root of the canonical bundle.
This does not stop us from placing the twisted theory on other complex surfaces, we just need to use another twisting homomorphism.
We will not belabor this point beyond this footnote, but it does play a role in the proper description of the symmetries an $\cN=4$ theory.} %
In addition to the $\boldsymbol{S}_a$'s there is a conserved current $\boldsymbol{R}$ generating a $U(1) \hookrightarrow SU(2)_R \times U(1)_r$ $R$-symmetry together with two supercurrents $\boldsymbol{G}_a$ and $\widetilde{\boldsymbol{G}}$.%
\footnote{When the $\cN = 2$ does not admit an action of $U(1)_r$ then a smaller, but nonetheless infinite-dimensional, symmetry persists.
We will not discuss it here; see \cite{BGWW} for more details.} %
In terms of holomorphic vector fields, $\boldsymbol{R}$ generates those of the form $f(z) \epsilon \partial_\epsilon$, $\boldsymbol{G}_a$ generates those of the form $f(z) \epsilon \partial_{z^a}$, and $\widetilde{\boldsymbol{G}}$ generates those of the form $f(z) \partial_\epsilon$, where $\epsilon$ is a choice of Grassmann odd coordinate on $\C^{2|1}$.
This is an infinite-dimensional enhancement of the twisted $\cN=2$ superconformal algebra, which is isomorphic to $\mathfrak{sl}(3|1)$.
In much the same way, the holomorphic twist of an $\cN = 4$ theory admits an action of the algebra of \emph{divergence-free} holomorphic vector fields on $\C^{2|3} - 0$ \cite{SaWi2}; the divergence-free condition is tied to the fact that the $\cN=4$ superconformal algebra is $\mathfrak{psl}(4|4)$ (rather than $\mathfrak{sl}(4|4)$) and so the twisted $\cN=4$ superconformal algebra is $\mathfrak{psl}(3|3)$.

\subsection{Deforming the minimal twist}

Just as with twisting a supersymmetric quantum field theory, we can use nilpotent symmetries of the holomorphic twist to deform the theory.
As seen above, the holomorphic twists of theories with extended supersymmetry naturally come with a wealth of nilpotent symmetries coming from the Grassmann odd vector fields on superspace arising from the residual supersymmetries of the underlying field theory.
As mentioned above, this provides a route to understanding aspects of more-topological field theories from the holomorphically-twisted theory; this is the avenue pioneered in \cite{EY}.

We will illustrate this process for $\cN=2$ theories and then briefly comment on $\cN=4$ theories, with a particular emphasis on the cases leading to Donaldson-Witten and Vafa-Witten theory.
Recall that equations of motion for the holomorphic field theory describing the minimal twist of an $\cN=2$ theory encodes the data of a $G$ bundle with holomorphic structure $\overline{\partial} + A$ together with a collection of matter fields that give rise to holomorphic sections of an associated $\mathcal{R}$-bundle (we will restrict to cases with $\mathcal{R} = U \oplus U^\vee$ for simplicity); the twisting homomorphism relevant for us will twist these sections by $K^{1/2}$.
Explicitly, these equations of motion are encoded in the BV/BRST differential as
\begin{equation}
\begin{aligned}
	\mathcal{Q} \boldsymbol{A} & = \overline{\partial} \boldsymbol{A} + \tfrac{1}{2}[\boldsymbol{A}, \boldsymbol{A}] & \qquad \mathcal{Q} \boldsymbol{\Lambda} & = \overline{\partial}_{\boldsymbol{A}} \boldsymbol{\Lambda} + \boldsymbol{\mu}(\boldsymbol{X}, \boldsymbol{Y})\\
	\mathcal{Q} \boldsymbol{X} & = \overline{\partial}_{\boldsymbol{A}} \boldsymbol{X} & \qquad \mathcal{Q} \boldsymbol{Y} & = \overline{\partial}_{\boldsymbol{A}} \boldsymbol{Y}\\
\end{aligned}
\end{equation}
where $\boldsymbol{\Lambda}$ is the fermionic partner of the adjoint chiral multiplet sitting in the $\cN=2$ vector multiplet, $\boldsymbol{X}$ and $\boldsymbol{Y}$ are the bosonic superfields coming from the hypermultiplet and transforming in $U$ and $U^\vee$, respectively, and $\boldsymbol{\mu}$ is (the pull back of) the complex moment map for the $\mathfrak{g}$ action on $U \oplus U^\vee$.
These fields can be viewed as holomorphic maps from spacetime into the holomorphic symplectic quotient $U \oplus U^\vee /\!\!/ G$; the remaining fields form an odd cotangent lift of these equations of motion.
In particular, the moduli space of solutions to these equations of motion take the following form \cite{ElSaWi}:
\begin{equation}
	\textrm{EOM}(\cM) = T^*[-1] \textrm{Sect}(\cM_{\overline{\partial}}, K^{1/2} \otimes (U \oplus U^\vee)/\!\!/G)~,
\end{equation}
where $\cM_{\overline{\partial}}$ is the locally ringed space built from $\cM$ whose structure sheaf is the complex $(\Omega^{0,\bullet}(\cM), \overline{\partial})$ and $\textrm{Sect}(\cM_{\overline{\partial}}, \cF)$ denotes the (derived) moduli space of holomorphic sections of the holomorphic bundle $\cF$.

This moduli space has two natural, geometric deformations coming from an analogue of the Dolbeault to de Rham deformation.
We can either apply this directly to the moduli space itself, leading to a moduli space of solutions to the equations of motion given by the de Rham stack of the base
\begin{equation}
	\textrm{Sect}(\cM_{\overline{\partial}}, K^{1/2} \otimes (U \oplus U^\vee)/\!\!/G)_{\textrm{dR}}~,
\end{equation}
i.e. the locally ringed space whose structure sheaf is the (holomorphic) de Rham complex.
This deformation can be directly tied the BV/BRST differential with $\widetilde{\boldsymbol{G}}$ or, equivalently the vector field $\partial_\epsilon$; this corresponds to the deformation of the holomorphic twist to the Donaldson-Witten twist \cite{Witt88}, cf. \cite[Sec.~10.2.2]{ElSaWi}.
If one is interested in invariants of general complex surfaces, rather than complex surfaces equipped with a choice of $K^{1/2}$ (which is the holomorphic avatar of a spin structure), one can work with a slightly different choice of twisting homomorphism that includes flavor symmetries, cf. Lecture 1; this will described in more detail in Section \ref{sec:Lecture4JM}.

If we instead choose a rank 2 transverse holomorphic folations on $\cM$ and then deform the theory with, say, $\boldsymbol{G}_1$, i.e. the nilpotent vector field $\theta \partial_{z^1}$, and get a holomorphic-topological field theory; this requires a slight modification of the twisting homomorphism that allows for the identification $\theta \sim \textrm{d} z^1$.
This deformation leads to a moduli space of solutions to the equations of motion of the form
\begin{equation}
	T^*[-1]\textrm{Sect}(\cM_{\textrm{d}'}, \widehat{K}^{1/2} \otimes U/G)~,
\end{equation}
where $\cM_{\textrm{d}'}$ is the locally ringed space build from $\cM$ with structure sheaf given by the complex $(\Omega^{(0);\bullet}(\cM), \textrm{d}')$ and $\widehat{K}$ is the pull-back of canonical bundle on the space of leaves to $\cM$.
This moduli space captures the equations of motion in the Kapustin twist \cite{Kapustin:2006hi} of the underlying supersymmetric field theory \cite[Sec~10.2.3]{ElSaWi}.

A similar story plays out in the $\cN = 4$ case, but now there are more possibilities.
We will not attempt to describe these additional deformations in any real detail, instead choosing to focus on the Donaldson-Witten and Vafa-Witten twists.
Although these twists come from the same supercharge, they make use of different twisting homomorphisms.
Namely, the Donaldson-Witten twist is described by the above by simply taking $U = U^\vee = \mathfrak{g}_\C$.
The twisting homomorphism used for Vafa-Witten theory instead makes the superfields $\boldsymbol{X}$, $\boldsymbol{Y}$ a scalar and a section of the canonical bundle $K$, respectively.
In this context, we see that the above equations of motion assert that $\overline{\partial} + A$ defines a holomorphic $G$ bundle $E$, the field $X$ is a holomorphic section of its adjoint bundle $\textrm{ad}(E)$, the field $Y$ is a holomorphic section of the adjoint bundle twisted by the canonical bundle $\textrm{ad}(E) \otimes K$; the field $\Lambda$ enforces the constraint that $X$ and $Y$ commute with one another:
\begin{equation}
\begin{aligned}
		(\overline{\partial} + A)^2 & = 0 & \qquad [X, Y] & = 0\\
		\overline{\partial}_A X & = 0 & \qquad \overline{\partial}_A Y & = 0
\end{aligned}
\end{equation}
These are precisely the holomorphic part of the Vafa-Witten equations, i.e. the equations prior to imposing suitable stability conditions that give rise to genuine Vafa-Witten theory.
The full moduli space of solutions to the equations of motion is thus the de Rham space of the moduli space of solutions to these equations of motion.

\begin{remark}
\begin{enumerate}
	\item As we just mentioned, the perspective we have been describing in this set of lectures entirely ignores the stability conditions one typically imposes to make the moduli space of solutions well-behaved.
	These stability conditions are obfuscated in taking the shortcut through the holomorphic twist; a proper treatment of the holomorphic twist should also take stability conditions into account.
	\item We have largely focused on the moduli space of solutions to the classical equations of motion; to study the quantum theory would require (geometric) quantization of these moduli spaces.
	To make contact with the invariants studied in the later sections, we note that quantization of these de Rham space leads to path integrals computing the Euler character of the underlying space.
	The standard example of this is topological quantum mechanics coming from twisting an $\cN=2$ quantum mechanical sigma model with target $X$ \cite{WittenMorse}; the moduli space of solutions to the equations of motion is
	\begin{equation}
		\textrm{Maps}(\R_{\textrm{dR}}, X)_{\textrm{dR}}
	\end{equation}
	and its path integral reproduces the Euler of $X$.
	In the context of Donaldson-Witten and Vafa-Witten theory, the path integral precisely computes the Euler character of the (stable parts of) the above moduli spaces.
	\item There are many other nilpotent symmetries in the holomorphic twist of theories with extended supersymmetry, some of which fill out the twists appearing in Table \ref{tab:twists}.
	In fact, even in the $\cN=2$ case there are infinitely many such deformations \cite{SaWi2}; beyond the ones described above (and rotations thereof), only one (and its rotations) has (at present) a natural interpretation from the underlying physical supersymmetric field theory. Deforming by the vector field $z^2 \partial_\epsilon$ leads to a holomorphic field theory localized on the $z^2 = 0$ plane.%
	\footnote{Infinitely many other nilpotent vector fields come from choosing a holomorphic function $f(z^1,z^2)$ and considering the vector field $f \partial_\epsilon$; this results in a holomorphic field theory localized on the curve $f = 0$.} %
	Lifting this vector field to the underlying physical theory, it can be identified with a conformal supercharge and the resulting deformation of the holomorphic twist recovers the construction of \cite{BLLPRvR}.
	Moreover, the $\lambda$-brackets of this vertex algebra can be extracted from the four-dimensional $\lambda$-brackets \cite{BGWW}.
\end{enumerate}
\end{remark}

\chapter[Topological correlators of \texorpdfstring{$\cN=2$}{N=2} supersymmetric Yang-Mills theory]{Topological correlators of \texorpdfstring{$\cN=2$}{N=2} supersymmetric Yang-Mills theory and four-manifold invariants}\label{chap:ySYM}
\vspace{-.4cm}
\begin{center}
\it{Jan Manschot}
\end{center}
\vspace{.4cm}
The aim of these lectures is to given an overview of topological twists of $\mathcal{N}=2$ supersymmetric Yang-Mills theories, and the connections to invariants of moduli spaces of instantons. 

\section[Lecture 1]{Lecture 1: Introduction and $\cN=2$ supersymmetric quantum field theory}
\label{sec:JMLect1}
The purpose of this first lecture is to introduce relevant aspects of $\cN=2$ supersymmetric quantum field theory. We consider space-time $X$ in Euclidean signature, either $X=\mathbb{R}^4$ or a smooth, compact without boundary four-manifold $X$. Let $G$ be a simply connected Lie group with Lie algebra $\frak{g}$, typically $G=SU(2)$ or $SO(3)$. The most important field during these lectures is the gauge potential $A$ (or connection on the principal $G$-bundle $P$),
\begin{equation}
A=A_\mu^a t^a_r dx^\mu \in \Omega^1(P,\frak{g}),
\end{equation}
where $a$ labels the generators of $\frak{g}$, and $t^a_r$ are generators in representation $r$. 
The field strength reads
\begin{equation}
F^a_{\mu\nu}=\partial_\mu A^a_\nu - \partial_\nu A^a_\mu+f^{abc} A_\mu^b A_\nu^c,
\end{equation}
with $f^{abc}$ the structure constants of the Lie algebra. A gauge transformation $U: X\to G$ acts on $A$ and $F$ as,
\begin{equation}
A_\mu\mapsto A^U_\mu=UA_\mu U^{-1}-(\partial_\mu U)\,U^{-1}, \qquad
F_{\mu\nu}\mapsto F^U_{\mu\nu}=UF_{\mu\nu}U^{-1}. 
\end{equation}
The set of all gauge transformations $U$ forms the group $\mathcal{G}$.

The Yang-Mills action reads
\begin{equation}
\begin{split}
    S&=\frac{1}{4g^2} \int_X d^4x\, (F_{\mu\nu}^a)^2 +\frac{i\theta}{64\pi^2} \int_X \epsilon_{\mu\nu\rho\sigma} F_{\mu\nu}^a F^a_{\rho\sigma}\\
    &=\frac{1}{g^2} \int_X  {\rm Tr}_{\bf 2}[F\wedge *F] +\frac{i\theta}{8\pi^2} \int_X {\rm Tr}_{\bf 2}[F\wedge F],
\end{split}
\end{equation}
where $g$ is the coupling constant and $\theta$ is the theta angle. These parameters determine the complexified coupling constant
\begin{equation}
    \tau=\frac{\theta}{2\pi}+\frac{4\pi i}{g^2}\in \mathbb{H}.
\end{equation}
For $G=SU(2)$, the action is minimized by the anti-self-dual (asd) connection 
\begin{equation}
\label{eq:DefInst}
F=-*F,
\end{equation}
or instanton. It is characterized by the instanton number
\begin{equation}
\label{eq:InstNum}
k=-\frac{1}{8\pi^2} \int_X {\rm Tr}_{\bf 2}[F\wedge F],
\end{equation}
and for $G=SO(3)$ the second Stiefel-Whitney class $w_2(P)\in H^2(X,\mathbb{Z}_2)$. If $w_2(P)$ is non-vanishing, $P$ cannot be lifted to a principal $SU(2)$ bundle. We let $E$ be the rank 2 bundle associated to the 2-dimensional projective representation of $SO(3)$. If $w_2(P)\neq 0$, this bundle does not exist globally. A globally well-defined rank 2 bundle $\tilde E$ can be obtained by multiplication of $E$ with a square root of a line bundle $\mathcal{L}$ with $c_1(\mathcal{L})=w_2(P) \mod H^2(X,2\mathbb{Z})$. The first Chern class $c_1(\tilde E)$ satisfies,
\begin{equation}
c_1(\tilde E)=w_2(P) \mod H^2(X,2\mathbb{Z}).
\end{equation}
For gauge group $SU(2)$ and $SO(3)$, we introduce furthermore
\begin{equation}
\label{eq:defmu}
\bfmu=\frac{1}{2} \bar w_2(P)\in L/2,
\end{equation}
with $\bar w_2(P)$ a lift of $w_2(P)$ to the lattice $L\simeq H^2(X,\mathbb{Z})$ modulo torsion. 

The exponentiated action of such solutions evaluates to 
\begin{equation}
e^{-S}=q^k, \qquad q=e^{2\pi i\tau}.
\end{equation}
The anti-self-duality equation (\ref{eq:DefInst}) gives rise to a
non-trivial space of solutions modulo gauge transformations, or moduli
space $M_k$,
\begin{equation}
M_k=\left\{ A_\mu\, dx^\mu\in \Omega^1(P,\mathfrak{g})\,\left\vert\, F=-*F;\,\,k=-\frac{1}{8\pi^2} \int_X {\rm Tr}_{\bf 2}[F\wedge F] \right.\right\}\biggm/ \mathcal{G} .
\end{equation}
These spaces will play a major role during these lectures. 
The Donaldson-Uhlenbeck-Yau theorem \cite{UhYa, Donaldson:1987} and Hitchin-Kobayashi \cite{Kobayashi:1982}
correspondence provide a strong connection between the instanton
solutions and mathematics. This will be discussed in detail in lectures by Y. Tanaka \ref{sec:Lecture1YT} and M. Kool \ref{sec:Lecture1MK}. 

To localize in quantum field theory to the anti-self-dual solutions, we consider
an $\mathcal{N}=2$ supersymmetric extension of Yang-Mills
theory. Contributions from states above the minimum energy bound will
vanish in such theories due to cancellations between bosonic and
fermionic degrees of freedom. There are many good texts on
supersymmetric field theory, such as the books by S. Weinberg \cite{Weinberg:2000cr} and Y. Tachikawa \cite{Tachikawa:2013kta}.

To introduce the field content of these theories, recall that $X=\mathbb{R}^4$ is a spin manifold. We denote the chiral and anti-chiral spin bundles by $S^\pm$.\footnote{Lecture 2 will discuss Spin and Spin$^c$ structures in more detail.} The global symmetry group of these theories include the $G_R=SU(2)_R$  R-symmetry group for $\cN=2$ supersymmetry. The field content of $\cN=2$ supersymmetric Yang-Mills theory is organized in terms of multiplets, which come in two types, the vector multiplets and hypermultiplets. We list the various fields together with their $SU(2)_+\times SU(2)_-\times SU(2)_R$ representation below: 
\begin{enumerate}
\item Vector multiplet:
\begin{itemize}
\item $A\in \Omega^1(X,{\rm End({\rm ad}}(P)))$, $({\bf 2},{\bf 2},{\bf 1})$,
\item $\phi\in \Gamma({\rm ad}(P))$, $({\bf 1},{\bf 1},{\bf 1})$,
\item $\Psi^I_\alpha\in\Gamma(S^+\otimes {\rm ad}(P))$, $({\bf 2},{\bf 1},{\bf 2})$,
\item $\bar \Psi^I_{\dot \alpha}\in \Gamma(S^-\otimes {\rm ad}(P))$, $I=1,2$, $({\bf 1},{\bf 2},{\bf 2})$,
with ${\rm ad}(P)$ the bundle associated to $P$ for the adjoint representation.
\end{itemize}
\item Hypermultiplet: 
\begin{itemize}
\item $Q,\tilde Q^\dagger \in \Gamma(R)$, $({\bf 1},{\bf 1},{\bf 2})\oplus ({\bf 1},{\bf 1},{\bf 2})$,
\item $\lambda_\alpha$, $\chi_\alpha\in \Gamma(S^+\otimes R)$,  $({\bf 2},{\bf 1},{\bf 1})\oplus ({\bf 2},{\bf 1},{\bf 1})$
\item $\bar \lambda_{\dot
    \alpha}$, $\bar \chi_{\dot \alpha}\in \Gamma(S^-\otimes R$), $({\bf 1},{\bf 2},{\bf 1})\oplus ({\bf 1},{\bf 2},{\bf 1})$.
\end{itemize} where $R$ is the bundle associated to $P$ for the quaternionic representation $r$ of the gauge group $G$ under which the hypermultiplet transforms. In these lectures, $R$ is either ${\rm ad}(P)$, or the bundle $E$ associated to the 2-dimensional, fundamental representation of $SO(3)$.
\end{enumerate}
A non-vanishing mass $m_j$, $j=1,\dots,N_f$ can be included for each hypermultiplet. The flavor symmetry group $G_F$ is $U(1)^{N_f}$ for $N_f$ distinct masses, while $G_F=SO(2N_f)$ for equal masses.

There is a standard $\cN=2$ supersymmetric action for these fields \cite{Weinberg:2000cr, Tachikawa:2013kta}. For $N_f=0$, the D-term superpotential reads
\begin{equation}
\label{eq:Vphi}
V(\phi)=\frac{1}{g^2}\,{\rm Tr}[\phi,\phi^\dagger]^2,
\end{equation}
minimized by $[\phi,\phi^\dagger]=0$. 

We specialize to $G=SU(2)$ or $SO(3)$ in the following. All solutions for minimizing $V(\phi)$ (\ref{eq:Vphi}) are then gauge equivalent to
\begin{equation}
\label{eq:Cphi}
\phi=a\, \sigma_3=\left(\begin{array}{cc} a & 0 \\ 0 & -a \end{array}\right),
\end{equation}
with $a$ a massless degree of freedom. This phase where the gauge symmetry is broken to $U(1)$, more generally the maximal Abelian subgroup, is known as the Coulomb branch of the theory. A separate branch of vacua can exist in theories with hypermultiplets. This is the Higgs branch where $\phi$ vanishes, but hypermultiplet scalars are non-vanishing.

Returning to the Coulomb branch, the gauge invariant order parameter on this branch is
\begin{equation}
\label{eq:Defu}
u=\left< {\rm Tr}[\phi^2] \right>.
\end{equation}
The $SU(2)$ gauge field becomes
\begin{equation}
A_\mu=\left(\begin{array}{cc} A^{U(1)}_\mu & W^+_\mu \\ W^-_\mu & -A^{U(1)}_\mu \end{array}\right).
\end{equation}
The off-diagonal components become massive $W$-bosons, and form together with the off-diagonal components of $\phi$ (\ref{eq:Cphi}) a massive vectormultiplet. This can be integrated out to obtain a low energy effective action with $U(1)$ gauge symmetry.

The theory contains states which preserve half the supersymmetry. These states are
\begin{itemize}
\item $W$-bosons with electric-magnetic charge $(m,e)=(0,2)$ and semi-classical central charge $Z=2a$,
\item monopoles with electric-magnetic charge $(m,e)=(1,0)$ and semi-classical central charge $Z=\tau a=a_D$,
\item dyons with electric-magnetic charge $(m,e)=(\pm 1,2\mathbb{Z})$ and semi-classical central charge $Z= m\, a_D+e\,a$,
\item instantons with instanton number $k$ (\ref{eq:InstNum}).
\end{itemize}

Quantum field theory processes give rise to an energy dependence of the effective coupling. In terms of $\cN=1$ multiplets the $\beta$-function reads,
\begin{equation}
\beta(E)=E\,\frac{dg(E)}{dE}=-\frac{g^3\,c}{(4\pi)^2}+O(g^4),
\end{equation}
with $c=3\,C({\rm adj})-C(r)$, where $r$ denotes the $G$-representation of $\cN=1$ (anti)chiral multiplets. The values of $C$ for adjoint and fundamental representation of $SU(N)$ are $C(\text{adj})=N$ and $C(\text{fund})=1/2$. The $\cN=2$ hypermultiplet consists of an $\cN=1$ chiral + anti-chiral multiplet which both transform under the same representation.

The theory is asymptotically free for $c>0$, and conformal for $c=0$. In the effective field theory on the Coulomb branch, the energy scale is set by the parameter $a$, such that the complexified gauge coupling reads,
\begin{equation}
\tau(a)=\tau_{\rm UV}-\frac{c}{2\pi i}\log\left(\frac{a}{\Lambda_{\rm UV}} \right)+\dots=-\frac{c}{2\pi i}\log\left(\frac{a}{\Lambda} \right)+\dots.
\end{equation}
with $\Lambda$ the complexified scale
\begin{equation}
\Lambda=\Lambda_{\rm UV}\,e^{2\pi i \tau_{\rm UV}/c},\qquad c>0.   
\end{equation}
As a result of supersymmetry, the effective action is determined by a holomorphic prepotential $\cF(a;\Lambda)$, which is homogeneous of degree 2. The effective coupling reads
\begin{equation}
\tau(a)=\frac{\partial^2 \cF(a;\Lambda)}{\partial a^2}.
\end{equation}
The prepotential can be expanded for small $\Lambda$ as,
\begin{equation}
\begin{split}
&\cF(a;\Lambda)=\frac{2i}{\pi}a^2\log(4a^2/\Lambda) \\
& \quad -\frac{i}{4\pi} \sum_{j=1}^{N_f} \left(a+\frac{m_j}{\sqrt{2}}\right)^2\log\left( \left(a+\frac{m_j}{\sqrt{2}}\right)/\Lambda\right)+\left(a-\frac{m_j}{\sqrt{2}}\right)^2\log\left( \left(a-\frac{m_j}{\sqrt{2}}\right)/\Lambda\right)\\
&\quad +\dots,
\end{split}
\end{equation}
with the logarithmic terms being due to the one-loop contributions. There are further subleading non-perturbative corrections due to instantons, which give rise to a power series in $\Lambda/a$. Seiberg and Witten \cite{SeWi, Seiberg:1994aj} determined these corrections to all orders in terms of a family of auxiliary elliptic curves parametrized by $u$ which have the effective coupling $\tau(a)$ as complex structure.

One formulation of this curve is as a family of complete intersections in $\mathbb{CP}^2$ parametrized by $u$. For $N_f=0$, this family is
\begin{equation}
\label{eq:SWcurve}
\Sigma_u=\left\{\left.(x,y)\in \mathbb{CP}^2 \right|\, y^2=x(x^2-u\,x+\tfrac{1}{4}\Lambda^4)\right\},
\end{equation}
with $a$ and $a_D$ given as the contour integrals over two homology cycles $A$ and $B$ such that the intersection $A\cap B=1$,
\begin{equation}
a=\oint_A \lambda,\qquad a_D=\oint_B \lambda,
\end{equation}
with $\lambda$ the Seiberg-Witten differential.

\begin{figure}[ht!]\centering 
	\includegraphics[scale=0.5]{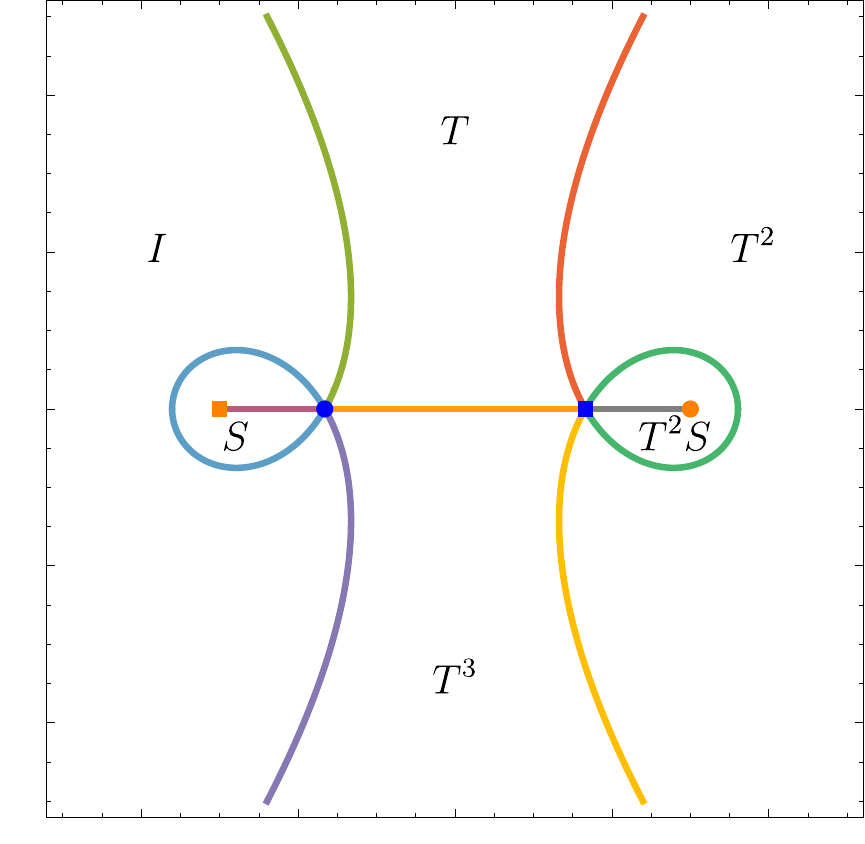}
	\caption{Plot of the $u$-plane for the $N_f=0$ theory with the singular points $u=\pm \Lambda^2$ indicated with the orange square and circle. The colored arcs are the images of the arcs in Fig. \ref{fig:fundgamma0(4)} under $u:\mathbb{H}/\Gamma^0(4)\to \mathbb{C}$. }\label{fig:uplane}
\end{figure}

The curve is singular for two values of $u$, $u=\pm \Lambda^2$. For these values of $u$, the low energy effective theory breaks down due to new massless particles. The domain for $u$ is plotted in Figure \ref{fig:uplane}.

The expression (\ref{eq:SWcurve}) can be brought to Weierstrass form, which is a convenient representation of $\Sigma_u$ for determining $u$ in terms of modular forms. Recall a {\it modular form} of weight $w\in \mathbb{Z}/2$ is a function $f:\mathbb{H}\to \mathbb{C}$ such that
\begin{equation}
f\!\left(\frac{a\tau+b}{c\tau+d}\right)=\varepsilon(\gamma)\,(c\tau+d)^w\,f(\tau),
\end{equation}
where $a,b,c,d\in \mathbb{Z}$ are entries of the matrix $\gamma$, 
\begin{equation}
\gamma=\left(\begin{array}{cc} a & b \\ c & d \end{array}\right)\in SL_2(\mathbb{Z}),
\end{equation}
or a congruence subgroup $\Gamma\subset SL_2(\mathbb{Z})$. Moreover, $\varepsilon(\gamma)$ is a 24'th root of unity depending on $\gamma$ and $f$. Classical examples of modular forms are the Dedekind eta function,
\begin{equation}
\eta(\tau)=q^{1/24}\prod_{n=1}^\infty(1-q^n),
\end{equation}
which is a modular form of weight $1/2$ for $SL_2(\mathbb{Z})$, and the three Jacobi theta functions $\vartheta_j$,
\begin{equation}
\label{eq:JacThetas}
\begin{split} 
&\vartheta_2(\tau)=\sum_{n\in \mathbb{Z}+1/2} q^{n^2/2},\qquad \vartheta_3(\tau)=\sum_{n\in \mathbb{Z}} q^{n^2/2},\qquad \vartheta_4(\tau)=\sum_{n\in \mathbb{Z}} (-1)^n\, q^{n^2/2}.  
\end{split}
\end{equation}
These are modular forms of weight $1/2$ for the congruence subgroup $\Gamma(2)$, ie the elements of $SL_2(\mathbb{Z})$ with $a,d=1\mod 2$ and $b,c=0 \mod 2$.

In terms of these functions, the order parameter $u$ \eqref{eq:Defu} is expressed as
\begin{equation}
\label{eq:utau}
\begin{split} 
\frac{u(\tau)}{\Lambda^2}&=\frac{\vartheta_2(\tau)^4+\vartheta_3(\tau)^4}{2\,\vartheta_2(\tau)^2\,\vartheta_3(\tau)^2}\\
&=\frac{1}{8}q^{-1/4}+\frac{5}{2}q^{1/4}+\dots.
\end{split}
\end{equation}
The function $u(\tau)$ is a modular form of weight 0 for the congruence subgroup $\Gamma^0(4)$,
ie the elements of $SL_2(\mathbb{Z})$ with $b=0\mod 4$. This subgroup is generated by two elements 
\begin{equation}
T^4=\left(\begin{array}{cc} 1 & 4 \\ 0 & 1 \end{array}\right)\,\quad \text{and}\quad  STS=\left(\begin{array}{cc} -1 & 0 \\ 1 & -1 \end{array}\right).
\end{equation}
As a result, there are
two strong coupling singularities, for $u=+\Lambda^2$ and
$u=-\Lambda^2$. Here a monopole respectively a dyon becomes
massless. Together with their superpartners, the monopole and dyon
form a hypermultiplet.

\begin{figure}[ht!]\centering 
	\includegraphics[scale=0.5]{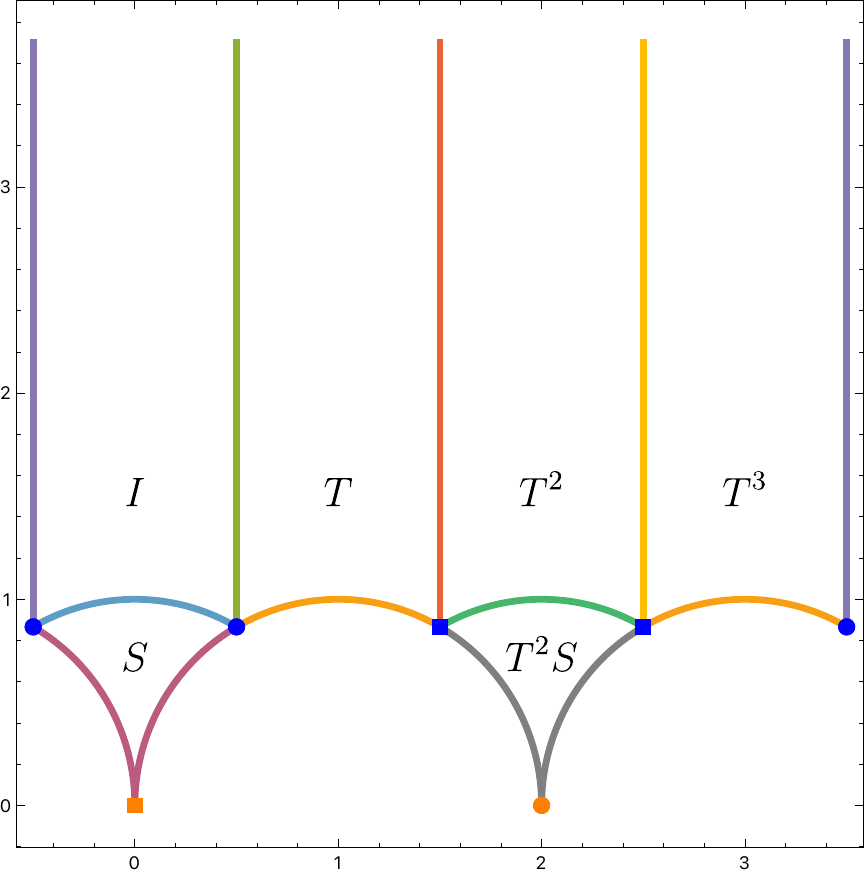}
	\caption{Fundamental domain of $\Gamma^0(4)$.  The two cusps
          on the real line correspond to the strong coupling
          singularities of the gauge theory, while the cusp at
          $\tau=i\infty$ corresponds to weak coupling. The boundaries with the same colour are identified.  }\label{fig:fundgamma0(4)}
\end{figure}

\newpage

\section[Lecture 2]{Lecture 2: Formulation on four-manifolds, topological
  twisting and topological correlators}
\label{sec:JMLect2}
This lecture introduces aspects of four-manifold and the notion of topologically twisting a supersymmetric quantum field theory.

Let $X$ be an oriented, smooth, compact four-manifold with Euler number $\chi$ and signature $\sigma=b_2^+-b_2^-$. We furthermore assume that $X$ is simply connected. We let $L$ be the uni-modular lattice naturally associated to $H^2(X,\mathbb{Z})$ with bilinear form $L\times L\to \mathbb{Z}$ given by $\int_X \alpha\wedge \beta$.

We summarize a few relevant structures:
\begin{itemize}
\item {\bf Smooth structure:}  Recall a smooth atlas is an atlas for which all transition functions are smooth. Two atlases are  smoothly equivalent if their union is smooth. A {\it smooth structure} is then defined as a collection of smoothly equivalent atlases. Two equivalent smooth structures are related by a diffeomorphism. A topological four-manifold may admit none, one or multiple, inequivalent smooth structures.

Note that a smooth structure is a prerequisite for the formulation of the equations of motion and the Lagrangian of a field theory.
\item {\bf Spin structure}: A Spin structure for a four-manifold $X$ is a principal Spin(4)$=SU(2)_-\times SU(2)_+$ bundle compatible with the tangent bundle $TX\to X$. A necessary requirement for $X$ to admit a Spin structure is that its second Stiefel-Whitney class vanishes, $w_2(X)=0$. 
\item {\bf Spin$^c$ structure}: Recall the definition of the group ${\rm Spin}^c(4)$,
\begin{equation}
   {\rm Spin}^c(4)=\{ (u_1,u_2)| {\rm det}(u_1)={\rm det}(u_2)\}\subset U(2)\times U(2),
\end{equation}
or equivalently as the exact sequence,
\begin{equation}
0\to S^1 \to {\rm Spin}^c\to SO(4)\to 0.
\end{equation}
A Spin$^c$ structure then corresponds to a principal Spin$^c(4)$ bundle compatible with action of the $SO(4)$ rotation group on the tangent bundle $TX$. Every smooth, orientable four-manifold admits a Spin$^c$ structure.
\end{itemize}

Our next subject is topological twisting twisting \cite{Witt88, VW, Kapustin:2006pk} of an $\mathcal{N}=2$ supersymmetric quantum field theory, in particular the theories discussed in Lecture \ref{sec:JMLect1}. N. Garner's lectures \ref{sec:NGLect1} and \ref{sec:NGLect2} discuss topological twists in detail. This notion corresponds to a transformation of the field content of the theory such that 
\begin{enumerate}
    \item the twisted theory can be formulated on an arbitrary smooth four-manifold, 
    \item the image of the supercharges $Q^I_\alpha$ in the twisted theory contains a scalar supercharge $\mathcal{Q}$.
\end{enumerate}

 This can be formulated most generally in terms of a homomorphism,
\begin{equation}
    \phi: {\rm Spin}^c(4)\to (G\times G_R\times G_F)/\mathbb{Z}_2.
\end{equation}
The fields then transform under the modified Lorentz group as
\begin{equation}
{\rm Spin}'(4)=(1\times \phi) {\rm Spin}(4)\subset {\rm Spin}(4)\times G_R. 
\end{equation}
The ${\rm Spin}'(4)$ representation of $Q^I_\alpha$, $\bar Q^I_{\dot \alpha}$ must contain a singlet. 

For the Donaldson-Witten twist, ${\rm Spin}'(4)=SU(2)_-\times SU(2)_+'$ with $SU(2)_+'$ the diagonal subgroup of $SU(2)_+\times SU(2)_R$. The ${\rm Spin}'(4)$ representation of $Q^I_\alpha$, $\bar Q^I_{\dot \alpha}$ is then $(\bf 1,1)\oplus (1,3)\oplus (2,2)$, or a scalar, self-dual 2-form and a vector. This is also the representation of the twisted fermions of the vector multiplet, which are respectively denoted by $\eta$, $\chi$ and $\psi$. The bosonic field content of the vector multiplet is unchanged and remains a vector and a scalar. The representation of hypermultiplets will be discussed in more detail in Lecture 4. See Section \ref{sec:Lecture4JM} below. Another useful perspective of topological twisting is as a reduction of the structure group \cite{Manschot:2021qqe, Moore:2024vsd}.

As a consequence of the nilpotent operator $\mathcal{Q}$, an observable $\cO$ is an element of one of the following three classes,
\begin{itemize}
\item $\cO$ such that $\{\mathcal{Q},\cO\}\neq 0$,
\item $\cO$ is $\mathcal{Q}$-exact, that is to say $\cO=\{\mathcal{Q},W\}$ for some $W$,
\item $\cO$ is an element of the $\mathcal{Q}$-cohomology, that is to say $\{\mathcal{Q},\cO\}=0$, but $\cO$ is not $\mathcal{Q}$-exact.
\end{itemize}
We then deduce that the vevs of $\mathcal{Q}$-exact observables formally vanish
\begin{equation}
\left< \{\mathcal{Q},\cO\} \right>=\int \D \mathcal{X}\,\{\mathcal{Q},\cO\} \, e^{-S[\mathcal{X}]}=0,
\end{equation}
where $\mathcal{X}$ stands for the collection of fields of the theory.
Crucially, the Lagrangian takes the following form
\begin{equation}
\cL=\frac{i}{8\pi}\tau\, {\rm Tr}[F\wedge F] + \{\mathcal{Q},W\},
\end{equation}
for some $W$. In particular, all metric dependence is $\mathcal{Q}$-exact. Without making reference to the explicit Langrangian, this follows from the structure of the energy momentum multiplet \cite{Moore:2024vsd}. As a result, correlation functions of $\mathcal{Q}$-closed observables are formally independent fo the metric
\begin{equation}
\delta_g\left< \cO \right>=0,\qquad {\rm if}\quad \{\mathcal{Q},\cO\}=0.
\end{equation}
We thus arrive at the hallmark of topologically twisted theories, namely that these are formally independent of the metric, while they do depend on the smooth structure discussed in the beginning of the lecture. In this way, they can have an impact on the classification of four-manifolds up to diffeomorphism. 

Examples of $\mathcal{Q}$-closed observables are
\begin{itemize}
\item the point observable for a point $p$ of $X$,
 \begin{equation}
     P(p)=2\,{\rm Tr}[\phi(p)^2],
 \end{equation}
\item the surface observable for $x\in H_2(X)$
\begin{equation}
 I(x)=\frac{1}{4\pi^2}\int_x {\rm Tr}[\psi\wedge \psi-\tfrac{1}{\sqrt{2}}\phi\,F].
\end{equation}
\end{itemize}
Such observables can be derived using the descent formalism \cite{Witt88, Baulieu:1988xs}, discussed in Section \ref{sec:descentlob}. The correlation functions of the $\mathcal{Q}$-closed observables are the physical counterpart of Donaldson invariants \cite{Witt88}, which are defined in mathematics as intersection numbers of the moduli space of instantons. See Section \ref{sec:Lecture2YT} for a detailed discussion. We thus arrive at
\begin{equation}
\left< P^r\,I(x)^s \right>=D_{r,s}\,\Lambda^{2r+s},
\end{equation}
with $D_{r,s}\in \mathbb{Q}$ a Donaldson invariant, which is an intersection number of $M_k$ with ${\rm vdim}_\mathbb{R}(M_k)=4r+2s$. 

In the following lectures, we will discuss the evaluation of the correlation functions through path integral methods. The path integral receives contributions from the Coulomb branch as well as a $\delta$-function support at the singular points, where some massive states becomes massless. Thus,
\begin{equation}
\left< \cO \right>=\Phi[\cO]+\left< \cO \right>_{\rm SW},
\end{equation}
where $\Phi[\cO]$ is the contribution from the Coulomb branch and 
$\left< \cO \right>_{\rm SW}$ is has delta-function support at the strong-coupling singularities of the $u$-plane. A monopole hypermultiplet becomes massless at these points whose  equation of motion are the Seiberg-Witten equations \cite{Witt94}. These equations will be discussed in Y. Tanaka's lectures. See Chapter \ref{sec:Lecture2YT}. The term $\left< \cO \right>_{\rm SW}$ involves the Seiberg-Witten invariants ${\rm SW}(c)$. 

\section[Lecture 3]{Lecture 3: Evaluation of the $u$-plane integral}
\label{sec:JMLect3}
We review in this lecture the formulation and evaluation of the $u$-plane integral. See Refs \cite{MoWi97, LoNeSha, Marino:1998bm, Marino:1998rg, Malmendier:2008db, Malmendier:2010ss, Korpas:2017qdo, Korpas:2019cwg, Manschot:2021qqe} for a selection of the literature. Ref. \cite{MoWi97} demonstrates that $\Phi[\cO]$ is only non-vanishing for $b_2^+\leq 1$. For $b_2^+=1$, the integral reduces to an integral over zero-modes of the various fields. This follows by combining the topological invariance, and a rescaling of the metric, $g_{\mu\nu}\to t^2 g_{\mu\nu}$, and careful treatment of the non-zero modes \cite{MoWi97}. We restrict in the following to gauge group $G=SU(2)$ or $SO(3)$. See \cite{Marino:1998bm} for the analysis with rank $>1$ gauge groups.

For the twisted $N_f=0$ theory with $G=SU(2)$ on a four-manifold with $b_1=0$, the path integral reduces to the finite dimensional integral,
\begin{equation}
\label{eq:Phia}
\Phi[\cO]=\sum_{U(1)\,\,{\rm fluxes}} \int da\wedge d\bar a\wedge dD\wedge d\eta\wedge d\chi\,A(u)^{\chi} B(u)^{\sigma} e^{-\int_X\cL_0} \cO,
\end{equation}
where $\chi$ in the exponent of $A$ is the Euler number of $X$, $\sigma$ the signature, and $\cL_0$ is the zero-mode Lagrangian,
\begin{equation}
\begin{split}
\cL_0&=\frac{i}{16\pi}\left(\bar \tau F_+\wedge F_++\tau F_-\wedge F_- \right)-\frac{y}{8\pi}D\wedge *D\\
&+\frac{i\sqrt{2}}{16\pi}\frac{d\bar \tau}{d\bar a} \eta\chi\wedge (F_++D),
\end{split}
\end{equation}
with $y={\rm Im}(\tau)$. The $U(1)$ fluxes are quantized as
\begin{equation}
\bfk=\frac{1}{4\pi} F\in L+\bfmu.
\end{equation}
Depending on the observable, contact terms can arise in the low energy theory \cite{LoNeSha, MoWi97}.

The Lagrangian $\cL_0$ depends on the metric through $F_\pm$. Explicit evaluation demonstrates that $\Phi$ is only locally constant as function of the metric, but may jump across walls. It depends on the metric through the period point $J$, that is the unique, normalized self-dual direction in $L$ oriented in the forward lightcone. We denote the integral by $\Phi^J_\bfmu[\cO]$. To evaluate the integral, it is useful to change variables from $a,\bar a$ to $\tau,\bar \tau$. After carrying out the integral over $\eta, \chi$ and $D$, the integral then takes the form
\begin{equation}
\label{eq:Phitau}
\Phi_\bfmu^J[\cO]=\int_{\mathbb{H}/\Gamma^0(4)} d\tau\wedge d\bar \tau\,\nu(\tau)\,\Psi^J_\bfmu(\tau,\bar \tau),
\end{equation}
with $\nu$ the (weakly) holomorphic function
\begin{equation}
\nu(\tau)=\frac{da}{d\tau}\,A^{\chi} B^{\sigma}=-\frac{i}{8} \frac{\vartheta_4(\tau)^{13-b_2}}{\eta(\tau)^9},    
\end{equation}
and the sum over fluxes
\begin{equation}
\Psi^J_\bfmu(\tau,\bar \tau)=\frac{1}{\sqrt{y}}\sum_{\bfk\in L+\bfmu} B(\bfk,J)\,(-1)^{B(\bfk,K)}\,q^{-\bfk_-^2}\bar q^{\bfk_+^2/2},
\end{equation}
with $K$ an integral lift of $w_2(X)$ to $L$, and $\bfk_\pm$ the projections parallel and orthogonal to $J$. The topological couplings $A$ and $B$ can be determined from R symmetry \cite{Witten:1995gf, MoWi97, Shapere:2008zf} or the equivariant partition function \cite{NY, Manschot:2019pog}. It is non-trivial to verify that the integrals \eqref{eq:Phia} and \eqref{eq:Phitau} are well-defined on the domain $\mathbb{H}/\Gamma^0(4)$.

Over the last years, much progress is made for the evaluation using ``mock modular forms" \cite{Malmendier:2008db, Malmendier:2010ss, Korpas:2017qdo, Korpas:2019cwg, Manschot:2021qqe}. 
The literature on mock modular forms is extensive, see for example \cite{Larry, MR2605321, ZwegersThesis}.
Rather then reviewing the theory of mock modular forms, I will illustrate the main features with an example. We specialize to $X=\mathbb{CP}^2$, which is simply connected with $b_2=b_2^+=1$, and the exponentiated point observable $\cO=e^{2pu}$. The sum over fluxes specializes to 
\begin{equation}
\Psi_\mu(\tau,\bar \tau)=\left\{\begin{array}{rr} 0,&\qquad \mu=0,\\ -\frac{i}{2\sqrt{2y}}\overline{\eta(\tau)^3},&\qquad \mu=1/2,\end{array} \right.
\end{equation}
We can evaluate the integral using Stokes' theorem using the following example of a mock modular form,
\begin{equation}
F(\tau)=-\frac{1}{\vartheta_4(\tau)}\sum_{n\in \mathbb{Z}} \frac{(-1)^n\, q^{n^2/2-1/8}}{1-q^{n-1/2}}=2\,q^{3/8}\,(1+3\,q^{1/2}+7\,q+\dots).
\end{equation}
This function does {\it not} transform as a modular form for $\Gamma^0(4)$. However, its non-holomorphic completion $\widehat F$, defined as
\begin{equation}
\widehat F(\tau,\bar \tau)=F(\tau)-\frac{i}{2} \int_{-\bar \tau}^{i\infty} \frac{\eta(w)^3}{\sqrt{-i(w+\tau)}}dw,
\end{equation}
does transform as a modular form of weight $1/2$ for $\Gamma^0(4)$. This is the characteristic property of mock modular forms. 

We then evaluate
\begin{equation}
\Phi_{1/2}[e^{2pu}]=\lim_{Y\to \infty} \int_{(\mathbb{H}/SL_2(\mathbb{Z}))_Y} d\tau\,\nu(\tau)\,\widehat F(\tau,\bar \tau)\,e^{2pu},
\end{equation}
where $(\mathbb{H}/SL_2(\mathbb{Z}))_Y$ is the truncated fundamental domain with ${\rm Im}(\tau)\leq Y$, and a similar cut-off near the cusps at $\tau=0$ and $\tau=2$. This evaluates to
\begin{equation}
\Phi_{1/2}[e^{2pu}]=4\left[ \nu(\tau)\,F(\tau)\,e^{2pu} \right]_{q^0} + \left[ \{\tau\to -1/\tau\}\right]_{q^0} +\left[ \{\tau\to (2\tau-1)/\tau\}\right]_{q^0},
\end{equation}
where the first term on the rhs is from the width 4 cusp for $\tau\to i\infty$, while the other two terms are from the width 1 cusps for $\tau\to 0$ and $\tau\to 2$. The brackets $[\dots]_{q^0}$ means the constant term of the expression in the brackets. This matches with results for Donaldson invariants by Ellingsrud--G\"ottsche \cite{ellingsrud1995wall, Gottsche:1996} and Moore--Witten \cite{MoWi97}.

This approach can also be applied for manifolds with $b_2^+=1$, $b_2^->0$ \cite{Korpas:2017qdo, Korpas:2019cwg}, which demonstrates explicitly the dependence on $J$. For these lectures, we restrict to the wall-crossing dependence,
\begin{equation}
\label{eq:WCTheta}
\begin{split}
\Phi_\bfmu^J[\cO]-\Phi_\bfmu^{J'}[\cO]=& 4\left[ \nu(\tau)\,\Theta_\bfmu^{JJ'}(\tau)\,\cO\right]_{q^0} \\
& +\left[ \{\tau\to -1/\tau\} \right]_{q^0} +\left[ \{\tau\to (2\tau-1)/\tau\}\right]_{q^0},
\end{split}
\end{equation}
with
\begin{equation}
\Theta_\bfmu^{JJ'}(\tau)=\tfrac{1}{2}\sum_{\bfk\in L+\bfmu} \left( {\rm sgn}(B(\bfk,J))-{\rm sgn}(B(\bfk,J'))\right)\,(-1)^{B(\bfk,K)}\,q^{-\bfk^2/2}.
\end{equation}
The wall-crossing of the SW contributions $\langle \mathcal{O} \rangle$ due to the jump of SW invariants,
\begin{equation}
{\rm SW}(c;J^+)-{\rm SW}(c;J^-)=-(-1)^{n(c)},\qquad n(c)=(c^2-2\chi-3\sigma)/8.
\end{equation}
is physically expected to cancel the terms due to strong coupling in Eq. \eqref{eq:WCTheta}. Making use of topological invariance of the theory, the SW contributions can be derived from the $u$-plane integral for $b_2^+>1$ \cite{MoWi97}. The result takes the form
\begin{equation}
\label{eq:UniZSWcont}
Z_{\rm SW, \bfmu}=\sum_{j=1,2} \sum_c {\rm SW}(c)\,\mathcal{A}_j^\chi\,\mathcal{B}_j^\sigma\,\mathcal{E}_j^{c^2}\,\mathcal{F}_{j,\bfmu},
\end{equation}
where $\mathcal{A}_j$, $\mathcal{B}_j$, $\mathcal{E}_j$ and $\mathcal{F}_j$ are universal, that is to say independent of the four-manifold. This provides a physical derivation of Witten's conjecture \cite{Witt94}, and physical understanding of the Kronheimer-Mrowka structure theorem \cite{KrMr}.
\vspace{.5cm}\\
{\bf Remark:} If the four-manifold $X$ is a complex toric surface, path integrals can be evaluated using equivariant localization in the $\Omega$-background associated to the $U(1)\times U(1)$ symmetry \cite{Nekr03}. The $\Omega$-background is characterized by two equivariant parameters $\varepsilon_1$ and $\varepsilon_2$, which were introduced in Refs \cite{Moore:1997dj, Moore:1998et, Lossev:1997bz}, and used in Ref. \cite{Moore:1998et} to regularize integrals over moduli spaces of instantons on $\mathbb{R}^4$. 
For a complex toric surface, such as $\mathbb{CP}^2$, the equivariant integral takes the form of a sum over the torus fixed points in the moduli space of semi-stable vector bundles with the summand expressed as a product of equivariant partition functions \cite{Nekrasov:2003vi, GNY1}. This has been applied to various physical theories \cite{ Bonelli:2020xps, Kim:2025fpz, KPT, Dennemann:2026idg}.

\section[Lecture 4]{Lecture 4: Hypermultiplets versus Euler numbers, Segre numbers and K-theoretic invariants}
\label{sec:Lecture4JM}

The previous lecture discussed the evaluation of the partition function of the ``pure" SU(2) without hypermultiplets, and Donaldson invariants. This fourth lecture discusses theories with hypermultiplets as well as the five-dimensional theory. These evaluate to different invariants of moduli spaces.

\subsection{Hypermultiplets}
\label{sec:Hypers}
We have seen in Lecture 2 that the topologically twisted vector multiplet can be formulated on any smooth, oriented four-manifold. This lecture will consider the hypermultiplets. The ${\rm Spin}'(4)$ representation of the bosonic scalars of the topologically twisted hypermultiplet is $(\bf 1, 2)\oplus (\bf 1, 2)$. Crucially, this involves the spinor representation and which cannot be formulated globally on a non-spin four-manifold. To formulate the twisted hypermultiplets, we couple these to a non-trivial background line bundles $\mathcal{L}_j$ for the flavor group \cite{Hyun:1995mb, Hyun:1995hz, Manschot:2021qqe, Aspman:2022sfj}. The quantization of the corresponding field strengths depends on the flavor representation, a single adjoint ($\cN=2^*$) or $N_f$ fundamental ($\cN=2$ SQCD) hypermultiplets. The condition for the adjoint hypermultiplet is \cite{Manschot:2021qqe}
\begin{equation}
\label{eq:FB}
\left[\frac{F_B}{2\pi}\right]\in w_2(X)+2L.
\end{equation}
while the condition for the fundamental hypermultiplet $j$ is \cite{Aspman:2022sfj}
\begin{equation}
\left[\frac{F_j}{2\pi}\right]+2\bfmu+ w_2(X)\in 2L.
\end{equation}
We denote $\left[\frac{F_B}{2\pi}\right]$ and $\left[\frac{F_j}{2\pi}\right]$ by $c_{\rm uv}$ respectively $c_j$.

A subset of the $\mathcal{Q}$-fixed equations of these theories are \cite{Hyun:1995mb, Labastida:1995zj, Hyun:1995hz, Labastida:1995gp, Labastida:1996tz}\footnote{A closely related set of equations, the “SO(3) monopole equations”, was studied in the mathematical literature \cite{Pidstrigach:1995, Bradlow:1996alg, Teleman:1996, okonek1996recent, Feehan:1997gj}.}
\begin{equation}
\label{eq:NonAbMonOp}
\begin{split} 
&F^{a+}_{\dot \alpha \dot \beta}+\frac{i}{2} \sum_{j=1}^{N_f} \bar M^{(j)}_{(\dot \alpha}T^a M^{(j)}_{\dot \beta)}=0,\\
&{\slashed{D}_j}M^{(j)}=0,
\end{split} 
\end{equation}
with $T^a$ a generator of the Lie algebra of the gauge group in the representation $R$ of the hypermultiplets, and $\slashed D_j$ the Dirac operator acting on $M^{(j)}$. For the adjoint hypermultiplet, the first equation of \eqref{eq:NonAbMonOp} can be written as \cite{Labastida:1998sk}
\begin{equation}
\label{eq:F+Adj}
F^+_{\mu\nu}+\frac{1}{2} (\bar \sigma_{\mu\nu})^{\dot \alpha \dot \beta} \left[ \bar M_{(\dot \alpha}, M_{\dot \beta)} \right]=0,
\end{equation}
where the matrix $\bar \sigma_{\mu\nu}$ transforms between indices $\mu\nu$ and $\dot \alpha\dot \beta$.

We denote the moduli space of solutions to these equations modulo gauge transformations by $M^Q_{k,c_{\rm uv}/c_j}$. The dimensions of the moduli spaces are
\begin{equation}
\begin{split}
&{\rm Adjoint:}\qquad \qquad \,\, {\rm vdim}_{\mathbb{R}}(M^Q_{k,c_{\rm uv}})=\frac{3}{2} (c_{\rm uv}^2-2\chi-3\sigma),\\
&{\rm Fundamental:}\qquad{\rm vdim}_{\mathbb{R}}(M^Q_{k,c_j})=2(4-N_f)k+\\
&\hspace{5.5cm}+\frac{1}{2}\left(-3\chi-(3+N_f)\sigma+\sum_{j=1}^{N_f}c_j^2\right).
\end{split}
\end{equation}
Given an almost complex structure on $X$ with canonical class $K_X$, there is a canonical
Spin$^c$ structure with characteristic class $c_{\rm uv}=-K_X$, $c_{\rm uv}^2=2\chi+3\sigma$. For
such a choice of Spin$^c$ structure, the chiral spin bundles $S^\pm \otimes \mathcal{L}^{1/2}$
correspond to differential forms,
\begin{equation}
W^+=\Lambda^{0,0}\oplus \Lambda^{0,2},\qquad  W^-=\Lambda^{0,1}.
\end{equation}
For this choice of $c_{\rm \mu\nu}$, the monopole fields $M$ being sections of $W^+$ are therefore  equivalent to the $C$ and $B^+$  fields of VW theory \cite{VW}, discussed in Y. Tanaka's and M. Kool's lectures. Note that the VW twist of
$\mathcal{N}=4$ Yang-Mills is distinct from the Donaldson-Witten
twist. The $\bf 4$ representation of the $SU(4)$ R-symmetry transforms
for the VW twist as $({\bf 1}, {\bf 2})\oplus ({\bf 1}, {\bf 2})$ of
$SU(2)_+\times SU(2)_-$,
while for the DW twist the $\bf 4$ tranforms as $({\bf 1}, {\bf 2})\oplus ({\bf 1}, {\bf
  1})\oplus ({\bf 1}, {\bf
  1})$. This results in a different field content in the absence of
background fluxes for the flavor group. 

For a canonical Spin$^c$ structure, $c_{\rm uv}=K_X$, Ref. \cite{Manschot:2021qqe} argues that the non-Abelian monopole equations (\ref{eq:NonAbMonOp}) are a compact deformation of the Vafa-Witten equations. These equations result from a different topological twist, the Vafa-Witten twist \cite{VW}, and are discussed in Lecture 4 by Y. Tanaka, Chapter \ref{sec:Lecture4YT}. As a result, the path integral with the adjoint hypermultiplet is therefore identical to that of VW theory for $m\to 0$. Other choices of $c_{\rm uv}$ correspond mathematically to L-twisted VW theory \cite{TT1, GW}.

The path integral of this theory including the $N_f$ hypermultiplets gives rise to the insertion of the Euler class of the in the integral over $M^Q_{k,c}$ in addition to observables \cite{Cordes:1994fc},
\begin{equation}
\label{eq:<O>MQ}
\left<\cO \right>=\sum_{k} \Lambda_{N_f}^{{\rm vdim}(M^Q_{k,c})}\int_{M^Q_{k,c}} {\rm Eul}({\rm Coker}({\slashed D}))\,\omega_{\cO},
\end{equation}
where ${\rm Coker}({\slashed D})\to \cM^Q_{k,c}$ is the total cokernel bundle for the $N_f$ Dirac operators ${\slashed D}_j$.

Using the flavor symmetry $U(1)^{N_f}$, $M^{(j)}\to e^{i\alpha}M^{(j)}$, integrals over the solution $M^Q_{k,c}$ can be localized to integrals over the branches \cite{VW, Pidstrigach:1995}:
\begin{itemize}
\item the instanton branch $M^{\rm i}_k$: $F^+=0$, $M_{\dot \alpha}^{(j)}=0$,
\item the monopole or Abelian branch $M^{\rm a}_k$: the connection $A_\mu$ is reducible. For the adjoint representation, $M_{\dot \alpha}$ is upper-triangular, while for the fundamental representation one of the two gauge components vanishes.
\end{itemize}
For localization to $M^{\rm i}_k$, the normal directions to $M^{\rm i}_k$ in $M^Q_{k,c}$ are the kernels of the Dirac operators. Hence, Atiyah-Bott-Berline-Vergne localization \cite{AB, BV82} gives rise to integration of the equivariant Euler class of the index bundle $W_k^{\rm i}\to M^{\rm i}_k$. Depending on the representation, the rank of the index bundle is
\begin{equation}
\label{eq:ranksIndex}
\begin{split} 
&{\rm Adjoint}:\qquad \qquad \,\,{\rm rk}(W_k^{\rm i})=-4k+\frac{3}{4}(c_{\rm uv}^2-\sigma),\\
&{\rm Fundamental}:\qquad {\rm rk}(W_k^{\rm i})=-N_f\,k+\frac{1}{4}\sum_{j=1}^{N_f} (c_j^2-\sigma).
\end{split} 
\end{equation}
Note that as a formal difference of ranks of the kernel and co-kernel bundle, the rank of the index bundle is negative for sufficiently large $k$. Localization to the monopole branch is more involved, but also reduces to integrals of Chern class over the Abelian branch. 

The correlation function (\ref{eq:<O>MQ}) reduces thus to integrals over the instanton and monopole branches \cite{VW, Pidstrigach:1995, LoNeSha, Dedushenko:2017tdw, Manschot:2021qqe, Aspman:2022sfj}, which take the following schematic form
\begin{equation}
\label{eq:HyperCorr}
\left<\cO\right>=\sum_k \Lambda_{N_f}^{{\rm vdim}(M^Q_{k,c})} \int_{M^{\rm i}_k \cup M^{\rm a}_k} \sum_l m^{{\rm rk}(-W_k^{\rm i/a})} \,\frac{c_l}{m^l}\,\omega_\cO,
\end{equation}
with $c_l$ Chern classes of the matter bundle over the fixed loci. 

Such correlation functions have been studied in detail over the passed years. We will first review the case of the adjoint representation (or $\cN=2^*$), followed by the fundamental representation (or $\cN=2$ SQCD).

\subsection{$\cN=2^*$} \label{sec:N=2*}
For this representation, the index bundle corresponds to the tangent bundle of the moduli space, such that Eq. \eqref{eq:HyperCorr} shows that partition function for this theory becomes the Euler characteristic of the moduli space. Somewhat schematically we arrive at
\begin{equation}
Z_{\bfmu,c_{\rm uv}}(\tau_{\rm uv})=\sum_k \chi(M_k^{\rm i})\,q_{\rm uv}^k+\sum_k \chi(M_k^{\rm a})\,q_{\rm uv}^k.
\end{equation}
The Euler numbers are actually quite intricate to define formally. See the lectures by M. Kool and Y. Tanaka.

Topological correlators can be determined using the low energy approach as for the $N_f=0$ theory discussed in Lecture 2. The SW curve for $\cN=2^*$ is determined in Ref. \cite{Seiberg:1994aj}. The $u$-plane contains an extra singularity due to the hypermultiplet. The corresponding modular fundamental domain is $\mathbb{H}/\Gamma(2)$ with an additional puncture at $\tau=\tau_{\rm uv}$. The order parameter $u$ is now an example of a bi-modular form \cite{Labastida:1998sk, Huang:2011qx, Bonelli:2019boe},
\begin{equation}
u(\tau,\tau_{\rm uv})=\frac{m^2}{4}\frac{\vartheta_2(\tau)^4\vartheta_3(\tau_{\rm uv})^4e_2(\tau_{\rm uv})-\vartheta_3(\tau)^4\vartheta_2(\tau_{\rm uv})^4 e_1(\tau_{\rm uv} )}{\vartheta_2(\tau)^4\vartheta_3(\tau_{\rm uv})^4-\vartheta_3(\tau)^4\vartheta_2(\tau_{\rm uv})^4},
\end{equation}
where $e_j$ are the half-periods, $e_1=\frac{1}{3}(\vartheta_3^4+\vartheta_4^4)$ and $e_2=-\frac{1}{3}(\vartheta_2^4+\vartheta_3^4)$. This function transforms as a modular form of weight 0, respectively weight 2, for $\Gamma(2)$ transformations of $\tau$, respectively $\tau_{\rm uv}$, while the modular group is the full $\SL_2(\mathbb{Z})$ under simultaneous transformations of $\tau$ and $\tau_{\rm uv}$.

\begin{figure}[ht!]\centering 
	\includegraphics[scale=0.7]{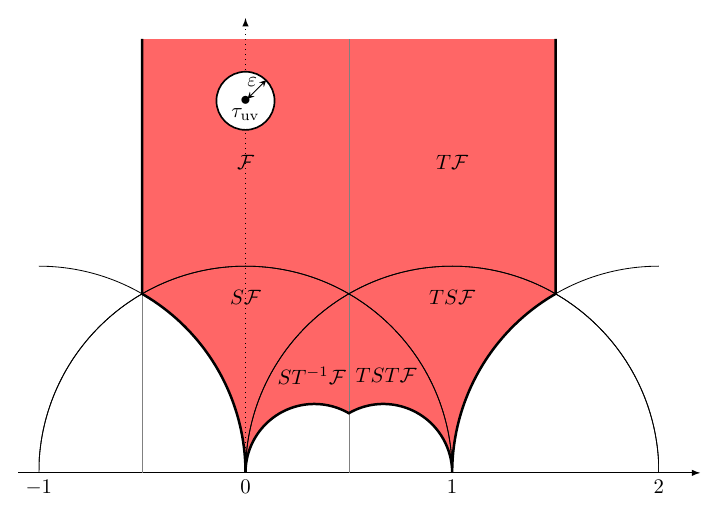}
	\caption{Fundamental domain for $\cN=2^*$. }\label{fig:fundgammaN=2*}
\end{figure}

The fundamental domain is depicted in Figure \ref{fig:fundgammaN=2*}. The contribution of the instanton branch corresponds to the contributions from the strong coupling cusps at $\tau=0,1$ as well as $\tau=\tau_{\rm uv}$. The contribution from the monopole branch is from the cusp at $\tau=i\infty$. 

To consistently couple the theory to the Spin$^c$ structure \eqref{eq:FB}, we weakly gauge the $U(1)_B$ symmetry and introduce a non-dynamical vector multiplet $(A_{B\mu},a_B=m,\chi_B,\eta_B)$ \cite{Manschot:2021qqe}.\footnote{For $c_{\rm uv}=0$, this theory was studied by \cite{Labastida:1998sk}.
} The LEET couplings are determined by the prepotential $\cF$, for example 
\begin{equation}
\frac{\partial^2 \cF}{\partial m^2} F_B^2,\qquad \frac{\partial^2 \cF}{\partial a\partial m} F\,F_B,
\end{equation}
and can be expressed in terms of modular forms.
The correlation functions decompose as
\begin{equation}
\left< \cO \right>_{\bfmu, c_{\rm uv}}=\Phi_{\bfmu,c_{\rm uv}}[\cO]+\left< \cO \right>_{{\rm SW}, \bfmu, c_{\rm uv}}
\end{equation}
and are functions of $\tau_{\rm uv}$. The partition function transforms as a modular form under $S$-duality transformations of $\tau_{\rm uv}$. $Z_{\bfmu,c_{\rm uv}}$ forms a vector-valued representation of $\SL_2(\mathbb{Z})$ as function of $\bfmu \in (L/2)/L$. The transformations under the two generators of $\SL_2(\mathbb{Z})$ are,
\begin{equation}
\begin{split}
&S:\qquad Z_{\bfmu,c_{\rm uv}}(-1/\tau_{\rm uv})=2^{-b_2} e^{\pi i (c_{\rm uv}^2-\sigma)/8}(-i\tau_{\rm uv})^{-\chi/2-4\ell}\\
&\hspace{4cm}\quad \times \sum_{\mathbf{\nu}\in (L/2)/L}e^{4\pi i B(\bfmu,\mathbf{\nu})}\,Z_{\mathbf{\nu},c_{\rm uv}}(\tau_{\rm uv}), \\
&T:\qquad  Z_{\bfmu,c_{\rm uv}}(\tau_{\rm uv}+1)=e^{-2\pi i(\bfmu^2+(\chi+2\ell)/12)}Z_{\bfmu,c_{\rm uv}}(\tau_{\rm uv}),
\end{split}
\end{equation}
with $\ell=(c_{\rm uv}^2-2\chi-3\sigma)/8$. The duality transformations of the partition functions of the SU(2), SO(3)$_+$ and SO(3)$_-$ can be deduced from these. 

For four-manifolds with $b^+_2(X)=1$, the results demonstrate that the correlations involve a continuous metric dependence, while depending on $\bar \tau_{\rm uv}$ at the same time \cite{MoWi97, Manschot:2021qqe}. This phenomenon is known as the {\it holomorphic anomaly} theory, and shows that for reducible connections certain $\mathcal{Q}$-exact terms of the Lagrangian do not decouple from the partition function.

For a canonical choice of the Spin$^c$ structure $c_{\rm uv}$, the partition function exhibits the properties expected for the partition functions of VW theory, ie S-duality of the correlation functions, and agreement with explicit results for $b_2^+=1$ \cite{Klya, Yos1, Yosh2} including the holomorphic anomaly \cite{VW, Manschot:2011dj, Dabholkar:2020fde}. For example, the $SO(3)$ partition function $Z_{\mu,K_{\mathbb{CP}^2}}$, $\mu=0,\tfrac{1}{2}$, for $X=\mathbb{CP}^2$ is proportional to the modular completion of the generating function of Hurwitz class numbers $H(n)$ \cite{Zag},
\begin{equation} \label{eqn:completedZagier}
\begin{split}
&Z_{\mu,K_{\mathbb{CP}^2}}(\tau_{\rm uv},\bar \tau_{\rm uv})=\\
&\qquad \frac{3}{\eta(\tau_{\rm uv})^6}\left(\sum_{n\in \mathbb{N}} H(4n-2\mu) \, q_{\rm uv}^n-\frac{i}{4\sqrt{2}\pi} \int^{i\infty}_{-\bar \tau_{\rm uv}} \frac{\vartheta_{3-2\mu}(2u)}{(-i(u+\tau))^{3/2}}\,du\right),
\end{split}
\end{equation}
with $\vartheta_j$ as introduced in Eq. (\ref{eq:JacThetas}). This is an example of a mock modular form discussed in the previous lecture. See Kool's lectures, Section \ref{sec:Kly}, for the derivation of the holomorphic part of this partition function using the toric geometry of $\mathbb{CP}^2$. 

For four-manifolds with $b_2^+\geq 1$, the universal Seiberg-Witten contributions can be derived using the approach discussed around Eq. \eqref{eq:UniZSWcont}. The universal couplings are universal functions of $\tau_{\rm uv}$, for example, $\mathcal{A}_1=\eta(\tau_{\rm uv})^3/\eta(2\tau_{\rm uv})^2$, and the complete partition function is a weakly holomorphic as function of $\tau_{\rm uv}$. Agreement is established between the universal functions of the $\mathcal{N}=2^*$ theory and those of VW theory \cite{VW, DPS, GK1} for $c_{\rm uv}=K_X$.

Another important aspect of the partition functions is that these reproduce the results of the $N_f=0$ theory in the limit where the hypermultiplet decouples, independent of the choice of $c_{\rm uv}$. This decoupling limit corresponds to $m\to \infty$ with $4m^4q_{\rm uv}=\Lambda_0^4$ held fixed, where $\Lambda_0$ is the scale of the $N_f=0$ theory. The $\mathcal{N}=2^*$ theory thus provides a one-parameter deformation connecting Donaldson theory with Vafa--Witten theory. This clarifies the puzzling aspect of Vafa-Witten theory that the Euler character is not a bordism invariant.

\subsection{$\cN=2$ SQCD} 
\label{sec:SQCD}
Considering hypermultiplets in the fundamental representation provides also interesting results for mathematics and physics. The structure of the correlation function is given by \eqref{eq:HyperCorr} in terms of Chern classes of index bundles over $M^{\rm i}_k$ and $M^{\rm a}_k$, with the rank of the index bundle $W^{\rm i}_k$ given by. To state the mathematical counterpart for the integral over $M^{\rm i}_k$ within algebraic geometry, we let $\alpha\in K(X)$ be a K-theory class for an algebraic surface $X$ with rank $s$, and $\alpha_M\in K(M_k^{\rm i})$ be the image of $\alpha$ under the Donaldson $\mu_D$-map. Then the contributions of the physical partition function for the instanton component correspond to \cite{GK4}
\begin{equation} 
\sum_k z^{{\rm vdim}(M_k)} \int_{[M_k]^{\rm vir}} c(\alpha_{M_k})\,e^{\mu_D(p)+\mu_D(x)},
\end{equation} 
In terms of the physical variables, we have $s=N_f$ and $z=\Lambda_0^2/2m^2$ with $m$ the mass of all hypers, $m_j=m$ for all $N_f$, and $\Lambda_0$ the scale of the $N_f=0$ theory which is related to the scale $\Lambda_{N_f}$ of the theory with $N_f$ hypers as $\Lambda_{0}^4=m^{N_f}\Lambda_{N_f}^{4-N_f}$.

The theory with equal masses gives rise to 3 singularities $u_1^*$, $u_2^*$ as before and a 3rd  singularity $u_3^*$ where $N_f$ monopoles becomes massless. Topological correlators are studied for vanishing masses by \cite{MoWi97, Kanno:1998qj, Malmendier:2008db}. We deduce from Eq. \ref{eq:HyperCorr} that the sum over $l$ then reduces to the contribution from the top Chern class $c_{\rm top}$, while in the limit $m\to \infty$ the dependence on the Chern classes disappears. 

Explicit results are derived for non-vanishing masses by \cite{Aspman:2022sfj, Aspman:2023ate}. The integration domains are more intricate then the ones shown above, typically also including branch points and cuts. The domains are moreover  different for different choices of the mass \cite{Aspman:2021vhs}, such as large or small. Some explicit results are
\begin{itemize}
\item Determination $\left< \cO \right>$ for $X=\mathbb{P}^2$, $N_f=1,2,3$ and vanishing background fluxes $c_j$ using the $u$-plane approach. The contribution from the strong coupling singularities $u_j^*$ vanish in this case. As an explicit example we give for $N_f=2$,
\begin{equation}
\left< u^2 \right>=\frac{19}{2^6} m^4+\frac{23}{2^7}\Lambda_2^2m^2+\frac{53}{2^{18}}\Lambda_2^4.
\end{equation}
The first term is the leading term in the decoupling $m\to \infty$ limit and agrees with the Donaldson invariant for this observable.
 Agreement is found between the calculations with both large and small mass expansions. In more recent work \cite{Dennemann:2026idg}, agreement is also established with the alternative approach of equivariant localization.
\item The $u$-plane integral is also evaluated in some cases for non-vanishing background fluxes $c_j\neq 0$ \cite{Aspman:2023ate}. 
\item For $b_2^+>1$, there are contributions from the singularities $u_j^*$, $j=1,2,3$ if the SW invariants are non-vanishing. The universal functions $\cA_j$, $\cB_j$, $\dots$ are now functions of $\Lambda_{N_f}/m$. The expressions for $\left<\cO\right>$ are Laurent polynomials in the masses if the contributions of {\it all} three singularities are included \cite{Aspman:2023ate, FMM:future}. For example for $X=K3$, the partition function for $N_f=1$ with $\bfmu=0$ and flavor background flux $c_1=0$ is \cite{Aspman:2023ate},
\begin{equation}
\label{eq:Z0K3}
Z_0=Z_0^{\rm i}+Z_0^{\rm m}=\frac{64\,m^2}{\Lambda_1^2},
\end{equation}
with $Z_0^{\rm i}$ the contribution from the instanton branch, singularities $u_1^*$ and $u_2^*$, and $Z_0^{\rm m}$ the singularity of the monopole branch, or singularity $u_3^*$.
\item If the contributions from the instanton branch, $u_{1}^*$ and $u_{2}^*$, are singled out, agreement is found in \cite{Aspman:2022sfj, Furrer:2026byd} with a number of the universal functions for Segre numbers derived in algebraic geometry  \cite{GK4}. For the instanton branch,   all instanton numbers with non-empty $M_k$ contribute for $b_2^+>1$, which suggests that the index bundles over $M_k^{\rm i}$ are sheaves. For the example of Eq. (\ref{eq:Z0K3}), the first few terms of $Z^{\rm i}_0$ are,
\begin{equation}
Z_0^{\rm i}=-\frac{3}{4}\frac{\Lambda_1^4}{m^4}-\frac{5}{16}\frac{\Lambda_1^7}{m^7}-\frac{63}{512}\frac{\Lambda_1^{10}}{m^{10}}+\dots.
\end{equation}
As a result, the contributions from the instanton and abelian component cancel each other for instanton number $k$ larger than a minimum value.

Ref. \cite{Furrer:2026aea} derives universal functions for higher rank gauge groups from the physical perspective.
\end{itemize}

\subsection{K-theoretic invariants and five-dimensional theories}
\label{sec:K-theoretic}
In this last part of the lecture series, I will discuss some results of Ref. \cite{Kim:2025fpz} in collaboration with H. Kim, G. W. Moore, R. Tao and X. Zhang. 

In addition to the previously discussed invariants of instanton moduli spaces, another interesting family of invariants are K-theoretic invariants such as the index of the Dirac operator, Ind$(\slashed{D}_A;M_k)$ and the holomorphic Euler characteristic $\chi_{\rm hol}(M_{k}^{\rm i})$. In the following we will consider generating functions of these invariants,
\begin{equation}
\label{eq:ZIndD}
    Z_{\bfmu}(\ccR)=\sum_k {\rm Ind}(\slashed{D}_A;M_k)\,\ccR^{4k},
\end{equation}
and their interpretation as the partition function of five-dimensional $\cN=1$ supersymmetric Yang-Mills on $X\times S^1$.

Five-dimensional Yang-Mills theories with $\cN=1$ supersymmetry are well studied since the mid-90's \cite{Seiberg:1996bd, Douglas:1996xp, Intriligator:1997pq}. The five-dimensional vector multiplet has bosonic fields $(A_m, \sigma)$ with $m=1,\dots,5$ and $\sigma$ a real scalar field. We choose gauge group $G=SU(2)$ or $SO(3)$ as before. The theory has a dimensionfull gauge coupling $g_{\rm 5d}$, with $g_{\rm 5d}^2$ having dimension of length. The theories are thus non-renormalizable, but can be realized as a deformation of a strong coupling RG fixed point. The fixed point for the $SU(2)$ theory without hypermultiplets is the $E_1$ theory. The theory has global symmetry $SU(2)_R\times U(1)^{(I)}$, where $SU(2)_R$ is the $R$-symmetry group and $U(1)^{(I)}$ is ``instanton" symmetry with current $j^{(I)}=*{\rm Tr}_{\bf 2}[F\wedge F]$. As before for the flavor symmetry of the hypermultiplets, we weakly gauge the $U(1)^{(I)}$ symmetry by introducing a non-dynamical vector multiplet with bosonic fields $(A_m^{(I)},\sigma^{(I)})$, with flux
\begin{equation}
\label{eq:nI}
\bfn_I=\left[ \frac{F^{(I)}}{2\pi} \right].
\end{equation}

An important aspect of five-dimensional theories is that the Lagrangian can consider Chern-Simons terms \cite{Tachikawa:2004ur}. We are interested in the mixed CS-term. The supersymmetrized mixed-CS term takes the form, 
\begin{equation}
S_{\rm mixed CS}=\frac{i}{8\pi^2}\int_{\mathbb{R}^5} F^{(I)}\wedge {\rm Tr}_{\bf 2}\left[AdA+\frac{2}{3}A^3\right]+\frac{1}{8\pi^2} \sigma^{(I)} {\rm Tr}_{\bf 2}[F\wedge *F]+\dots,
\end{equation}
where the dots indicate further terms. If we set $\sigma^{(I)}=\frac{8\pi^2}{g_{5d}^2}$, we reproduce the standard kinetic term of the gauge fields. As a result, it suffices for us to consider $S_{\rm mixed CS}$.

We consider the theory on $X\times S^1$, with radius $R$ for $S^1$. The spectrum of the theory includes a tower of Kaluza-Klein states, arising from momenta around $S^1$. The complex scalar field $\phi$ of the four-dimensional theory is related to the 5d fields by
\begin{equation}
\phi=\sigma+\frac{i}{R}\oint_{S^1} A_m\, dx^m.
\end{equation}
We introduce the dimensionless parameter $\ccR$ by
\begin{equation}
\ccR^4=e^{-\frac{8\pi^2 R}{g_{5 \rm d}^2}+i\oint_{S_1}A^{(I)}}=R^4\Lambda^4,
\end{equation}
which appeared in Eq. \eqref{eq:ZIndD}.

We carry out a partial topological twist on $X$ of this theory, such that its ``topological observables" are formally independent of the metric on $X$ but do depend on the metric of $S^1$ in particular the radius $R$ and $\ccR$. There are two approaches to evaluate topological correlators, either through reduction of the degrees of freedom to $S^1$, or reduction to $X$.

To reduce the degrees of freedom to $S^1$, we expand the fields in collective coordinates $A(x)\mapsto A(x)+\delta_n A$, which satisfy the anti-self-duality equations and similarly for the fermions. As a result, one obtains a supersymmetric quantum-mechanical theory with target space $M_k$. The path integral of this theory evaluates to the Witten index, 
\begin{equation}
{\rm Tr}_{\cH} (-1)^F\,e^{-\beta H}={\rm Ind}(\slashed{D}),
\end{equation}
{\bf NB} Note that $\slashed{D}$ and hence ${\rm Ind}(\slashed{D})$ are not well-defined if $M_k$ is not a spin manifold! In particular, the sign of Pfaff($\slashed{D}$) is not well-defined for non-spin manifolds. Ref. \cite{Kim:2025fpz} conjectures that:
\begin{itemize}
\item $M_k$ is a spin-manifold if 
\begin{equation}
B(w_2(P),w_2(X))=0\mod 2.
\end{equation}
\item the sign ambiguity of Pfaff($\slashed{D}$) is resolved by choosing the flux $\bfn_I$ \eqref{eq:nI} such that 
\begin{equation}
\label{eq:CondNonSpin} 
B(w_2(P),w_2(X)+\bfn_I)=0\mod 2,
\end{equation}
and that in this case, the Dirac operator $\slashed D$ is coupled to a Spin$^c$ line bundle $\cL_I \to M_k$ such that $\slashed D$ is well-defined. 
\end{itemize}
These aspects are proven by D. S. Freed, M. J. Hopkins, G. W. Moore \cite{Freed:toappear}.

\begin{figure}[ht!]\centering 
	\includegraphics[scale=1]{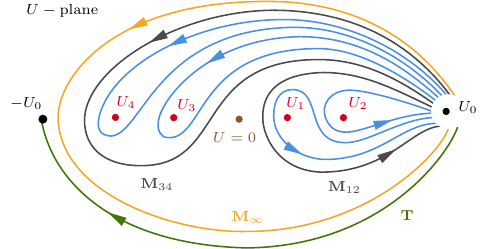}
	\caption{Schematic representation of the $U$-plane with singularities $U_j$ and monodromies starting from the base point $U_0$. $\mathbf{T}$ corresponds to the action of the $\mathbb{Z}_2$ one-form symmetry (\ref{eq:Toneform}).}\label{fig:Uplane}
\end{figure}

\begin{figure}[ht!]\centering 
	\includegraphics[scale=.35]{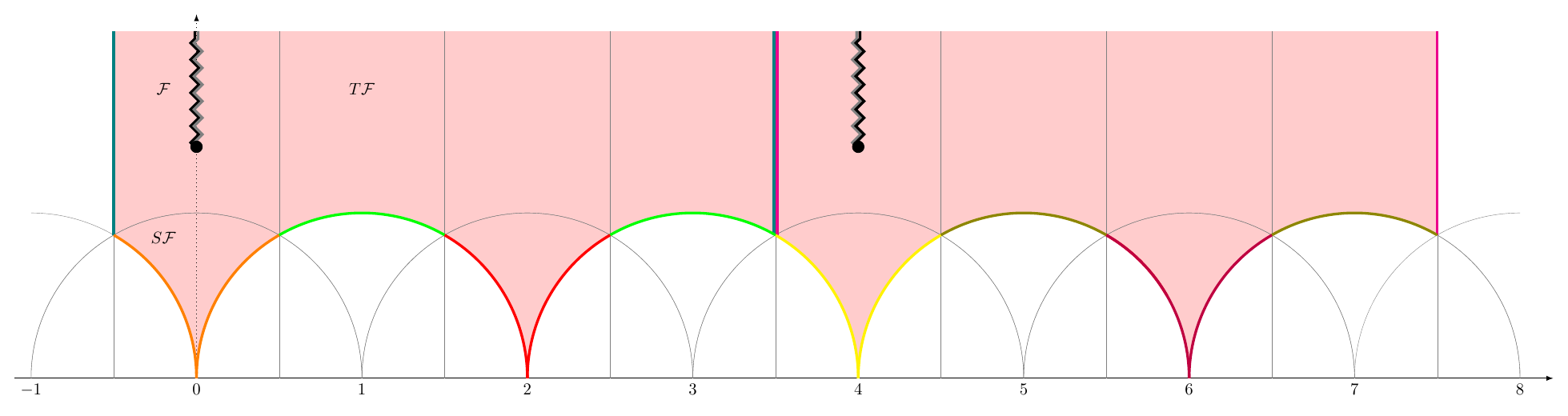}
	\caption{Fundamental domain for the effective coupling $\tau$, with the black dots being the branch points of the double cover. }\label{fig:FRCG}
\end{figure}

We continue by discussing the five- and four-dimensional perspective. The order parameter on the Coulomb branch for the five-dimensional theory on $\mathbb{R}^4\times S^1$ is the Wilson line,
\begin{equation}
\left< W_F \right>=\left<{\rm Tr}_F {\rm P} \exp\left(\oint_{S_1} (\sigma +i A_5)dx_5\right) \right>,
\end{equation}
where $F$ indicates the representation of the gauge group and ${\rm P}$ path ordering of the exponential. This observable is traditionally denoted by $U$, and the leading terms in the $\ccR$ expansion are,
\begin{equation}
U=e^{Ra}+e^{-Ra}+O(\ccR^4).
\end{equation}
The LEET of this theory is solved in terms of the prepotential $\cF(a,\ccR)$ which gives the effective coupling $\tau$. The $U$-plane with the singularities is displayed in Figure \ref{fig:Uplane}. 
$U$ is given in terms of modular forms as
\begin{equation}
U(\tau)^2=-8\ccR^2 u(\tau)/\Lambda^2+4\ccR^4+4,
\end{equation}
with $u$ as in Eq. (\ref{eq:utau}). The fundamental domain in $\tau$-space is a branched double cover of $\mathbb{H}/\Gamma^0(4)$, sketched in Figure \ref{fig:FRCG}.

As suggested by these figures, the theory includes an $\mathbb{Z}_2$ symmetry $\bf T$. This is the electric $\mathbb{Z}_2$ one-form symmetry (or center symmetry) multiplies the holonomy of $A_\mu$ by a $Z(G)$-valued character of the fundamental group of space-time. For $G=SU(2)$, this symmetry acts on $U$ and $a$ as
\begin{equation}
\label{eq:Toneform}
{\mathbf T}:\quad U\mapsto -U, \qquad a\mapsto a+\frac{\pi i}{R}.
\end{equation}
There is a potential anomaly for this symmetry if $w_2(P)\neq 0$ \cite{BenettiGenolini:2020doj, Apruzzi:2021nmk}. Namely the path integral changes by a sign,
\begin{equation}
\label{eq:TZ}
{\bf T}:\qquad Z_{\bfmu,\bfn_I}\mapsto (-1)^{B(2\bfmu,w_2(X)+\bfn_I)}\,Z_{\bfmu,\bfn_I},
\end{equation}
such that the theory is anomalous if the condition in Eq. (\ref{eq:CondNonSpin}) is not satisfied. 

We formulate the $U$-plane integral $\Phi_{\bfmu,\bfn_I}(\ccR)$ for this theory. This reads 
\begin{equation}
    \Phi_{\bfmu,\bfn}(\ccR)=\frac{1}{2} \int_{\cF_\ccR} d\tau\wedge d\bar
\tau\,\nu_\ccR(\tau)\,C^{\bfn^2}\,\Psi_\bfmu^J(\tau,\bar
\tau,v\bfn,\bar v\bfn),
\end{equation}
where
\begin{equation}
\nu_\ccR=\frac{\vartheta_4(\tau)^{13-b_2}}{\eta(\tau)^9}\frac{1}{U},\qquad C=\frac{\vartheta_4(\tau,v)}{\vartheta_4(\tau)}, \qquad
   v=-\frac{1}{2\pi i}\partial_a\partial_{\log(\ccR)}\cF.
\end{equation}

Upon evaluation the integral takes the form
\begin{equation}
\Phi_{\bfmu,\bfn_I}(\ccR)=2(1+(-1)^{B(2\bfmu,w_2(X)+\bfn_I)})\,{\rm Coeff}_{q^0}{\rm Ser}_{\ccR}\left[\nu_\ccR(\tau)\,C^{\bfn_I^2} G_{\bfmu,\bfn_I}(\tau,v) \right].
\end{equation}
Thus the path integral in fact vanishes if the condition (\ref{eq:CondNonSpin}) is not satisfied, and thus satisfies \eqref{eq:TZ} trivially. This is similar to the $\pi_4(SU(2))\simeq \mathbb{Z}_2$ anomaly of Ref.  \cite{Witten:1982fp}. 

For the evaluation, we make a small $\ccR$ expansion and then determine the
$q^0$ term. For example for $X$ with $b_2(X)=1$, for example $\mathbb{P}^2$, we obtain
$$
\Phi_{1/2,n}(\ccR)=\left\{\begin{array}{ll} 
1+O(\ccR^{13}),& n=\pm 1, \\
                               1+\ccR^4+\ccR^8+\ccR^{12}+\dots , & n=\pm 3,\\ 
1+6\,\ccR^{4}+21\,\ccR^{8}+56\,\ccR^{12}+\dots, & n=\pm 5, \\
1+21\,\ccR^{4}+210\,\ccR^{8}+1401\,\ccR^{12}+\dots, & n=\pm 7,\\ 
                               1+55\,\ccR^{4}+1310\,\ccR^{8}+19432\,\ccR^{12}\dots, & n=\pm 9.
                                 \\  \end{array}\right.
                             $$
and 0 for even $n$. In agreement with results by G\"ottsche, Nakajima, Yoshioka \cite{GoNaYo2} for $\mathbb{P}^2$.                             
Similarly,  this approach gives a physical derivation of the universal functions determined by G\"ottsche, Kool, Williams \cite{Gottsche:2019vbi} for algebraic surfaces. 

Together with the results for the other discussed theories, the remarkable agreement of the physical approach using path integrals with the formal mathematical approach stimulates further research on both sides.


\chapter{From the Donaldson invariants to the Vafa-Witten theory}
\label{chap:fromDtoVW}
\vspace{-.4cm}
\begin{center}
\it{Yuuji Tanaka}
\end{center}
\vspace{.4cm}

The aim of these lectures is to give an introduction to the differential and algebro-geometric approaches to gauge-theoretic invariants of four-manifolds, from Donaldson theory to Vafa--Witten theory. We begin with moduli spaces of anti-self-dual instantons and their Uhlenbeck compactifications, leading to the definition of Donaldson invariants. We then explain, via the Donaldson--Uhlenbeck--Yau theorem, how instanton moduli spaces on K\"ahler surfaces are related to moduli spaces of semistable sheaves. Finally, we introduce Vafa--Witten theory and its algebro-geometric formulation in terms of moduli spaces of semistable Higgs sheaves and perfect obstruction theories. 

\section[Lecture 1]{Lecture 1: Anti-self-dual instanton moduli spaces}
\label{sec:Lecture1YT}

In this lecture, we study the moduli space of anti-self-dual (ASD) instantons (connections) on principal $G$-bundles over a closed, oriented, smooth four-manifold. 
The moduli spaces lie at the heart of the Donaldson theory. 
Standard references for the subject are \cite{FrUh} and \cite{DoKr}. 

\subsection{Anti-self dual instantons, gauge transformations, and the moduli spaces}

Let $X$ be a closed, oriented, smooth 4-manifold, and let $P \to X$ be a principal $G$-bundle over $X$, where $G$ is a compact Lie group. 
In most applications, we take $G=SU(2)$ or $SO(3)$; these are the cases most relevant for the Donaldson theory and also the cases in which the transversality results described below are classically formulated.
We denote by $\text{ad} (P) := P \times_{ad} \mathfrak{g}$ the adjoint bundle of $P$, where $\mathfrak{g}$ is the Lie algebra of $G$.

Fix a Riemannian metric $g$ on $X$. This defines the Hodge star operator $*_g : \Lambda^p_X \to \Lambda^{4-p}_X$ for $p= 0, \dots, 4$, where $\Lambda^{p}_X := \Lambda^p T^* X$. For $p=2$, this map is $\Lambda^{2}_X \to \Lambda^2_X$ and it is easy to check $(*_g)^2 = 1$. 
From this, $\Lambda^2_X$ decomposes into self-dual and anti-self-dual parts:
\[
\Lambda^2_X = \Lambda^+ \oplus \Lambda^- ,
\]
where $\Lambda^\pm $ are vector bundles of rank 3 over $X$, whose fibres at $x \in X$ are the $\pm1$-eigenspaces of the operator $*_g|_{x}$, respectively.

\begin{definition}[Anti-self-dual (ASD) connections]
Let $A$ be a connection on $P$ with curvature $F_A$. We say $A$ is \emph{anti-self-dual (ASD) connection}, or \emph{anti-self-dual (ASD) instanton}, if
\[
F_A^+ := \pi^+(F_A) = 0,
\]
where $F_A$ is the curvature of $A$ and $\pi^+$ is the projection onto $\Gamma (\Lambda^+  \otimes \text{ad}(P))$. 
\end{definition}

\paragraph*{Gauge group action.}  
Let $u: P \to G$ be an equivariant map, namely, $u$ satisfies  
\[
u(pg) = g^{-1}u(p)g, \quad \forall p \in P, \forall g \in G .
\]
We call such a $u$ \textit{gauge transformation}. 
It corresponds to a fibre-preserving bundle automorphism $\varphi: P \to P$ satisfying $\varphi(pg) = \varphi(p)g$ for $p \in P$ and $g \in G$.

Denote by $\mathcal{A}_P$ the space of all connections on $P$, and by $\mathcal{G}_P$ that of all gauge transformations. We call $\mathcal{G}_P$ 
the {\it gauge group}.

The gauge group $\mathcal{G}_P$ acts on $\mathcal{A}_P$ by 
$A + (d \varphi) \varphi^{-1}$ 
for $A \in \mathcal{A}_P$ and $u \in \mathcal{G}_P$. 
Locally, on an open subset $U \subset X$, the gauge transformation on a connection $A$ is written as 
$$ \varphi^{-1}_{U} A_{U} \varphi_{U} 
+ \varphi_{U}^{-1} d \varphi_{U},$$
whre $A_{U} := A|_{U}, A \in \mathcal{A}_P$, 
$\varphi_{U} := \varphi|_{U}, \varphi \in \mathcal{G}_{P}$.

\subsection{Instanton moduli spaces}

The {\it instanton moduli space} is defined as:
\[
\mathcal{M}_X(P,g) := \{ A \in \mathcal{A}_P \mid F_A^+ = 0 \} / \mathcal{G}_P.
\]

At this stage, this is merely a set, but one can equip a manifold structure on it, namely,  
$\mathcal{M}_X (P , g)$ becomes an oriented smooth manifold of expected dimension for a generic choice of Riemannian metrics $g$ on $X$, as we see below.

\subsection{Sobolev spaces and elliptic theory}

To define smooth structures on the moduli space, one has to work with suitable Sobolev completions $L^2_k$ of the space of connections $\mathcal{A}_{P}$ and gauge transformations $\mathcal{G}_P$.

Let $L^2_k(X, E)$ denote the Sobolev completion of smooth sections of a vector bundle $E$ over $X$ with respect to the Sobolev $L^2_k$-norm.

We define:   
\[
\mathcal{A}_P^k := \{ A \in \mathcal{A}_P \mid A \text{ is } L^2_k \},
\quad \mathcal{G}_P^{k+1} := \{ u \in \mathcal{G}_P \mid u \text{ is } L^2_{k+1} \}.
\]

If $k> 2$, we have $L^2_{k} \to C^{0}$ from the Sobolev embedding theorem. 
Thus, $\varphi \in L^2_k \, (k >2)$ is continuous. 
Hence, we work on $L^2_k \, (k > 2)$ gauge transformations, namely, $L^2_k$-bundles $(k >2)$ which are topologically well-defined bundles. 

Then the completed moduli space is then formed as
\[
\mathcal{M}_X(P,g) := \{ A \in \mathcal{A}_P^k \mid F_A^+ = 0 \} / \mathcal{G}_P^{k+1}.
\]
In practice, one can prove that the resulting smooth structure is independent of the Sobolev indices, 
once $k$ is takent to be sufficiently large.

\subsection{Atiyah--Hitchin--Singer complex and the Kuranishi model}

The infinitesimal deformation theory of ASD connections is governed by the following elliptic complex:
\[
0 \to L^2_3 (X,  \Lambda^0_X \otimes \text{ad}(P)) \xrightarrow{d_A} 
 L^2_2 (X,  \Lambda^1_X \otimes \text{ad}(P))
\xrightarrow{d^+_A} 
 L^2_1 (X,  \Lambda^+ \otimes \text{ad}(P))\to 0,
\]
where $A$ is an ASD connection, $d^+_A = \pi_+ \circ d_A$ is the composition of exterior covariant derivative $d_A$ and the projection $\pi$ onto self-dual 2-forms.

Denote the cohomology of the above complex by  $\mathbb{H}^0_A := \ker d_A$, $\mathbb{H}^1_A := \ker d^+_A / \text{Im}(d_A)$, and $\mathbb{H}^2_A := \text{coker}(d^+_A)$.

The following gives a criteria when we have a smooth structure on the moduli space $\mathcal{M}_{X}(P,g)$. 
\begin{theorem}[Atiyah--Hitchin--Singer \cite{AHS}] 
If $\mathbb{H}^0_A = 0$ and $\mathbb{H}^2_A = 0$, then the moduli space $\mathcal{M}_{p}(X,g)$ is smooth around $[A]$ of the expected dimension, and the Zariski tangent space is given by $\mathbb{H}^1_A$.
\end{theorem}

Note that $\mathbb{H}^0_A$ is the Lie algebra of the gauge group, and $\mathbb{H}^2_A$ is the {\it obstruction space} to deform the ASD connection $A$. 

The index of the complex can be calculated via the Atiyah--Singer index theorem, and it is:
\[
\text{index} = - 8 c_2(E). [X] + 3(1 - b^1 + b^+)
\]
for $G=SU(2)$, where $E$ is the associated $SU(2)$ vector bundle over $X$, $c_2 (E)$ is the second Chern class of $E$, and $[X]$ is the fundamental class of $X$. For the $G=SO(3)$ case, we have
\[
\text{index} = 2 p_1(E) . [X] + 3(1 - b^1 + b^+) , 
\]
where $E$ is the associated $SO(3)$ vector bundle over $X$ and $p_1 (E)$ is the first Pontryagin class of $E$. 
The number $- \text{index}$ is called the {\it expected} or {\it virtual} dimension of the moduli space.   

We introduce the notion of {\it irreducible} and {\it reducible} connections.
For $A \in \mathcal{A}_{P}$, define the stabiliser group at $A$ by 
$$ \Gamma_{A} := \{ \varphi \in \mathcal{G}_{P} \, | \, \varphi (A) = A  \} .$$

\begin{definition}
    A connection $A$ is said to be {\it irreducible}, if the $\Gamma_{A}$ coincides with the centre of $G$, and {\it reducible} otherwise. 
\end{definition}

Denote by $\mathcal{A}_{P}^* \subset \mathcal{A}_P^k$ the space of all irreducible connections, and by $\mathcal{M}_{X}^* (P, g)$ the moduli space of irreducible ASD connections. 

As in the theory of deformation of complex structures on complex manifolds, we have the following Kuranishi type theorem: 

\begin{theorem}[Atiyah--Hitchin--Singer \cite{AHS}]
Let $[A] \in \mathcal{M}_{X}^{*} (P,g)$. Then there exists a neigbourhood $U$ of $0$ in $\mathbb{H}^1_{A}$ such that there exists a real analytic map 
$ \kappa_{A} : U \to \mathbb{H}^2_{A} $ 
with $\kappa_{A} (0) =0$ and the first derivative vanishing, and the moduli space $\mathcal{M}_{X}^{*} (P,g)$ around $[A]$ is locally modelled on the zero set of $\kappa_{A}$. 
\end{theorem}

Note that for a reducible ASD connection $A$,  $\mathbb{H}^1_A$ and $\mathbb{H}^2_A$ are $\Gamma_A$-equivariant, and the moduli space $\mathcal{M}_{P} (X,g)$ is locally modelled on $\kappa^{-1}(0) / \Gamma_A$ around $[A]$.  

\subsection{The Freed--Uhlenbeck theorem}

We are now interested in when the condition $\mathbb{H}^0_A = 0$ and $\mathbb{H}^2_A = 0$ can be achieved. 
The basic idea is that,
after varying the Riemannian metric $g$ generically, the instanton moduli spaces become smooth.

For $\mathbb{H}^2_A =0$, we achieved it by choosing a generic Riemannian metric $g$ on $X$. 
To sate it, we denote by $\mathcal{C} = \text{Riem}(X)/\text{Conf}(X)$ be the space of all conformal structures on $X$, where $\text{Riem}(X)$ denotes the space of all Riemannian metrics on $X$, and $\text{Conf}(X)$ the space of all conformal transformations on $X$. 

\begin{theorem}[Freed--Uhlenbeck \cite{FrUh}]
Let $G = SU(2)$ or $SO(3)$. Assume that $\pi_1 (X)$ is trivial. Then there exists an open dense subset $\mathcal{D} \subset \mathcal{C}$ such that for all $[g] \in \mathcal{D}$ and all $[A] \in \mathcal{M}_X (P, g)$, we have $\mathbb{H}^2_A = 0$. 
\end{theorem}

Thus, for generic metrics, the obstruction space vanishes.

\begin{remark}
There are other ways to achieve the transversality, see e.g. \cite{Dona83}, \cite{Kron}.
\end{remark}

For the vanishing of $\mathbb{H}^0_A$, we have the following: 
\begin{theorem}[Freed--Uhlenbeck \cite{FrUh}]
Let $G = SU(2)$ or $SO(3)$. Assume that $\pi_1 (X)$ is trivial. Then, if $b^+ >0$, there exists an open dense subset $\mathcal{D} \subset \mathcal{C}$ such that for all $[g] \in \mathcal{D}$ and all $[A] \in \mathcal{M}_X(P, g)$, we have $\mathbb{H}^0_A = 0$. 
\end{theorem}

\subsection{Orientations}

For a finite-dimensional smooth manifold $M$ of real dimension $n$, $M$ is orientable if and only if the determinant line bundle  $\Lambda^n TM$ of the tangent bundle $TX$ is trivial, and an orientation is given by a choice of trivialisations.

For the instanton moduli space $\mathcal{M}_X (P,g)$, we consider the linearised operator of the ASD instanton equations:
$$ \delta_{A} := (d_A^{*} , d_{A}^+) :
L^2_2 ( \Lambda^1 \otimes \text{ad} \, (P) ) \to L^2_3 ( \Lambda^0 \otimes \text{ad} \, (P) ) \oplus L^2_1 ( \Lambda^+ \otimes \text{ad} \, (P) )  $$
for $[A] \in \mathcal{A}_{P}$. This is a family of elliptic operators. 
We then consider
\begin{equation*}
    \begin{split}
    \mathcal{L}_{A} &:= \det ( \text{ind}\, \delta_A ) \\
    &= \det ( \text{ker}\, \delta_A) \otimes \det ( \text{coker}\, \delta_A )^{*}. 
    \end{split}
\end{equation*}
This defines a line bundle $\mathcal{L}$ over $\mathcal{A}_P$, and it in fact descends to $\mathcal{B}_{P}$. 

If $\mathbb{H}^0_A = \mathbb{H}^2_A =0$ for all $[A] \in \mathcal{M}_{X} (P,g)$, $\mathcal{L}|_{\mathcal{M}_{X}(P,g)}$ is the determinant line bundle of the tangent bundle of $\mathcal{M}_{X}(P,g)$. 

\begin{theorem}[Donaldson \cite{Dona87}, Donaldson--Kronheimer \cite{DoKr}]
Let $X$ be a compact, oriented, smooth 4-manifold, and let $P \to X$ be a principal $G$-bundle over $X$. Assume either
\begin{enumerate}
    \item[$(1)$] $G$ is $U(n)$ or $SU(n)$. 
    \item[$(2)$] $\pi_1 (X) = \{ 1 \}$
 and $G$ is a simply-connected simple Lie group. 
\end{enumerate}
Then, 
\begin{itemize}
\item $\mathcal{L} \to \mathcal{B}_P$ is trivial, hence, the smooth part of the moduli space $\mathcal{M}_X (P, g)$ is orientable. 
\item An orientation can be given by choosing an orientation on $H^1 (X) \oplus \mathbb{H}^{+} (X)$. 
\end{itemize}
\end{theorem}

There is a refinement by Joyce, Upmeier, and fourth-named author: 

\begin{theorem}[\cite{JTU}]
Let $X$ be a compact, oriented, smooth 4-manifold, and let $P \to X$ be a principal $G$-bundle over $X$. 
\begin{itemize}
    \item Choose an orientation on $H^1 (X) \oplus \mathbb{H}^{+} (X)$ and a $Spin^c$-structure on $X$. Then one can construct an orientation on the smooth part of the moduli space $\mathcal{M}_X (P, g)$. 
    \item If $G$ is simply-connected, or $G=U(n)$, then the orientation on the moduli space $\mathcal{M}_X (P, g)$ in the above is independent of the choice of $Spin^c$-structure. 
\end{itemize}
\end{theorem}

\subsection{Example: charge 1 ASD instanton moduli space on $S^4$.}

We describe an example of the moduli space $\mathcal{M}_P(X,g)$ for the case $X= S^4$ with the standard Riemannian metric $g_0$, $G=SU(2)$ and $c_2(E) =1$, where $E$ is the associated vector bundle. (See e.g. \cite{Atiy} for more details.)  

We use quaternionic coordinates $x = x^1 + x^2 i + x^3 j + x^4 k \in \mathbb{H}$, where $x^1, \dots ,x^4 \in \mathbb{R}$, and $i^2 = j^2 = k^2 = ijk = -1$. 
We define $\text{Im}\,  x := x^2 i + x^3 j + x^4 k , \, \bar{x} := x^1 - x^2 i - x^3 j - x^4 k, \, |x|^2 := x \bar{x}$.

With the above notations, 
\begin{equation*}
\begin{split}
d \bar{x} \wedge d x &= 2 ( (dx^1 \wedge  dx^2 - dx^3 \wedge dx^4) i
+  (dx^1 \wedge  dx^3 - dx^4 \wedge dx^2) j \\
& \qquad \qquad \qquad +  (dx^1 \wedge  dx^4 - dx^2 \wedge dx^3) k) \\ 
\end{split}
\end{equation*}
is an $SU(2) ( \cong \text{Im} \, \mathbb{H})$-valued anti-self-dual two-form, and similarly, 
\begin{equation*}
\begin{split}
d {x} \wedge d \bar{x} &= - 2 ( (dx^1 \wedge  dx^2 + dx^3 \wedge dx^4) i
+  (dx^1 \wedge  dx^3 + dx^4 \wedge dx^2) j \\
& \qquad \qquad \qquad +  (dx^1 \wedge  dx^4 + dx^2 \wedge dx^3) k) \\ 
\end{split}
\end{equation*}
is an  $SU(2)$-valued self-dual two-form.

Consider the following $SU(2)$-connection on $\mathbb{R}^4$:   
$$ 
A(x) := \text{Im}\left( \frac{\bar{x} d x}{(1 + |x|^2} \right).$$ 
Then, the curvature becomes 
$$ F_{A} = \frac{d \bar{x} \wedge d x }{( 1 + |x|^2 )^2} .$$
Thus, this connection $A$ is anti-self-dual.

We then consider the scale transformation $x \mapsto \lambda x$. Then we have
$$ A_{\lambda} := \lambda^* A = \frac{\lambda^2 \, \text{Im} \, (\bar{x} \wedge dx)}{1 + \lambda^2 |x|^2}, $$
and 
$$ F_{A_{\lambda}} =  \frac{\lambda^2  d \bar{x} \wedge dx)}{(1 + \lambda^2 |x|^2)^2}. $$
Note that, when $\lambda \to \infty$, the curvature density function  $|F_{A_{\lambda}}|^2 = \frac{\lambda^4}{( 1 + \lambda^2 |x|^2 )^4}$ becomes $\delta$-function at $x=0$. This phenomenon is called {\it bubbling}. 

We combine the translation to the above scale transformation, namely, for $x_0 \in \mathbb{H}$ and $\lambda >0$ we consider the following connection:
$$ A_{(x_0 , \lambda)} (x)  = \frac{\lambda^2 \, \text{Im} \, (\overline{(x - x_0)} \wedge dx)}{1 + \lambda^2 |x -x_0|^2}. $$
This $ A_{(x_0 , \lambda)}$ is also an anti-self-dual instanton. 
We call $\lambda$ the {\it strength} and $x_0$ the {\it centre} of the connection $ A_{(x_0 , \lambda)}$.  
Note that the scaling gives a 1-parameter family, and the translations give four more parameters.

Define the quaternionic projective space by 
\[
\mathbb{HP}^1 := (\mathbb{H}^2 \setminus \{0\})/\mathbb{H}^* \cong S^4.
\]
Consider 
$$ S^7 := \{ (x,y) \in \mathbb{H}^2 \setminus \{0 \} \, | \, |x|^2 + |y|^2 = 1 \} . $$
This can be thought of as 
a principal $SU(2) = Sp(1) := \{ x \in \mathbb{H} \, | \, |x| =1 \} \cong S^3$-bundle over $\mathbb{H} \mathbb{P}^1 = S^4$. 
(Note that this is the quaternionic analogue of the Hopf fibration $S^3 \to S^2$.)
Denote by $E$ the associate $SU(2)$-vaector bundle over $S^4$. Then one can check that $c_2(E) =1$. 

In fact, the previously defined instanton $A_{(x_0,\lambda)}$ can be extended to a connection on this bundle by the removal singularity theorem by Uhlenbeck \cite{Uhle1}.
Moreover, the following holds:

\begin{theorem}[Atiyah--Hitchin--Singer \cite{AHS}]
All ASD connections on the $SU(2)$-vector bundle $E \to S^4$ with $c_2 (E)=1$ are given by the extension of $A_{(x_0, \lambda)}$'s. 
\end{theorem}

Here, the number of the parameters is 
$$ \text{the centre (4 parameters)} + \text{the scaling (1 parameter)} = 5 .
$$
This coincides with the expected dimension $8 c_2(E) - 3( 1 -b^1 +b^+)= 8 -3 =5$. 
In fact, the instanton moduli space $\mathcal{M}_{S^4}(E, g_0)$ is identified with the five-dimensional open ball $B^5$. 
Note that the boundary of $B^5$ is the $S^4$ itself.

\subsection*{Exercises}

\begin{enumerate}
  \item Prove that the instanton $A(x)$ on $\mathbb{R}^4$ extends smoothly to a connection on the principal $SU(2)$ bundle over $S^4$.

  \item Let $E \to S^4$ be the associated rank 2 complex vector bundle of the principal $SU(2)$-bundle $S^7$ over $S^4$ via the standard representation $SU(2) \to GL(2, \mathbb{C})$. Prove that the second Chern class $c_2(E) = 1$.

  \item Prove that all ASD instantons on $E \to S^4$ with $c_2(E) = 1$ are gauge equivalent to a translated and rescaled version of the basic instanton $A_{(x_0, \lambda)} (x)$ for some $x_0 \in S^4$ and $\lambda >0$.

  \item Compute the dimension of the moduli space $\mathcal{M}_{S^4}(E, g_0)$ with $c_2(E) =1$.
\end{enumerate}

\section[Lecture 2]{Lecture 2: The Uhlenbeck compactification and the Donaldson invariants}
\label{sec:Lecture2YT}

This lecture describes the {\it Uhlenbeck compactification} of the ASD instanton moduli spaces and the definition of {\it Donaldson invariants}. The readers are referred to \cite{FrUh}, \cite{DoKr} again for further details on the materials in this section. For the Seiberg--Witten invariants, see e.g. \cite{Mor} for more details.  

\subsection{Uhlenbeck compactness and the Uhlenbeck compactification}

Let $X$ be a closed, oriented, smooth four-manifold, and let $P \to X$ be a principal $SU(2)$-bundle. 
We fix a Riemannian metric $g$ on $X$. 

The instanton moduli space $\mathcal{M}_{X} (P,g)$ is not compact; sequences of ASD-instantons can develop curvature concentration at infinitely many points. 
This phenomenon is called {\it bubbling}. 
The following Uhlenbeck compact theorem gives a precise description of this loss of compactness: 

\begin{theorem}[Uhlenbeck compactness \cite{Uhle2}, \cite{Sedl}]
Consider a sequence $\{A_n\}$ of $L^2_2$ ASD connections on $P$. Then, after passing to a subsequence, we obtain:

\begin{itemize}
    \item a finite set of bubbling points $\{p_1, \dots, p_k\} \subset X$;
    \item $\{ m_1 , \dots , m_k \} \subset \mathbb{Z}_{>0}$; 
    \item a smooth $SU(2)$-bundle $P' \to X$ with $c_2(P') = c_2 (P) - \sum_{j=1}^{k} m_j$; 
    \item an $L^2_2$ ASD connection on $P'$; and 
    \item a sequence of $L^2_3$ gauge transformations $u_n : P'|_{X \setminus  \{ p_1, \dots, p_k\}} \xrightarrow{\cong} P'|_{X \setminus \{p_1, \dots, p_k\}}$ 
\end{itemize}
such that:  
\begin{enumerate}
    \item $u_n(A_n) \to A'$ in $L^2_{2}$ on compact subsets of $X \setminus \{p_1, \dots, p_k\}$.
    \item $|F_{A_n}|^2 \, d\text{vol}_g \rightharpoonup |F_{A'}|^2 \, d\text{vol}_g + 8\pi^2 \sum m_j \delta_{p_i}$ (weakly as measures), where $\delta_{p_j}$ is the $\delta$-measure.
\end{enumerate}
\end{theorem}

The theorem says that all lost energy is concentrated at finitely many points, and that the
amount of energy lost at each point is a multiple of $8 \pi^2$. The integer $m_j$ is the charge
lost at the point $p_j$. Analytically, one obtains convergence away from the bubbling points after
choosing suitable gauges; topologically, the limiting bundle has smaller second Chern class.
This result motivates the following definition. For each integer $l$ with $0 \leq l \leq c_2(P)$, choose
a principal $SU(2)$-bundle $P_l \to X$ satisfying $c_2(P_l) = c_2(P) - l$,
with $P_0 = P$. Let $\text{Sym}^l(X)$ denote the $l$-th symmetric product of $X$, parametrising unordered
$l$-tuples of points on $X$, counted with multiplicity.

\begin{definition}[Uhlenbeck compactification]
The {\it Uhlenbeck compactification} of the moduli space $\mathcal{M}_X (P,g)$ is the stratified space 
\[
\overline{\mathcal{M}}_X(P,g) := \bigsqcup_{l=0}^{c_2(P)} \mathcal{M}_{X}(P_l ,g)  \times \text{Sym}^l(X),
\]
where $P_l$ is a principal $SU(2)$-bundle over $X$ with $c_2 (P_l) = c_2 (P) - l, (l = 0, \dots , c_2(P))$ and $P_0 =P$, $\text{Sym}^l (X)$ is the $l$-th symmetric product of $X$.
\end{definition}

The topology on
$\overline{\mathcal{M}}_X(P,g) $ is defined so that a sequence of instantons converges precisely
when it converges in the sense described by Uhlenbeck compactness. Thus a limiting point
consists of a genuine instanton on a lower-charge bundle plus a zero-dimensional cycle recording
where the missing energy has concentrated.

\subsection{Stable range and fundamental class}
To define Donaldson invariants by integration, one needs a fundamental class of the compactified moduli space. For this purpose, it is important that all lower strata of the Uhlenbeck
compactification have sufficiently high codimension.

Choose a generic Riemannian metric on $X$ to ensure smoothness of moduli spaces $\mathcal{M}_{X}(P_l,g)$ in the Uhlenbeck compactification $\overline{\mathcal{M}}_X(P,g)$. 
We want all the stratum other than $\mathcal{M}_X (P, g)$ in $\overline{\mathcal{M}}_X(P,g)$ have the dimensions at least two less than that of $\mathcal{M}_{X} (P,g)$. 

For $0 < l < c_2 (P)$, 
$$ \dim (\mathcal{M}_{X} (P_l, g) \times \text{Sym}^{l} (X) ) = d ( \mathcal{M}_{X} (P, g) )- 4 l ,$$
where $d (\mathcal{M}_{X} (P, g) )$ is the expected dimension of $\mathcal{M}_{X} (P, g)$, namely, it is $-$ (the index of Atiyah--Hitchin--Singer complex). So, in this case, the above requirement is automatically satisfied. 

For the component $\mathcal{M}_{X} (P_{c_2(P)}, g) \times \text{Sym}^{c_2 (P)} X \cong  \text{Sym}^{c_2 (P)}$, we have
\begin{equation*}
\begin{split}
 \dim ( \text{Sym}^{c_2 (P)} X ) &= 4 c_2 (P) \\
 &= d( \mathcal{M}_{X}(P,g)) + 3 (1 + b^{+}(X) - 4c_2 (P) .
\end{split}
\end{equation*}
On the other hand, we want
$$ d (\mathcal{M}_{X} (P,g)) -2 \geq \dim ( \text{Sym}^{c_2 (P)} X ) ,$$
so, we want 
$$  d (\mathcal{M}_{X} (P,g) \geq 
 d( \mathcal{M}_{X}(P,g)) + 3 (1 + b^{+}(X) - 4c_2 (P) .$$
 In other words, we require 
 $$ 4 c_2 (P) \geq 5 + 3 b^{+} (X) .$$
 We call the above range for $c_2 (P)$ the {\it stable range}. 
 In this range, the lower strata of the Uhlenbeck compactification have codimension at least two, and the compactified moduli space $\overline{\mathcal{M}}_X(P,g) $ has a well-defined fundamental class
$[\overline{\mathcal{M}}_X(P,g) ]$. 

\subsection{Donaldson invariants}

We now define Donaldson invariants in the stable range. 

Let $X$ be a closed, oriented, smooth four-manifold with $b_2^+(X) > 0$ and odd. Let $P \to X$ be a principal $SU(2)$-bundle.
Consider the case $\pi_1 (X) =  \{ 1 \}$ and $G=SU(2)$. 

Denote by $\hat{\mathcal{G}}_P := \mathcal{G}_{P}/Z(G)$, where $Z(G)$ is the centre of $G$. 
Since $\mathcal{A}^{*}_{P} \to \mathcal{B}^{*}_{P} / \hat{\mathcal{G}}_P $ is a principal $\hat{\mathcal{G}}_P$-bundle.  
We then define a \textit{universal bundle} $\mathcal{U} \to \mathcal{B}^*_P \times X$ by 
\[
\mathcal{U} : = \mathcal{A}^*_P \times_{\hat{\mathcal{G}}_P} P \to \mathcal{B}^*_P \times X
\]
This is a principal $SO(3)$-bundle.

We then consider \textit{$\mu$-map:}
\[
\mu: H_2(X, \mathbb{Z}) \to H^{2}(\mathcal{B}_P^*, \mathbb{Q}),
\]
defined using the slant product  $/ : H_{p} (X, \mathbb{Q} ) \times H^{q} (\mathcal{B}_{P}^{*} \times X, \mathbb{Q}) \to H^{q-p} (\mathcal{B}^*_P , \mathbb{Q})$ with $p_1(\mathcal{U})$, namely, $\mu ( [\Sigma]) := - \frac{1}{4} p_1 (\mathcal{U} )/ [\Sigma]$ for $[\Sigma ] \in H_2 (X, \mathbb{Q})$. 
Then by restricting the $\mu$-map on the moduli space $\mathcal{M}_{X}^* (P,g)$, we obtain 
$$ \mu : H_2 (X, \mathbb{Q}) \to H^2 (\mathcal{M}_{X}^{*} (P, g) , \mathbb{Q} ).$$

\begin{proposition}
The $\mu$-map extends over the compactified moduli space $\overline{\mathcal{M}}_X(P,g)$.  We denote it by $\bar{\mu} : H_2(X, \mathbb{Q}) \to H^2 ( \overline{\mathcal{M}}_X(P,g) , \mathbb{Q})$. 
\end{proposition}

Remember in the stable range: $ 4 c_2 (P) \geq 3 b^{+} (X) + 5$, the fundamental class $[ \overline{\mathcal{M}}_X(P,g) ]$ is well-defined. We also assume that $b^{+} (X)$ odd so that 
$$ d:= 4 c_2(P) - \frac{3}{2} ( 1 + b^{+} (X) ) \in \mathbb{Z} , $$
where $2 d = $ the expected dimension of $\mathcal{M}_X (P,g)$ ($= -$ the index of the Atiyah--Hitchin--Singer complex). 
We write $k:= c_2 (P)$.

We then define \textit{Donaldson (polynomial) invariant} $q_{k} (X) : H_(X) \otimes \cdots \otimes H_2 (X) \to \mathbb{R}$ by 
\begin{equation*}
\begin{split}
q_k (X) (\Sigma_1, \dots, \Sigma_d) &= \langle \bar{\mu} (\Sigma_1) \cup \dots \cup \bar{\mu}(\Sigma_d), [\overline{\mathcal{M}}_X(P,g) ] \rangle  \\
&= \int_{[\overline{\mathcal{M}}_X(P,g) ]} \bar{\mu}(\Sigma_1) \cup \dots \cup \bar{\mu}(\Sigma_d), 
\end{split}
\end{equation*}
where $[\Sigma_1], \dots, [\Sigma_d] \in H_2(X ,\mathbb{Q})$. 

\begin{theorem}[Donaldson \cite{Dona90}]
If $b_2^+(X) > 1$, then $q_d (X)$ is independent of the choice of Riemannian metrics.
\end{theorem}

\begin{remark}
There are \textit{wall-crossing phenomena},  if $b_2^+(X) = 1$. In this case,  $q_d (X)$ can change under a metric variation. See e.g. \cite{Kots}, \cite{KoMo}, \cite{MaWe}, \cite{ElGo}, \cite{FrQi}, \cite{Gottsche:1996}, \cite{GoZa}, \cite{MoWi97}, \cite{GNY1}. 
\end{remark}

\subsection{Donaldson invariants in the unstable range}

The definition above used the stable range condition. If the desired degree lies outside the stable
range, Donaldson invariants can still be defined by means of blow-up formulae.

\begin{proposition}[Freedman--Morgan \cite{FrMo}]
Suppose $k$ is in the stable range for $X$. 
If $X'$ is the blowup in a point in $X$, in other words, $X' = X \# k \overline{\mathbb{CP}}^2$, 
then $d+1$ is in the stable range for $X'$, and 
$$ q_{k+1} (X) (\Sigma_1, \dots, \Sigma_d, e, e, e, e) = - 2 q_k ( \Sigma_1, \dots, \Sigma_d) , $$
where $ e \in H_2 (X, \mathbb{Z})$ is the class represented by the exceptional curve in $X$. 
\end{proposition}

\begin{proposition}[Friedman--Morgan \cite{FrMo}]
Fix $k \in \mathbb{Z}$. Suppose $b^+ (X) > 1$. Then there exists $b>0$ such that $k+b$ is in the stable range for $b$-times blowups $X'$ of $X$. 
\end{proposition}

We then define the Donaldson invariant for $k$ in the unstable range by 
$$ q_k (X) (\Sigma_1 , \dots , \Sigma_d) := \left( - \frac{1}{2} \right)^b q_{k+b} (X') 
 (\Sigma_1, \dots, \Sigma_d, \underbrace{e_1 , e_1 , e_1 ,e_1,  \dots, e_b, e_b, e_b , e_b}_{\text{each } e_i \text{ occurs 4 times here}}) . $$

\subsection{The structure theorem for the Donaldson series by Kronheimer--Mrowka}

More generally, for 
$$ \mathbb{A} (X) := \underbrace{H_2 (X) \otimes \cdots \otimes H_2 (X)}_{a \text{ times}}  \otimes \underbrace{H_0 (X) \otimes \cdots \otimes H_0 (X)}_{b \text{ times}}, $$ 
we define a Donaldson (polynomial) invariant 
$q_d (X) :  \mathbb{A} (X) \to \mathbb{Q}$
by 
$$ \langle \bar{\mu} (\Sigma_1) \cup \dots \cup \bar{\mu} (\Sigma_a), \bar{\nu}(p_1) \cup \dots \cup \bar{\nu} (p_b) , [\overline{\mathcal{M}}_X(P,g) ] \rangle  , $$
where $\bar{\nu} : H_0 (X) \to H^4 (\overline{\mathcal{M}}_X(P,g) )$ is defined in a similar way to $\bar{\mu}$, and $a + 2b =d$. 

\begin{definition}
A four-manifolds $X$ with $b^+ >0$ and $b^{+} - b^1$ odd is said to be {\it of simple type}, if 
$q_k (X) (z p^2 ) = 4 q_k (X) (z)$
for all $z \in \mathbb{A} (X)$, where $p \in H_0 (X)$.
\end{definition}

We define the {\it Donaldson series} by 
\[
\mathcal{D}_X(h) := \sum_{k=0}^\infty \frac{1}{d!} q_k (X) (h^d) 
+ \frac{1}{2}  \sum_{k=0}^\infty \frac{1}{d!} q_k (X) (ph^d)
\]
for $h \in H_2(X)$. 
We think of this as a formal power series. 

\begin{theorem}[Structure theorem by Kronheimer and Mrowka \cite{KrMr}]
If $X$ is of simple type and $b^1 (X) = 0$, then there exist $K_1 \dots, K_S \in H^2 (X , \mathbb{Z}) $ and non-zero rational numbers $\alpha_1 , \dots , \alpha_s$ such that
\[
\mathcal{D}_X(h) = \exp(Q(h)/2) \sum_{i=1}^s \alpha_i e^{K_i(h)} , 
\]
where $Q$ is the intersection form of $X$. 
\end{theorem}

We call $K_i$ a {\it basic class}.  
This theorem shows that, for manifolds of simple type, all Donaldson invariants are encoded
by finitely many cohomology classes and rational coefficients. 

\subsection{Seiberg--Witten invariants and Witten's conjecture}

We now recall the Seiberg--Witten invariants and their relationship with Donaldson invariants.

Let $X$ be a closed, oriented, smooth four-manifold. 
We take a Spin$^c$ structure $\mathfrak{s}$ on $X$ (note that it always exists on an oriented four-manifold).
We then have the spinor bundle $S= S^+ \oplus S^-$ and the determinant line bundle $\mathcal{L} := \det S^{+}$, where $S^{\pm}$ are complex vector bundles of rank $2$.

Take a connection $A$ on the line bundle $\mathcal{L}$. Then, this connection with the Levi-Civita connection on $TX$ induces a connection $\nabla_{\tilde{A}}$ on the spinor bundle $S$.  
We then define the {\it Dirac operator} $D_A$ by 
$$ 
D_A : \Gamma (S) \xrightarrow{\nabla_{\tilde{A}}} 
 \Gamma (S \otimes T^*X ) \xrightarrow{\rho} \Gamma (S), $$
where $\rho$ is a map defined by the Clifford multiplication. 
This $D_A$ swap $S^+$ and $S^{-}$ in four dimensions (in fact in even dimensions), namely, $D_A : \Gamma (S^{\pm}) \to \Gamma (S^{\mp})$. 

Define a map $ q : \Gamma (S^{+}) \to \text{End}_0 (S^+) $ by $q (\psi) := \psi^{*} \otimes \psi - \frac{1}{2} | \psi |^2 \text{Id}$ for $\psi \in \Gamma (S^+)$. 
We then consider the following equations for $(A , \psi ) \in \mathcal{A}_{\mathcal{L}} \times \Gamma (S^+)$:
\[
D_A \psi = 0, \quad c ( F_A^+ ) = \frac{1}{2} q (\psi) , 
\]
where $c$ is a map $i \Gamma (\Lambda^+ ) \to \text{End}_0 (S^+)$ defined by the Clifford multiplication. 
We call the above equations the {\it Seiberg--Witten equations}.

Denote the configuration space by $\mathcal{C} := \mathcal{A}_{\mathcal{L}} \times \Gamma (S^+)$, where $\mathcal{A}_{\mathcal{L}}$ denotes the space of all connections on $\mathcal{L}$. 
We define the {\it gauge group} $\mathcal{G}$ by $\mathcal{G} := \{ \gamma \, | \, X \to S^1 \} $. 
This acts on the configuration space $\mathcal{C}$ by 
$(A - 2 (d \gamma ) \gamma^{-1} , \gamma \psi) $ for $(A, \psi) \in \mathcal{C}$ and $\gamma \in \mathcal{G}$. 
We then denote by $\mathcal{M}^{SW}_{X} (\mathfrak{s} ) $ the moduli space of solutions to the Seiberg--Witten equations, namely, 
$$ \mathcal{M}^{SW}_{X} (\mathfrak{s}) 
:= \left\{ (A, \psi ) \in \mathcal{C} \, | \, 
D_A \psi = 0, \, c ( F_A^+ ) = \frac{1}{2} q (\psi) \right\} / \mathcal{G}. $$
The expected dimension is given by 
$ \frac{1}{4} c_1^2 ( \mathcal{L}) . [X] - \frac{2 \chi + 3 \sigma}{4}    $.

One of the remarkable properties of the moduli space $\mathcal{M}^{SW}_{X} (\mathfrak{s} )$ is the following: 
\begin{theorem}
The moduli space $\mathcal{M}^{SW}_{X} (\mathfrak{s}) $ is compact.
\end{theorem}

For the Seiberg--Witten equations, it is more convenient to add a self-dual two-form to perturbed the equations, and after a generic choice of these perturbation term and and a generic choice of Riemannian metrics, one obtains a smooth structure on $\mathcal{M}^{SW}_{X} (\mathfrak{s})$. 
Also, as in the case of the ASD instanton moduli spaces, an orientation on $H^1 (X) \oplus \mathbb{H}^{+} (X)$ determines an orientation of the moduli space $\mathcal{M}^{SW}_{X} (\mathfrak{s})$.

Fix $x_0 \in X$, we define
$$ \mathcal{G}^0 := \{ \gamma \in \mathcal{G} \, | \, \gamma (x_0) =1 \in S^1 \} .$$
We then denote $\mathcal{B}^0 := \mathcal{C}^{*} / \mathcal{G}^0$, where $\mathcal{C}^*$ is the set of all irreducible $(A, \psi)$. 
Here $(A, \psi)$ is called {\it irreducible}, if $\psi \neq 0$. 
This $\mathcal{B}^{0}$ is an $S^1$-bundle over $\mathcal{B}:= \mathcal{C} / \mathcal{G}$. We denote this bundle by $\mathcal{U}$. 

Suppose the expected dimension is even, and write it by $2 d$. 
We define the {\it Seiberg--Witten invariant} by 
$$ SW (\mathfrak{s}) := c_1 ( \mathcal{U} )^d . [\mathcal{M}^{SW}_{X} (\mathfrak{s})] \in \mathbb{Z} . $$
This is independent of the choice of Riemannian metrics on $X$, if $b^{+} > 1$. 
(If the expected dimension is odd, we define the invariant to be zero.)

\begin{conjecture}[Witten’s conjecture \cite{Witt94}] Donaldson invariants are determined by Seiberg--Witten invariants. More concretely, the basic classes by Kronheimer and Mrowka coincide with the classes where the Seiberg--Witten invariants do not vanish. 
\end{conjecture}

Proofs for the above Witten conjecture were established for 
\begin{itemize}
    \item projective surfaces by G\"ottsche--Nakajima--Yoshioka \cite{GNY3} via Mochizuki’s formula \cite{Moc} and a solution to the Nekrasov conjecture \cite{Nekr03, Moore:1997dj, Moore:1998et, Lossev:1997bz}, \cite{NaYo05,NeOk, BrEt}; and 
    \item general four-manifolds by Feehan--Leness via $PU(2)$-monopoles, see \cite{FeLe1}, \cite{FeLe2} as well as the references therein.   
\end{itemize}

Note that, taking advantage of the a priori compactness of the Seiberg--Witten moduli spaces, Furuta \cite{Furu} proved the {10/8 theorem} using a finite-dimensional approximation of the Seiberg--Witten map. This approach was subsequently refined into the Bauer--Furuta invariant \cite{BaFu}, which takes values in the equivariant stable cohomotopy group. It is one of the strongest invariants currently available for detecting subtle phenomena in the smooth structures on four-manifolds.

\subsection*{Exercises}

\begin{enumerate}
    \item Prove that energy lost in the Uhlenbeck compactification is $8\pi^2$ per bubbling point.
    \item Prove that the Seiberg--Witten moduli space is always compact.
\end{enumerate}

\section[Lecture 3]{Lecture 3: The Donaldson--Uhlenbeck--Yau theorem and algebro-geometric formulations}

In this lecture we explain how anti-self-dual instantons on compact K\"ahler surfaces can be described in algebro-geometric terms. The central result is the Donaldson--Uhlenbeck--Yau theorem, which identifies Hermitian-Einstein connections with stable holomorphic bundles. This theorem provides the bridge between the differential-geometric moduli spaces appearing in Donaldson theory and moduli spaces of stable sheaves in algebraic geometry. 
Standard references for the materials in this section are \cite{Koba}, \cite{LuTe} as well as Donaldson's original papers \cite{Dona85, Donaldson:1987} and the work of Uhlenbeck--Yau \cite{UhYa}.   

\subsection{Hermitian--Einstein connections}

Let $X$ be a compact Kähler surface with Kähler form $\omega$. 
In this setting, by using the complex structure on $X$, we have
$$\Lambda^2_X \otimes \mathbb{C} = \Lambda^{2,0} \oplus \Lambda^{0,2} \oplus \Lambda^{1,1} .$$
Furthermore, we have 
\begin{equation*}
\begin{split}
\Lambda^+  \otimes \mathbb{C} &= \Lambda^{2,0} \oplus \Lambda^{0,2}  \oplus \mathbb{C} \omega, \\
\Lambda^- \otimes \mathbb{C} &= \Lambda^{1,1}_{0},
\end{split}
\end{equation*}
where $\Lambda^{1,1}_0$ is the orthogonal complement of $\mathbb{C} \omega$ in $\Lambda^{1,1}$. 

Consider an $SU(r)$-bundle $P$, and its associated  complex vector bundle $E$ of rank $r$ over $X$.

\begin{proposition}
A connection $A$ is an anti-self-dual (ASD) connection if and only if 
\begin{equation}
F_A^{2,0} = F_A^{0,2} = 0,\quad \Lambda F_A = 0,  
\label{eq:HE}
\end{equation}
where $\Lambda =(\wedge \omega)^*$. 
\end{proposition}

\begin{remark}
If $F_A^{0,2} =0$, then $A$ determines a holomorphic structure on $E$.   
\end{remark}

A connection $A$ satisfying the above equations \eqref{eq:HE} is called a \textit{Hermitian-Einstein connection} (or called a {\it Hermitian–Yang–-Mills connection}).
This is defined in an arbitrary complex dimension and also for $U(r)$ bundles by considering the following:
\[
F_A^{2,0} = F_A^{0,2} = 0,\quad \Lambda F_A =  c_E \cdot \text{Id}_E , 
\]
where $ c_E := \frac{2\pi \cdot \deg(E)}{\text{rank}(E) \cdot \text{Vol}(X)}$.

\subsection{Slope (Mumford–Takemoto) stability}

Let $X$ be a compact K\"ahler manifold of dimension $n$ with K\"ahler form $\omega$, and let $E \to X$ be a holomorphic vector bundle.

\begin{definition} 
$E$ is \emph{slope-stable} if for any proper coherent subsheaf $F \subset E$ with $0 < \text{rank}(F) < \text{rank}(E)$, the following inequality holds: 
\[
\mu(F) < \mu(E) , 
\]
where the {\it slope} $\mu(E)$ is defined as:
\[
\mu(E) = \frac{1}{\text{rank}(E)} \int_X c_1(E) \wedge \omega^{n-1}  
\]
We call $E$ {\it slope semistable} if $\mu (F) \leq \mu (E)$ holds in the above. 
\end{definition}

\subsection{The Donaldson--Uhlenbeck--Yau theorem}
The following theorem is one of the central bridges between Gauge Theory and Algebraic Geometry. 

\begin{theorem}[Donaldson--Uhlenbeck--Yau \cite{Dona85, Donaldson:1987}, \cite{UhYa}] 
Let $E$ be an irreducible holomorphic vector bundle. Then $E$ is slope stable if and only if $E$ admits a Hermitian-Einstein connection.
\end{theorem}

In complex dimension two, and especially for $SU(r)$ bundles, this theorem says that ASD
connections on a K\"ahler surface are equivalent to stable holomorphic bundles equipped with
Hermitian-Einstein metrics. Thus, the instanton moduli space can be identified, on its stable
locus, with a moduli space of stable holomorphic bundles.
The theorem has two directions. The direction from stability to existence is an analytic
existence theorem. Donaldson \cite{Dona85, Donaldson:1987} proved it for the projective cases using a parabolic flow and functional
methods; Uhlenbeck and Yau \cite{UhYa} proved the general result by a continuity method. The direction
from existence to stability is more elementary and follows from a vanishing theorem. 

\subsection{Hermitian metric description}

Let $X$ be a compact Kähler manifold with Kähler form $\omega$, and let $E$ be a complex vector bundle over $X$ with a holomorphic structure $\bar{\partial}_E$.

In this setting, we have the decomposition: 
$$ \Lambda^{1} \otimes \mathbb{C} = \Lambda^{1,0} \oplus \Lambda^{0,1} .$$ 
From this, a covariant derivative $\nabla_{A}$ of $E$ decomposes as 
$$ \nabla_{A} = \partial_{A} + \bar{\partial}_{A} , 
$$
where $\partial_{A} : \Gamma (E) \to \Gamma (E \otimes \Lambda^{1,0}) , \bar{\partial}_{A} : \Gamma (E) \to \Gamma (E \otimes \Lambda^{0,1})$. (Note that similar decompositions hold for covariant exterior derivatives $D_A$.)

Let $h$ be a Hermitian metric on $E$. Then there exists a unique compatible connection $A_h$, called the {\it Chern connection},  such that $\bar{\partial}_{A_h} = \bar{\partial}_{E}$. 

\begin{definition}
A hermitian metric $h$ on $E$ is said to be a \emph{Hermitian–Einstein metric}, if its Chern  connection $A = A_h$ satisfies 
\[
F_{A_h}^{0,2} = 0, \quad \text{and} \quad i \Lambda  F_{A_h} = c_E \, \text{Id}_E .
\]
\end{definition}
The first equation is automatically satisfied for the Chern connection of a holomorphic
bundle, so the essential condition is the second equation. 

\subsection{Vanishing theorem}

The stability of $E$ under the existence of a Hermitian-Einstein metric is essentially from the following vanishing theorem:

\begin{theorem}
If a holomorphic vector bundle $E$ over $X$ admits a Hermitian–Einstein metric $h$ with
\[
i\Lambda F_{A_h} = \lambda \cdot \text{Id}_E \,\text{ and } \lambda < 0,
\]
then there is no non-trivial holomorphic section of $E$.
\end{theorem}

With this in hand, a proof of the stability of $E$ under the existence of a Hermitian-Einstein metric goes in the following way: 
Let $\mathcal{F} \subset E$ be a subsheaf with $r' := \text{rank}\, \mathcal{F} < \text{rank}\, E$. 
Then the inclusion map $i : \mathcal{F} \to E$ induces a homomorphism:
$$ \det i : \det \mathcal{F} ( = ( \Lambda^{r'} \mathcal{F})^{**} ) \to (\Lambda^{r'} E)^{**} = \Lambda^{r'} E . $$
This is, in fact, injective, thus, we obtain 
$$ s : \mathcal{O}_{X} \to \Lambda^{r'} E \otimes ( \det \mathcal{F} )^{*} ,  $$
namely, a holomorphic section of $\Lambda^{P} E \otimes ( \det \mathcal{F} )^{*}$. 

Now, from the existence of a Hermitian-Einstein metric h on $E$, we have 
$$ 
i\Lambda F_{A_h} = c_{E} \cdot \text{Id}_E, $$
where 
$$ c_{E} = \frac{2 n \pi \mu (E)}{n! \text{vol}_X}. $$
On the other hand, there is also a Hermitian-Einstein metric on the line bundle $\det \mathcal{F}$ with 
$$ c_{\det \mathcal{F}} = \frac{2 n \pi \mu (\det \mathcal{F}) }{n! \text{vol}_{X}} 
= \frac{2 n \pi r' \mu (\mathcal{F}) }{n! \text{vol}_{X}} . $$
Hence, $\Lambda^{r'} E \otimes (\det \mathcal{F})^{**}$ admits a Hermitian--Einstein metric with $ r' c_E - c_{\det \mathcal{F}}$. 
Since $\Lambda^{r'} E \otimes (\det \mathcal{F})^{**}$ has a non-trivial section, the above vanishing theorem implies 
$$ r' c_{E} - c_{\det \mathcal{F}} \geq 0. $$
This is equivalent to $\mu ( \mathcal{F} ) \leq \mu ( \mathcal{E} )$. 
Thus the existence of a Hermitian–Einstein metric implies slope semistability. A refinement of
the argument gives stability when the connection is irreducible.

\subsection{Existence of Hermitian–Einstein metrics}

Let $E \to X$ be a holomorphic vector bundle over a K\"ahler manifold $X$ with K\"ahler form $\omega$ of $\dim_{\mathbb{C}} X =n $. 

Denote by $\text{Herm}^+ (E)$ the space of all hermitian metrics on $E$. 
We consider a functional $L : \text{Herm}^+ (E) \to \mathbb{R}$, whose gradient is:
$$\text{grad} \, L = ( i \Lambda F_{A_h} - c \cdot  \text{Id}_E) h,$$ 
where $ c:= 2 n \pi \int_{X} c_1 (E) \wedge \omega^{n-1} / r \int_{X} \omega^n . $

To define $L$, we set 
$$ L(h ,k ) := \int_{X} 
\left( Q_2 (h , k) - \frac{c}{n} Q_1 (h ,k ) \omega ) \wedge \frac{\omega^{n-1}}{(n-1)!} \right), $$
for $h, k \in \text{Herm}^+ (E)$, where 
\begin{equation*}
    \begin{split}
  Q_1 (h ,k ) &= \text{log} \, (\det k^{-1} h ) ,  \\       
  Q_2 (h, k) &= i \int_{0}^{1} \text{tr}\, (v_t \cdot F_t ) dt . \\
    \end{split}
\end{equation*}
Here, $v_t := h_t^{-1} \partial_t h_t \in \Gamma (\text{End} \, (E) )$, where $h_t \, (0 \leq t \leq 1)$ is a curve in $\text{Herm}^+ (E)$ with $k= h_0$ and $h = h_1$, and $F_t := F_{A_h} \in \Gamma (\Lambda^{1,1}_{X} \otimes \text{End}\, (E))$. 

Fix $k \in \text{Herm}^+ (E)$ and define $L(h) := L (h ,k )$ for $h \in \text{Herm}^{+} (E)$. 
We then consider the evolution equation:
\begin{equation}
    \frac{d h_t}{dt} = - \text{grad}\, L 
    = - ( i \Lambda F_{A_h} - c \cdot \text{Id}_E) h_t. 
    \label{eq:evoeq}
\end{equation}

A stationary point of this flow is precisely a Hermitian–Einstein metric.
Donaldson proved:
\begin{itemize}
  \item Short time existence of a solution to \eqref{eq:evoeq} and its regularity. 
\end{itemize}
Moreover, he proved solutions exist for all time. 

Now, the problem is whether a solution converges or not. 
This follows from 
\begin{itemize}
\item $\max_{X} |i\Lambda F_{A_h} - c \cdot \text{Id}_E|$ decreases in $t$, and 
\item The functional $L$ is bounded below.
\end{itemize}

In fact, Donaldson proved: 
\begin{theorem}[Donaldson \cite{Donaldson:1987}]
\begin{equation*}
        L_X (h , k ) 
        \geq \frac{1}{n-1} L_Y (h , k) 
        - C ( \max_{X} |i\Lambda F_{A_h} - c \cdot \text{Id}_E| ) -C' , 
\end{equation*}
where $Y$ is a non-singular hypersurface of $X$, and $C, C' >0$ are constants which are independent of $h_t$. 
\end{theorem}

On Riemann surfaces, we have the following: 
\begin{theorem}
    Let $E$ be a semistable vector bundle over a Riemann surface $\Sigma$.  Then for a fixed $k \in \text{Herm}^{+} (E)$, $L(h,k)$ is bounded from below on $\Sigma$. 
\end{theorem}

\begin{remark} Uhlenbeck--Yau's proof \cite{UhYa} of the existence theorem of Hermitian--Einstein metrics for general compact K\"haler manifolds uses the method of continuity. They prove this by contradiction, showing that if the method of continuity were to fail, their analysis would produce a destabilising subsheaf, contradicting the stability assumption.
\end{remark}

\subsection{Gieseker stability and compactification}
We now move from differential geometry to algebraic geometry.  
Let $X$ be a smooth projective variety,  and let $H$ be an ample line bundle.

The \emph{Hilbert polynomial} of a torsion-free coherent sheaf $\mathcal{E}$ is:
\[
p_{\mathcal{E}}(m):= \chi(\mathcal{E}\otimes \mathcal{O}_X (mH)) ,
\]
where $\chi (\mathcal{E}) := \sum (-1)^i \dim H^{i} (X , \mathcal{E})$. 

\begin{definition} 
A torsion-free coherent sheaf $\mathcal{E}$ is \emph{Gieseker semistable}, if for all proper subsheaves $\mathcal{F} \subset \mathcal{E}$,
\[
    \frac{p_{\mathcal{F}}(m)}{\mathrm{rank}(\mathcal{F})} \leq \frac{p_{\mathcal{E}}(m)}{\mathrm{rank}(\mathcal{E})} \quad \text{for } m \gg 1 . 
\]
It is called {\it Gieseker semistable},  if the strict inequality holds in the above. 
\end{definition}

Consider the case $[\omega] = c_1 (H)$, then by the Riemann--Roch formula, we have
\begin{equation*}
\begin{split}
\chi ( \mathcal{E} & \otimes \mathcal{O} (m H) )
 = \text{rank} (\mathcal{E}) \frac{m^n}{n!} \int_X c_1 (H)^n \\ & + \frac{m^{n-1}}{(n-1)!} \int_X c_1 (H)^{n-1} \left( c_1 (\mathcal{E}) + \frac{\text{rank}(\mathcal{E})}{2} c_1 (X) \right) + \cdots .
\end{split}
\end{equation*}
Thus, we obtain 
$$p_{\mathcal{E}} (m) - p_{\mathcal{F}} (m) = \frac{m^{n-1}}{(n-1)!} (\mu (\mathcal{E}) - \mu (\mathcal{F})) + \cdots . $$
Hence, we have: 

\begin{proposition}
\[
    \text{slope-stable} \Rightarrow \text{Gieseker-stable} \Rightarrow \text{Gieseker-semistable} \Rightarrow \text{slope-semistable}. 
\]
\label{prop:ss}
\end{proposition}

\subsection{Jordan--H\"older filtration}
If $\mathcal{E}$ is Gieseker or slope semistable, there exists a filtration:
\[
    0 = \mathcal{E}_0 \subset \mathcal{E}_1 \subset \cdots \subset \mathcal{E}_l = \mathcal{E}
\]
with $\mathrm{gr}_i(E) = E_i / E_{i-1}$ stable and 
$$\frac{\chi (\mathrm{gr}_i( \mathcal{E}) \otimes \mathcal{O} (mH))}{\text{rank} (\mathrm{gr}_i( \mathcal{E}) )} = 
\frac{\chi( \mathcal{E} \otimes \mathcal{O} (mH) )}{\text{rank} (\mathcal{E})} $$
for all $i$.

We then write $\mathrm{gr} (\mathcal{E}) := \oplus_{i} \mathrm{gr}_i( \mathcal{E}). $

\begin{definition} Two semistable sheaves $\mathcal{E}_1, \mathcal{E}_2$ or rank $r$ with the same Hilbert polynomial are said to be \emph{S-equivalent} if:
\[
    \mathrm{gr}(\mathcal{E}_1) \cong \mathrm{gr}(\mathcal{E}_2) . 
\] 
\end{definition}

We denote by $\mathcal{M}^{ss}_{H}(p, r)$ the moduli space of S-equivalence classes of Gieseker semistable torsion-free sheaves of rank $r$ with Hilbert polynomial $p$ on $X$.

\begin{theorem}[Gieseker, Maruyama, Simpson] $\mathcal{M}^{ss}_{H}(p, r)$ is a projective scheme (in particular, it is compact).
\end{theorem}

Denote by $\mathcal{M}^{\text{stable}}_{H}(p, r)$ the moduli space of Gieseker stable sheaves of rank $r$ and with Hilbert polynomial $P$ on $X$; and by $\mathcal{M}^{\mu-\text{stable}}_{H} (p, r)$ the moduli spaces of slope stable sheaves of rank $r$  with Hilbert polynomial $p$ on $X$. 
We then have the following open inclusions by Proposition \ref{prop:ss}:
$$ \mathcal{M}^{\mu-\text{stable}}_{H} (p, r) \subset \mathcal{M}^{\text{stable}}_{H}(p, r) \subset \mathcal{M}_{H}^{ss} (p, r) . $$
By the Donaldson--Uhlenbeck--Yau theorem, $\mathcal{M}^{\mu-\text{stable}}_{H} (p, r)$ is the same as the instanton moduli space $\mathcal{M}_{P} (X ,g )$. Thus, the Gieseker moduli space $\mathcal{M}_{H}^{ss} (p, r)$ gives a compactification of $\mathcal{M}_{P} (X ,g )$.

\subsection{Li--Morgan's Map}
Let $X$ be a smooth projective surface with $\pi_1 (X) = \{ 1 \}$.

Take 
$[ \mathcal{E} ]\in \mathcal{M}_{X,H}^{ss} (P,r)$. As $\mathcal{E}$ is Gieseker semistable, it is slope stable. 
Take the Jordan--H\"older filtration for slope stability, namely, consider $\mathrm{gr}_{\mu}(\mathcal{E})$. 
This is the direct sum of slope stable sheaves. 
Take the double dual $(\mathrm{gr}_{\mu}(\mathcal{E}))^{**}$, this is reflexive,  thus locally-free (i.e. a vector bundle) on $X$. Thus, by the Donaldson--Uhlenbeck--Yau theorem, there exists a Hermitian-Einstein metric on it.  
The quotient
$\frac{(\mathrm{gr}_{\mu}(\mathcal{E}))^{**}}{\mathrm{gr}_{\mu}(\mathcal{E})}$ 
is supported in dimension zero. Hence, its support determines a point in a symmetric product
of $X$, recording the bubbling points in the Uhlenbeck compactification.
Then, consider a map 
$    \mathcal{C}: \mathcal{M}^{ss}_{X, H} (p, r) \to \mathcal{M}_X (P,g)
$ defined by 
$$ \mathcal{C} ( [ \mathcal{E} ]) := 
\left( (\mathrm{gr}_{\mu}(\mathcal{E}))^{**} , \text{supp} \left( \frac{(\mathrm{gr}_{\mu}(\mathcal{E}))^{**}}{\mathrm{gr}_{\mu}(\mathcal{E})} \right) \right) .$$

\begin{theorem}[Jun Li \cite{Liju93}, Morgan \cite{Morg93}]
    The map $\mathcal{C}$ is well-defined and continuous, and it is an isomorphism outside the lower strata. 
    Hence, 
    $$ \mathcal{C}_{*} [ \mathcal{M}^{ss}_{X, H} (p, r)] = [ \mathcal{M}_X (P, g) ] .$$
\end{theorem}

This theorem explains precisely how the Gieseker compactification and the Uhlenbeck compactification are related. The locally free part of a sheaf records the limiting instanton, while
the quotient by its double dual records the points where curvature has concentrated. 

\subsection{Algebro-Geometric Donaldson invariants}

If the Chern classes of $E$ are large enough, the dimension of the Gieseker moduli space coincides with the expected one, and the lower strata have codimension $\geq 2$. 
In this case, Donaldson invariants are defined by using the Gieseker moduli spaces. 
The Li--Morgan theorem
ensures that the resulting fundamental classes agree under the comparison map.

\subsection{Donaldson–Mochizuki invariants on projective surfaces}

Using virtual techniques in \cite{BF}, Mochizuki constructs a perfect obstruction theory on the moduli space of semistable sheaves on projective surfaces, and defines {\it Donaldson–Mochizuki invariants} by using the virtual fundamental classes of the moduli spaces, even when singularities are present \cite{Moc}.

More precisely, Mochizuki 
\begin{itemize}
\item[$\bullet$] formulates the invariants where the moduli space may have strictly semistable sheaves;  
\item[$\bullet$] proves a weak wall-crossing formula for his invariants; and 
\item[$\bullet$] expresses them in terms of Seiberg--Witten invariants.  
\end{itemize}

These lead to the determination of the wall-crossing terms and a resolution of Witten's conjecture $D=SW$ on complex projective surfaces, both by G\"ottsche--Nakajima--Yoshioka \cite{GNY1, GoNaYo2, GNY3}, both with analysis on the Nekrasov partition function \cite{Nekr03, Moore:1997dj, Moore:1998et, Lossev:1997bz}, \cite{NaYo05, NeOk, BrEt}.

\subsection*{Exercises}
\begin{enumerate}
    \item Let $E = \mathcal{O}_{\mathbb{P}^2}(1) \oplus \mathcal{O}_{\mathbb{P}^2}(-1)$. Is $E$ slope stable?
    \item Prove that if $\mu(F) < \mu(E)$ for all $F \subset E$, then $E$ is indecomposable.
    \item Compute the Hilbert polynomial for $E = \mathcal{O}_X(H)^{\oplus 2}$.
    \item Show that for a torsion-free sheaf $E$ on a surface, $E^{**}$ is locally free. 
\end{enumerate}

\section[Lecture 4]{Lecture 4: Vafa--Witten theory in algebraic geometry}
\label{sec:Lecture4YT}

This lecture gives an introduction to the theory of Vafa--Witten from the mathematical side. 
The goal is to explain how the physical prediction of Vafa and Witten \cite{VW} is reflected in the geometry of moduli spaces of sheaves and Higgs sheaves on projective surfaces. In particular, we will see how Euler characteristics, virtual Euler characteristics, perfect obstruction theories, and modular forms naturally enter the story. 

The discussion begins with moduli spaces of bundles on curves, where the obstruction theory is comparatively simple, and then moves to moduli spaces of sheaves on surfaces, where obstruction spaces appear and virtual techniques become indispensable. We then recall the {\it Vafa–Witten equations}, describe their algebro-geometric counterpart in terms of Higgs pairs, and finally explain how virtual localisation leads to the definition of Vafa–Witten invariants on projective surfaces
\subsection{Betti numbers of the moduli spaces of bundles/sheaves}

Before introducing the Vafa--Witten theory in four dimensions, let us mention some work in two dimensions. 

Let $C$ be a complex curve, and 
consider $\mathcal{M}_{C}^{ss} (n ,r)$ the moduli space of semistable vector bundles of rank $r$ with degree $n$ over $C$. For
simplicity, assume $n$ and $r$ are coprime. Then semistability coincides with stability, and the moduli space $\mathcal{M}_{C}^{ss} (n ,r)$ is smooth.  
The geometry and topology of the moduli spaces were studied by several
authors.

\begin{itemize}
\item {Harder--Narasimhan} \cite{HaNa}
obtained an inductive formula of and hence computed the Poincar\'e polynomial of $\mathcal{M}_{C}^{ss} (n,r)$ via the Weil conjecture by counting in a finite field. 
\item {Atiyah--Bott} \cite{AtBo} reproved and also improved the result by Harder--Narashimhan via the infinite dimensional Morse theory.  
\item {Kirwan} \cite{Kirw} obtained the Poincar\'e polynomial of GIT quotient as a finite-dimensional version of the result by Atiyah--Bott.  
\end{itemize}

A key reason why the theory is relatively well behaved on curves is that the obstruction
space
$\text{Ext}^2 (E, E)_0$  
vanishes automatically for dimensional reasons. Thus, the deformation theory of stable bundles on
curves is unobstructed. 
Note that this is no longer true on surfaces

\subsection{Euler characteristics of the instanton moduli spaces}

In the 1990s, researchers began to study the geometry of moduli spaces of semistable sheaves
on projective surfaces. In general, these moduli spaces are highly singular, even if one imposes
coprimality assumptions analogous to those in the curve case. The essential new phenomenon is
the appearance of the obstruction space $\text{Ext}^2 (E, E)_0$. 

However, for del Pezzo surfaces, the obstruction space $\text{Ext}^2 (E, E)_0$ vanishes, thus, one obtains smooth moduli spaces. In fact, the following people computed the Betti numbers (and the Euler characteristics) of the moduli spaces for those cases: 
\begin{itemize}
    \item Klyachko \cite{Klya}, Yoshioka \cite{Yos1}: $\mathbb{P}^2$
    \item Yoshioka: ruled surfaces \cite{Yosh2}, rational elliptic surfaces \cite{Yosh3}. 
\end{itemize}

Before the above results, G\"ottsche \cite{Got} computed the Euler characteristics of the Hilbert schemes of points $\mathrm{Hilb}^n(X) = X^{[n]}$ on projective spaces (which can be thought of as the moduli spaces of semistable sheaves of rank one on projective surfaces).

On projective surfaces, one can consider a {\it generating series} of e.g. the Euler characteristics of the moduli spaces (by varying $c_2$, which one cannot do that on curves). 
Remarkably, one obtains modular forms, namely, those generating series are written in terms of modular forms. 

In fact, what G\"ottsche \cite{Got} obtained was the following beautiful formula:     
    \[
        \sum_{n=0}^{\infty} e(X^{[n]}) q^n = \left(\prod_{n=1}^{\infty} \frac{1}{1 - q^n}\right)^{e(X)}
    \]
    where  $\eta (q) := q^{\frac{1}{24}} \prod_{n=1}^{\infty} (1 - q^n)$ is the Dedekind eta function and $e(X)$ is the topological Euler characteristic of $X$. See Lecture 2 of Martijn Kool's lecture notes for how to deduce the above G\"ottche's formula. 

Later developments by Grojnowski \cite{Groj} and Nakajima \cite{Naka3} explain this formula via the representation theory of Heisenberg (super) algebras acting on the direct product of homology groups of $X^{[n]}$.

\begin{remark}
    Nakajima's pioneering work \cite{Naka1} on the moduli spaces of ASD instantons on ALE spaces, and their relation to representations of affine Lie algebras and the modularity of the generating series.
\end{remark}

\subsection{Donaldson’s theory and $\mathcal{N} = 2$ super Yang--Mills theory}

Witten \cite{Witt88} describes Donaldson's theory in terms of a topological twist of $\mathcal{N}=2$ super Yang--Mills theory. 
In this interpretation, the Donaldson invariants arise as correlation functions in
a topological quantum field theory. 

Later, Seiberg and Witten \cite{SeWi} found an exact solution to the low-energy effective theory of $\mathcal{N}=2$ super Yang--Mills theory. 
This uses a certain family of elliptic curves (the {\it Seiberg--Witten curves}) to determine the effective theory. 
The Seiberg--Witten
invariant that arise from this analysis provides a much simpler way to describe many four-manifolds
invariants.

Moreover, the modular properties of the wall-crossing term and blowup formula for Donaldson invariants are explained by using {\it $u$-plane integral} \cite{MoWi97}, while this approach also provides a physical derivation of Witten's conjecture.  See Lectures \ref{sec:JMLect2} and \ref{sec:JMLect3} of Jan Manschot's lecture notes. 
This is one of the first strong indications
that modularity is not accidental but is built into the structure of four-dimensional gauge theory.

\subsection{$\mathcal{N} = 4$ Super Yang--Mills theory and S-duality conjecture}

Vafa and Witten \cite{VW} looked into a more symmetric model, which is a topological twist of $\mathcal{N}=4$ super Yang--Mills theory to study {\it $S$-duality}. 

Let $X$ be a closed, oriented, smooth four-manifold.  
Let $P \rightarrow X$ be a principal $G$-bundle over $X$ ($G$ is a compact Lie group).  
Fix a Riemannian metric $g$.

For $(A, B, C ) \in \mathcal{A}_{P} \times \Gamma (\text{ad}\, (P)  \otimes \Lambda^{+} )  \times \Gamma (\text{ad}\, (P) \otimes \Lambda^0_{X} )$, consider:

\[
d_A^* B + d_A C = 0 ,
\]
\[
F_A^+ + [B. B] + [B, C] = 0, 
\]
where $[B. B] \in \Gamma ( \text{ad}\, (P) \otimes \Lambda^{+}_{X} )$ denotes the natural quadratic expression obtained from the Lie bracket on $\text{ad}\,  (P)$ and the wedge
product of self-dual forms.  
We call the above equations the {\it Vafa--Witten equations}.

Roughly speaking, what Vafa and Witten considered was the following signed counting of solutions to the Vafa--Witten equations: 
$$ \mathcal{M}^{VW}_X (P,g) := \chi (  \overline{\mathcal{M}}_X (P, g) ) + \sum_{\xi \in \mathcal{M}^{VW}_X (P,g) \setminus \mathcal{M}_X (P, g) } \varepsilon (\xi) , $$
where $\chi$ is the ``Euler characteristic'', $\overline{\mathcal{M}}_X (P, g )$ is the Uhlenbeck compactification of the ASD instanton moduli space ${\mathcal{M}}_X (P, g) $, $\mathcal{M}^{VW}_X (P,g)$ is the moduli space of solutions to the {\it Vafa--Witten equations}, and 
$\varepsilon( \cdot )$ is a ``signed counting'' of points in ${\mathcal{M}^{VW}_X (P,g) \setminus \mathcal{M}}_X (P, g )$; and 
they conjectured the generating series of this invariant could be written in terms of modular forms.

Note that for del Pezzo surfaces, K3, and ALE spaces where $B = C = 0$, the moduli spaces are unobstructed. 

For studies from analytical perspectives on the moduli space of solutions to the Vafa--Witten equations, see e.g. \cite{Chen}, \cite{DaGu}, \cite{Guan}, \cite{Guan2}, \cite{Tana2}, \cite{Tana3}, \cite{Tana4}, \cite{Taub1}. \cite{Taub2}, \cite{Taub3}. Moreover, the Vafa--Witten equations are a compact deformation of the non-Abelian adjoint monopole equations \eqref{eq:NonAbMonOp}, \eqref{eq:F+Adj} in Section \ref{sec:Hypers} in Jan Manschot's lecture notes (see \cite{Manschot:2021qqe} for more detail).

\subsection{Algebro-geometric counterpart}

We consider the case for general projective surfaces (namely, the obstructed case). 
Let $X$ be a projective surface, and let 
$E$ be a holomorphic vector bundle over $X$. 

In this setting the Vafa--Witten equations become: 
\begin{equation}
\bar{\partial}_A \varphi = 0, \quad F_A^{0,2} =0, \quad  F^{1,1}_A \wedge \omega + [\varphi, \varphi^*] = 0, 
\label{eq:VWK}
\end{equation}
where $\varphi \in \Gamma (\text{End}\, (E) \otimes \Lambda^{2,0})$. 

The condition $F_A^{0,2} =0$
 says that $A$ defines a holomorphic structure on $E$, while the first equation
says that $\varphi$ is holomorphic with respect to this holomorphic structure.

As in the case of the anti-self-dual equations, via the Donaldson--Uhlenbeck-Yau type theorem, solutions to the Vafa--Witten equations \eqref{eq:VWK} on a projective surface correspond to stable Higgs bundles $(E , \phi)$, where 
$\phi: E \to E \otimes K_X$,  and 
$K_X$ is the canonical line bundle of $X$ \cite{AlGa03, Tana}.

\subsection{cf. Hitchin's equations on Riemann surfaces}

Let $\Sigma$ be a compact Riemann surface of genus $2$, and let $E$ be a hermitian vector bundle over $\Sigma$. 

For $ (A , \Phi ) \in \mathcal{A}_{E} \times \Gamma ( \text{End}\, (E) \otimes \Lambda^{1,0})$, we consider the following equations:
$$ \bar{\partial}_{A} \Phi =0, \quad F_A + [\Phi , \Phi^* ]=0 .$$
These equations are called {\it Hitchin's equations} \cite{Hitc}. 

Solutions to Hitchin's equations correspond to stable Higgs bundles $(E , \phi)$, where $\phi : E \to E \otimes K_{\Sigma}$, and $K_{\Sigma}$ is the canonical line bundle of $\Sigma$. 

Note that, on Riemann surfaces, $\Lambda^{1,0} =  K_{\Sigma} = \Omega_{\Sigma}^1 $, where $\Omega^1_{\Sigma}$ is the holomorphic cotangent bundle of $\Sigma$. 
Simpson generalised this theory by considering semistable Higgs pairs $(E, \phi )$, where $\phi : E \to E \otimes \Omega_{X}^1$ on smooth projectile variety $X$ in arbitrary dimension. The Vafa--Witten case corresponds to a generalisation of the case for $\phi : E \to E \otimes K_{X}$.  
Note that the former also appears in the Kapustin--Witten theory \cite{KaWi}, which is obtained from a different topological twist of $\mathcal{N}=4$ super Yang--Mills theory on a Kähler surface \cite{Naka4}, \cite{Tana}, \cite{LiRaTa}, \cite{KrTaYa}.

\subsection{Perfect obstruction theory}

In Algebraic Geometry, there is a way to define a {\it virtual fundamental class} of the moduli spaces developed by Li--Tian \cite{LT} and Behrend--Fantechi \cite{BF}.

Denote by $\mathcal{M}$ the moduli space of semistable sheaves on a projective surface $X$.
We consider the case that semistability implies stability. 
Then, locally, $\mathcal{M}$ is the zero set of the Kuranishi map $\kappa: \text{Ext}^1 (E ,E)_0 \to \text{Ext}^2 (E, E)_0 $ around $[E] \in \mathcal{M}$.

A moral here is to consider the following local model attached to it:

$$\xymatrix@R=16pt@C=0pt{
& \mathcal{V} \ar[d] \\
M : = Z(s) \subset & \mathcal{X}, \ar@/_/[u]_s}
$$
where $\mathcal{X}$ is an ambient space, $\mathcal{V}$ is a bundle on it, and the moduli space is locally the zero set $Z(s)$ of a section $s$ of $\mathcal{V}$.

This local model in fact yields a two-term locally-free resolution of the Kuranishi map, and by dualising the above geometric description one may package it into the following diagram:

$$
\xymatrix@R=20pt@C=20pt{
\mathcal{V}^{\vee}|_{M} \ar@{->>}[d]^{s^{\vee}|_{M}} \ar[r]^{d s^{\vee}|_{M}} & \Omega_{\mathcal{X}}|_{M} \ar[d]^{id} \\
\mathcal{I}_{M/\mathcal{X}} / \mathcal{I}^2_{M/\mathcal{X}} \ar[r] & \Omega_{\mathcal{X}}|_{M} . }
$$

Then Behrend and Fantechi consider a globalisation of this. 
They define a {\it perfect obstruction theory} on a Deligne--Mumford stack $\mathbb{M}$ as a complex $E^{\bullet} \in D^{b} (\mathbb{M})$, 
locally quasi-isomorphic to a complex of locally-free sheaves supported in degree $[-1, 0]$ with a morphism $\phi : E^{\bullet} \to \mathbb{L}_{\mathbb{M}}^{\geq 1}$, where $\mathbb{L}_{\mathbb{M}}^{\geq 1}$ is $[-1, 0]$-truncation of the cotangent complex, so that $h^{0} (\phi)$ is isomorphism and $h^{-1}(\phi)$ is surjective.

From these data, they construct a {\it virtual fundamental class} in the Chow group of $\mathbb{M}$ of the right degree and prove that it is a deformation invariant.

\subsection{Symmetric obstruction theory on the Vafa--Witten moduli spaces, their $\mathbb{C}^*$-fixed loci, and Vafa--Witten invariants on projective surfaces}

Denote by $\mathcal{N}$ the moduli space of semistable Higgs sheaves on a projective surface $X$. 
To define deformation invariants out of the moduli space, we need:
\begin{itemize}
    \item transversality of the moduli space, and 
    \item compactness. 
\end{itemize}

In the Vafa--Witten theory on projective surfaces, the higher obstruction space is $\text{Ext}^2 (E, E \otimes K_X)_0$, which vanishes when there are no strictly semistable sheaves in the moduli space, since $\mathrm{Ext}^2(E, E \otimes K_X)_0 \cong \mathrm{Ext}^0(E, E)_0^\vee$ by the Serre duality. 
Moreover, the obstruction spaces $\text{Ext}^2 (E, E)_0$ and $\text{Ext}^1 (E, E \otimes K_X)_0$ are dual to the {\it virtual tangent spaces} $\text{Ext}^1 (E, E)_0$ and $\text{Ext}^0 (E, E \otimes K_X)_0$ of the moduli space, again, by the Serre duality. 
In fact, Richard Thomas and the fourth-named author \cite{TT1} constructed a symmetric perfection obstruction theory on the moduli space.

On the other hand, the moduli space is not proper. However, there is a $\mathbb{C}^*$-action on the moduli space, which is induced from the multiplication on the Higgs fields, and the fixed loci are compact.\footnote{From the perspective of the physical $\mathcal{N}=2^*$ theory discussed in Lecture \ref{sec:Lecture4JM}, the $\mathbb{C}^*$-action corresponds to the action of the $U(1)$ flavor group. The fixed point locus of this action consists of the instanton branch and monopole branch.} 
The fixed loci consist of:
\begin{itemize}
    \item Gieseker semistable sheaves
    \item Nilpotent Higgs fields
\end{itemize}
Namely, we have
$$ \mathcal{N}^{\mathbb{C}^*} \cong \mathcal{M}_{X}^{ss} (p,r, L) \sqcup \mathcal{M}^{\mathrm{Higgs}} , $$
where $\mathcal{M}_{X}^{ss} (P,r, L) $ is the moduli space of semistable sheaves of rank $r$ with fixed Hilbert polynomial $P$ and fixed determinant $L$ on $X$, and $\mathcal{M}^{\mathrm{Higgs}}$ is the moduli space of nilpotent trace-free Higgs fields, if $r$ is prime. 

Using $\mathbb{C}^*$-action and fixed loci, we define 
$$ VW := \int_{[\mathcal{N}^{\mathbb{C}^*}]^{\mathrm{vir}}} \frac{1}{e^{\mathbb{C}^{*}} (N_{\mathcal{N}^{\mathbb{C}^{*}}}^{\mathrm{vir}})} ,$$
where $N^{\text{vir}}_{\mathcal{N}^{\mathbb{C}^{*}}}$ is the virtual normal bundle of $\mathcal{N}^{\mathbb{C}^{*}}$, $e^{\mathbb{C}^{*}}$ denotes the $\mathbb{C}^{*}$-equivariant Euler class, and $[ \cdot ]^{vir}$ indicates the virtual class constructed from the induced perfect obstruction theory on $\mathcal{N}^{\mathbb{C}^{*}}$. 

By a result of Jiang--Thomas \cite{JiTh}, one sees: 
\begin{equation*}
\begin{split}
VW & := \int_{[\mathcal{N}^{\mathbb{C}^*}]^{\text{vir}}} \frac{1}{e^{\mathbb{C}^{*}} (N_{\mathcal{N}^{\mathbb{C}^{*}}}^{\text{vir}})}  \\
& = \int_{[\mathcal{M}_{X}^{ss} (P,r, L) ]^{\text{vir}}} \frac{1}{e^{\mathbb{C}^{*}} (N_{\mathcal{M}_{X}^{ss} (P,r, L) }^{\text{vir}})} 
 +\int_{[\mathcal{M}^{\mathrm{Higgs}}]^{vir}} \frac{1}{e^{\mathbb{C}^{*}} (N_{\mathcal{M}^{\mathrm{Higgs}}}^{\text{vir}})}  \\
 & = \underbrace{\int_{[\mathcal{M}_{X}^{ss} (p,r, L) ]^{\text{vir}}} c_{\text{vd}} (T_{\mathcal{M}_{X}^{ss} (p,r, L) }^{\text{vir}})}_{\text{the instanton (horizontal) contribution}}
+\underbrace{\int_{[\mathcal{M}^{\mathrm{Higgs}}]^{\text{vir}}} \frac{1}{e^{\mathbb{C}^{*}} (N_{\mathcal{M}^{\mathrm{Higgs}}}^{\text{vir}})}}_{\text{the monopole (vertical) contribution }} ,
\end{split}
\end{equation*}
where $T^{\text{vir}}_{\mathcal{M}_{X}^{ss} (P,r, L)}$ denotes the virtual tangent sheaf 
on $\mathcal{M}_{X}^{ss} (p,r, L)$ and $\text{vd}$ is the virtual dimension, thus, the first term in the rightmost side is the {virtual Euler characteristic} of $\mathcal{M}_{X}^{ss} (P,r, L)$ introduced by {Ciocan-Fontanine--Kapranov} \cite{CiKa}, {Fantechi--G\"ottsche} \cite{FG}. 

\subsection{Virtual Euler characteristics of moduli spaces of semistable sheaves on projective surfaces}

The moduli space $\mathcal{M}_X^{ss} (p, r, L)$ carries a perfect obstruction theory due to Mochizuki \cite{Moc}, and one has the virtual tangent bundle: 
$$T_{\mathcal{M}^{ss}_X (p, r, L)}^{\text{vir}} = R \pi_{*} R \mathcal{H}om (\mathcal{E}, \mathcal{E} )_{0} [1] , $$ 
where $\mathcal{E}$ is a universal sheaf on $X \times \mathcal{M}^{ss}_X (p, r, L)$ and $\pi: X \times \mathcal{M}^{ss}_X (p, r, L) \to \mathcal{M}^{ss}_X (p, r, L)$ is the projection.

Fantechi and G\"ottsche  define the {\it virtual Euler characteristic} of $\mathcal{M}^{ss}_X (p, r, L)$ by
$$ e^{\text{vir}} (\mathcal{M}^{ss}_X (p, r, L)) := \int_{[\mathcal{M}^{ss}_X (p, r, L)]^{\text{vir}}} c_{\text{vd}} (T^{\text{vir}}_{\mathcal{M}^{ss}_X (p, r, L) }). $$
More generally, they define the {\it virtual $\chi_y$-genus} of $\mathcal{M}^{ss}_X (p, r, L)$ by 
$$ \chi_{-y}^{\text{vir}} (\mathcal{M}^{ss}_X (p, r, L)):= \int_{[\mathcal{M}^{ss}_X (p, r, L)]^{\text{vir}}} (1-y \, \text{ch} (\text{T}^{\text{vir}})^{-1}) \cdot \text{td} (\text{T}^{\text{vir}}) ) \in \mathbb{Q}[y] , $$
where $\text{T}^{\text{vir}}$ is the virtual tangen sheaf of $\mathcal{M}^{ss}_X (p, r, L)$. 
G\"ottsche--Kool \cite{GK1, GK3}, G\"ottsche--Kool--Laarakker \cite{GKL} express these as a Donaldson--Mochizuki invariant, and form a conjecture that the generating series of them 
could be written in terms of modular forms and Seiberg--Witten invariants. They verified it in many examples.

Note that the virtual Euler characteristic or the virtual $\chi_y$-genus of the moduli space of semistable sheaves is the {instanton part} of the Vafa--Witten invariant or a refinement of it. 

G\"ottsche--Kool \cite{GK1, GK3}, G\"ottsche--Kool--Laarakker \cite{GKL} conjecture the generating series of them could be written in terms of modular forms and Seiberg--Witten invariants.  
In particular, they anticipate that the virtual Euler characteristics $e^{\text{vir}} (\mathcal{M}_{X}^{ss} (r, c_1, c_2))$ is the coefficient of $q^{c_2 - \frac{r-1}{2r} c_1^2 - \frac{r}{2} \chi ( \mathcal{O}_X ) + \frac{r}{24} K_X^2}$ of 
\vspace{-0.2cm}
\begin{equation}
\begin{split}
& r^{2 +K_X^2 - \chi (\mathcal{O}_X)}  \left( \frac{1}{\Delta (q^{\frac{1}{r}})^{\frac{1}{2}} } \right)^{\chi (\mathcal{O}_{X})} 
\left( \frac{\Theta_{A^{\vee}_{r-1},0} (q) }{\eta (q)^r} \right)^{- K_X^2} 
\\ 
& \qquad \times D_0 (q)^{K_X^2} \sum_{\mathbf{\beta} \in H^2 (X, \mathbb{Z})^{r-1}} 
\prod_{i} \epsilon_r^{i \beta_i c_1} SW (\beta_i) \prod_{i \leq j} D_{ij} (q)^{\beta_i \beta_j}, 
\end{split}
\label{eq:GKL}
\end{equation}  
where $D_0$, $D_{ij}$'s are universal power series. See their paper for the notations etc. (the above is Conjecture 1.10 in in \cite{GKL}).

See Lecture 3 of Martijn Kool's lecture notes for more detailed accounts. 

\subsection{Modularity of the generating series}

The generating series of the Vafa--Witten invariants are expected to be modular forms.  
This was first verified in examples via the work by 
Yoshioka (for $\mathbb{P}^2$, ruled surfaces) \cite{Yos1, Yosh2} , Göttsche (for K3) \cite{Got}, and Nakajima (for ALE spaces) \cite{Naka1}.

In the semistable case, for example on K3,  Thomas and the fourth-named author \cite{TT2} proved the following, assuming a conjecture by {Yukinobu Toda} \cite{Toda} in Donaldson--Thomas theory (it was later proved by {Maulik} and {Thomas} 
 \cite{MaTh}, which also matches with the conjecture by Vafa and Witten: 
\vspace{-0.2cm}
\begin{equation} \sum_{c_2} VW_{r, c_2} q^{c_2} = \sum_{d | r} \frac{d}{r^2} \sum_{j=0}^{d-1} \eta \left( e^{\frac{2 \pi  i j}{d}} q^{\frac{r}{d^2}} \right)^{-24}. 
\label{K3}
\end{equation}
where $\eta(q)$ is the Dedekind eta function.

To be precise, the above is for the $SU (r)$ Vafa--Witten invariants, namely, we fix the first Chern class of sheaves and consider $\phi$ with $\text{tr} \phi =0$, let us denote them by $VW_{r,c_2}^{SU(r)}$. 

The {\it Vafa--Witten partition function} is defined to be: 
$$ Z_{VW}^{SU(r)} (q) := r^{-1} q^{\lambda} \sum_{c_2} q^{\frac{1}{2r} \text{vd}} (-1)^{\text{vd}} VW_{r, c_2}^{SU(r)} ,  $$
where $\lambda := - \frac{1}{2} \chi (\mathcal{O}_{X} ) + \frac{r}{24} K_{X}^2$, and $\text{vd}$ is the virtual dimension of the moduli space $\mathcal{M}_{X}^{ss} (p,r, L) $. 

Jiang and Kool \cite{JK} proved the {\it $S$-duality conjecture} for $K3$ surfaces by constructing the Langlands dual side of the story, when the rank $r$ is prime:
\vspace{-0.2cm}
$$
Z_{VW}^{SU(r)} (-1 / \tau) = (-1)^{(r-1) \chi(\mathcal{O}_{X})} \left( \frac{r \tau}{i} \right)^{- \frac{e(X)}{2}} Z_{VW}^{{}^L SU(r)} (\tau), 
$$
where $q = e^{2 \pi i \tau}$, and the ${}^{L} SU(r)$ Vafa--Witten partition function $Z_{VW}^{{}^L SU(r)}$ on the right-hand side is defined through the moduli spaces of  
twisted sheaves by Yoshioka.  

\subsection{Blowup formulae}

Let $p : \hat{X} \to X$ be the blowup at a point of a projective surface $X$. 
Fix a cohomology class $c:= r + c_1 + c_2$ of $X$ and consider $\hat{c} = p^{*} c - k e$ of $\hat{X}$  for some $k \geq 0$, where $e := \text{ch} (\mathcal{O}_C (-1))$ with $C$ the exceptional divisor. 

\begin{theorem}[\cite{KLT}]
\begin{equation*}
\begin{split}
q^{\bullet}  & \sum_{c_2}  e^{\rm{vir}} ( \mathcal{M}^{ss}_{\hat{X}} (r, \hat{c}_1 , c_2)  )) 
= \frac{\Theta_{A^{\vee}_{r-1},k } (q) }{\eta(q)^{r}}  
      \sum_{c_2}  e^{\rm{vir}} ( \mathcal{M}^{ss}_{X} (r,  c_1 , c_2) ))
\end{split}
\end{equation*}
where 
$\eta$ is the Dedekind eta function. 
\label{th:KLTblowup}
\end{theorem}

The above formula coincides with the conjectural formula by  G\"ottsche--Kool \cite{GK1, GK3} mentioned earlier. 
Similar formula for the virtual $\chi_y$-genera of the moduli spaces also holds. 
Note that G\"ottsche's conjectural formula for the topological $\chi_y$-genera of the moduli spaces \cite{Gott3} and the one by Vafa--Witten \cite{VW} have the same shape as ours.  

In addition, the above blowup formula determines some of  universal functions described by G\"ottsche--Kool \cite{GK1}. For instance, this blowup formula with the calculations on the $K3$ surface yields the full computations of the generating series of the instanton part of the Vafa--Witten invariants for the rank two case, provided the existence of the above universal functions. 
As for the existence, 
Joyce works on proving these conjectures for the above various virtual enumerative invariants with his wall-crossing formulae \cite{Joy} (cf. \cite{GJT}) derived from his reinvention of the vertex algebras on the homology of the moduli stacks of coherent sheaves \cite{Joyc1}.

Most recently, Arbesfeld--Kool--Laarakker \cite{AKL} determined some of universal functions that appear in Laarakker's formulae \cite{Laa1, Laa2} on the generating series of the {vertical component} of the Vafa--Witten invariants by using our blowup formula above. 
In particular,  they determined the generating series of the vertical Vafa-Witten invariants on a projective surface for the rank two case.

Theorem \ref{th:KLTblowup} is a consequence of a more general blowup formula proved by Kuhn and the fourth-named author in \cite{KuTa}, together with an explicit computation on $\mathbb{P}^2$. The blowup formula in \cite{KuTa} is expressed in terms of universal functions that are independent of the underlying surface. In other words, it depends only on the local geometry of the blowup (see \cite{KuTa} for details).

To state the precise statement, 
let $\mathcal{E}$ be a universal sheaf on $X \times \mathcal{M}^{ss}_{X} (c)$, and denote by 
 $\Phi (\mathcal{E})$ a Chow cohomology class of the moduli space, which depends only on the universal sheaf and cohomology classes of $X$ such as the insertions for the virtual enumerative invariants on a projective surface listed below.

\begin{theorem}[\cite{KuTa}]
Fix a Chern class $\hat{c} = p^*c-k e$ for some $k \geq 0$ on $\hat{X}$, where $c=r+c_1+c_2$ is that on $X$ and $e$ is the class defined by the exceptional divisor. 
		 	Then there exist universal power series $\Omega_n\in \mathbb{Q}[[\nu_2,\ldots,\nu_r]]$ depending only on $r,\Phi$ and $k$, which satisfy:\begin{equation*}\label{eq:structeq}
				\int_{[\mathcal{M}_{\hat{X}}^{ss}(\hat{c})]^{\rm{vir}}} \Phi(\mathcal{E}) = \sum_{n=0}^{\infty} \int_{[\mathcal{M}^{ss}_X(c+n[pt])]^{\rm{vir}}} \Phi(\mathcal{E}) \, \Omega_n(\mathcal{E}).
			\end{equation*}
            \label{th:blowupKT}

\end{theorem}

This result is a virtual analogue of the blowup formula of Nakajima--Yoshioka \cite{NaYo1, NaYo2, NaYo3}, which played a key role in the proof of the equality $D=SW$ for projective surfaces by G\"ottsche--Nakajima--Yoshioka \cite{GNY3}. One important application is a Friedman--Morgan type blowup formula for Donaldson--Mochizuki invariants, which provides a direct proof of the equivalence between the classical Donaldson invariants and the Donaldson--Mochizuki invariants for the rank two case. We note that this equivalence also follows from the aforementioned work of G\"ottsche--Nakajima--Yoshioka on $D=SW$ \cite{GNY3}.

Theorem \ref{th:blowupKT} may have many further applications to virtual enumerative invariants on projective surfaces, including:
\begin{itemize}
\item Donaldson--Mochizuki invariants;
\item virtual Euler characteristics;
\item virtual $\chi_y$-genera;
\item virtual Segre numbers (a version of Donaldson--Mochizuki invariants with fundamental matter); and
\item virtual Verlinde numbers (a generalisation of $K$-theoretic Donaldson invariants),
\end{itemize}
all of which are defined via the virtual classes of moduli spaces of semistable sheaves on a projective surface.

\chapter{Algebraic geometry and Vafa--Witten theory} \label{ch:MK}
\vspace{-.4cm}
\begin{center}
\it{Martijn Kool}
\end{center}
\vspace{.4cm}

Yang--Mills theories are an important class of quantum field theories. In particular, the Standard Model of particle physics involves Yang--Mills theory for gauge groups $\mathrm{U}(1), \SU(2), \SU(3)$ on the differentiable 4-manifold $\R^4$. 

In their seminal 1994 paper on $S$-duality, Vafa--Witten \cite{VW} considered $\mathcal{N}=4$ super-symmetric Yang--Mills theory for gauge group $\SU(r)$ (and its Langlands dual group $^{L}\SU(r)$) on a compact connected Riemannian 4-manifold $M$. Moreover, they carry out a topological twist which makes the observables of the theory metric independent. As a consequence, the partition function for gauge group $\SU(r)$ of the theory at strong (complexified) coupling $\tau \in \C$ equals the partition function for gauge group $^{L}\SU(r)$ at weak coupling $-1/\tau$ up to a modular factor. In mathematical terms, the partition functions for gauge groups $\SU(r)$ and $^{L}\SU(r)$ form the components of a vector-valued modular form. 

We consider the case when $M$ underlies a smooth projective surface $S$ over $\C$. Roughly speaking, there are three types of surfaces that we will study:
\begin{enumerate}
\item[(1)] Del Pezzo surfaces, i.e., smooth projective surfaces with ample anti-canonical divisor $-K_S$. 
\item[(2)] $K3$ surfaces, i.e., smooth projective surfaces satisfying $b_1(S) = 0$ and with trivial canonical divisor $K_S = 0$.
\item[(3)] Surfaces with non-zero holomorphic 2-form, i.e., $h^{2,0}(S) > 0$.\footnote{Typically, these surfaces have a non-zero effective divisor $C \in |K_S|$. When such a surface $S$ is minimal and $K_S^2 >0$, $S$ is called ``of general type''.}
\end{enumerate}

For $S \subset \PP^3$ defined by a homogeneous polynomial of degree $d$, we have that $d \leq 3$ leads to del Pezzo surfaces, $d=4$ gives $K3$ surfaces, and $d \geq 5$ yields surfaces of general type (with holomorphic 2-form).

\begin{exercise}
Prove the statements in the last sentence. Classify all surfaces in categories (1)--(3) that are obtained as complete intersections in products of projective spaces. 
\end{exercise}

As discovered by Tanaka--Thomas \cite{TT1} in 2017, for $S$ a smooth projective surface, Vafa--Witten's partition functions can be realized entirely within \emph{algebraic geometry}. More specifically, they can be defined as intersection numbers on certain moduli spaces of Gieseker stable sheaves (and Higgs pairs) on $S$. 

The goal of these lectures is to explain the algebro-geometric side of the story with a focus on computational techniques. For the physics aspects and the relevant differential geometry, I refer to the other lectures. The great advantages of using algebraic geometry in this context originate from the following observations:
\begin{enumerate}
\item Algebraic geometry is good at providing \emph{compactifications} of moduli spaces such as Hilbert schemes of points and, more generally, Gieseker--Maruyama--Simpson moduli spaces of rank $r$ stable sheaves \cite{HL}. Miraculously, these are the correct compactifications for applications to physics.\footnote{Concretely, this means that intersection numbers on these moduli spaces match the results obtained by path integrals and supersymmetric localization in physics.}
\item In special cases, such as for $r=1$ on any smooth projective surface, or any $r>0$ for del Pezzo surfaces, these moduli spaces are smooth. However, in general they are singular (they can have multiple irreducible components, which can be non-reduced and of unexpected dimension). However, it turns out they carry a \emph{virtual cycle} in the sense of Behrend--Fantechi \cite{BF} and Li--Tian \cite{LT}. These virtual cycles can be used to give algebro-geometric definitions of Donaldson, Seiberg--Witten, and Vafa--Witten invariants.
\item Although the general definition of a virtual cycle is quite involved, there are lots of tools for explicitly carrying out \emph{calculations} involving virtual cycles such as Graber--Pandharipande localization, cosection localization, and wall-crossing techniques (to name a few). Therefore, the partitions functions from physics can actually be determined explicitly in examples. Sometimes this leads to verifications, or in the best cases proofs, of predictions from physics. In other cases, it allows mathematicians to go beyond the original physics calculations (e.g.~higher rank gauge groups, refined invariants, adding insertions, ...).
\end{enumerate}

\section[Lecture 1]{Lecture 1: Moduli of sheaves and Hilbert schemes}
\label{sec:Lecture1MK}

\subsection{Moduli of sheaves} \label{sec:modulisheaves}

In this section, we review some basic facts on moduli of sheaves. For a detailed exposition, we refer to the book by Huybrechts--Lehn \cite{HL}, which contains proofs of most of the statements made in this section. 

Let $X$ be a smooth projective variety of dimension $d$ over $\C$ and let $H$ be an ample divisor on $X$. By the Donaldson--Uhlenbeck--Yau theorem, covered in Y.~Tanaka's lectures, we are interested in slope (semi-)stable and Gieseker (semi)stable coherent sheaves $E$ on $X$. We briefly repeat the definition. 

Recall that the Hilbert polynomial of a coherent sheaf $E$ on $X$ is defined by $P_E(m) = \chi(E \otimes \O_X(mH))$ which, by the Hirzebruch--Riemann--Roch theorem, is given by
\[
P_E(m) = \int_X e^{m H} \, \ch(E) \, \td(T_X),
\]
where $T_X$ denotes the (holomorphic) tangent bundle of $X$. This formula exhibits $P_E(m)$ as a polynomial in $m$. Denoting the scheme-theoretical support of $E$ by $\mathrm{supp}(E)$, we say that $E$ is $\delta$-dimensional when $\mathrm{supp}(E)$ is $\delta$-dimensional. Since $\ch_i(E) = 0$ for $i<d-\delta$ and $\ch_{d-\delta}(E)$ is Poincar\'e dual to the class $[\mathrm{supp}(E)] \in H_{2\delta}(X)$, it follows that the leading coefficient of $P_E(m)$ is $(\ch_{d-\delta}(E) H^{\delta}) (m^{\delta}/\delta !)$ which is positive since $H$ is ample (Nakai--Moishezon criterion).\footnote{We view Chern classes as elements of singular cohomology, i.e., $c_i(E) \in H^{2i}(X,\Z)$ and therefore $\ch_i(E) \in H^{2i}(X,\Q)$.} We define the reduced Hilbert polynomial of $E$ by 
$$
p_E(m) = \frac{P_E(m)}{\ch_{d-\delta}(E) H^{\delta}}.
$$

\begin{definition}
A coherent sheaf $E$ on $X$ is called torsion free whenever any subsheaf $0 \neq F \subset E$ has dimension $d$. We define its slope (with respect to $H$) as $\mu(E) := c_1(E) H^{d-1} / \rk(E)$. A torsion free sheaf $E$ on $X$ is called slope stable (with respect to $H$) when
\[
\mu(F) < \mu(E)
\]
for all subsheaves $F \subset E$ satisfying $0 < \rk(F) < \rk(E)$. Moreover, $E$ is called Gieseker stable (with respect to $H$) when
\[
p_F(m) < p_E(m), \quad \textrm{for all } m \gg 0,
\]
for all $0 \neq F \subsetneq E$. We say that $E$ is slope semistable (resp.~Gieseker semistable), when the same holds with ``$<$'' replaced by ``$\leq$''. Sheaves which are slope semistable and not slope stable are called strictly slope semistable and similarly for Gieseker stability.
\end{definition}

As already discussed in Y.~Tanaka's lectures, for a torsion free sheaf $E$ on $X$, we have a chain of implications
\[
    \text{slope-stable} \Rightarrow \text{Gieseker stable} \Rightarrow \text{Gieseker semistable} \Rightarrow \text{slope-semistable.}
\]

\begin{exercise}
Prove these implications. Pay special attention to subsheaves $F \subsetneq E$ with $\rk(F) = \rk(E)$. 
\end{exercise}

\begin{exercise} \label{exc:tensorlinebundle}
\hfill
\begin{enumerate}
\item[(a)] Suppose $E$ is slope (semi)stable and $L$ is a line bundle on $X$. Prove that $E \otimes L$ is also slope (semi)stable. 
\item[(b)] Suppose $E$ is Gieseker (semi)stable (with respect to $H$). Prove that $E \otimes \O_S(\ell H)$ is Gieseker (semi)stable (with respect to $H$) for all $\ell \in \Z$.
\end{enumerate}
\end{exercise}

In the next lecture, it will be crucial that a Gieseker stable sheaf $E$ on $X$ is simple, i.e., $\Hom_X(E,E) = \C$. Therefore, we are interested in settings where there are no stricty semistable sheaves. This is guaranteed by the following criterion: let $r > 0$ and $c_1 \in H^2(X,\Z)$ such that 
\[
\gcd(r, c_1 \cdot H) = 1,
\]
then there are no rank $r$ strictly slope semistable sheaves $E$ on $X$ with $c_1(E) = c_1$. 

\begin{exercise} 
\hfill
\begin{enumerate}
\item[(a)] Prove that a Gieseker stable sheaf $E$ is simple. As follows: (1) show that any non-zero $E \to E$ is an isomorphism using \cite[Proposition 1.2.7]{HL} and (2) use that any finite-dimensional division algebra over $\C$ is trivial.
\item[(b)] Prove that if $\gcd(r,c_1 \cdot H) = 1$ there are no rank $r$ strictly slope semistable sheaves with first Chern class $c_1$.
\end{enumerate}
\end{exercise}

\begin{theorem}[Gieseker, Maruyama, Simpson]
Let $X$ be a smooth projective variety of dimension $d$ with ample divisor $H$. Fixing $r>0$ and Chern classes $c_i \in H^{2i}(X,\Z)$ for all $i=1, \ldots, d$, there exists a quasi-projective scheme $M_X^H(r,c_1, \ldots, c_d)$ co-representing an associated natural modular functor. Its $\C$-valued points are precisely the isomorphism classes of rank $r$ Gieseker $H$-stable torsion free sheaves $E$ on $X$ with $c_i(E) = c_i$ for all $i$. When there are no rank $r$ strictly Gieseker $H$-semistable sheaves $E$ on $X$ with $c_i(E) = c_i$ for all $i$, the scheme $M_X^H(r,c_1, \ldots, c_d)$ is projective.
\end{theorem}

For the notion of co-representability and the relevant moduli functor, we refer to \cite{HL}. The crucial part of the theorem is the final sentence: in the absence of strictly semistable sheaves $M:=M_X^H(r,c_1, \ldots, c_d)$ is projective, which will allow us to do intersection theory on $M$ (in particular, there is a degree map $\int_M \colon H_{2*}(M) \to \Z$). Nonetheless, $M$ is generally highly singular with several (possibly non-reduced) irreducible components.\footnote{In fact, moduli of stable sheaves satisfy Murphy's Law by \cite{Vak}.} 

However, at a point $[E] \in M$, we have the following dimension estimates\footnote{When we write $[E] \in M$, we always mean that $[E]$ is a $\C$-valued point of $M$.}
\[
\dim \Ext^1_X(E,E) - \dim \Ext^2_X(E,E) \leq \dim_{[E]} (M) \leq \dim \Ext^1_X(E,E),
\]
where $T_M|_{[E]} \cong \Ext^1_X(E,E)$ is the Zariski tangent space of $M$ at $[E]$ and $\dim_{[E]}(M)$ denotes the dimension of $M$ at $[E]$. In particular, we see that if $\dim \Ext_X^2(E,E) = 0$ for all $[E] \in M$, then the Zariski tangent spaces do not jump, which suggests that $M$ is smooth, which is indeed the case.

In general, $M$ is only a coarse moduli space, meaning that it co-represents the relevant moduli functor and its $\C$-valued points are the isomorphism classes of the sheaves of interest. However, sometimes $M$ represents its moduli functor in which case there exists a universal sheaf $\cE$ on $M \times X$. It has the tautological property
\[
\cE|_{\{[E]\} \times X} \cong E.
\]

We are also interested in the determinant map
\[
\det \colon M \to \Pic^{c_1}(X), \quad [E] \mapsto [\det(E)],
\]
where $\Pic^{c_1}(X) \leq \Pic(X)$ denotes the subgroup of line bundles $L$ on $X$ satisfying $c_1(L) = c_1$. For a fixed $[L] \in \Pic^{c_1}(X)$, we consider the fibre
\[
M_X^H(r,L,c_2 \ldots, c_d) := \mathrm{det}^{-1}([L]) \subset M_X^H(r,c_1,c_2, \ldots, c_d).
\]

The Zariski tangent space to $[E] \in M_L := M_X^H(r,L,c_2 \ldots, c_d)$ has a nice description in terms of trace maps. We recall that there exist trace maps \cite[Section 10.1]{HL}
\[
\tr^i \colon \Ext^i_X(E,E) \to H^i(X,\O_X),
\]
for all $i \geq 0$ and we define $\Ext^i_X(E,E)_0 := \ker(\tr^i)$. It turns out that $\tr^i$ is surjective and has a section $H^i(X, \O_X) \to \Ext^i_X(E,E)$ induced by the identity map $\O_X \to \mathcal{E}{\it{nd}}_{X}(E)$. We have
\[
T_{M_L} |_{[E]} \cong \Ext_X^1(E,E)_0.
\]
Similar to above, at a point $[E] \in M_L$, we have 
\[
\dim \Ext^1_X(E,E)_0 - \dim \Ext^2_X(E,E)_0 \leq \dim_{[E]} (M_L) \leq \dim \Ext^1_X(E,E)_0,
\] 
and, when $\dim \Ext^2_X(E,E)_0 = 0$ for all $[E] \in M_L$, then $M_L$ is smooth. Note that if $b_1(X) = 0$, then $M_L = M$.

We end with a series of important observations. 

\begin{exercise}
Let $X = C$ be a smooth projective curve. Prove that $M_C^H(r,c_1)$ is smooth. 
\end{exercise}

\begin{exercise} \label{exc:modulionK3}
Let $X = S$ be a $K3$ surface. Prove that $M$ is smooth. \emph{Hint: Serre duality.} 
\end{exercise}

\begin{exercise} \label{exc:moduliondelPezzo}
Let $X = S$ be a del Pezzo surface. Prove that $M$ is smooth. \emph{Hint: For $[E] \in M$, prove that that $\Hom_S(E,E \otimes K_S) = 0$ (see \cite[Proposition 1.2.7]{HL}) and use Serre duality.} 
\end{exercise}

\begin{exercise} \label{exc:rank1tf}
\hfill
\begin{enumerate}
\item[(a)] Let $E$ be a torsion free sheaf on a smooth variety $X$. Then the natural map $E \to E^{**}$, where $E^* := \hom_X(E,\O_X)$, is injective and $E^{**} / E$ is supported in codimension $\geq 2$. For $X = S$ a smooth projective surface, use this fact to deduce that if $E$ is a rank 1 torsion free sheaf on $X$, we have $E \cong I_{Z/S} \otimes L$, where $[L] \in \Pic(S)$ and $I_{Z/S} \subset \O_S$ is the ideal sheaf of a 0-dimensional subscheme $Z \subset S$.
\item[(b)] For $X = S$ a smooth projective surface and $Z \subset S$ a 0-dimensional subscheme, denote by $I := I_{Z/S} \subset \O_S$ the corresponding ideal sheaf. For any line bundle $K$ on $S$, show that $\Hom_S(I,I \otimes K)_0 = 0$. Use this to show that $\Ext^2_S(I,I)_0 = 0$. Use (a) to conclude that $M_S^H(1,L,c_2)$ is a smooth projective variety. \emph{Hint: Let $U = S \setminus Z$. Use torsion-freeness of $I$ to deduce that the restriction $\Hom_S(I,I \otimes K) \rightarrow \Hom_S(I|_U, (I \otimes K)|_{U}) = \Gamma(U,K|_U)$ is an injection. Then use that, since $\mathrm{codim}(Z) = 2$, the restriction $\Gamma(S,K) \to \Gamma(U,K|_U)$ is an isomorphism.}
\end{enumerate}
\end{exercise}

\subsection{Hilbert schemes of points} \label{sec:hilb}

Let $X$ be a smooth quasi-projective variety of dimension $d$. For any $n \geq 0$, we denote by 
\[
X^{[n]} = \big\{ Z \subset X \, : \, Z \textrm{ is } 0\textrm{-dimensional } \dim H^0(X,\O_Z) = n \big\},
\]
the Hilbert scheme parametrizing 0-dimensional subschemes of length $n$ on $X$. Think of these as collections of $n$ points on $X$ that may cluster together and acquire non-trivial scheme structure. 
The precise statement, due to Grothendieck \cite{Gro}, is that there exists a quasi-projective scheme representing a natural moduli functor (involving appropriate flat families of closed schemes relative to a base scheme). Moreover, if $X$ is projective, then $X^{[n]}$ is projective. One difference compared to moduli of sheaves is that $X^{[n]}$ is a fine moduli space --- in particular there exists a universal subscheme $\cZ \subset X^{[n]} \times X$ satisfying (in particular)
\[
\cZ|_{\{Z\} \times X} = Z.
\]
It is well-known that the Zariski tangent space to $Z \subset X$ is given by \cite{Leh1}
\[
T_{X^{[n]}}|_Z \cong \Hom_X(I_{Z/X},\O_Z),
\]
where $I_{Z/X} \subset \O_X$ denotes the ideal sheaf of $Z \subset X$.

For background on the foundational aspects of Hilbert schemes, we refer to \cite{Leh1, HL}. Instead, we will mostly focus on intersection-theoretic properties.

For $X$ a smooth \emph{projective} variety of dimension $d$, we relate $X^{[n]}$ to a moduli space of rank 1 torsion free sheaves as follows. Let $c_i = 0$ for $i < d$ and $c_d := (-1)^{d} (d-1)! n$ (i.e.~$\ch = (1,0, \ldots, 0,-n)$), then Exercise \ref{exc:rank1tf} implies that for $d>1$ there is a bijection of sets (at the level of $\C$-valued points)
\begin{equation} \label{eqn:isoMandHilb}
M_X(1,\O_X,0, \ldots, 0,c_d) \to X^{[n]}, \quad [I_{Z/X}] \mapsto Z. 
\end{equation}
Note that there is no dependence on ample divisor $H$, since stability is automatic for $r=1$. Also note that we crucially used $d>1$. In fact, this map underlies an isomorphism of schemes.\footnote{This is not a trivial statement since it involves comparing two quite different moduli functors. The isomorphism follows from \cite[Lemma B.5.6]{KPS} or (a slight adaptation of) the arguments in \cite[Section 2]{PP} as spelled out in \cite[Section 3.3]{BKP}. MK thanks A.~Ricolfi for pointing out the first reference.} At the level of Zariski tangent spaces, we obtain an interesting isomorphism
\[
\Ext^1(I_{Z/X},I_{Z/X})_0 \cong \Hom_X(I_{Z/X},\O_Z),
\]
which is induced by the short exact sequence $0 \to I_{Z/X} \to \O_X \to \O_Z \to 0$.

Back to a smooth quasi-projective variety $X$ of dimension $d$, we denote by $$\Sym^n(X) := X^{\times n} / \mathfrak{S}_n$$ the $n$th symmetric product of $X$, where $\mathfrak{S}_n$ acts on $X^{\times n}$ by permuting the factors. Then there exists a \emph{proper} morphism
\[
\mathsf{hc} \colon X^{[n]} \to \Sym^n(X), \quad Z \mapsto \sum_{P \in Z} \mu_P(Z) \cdot \{P\},
\]
where $\mu_P(Z) = \dim_{\C}(\O_{Z,P})$ denotes the multiplicity of $Z$ at $P$. This map is known as the Hilbert--Chow morphism. Here are some interesting facts:
\begin{itemize}
\item For $d=1$, i.e.~$X=C$ is a smooth curve, $\mathsf{hc}$ is an isomorphism and $\Sym^n(C)$ is smooth (essentially by the fundamental theorem on symmetric polynomials).
\item For $d=2$, i.e.~$X=S$ is a smooth surface, $\Sym^n(S)$ is generally singular and $\mathsf{hc}$ is a (crepant) resolution. In particular, $S^{[n]}$ is smooth of dimension $2n$. Granted the isomorphism \eqref{eqn:isoMandHilb}, smoothness follows from Exercise \ref{exc:rank1tf}.
\item For $d \geq 3$, $X^{[n]}$ is generally singular. (In fact, $\bigcup_n (\C^{16})^{[n]}$ is in some sense arbitrarily singular by a result of Jelisiejew \cite{Jel}.) However, it turns out that $X^{[n]}$ has a special structure for $d=3$ and $4$, which forms the subject of Donaldson--Thomas theory (e.g.~\cite{Tho1, BJ, OT} and references therein).
\end{itemize}

One can construct interesting tautological vector bundles on $X^{[n]}$ as follows. Let $E$ be a rank $r$ vector bundle on $X$ and consider the projections
\begin{displaymath}
\xymatrix
{
& \cZ \ar_{q}[dl] \ar^{p}[dr] & \\
X & & X^{[n]}
}
\end{displaymath}
Viewing $E$ as a locally free sheaf, one can define 
\[
E^{[n]} := p_* q^* E,
\]
which is a locally free sheaf of rank $rn$ on $X^{[n]}$. 

\begin{exercise}
Prove the last statement using Grauert's theorem \cite[Corollary III.12.9]{Har}. 
\end{exercise}

In terms of algebraic vector bundles, $E^{[n]}$ is a rank $rn$ vector bundle with fibre 
\[
E^{[n]}|_Z \cong H^0(X,E|_Z).
\]

We are interested in the case $X = S$ is a smooth projective surface. Then we have the following powerful theorem \cite{EGL}.
\begin{theorem}[Ellingsrud--G\"ottsche--Lehn] \label{thm:EGL}
Fix $n \geq 1$, let $S$ be a smooth projective surface, and let $E_1, \ldots, E_{\ell}$ be vector bundles on $S$ of ranks $r_1, \ldots, r_{\ell}$ respectively. Let $P$ be a polynomial expression, with rational coefficients, in the Chern classes of $E_i^{[n]}$ and $T_{S^{[n]}}$. Then there exists a polynomial expression $\widetilde{P}$, with rational coefficients, in the ranks $r_i$ and Chern classes of $E_i$ and $T_S$ such that
\[
\int_{S^{[n]}} P  = \int_S \widetilde{P}.
\]
Moreover, the polynomial expression $\widetilde{P}$ \emph{only} depends on $P$.
\end{theorem}

This result is called the universality theorem of Ellingsrud--G\"ottsche--Lehn. The type of expressions we have in mind for $P$ are
\begin{equation} \label{eqn:examplesP}
c_{2n}(T_{S^{[n]}}), \quad c_n(T_{S^{[n]}})^2 + \frac{1}{2} c_n(L^{[n]}) c_n(T_{S^{[n]}}), \quad \ldots
\end{equation}
By the Hopf index theorem, the first insertion calculates the topological Euler characteristic $e(S^{[n]}) = \sum_i (-1)^i b_i(S^{[n]})$ of the smooth variety $S^{[n]}$. 

Denote by $c(E) := 1 + c_1(E) + c_2(E) + \cdots$ the total Chern class of $E$. For $\ell=1$ and $E_1 = L$ a line bundle, some more examples of possible choices of $P$ in Theorem \ref{thm:EGL} include
\begin{equation} \label{eqn:moreexamplesP}
\Big\{ \frac{1}{c(L^{[n]})} \Big\}_{2n}, \quad \Big\{ \ch(\det(L^{[n]})) \cdot \td(T_{S^{[n]}}) \Big\}_{2n}, \quad \Big\{ c_n(L^{[n]}) \frac{c(T_{S^{[n]}})}{c(L^{[n]})} \Big\}_{2n},
\end{equation}
where the denominators are expanded by the geometric series, and $\{-\}_{2n}$ takes the degree $4n$ part of $(-)$ viewed as an element of $H^{\mathrm{even}}(S^{[n]},\Q)$. 

\begin{exercise} \label{exc:Verlinde+SymC}
The first insertion of \eqref{eqn:examplesP} is called the top Segre class $s_{2n}(L^{[n]})$. It has interesting enumerative meaning and we refer to $\int_{S^{[n]}} s_{2n}(L^{[n]})$ as a \emph{Segre number} \cite{Leh2}. 
\begin{enumerate}
\item[(a)] Argue that the second insertion in \eqref{eqn:moreexamplesP} calculates $\chi(S^{[n]},\det(L^{[n]}))$. This is an example of a \emph{Verlinde number}.
\item[(b)] Suppose the linear system $|L|$ has a smooth curve $C$. Show that the third insertion in \eqref{eqn:moreexamplesP} calculates the topological Euler characteristic $e(\Sym^n(C))$. \emph{Hint: Let $s_C \colon \O_S \to \O_S(C)$ be the section defining $C \subset S$. For any $Z \subset S$, we have an element $s_C|_Z \in H^0(S,L^{[n]}|_Z)$ which vanishes if and only if $Z \subset C$ (scheme theoretically). This defines a tautological section $s \in H^0(S^{[n]}, L^{[n]})$ with scheme theoretical zero locus $\iota \colon Z(s) = \Sym^n(C) \subset S^{[n]}$ and its fundamental class (therefore) satisfies $\iota_*[\Sym^{n}(C)] = c_n(L^{[n]}) \cap [S^{[n]}]$. Also use the short exact sequence $0 \to T_{Sym^n(C)} \to T_{S^{[n]}}|_{\Sym^n(C)} \to N_{\Sym^n(C)/S^{[n]}} \to 0$, where $N_{\Sym^n(C)/S^{[n]}} \cong L^{[n]}|_{\Sym^n(C)}$.}\footnote{There is a much easier topological stratification argument which shows that $\sum_{n\geq 0} e(\Sym^n(C)) q^n = (1-q)^{-e(C)}$, which is known as MacDonald's formula (e.g.~see \cite[Appendix A.2]{BK} for a quick derivation).}
\end{enumerate}
\end{exercise}

\begin{exercise} 
(More work.) In the special case $n=2$, there is a description of $S^{[n]}$ in terms of blow-ups. Denote by $\Delta \subset \Sym^2(S)$ the image of the diagonal $\Delta \subset S \times S \to \Sym^2(S)$. Then $S^{[2]} \cong \mathrm{Bl}_{\Delta}(\Sym^2(S))$. Moreover, there is a Cartesian diagram
\begin{displaymath}
\xymatrix
{
\mathrm{Bl}_{\Delta}(S \times S) \ar[r] \ar[d] & \mathrm{Bl}_{\Delta}(\Sym^2(S)) \ar[d] \\
S \times S \ar[r] & \Sym^2(S).
}
\end{displaymath}
Using general facts from intersection theory on blow-ups of a smooth projective variety along a smooth subvariety \cite[Section 15.4]{Ful}, derive Theorem \ref{thm:EGL} for $n=2$. 
\end{exercise}

The proof of Theorem \ref{thm:EGL} uses the nested Hilbert scheme $S^{[n,n+1]}$ parametrizing flags of 0-dimensional subschemes $Z_0 \subset Z_1 \subset S$ with $Z_0$ of length $n$ and $Z_1$ of length $n+1$. This leads to a diagram
\begin{displaymath}
\xymatrix
{
& S^{[n,n+1]} \ar[dl] \ar[dr] & \\
S^{[n+1]} & & S^{[n]} \times S
}
\end{displaymath}
where the map to $S$ is defined as the support of $Z_1 \setminus Z_0$. Remarkably, $S^{[n,n+1]}$ is smooth and Theorem \ref{thm:EGL} is proved by transporting the calculation from $S^{[n+1]}$ to $S^{[n]} \times S$, via the above correspondence, and then recursively all the way down to $S^{\times (n+1)}$. In fact, the recursive procedure is constructive and can be implemented on a computer, though the method explained later in Section \ref{sec:locHilb} is often a much more efficient way for calculating intersection numbers on $S^{[n]}$.

We will now discuss a remarkable corollary to Theorem \ref{thm:EGL}. In the setting of Theorem \ref{thm:EGL}, suppose we have a polynomial expression $P_n$ in the Chern classes of $E_1, \ldots E_{\ell}, T_{S^{[n]}}$ on $S^{[n]}$ for each $n \geq 1$. Then we can form the generating series
\[
\mathsf{Z}_{S,E_1, \ldots, E_{\ell}}(q) = 1 + \sum_{n=1}^{\infty} q^n \int_{S^{[n]}} P_n.
\]
Up to now, we (tacitly) worked with connected smooth projective surfaces $S$. However, Theorem \ref{thm:EGL} also holds for \emph{disconnected} smooth projective surfaces. We assume that the collection $\{P_n\}_{n=1}^{\infty}$ satisfies the following \emph{multiplicative property}: for any (possibly disconnected) smooth projective surfaces $S', S''$ with vector bundles $E_1', \ldots, E_\ell'$ of ranks $r_1, \ldots, r_{\ell}$ on $S'$, and vector bundles $E_1'', \ldots, E_\ell''$ of ranks $r_1, \ldots, r_{\ell}$ on $S''$, we have
\begin{equation} \label{eqn:multprop}
\mathsf{Z}_{S' \sqcup S'',E_1' \sqcup E_1'', \ldots, E_{\ell}' \sqcup E_{\ell}''}(q) = \mathsf{Z}_{S',E_1', \ldots, E_{\ell}'}(q) \cdot \mathsf{Z}_{S'',E_1'', \ldots, E_{\ell}''}(q).
\end{equation}
\begin{corollary} \label{cor:univ}
In the above setup, suppose the multiplicative property \eqref{eqn:multprop} holds. Then there exist formal power series 
\[
A(q), B(q), \{ C_i(q) \}_{i=1}^{\ell}, \{D_i(q)\}_{i=1}^{\ell}, \{E_{ij}(q)\}_{1 \leq i \leq j \leq \ell}, \quad \mathrm{in} \quad 1+q\Q[[q]] 
\]
\emph{only} depending on the choice of $\{P_n\}_{n=1}^{\infty}$ and $r_1, \ldots, r_{\ell}$ with the following property. For any smooth projective surface $S$ and vector bundles $E_1, \ldots, E_{\ell}$ on $S$ of ranks $r_1, \ldots, r_{\ell}$ respectively, we have\footnote{For a formal power series $A(q) \in 1+ q\Q[[q]]$ and $n \in \Z$, we have $A(q)^n := \exp(n \log(A(q)))$. For notational convenience, we omit the degree map in the powers, i.e., $c_1(E_i)c_1(E_j)$ stands for the intersection number $\int_S c_1(E_i)c_1(E_j)$. Later, for line bundles $L$, we will also often write $L^2$ instead of $c_1(L)^2$, $L K_S$ for $-c_1(L) c_1(T_S)$ etc.}
\[
\mathsf{Z}_{S,E_1, \ldots, E_{\ell}}(q) = A(q)^{c_1(T_S)^2} B(q)^{c_2(T_S)} \prod_i C_i(q)^{c_1(E_i) c_1(T_S)} \prod_i D_i(q)^{c_2(E_i)} \prod_{i \leq j} E_{ij}(q)^{c_1(E_i)c_1(E_j)}. 
\]
\end{corollary}

An informal way of stating this (which can be made precise) is that $\mathsf{Z}_{S,E_1, \ldots, E_{\ell}}(q)$ is a cobordism invariant for tuples $(S,E_1, \ldots, E_\ell)$. This result is extremely powerful; it implies that we ``only'' need to calculate $\mathsf{Z}_{S,E_1, \ldots, E_{\ell}}(q)$ on any \emph{finite} collection of tuples $(S,E_1, \ldots, E_{\ell})$ for which the corresponding vectors of Chern numbers
\begin{equation} \label{eqn:Chernvector}
\Big( c_1(T_S)^2, c_2(T_S), \{c_1(E_i)c_1(T_S)\}_{i=1}^{\ell}, \{c_2(E_i)\}_{i=1}^{\ell}, \{c_1(E_i)c_1(E_j)\}_{1 \leq i \leq j \leq \ell} \Big)
\end{equation}
form a $\Q$-basis --- taking $\log$ and performing Gaussian elimination then gives $A(q)$, $B(q)$, $C_i(q)$, $D_i(q)$, $E_{ij}(q)$. 

For example, suppose $\ell=1$ and $E_1 = L$ a line bundle, one could take the basis of $\Q^4$ determined by the four pairs
\[
(S,L) = (\PP^2,\O), (\PP^2,\O(1)), (\PP^2,\O(2)), (\PP^1 \times \PP^1,\O).
\]

\begin{exercise} \label{exc:listmult}
Show for the first insertion in \eqref{eqn:examplesP}, and all insertions in \eqref{eqn:moreexamplesP}, that the corresponding generating series satisfy the multiplicative property \eqref{eqn:multprop} by using properties of Chern classes. For the first insertion in \eqref{eqn:examplesP}, prove the multiplicative property also by using properties of topological Euler characteristics such as $e(X \sqcup Y) = e(X) + e(Y)$ and $e(X \times Y) = e(X)e(Y)$. 
\end{exercise}

\begin{exercise} 
For a smooth projective curve $C$, define $$\mathsf{Z}_C(q) = \sum_{n = 0}^{\infty} e(\Sym^n(C)) q^n.$$
\begin{enumerate}
\item[(a)] Suppose $C \in |L|$ is a smooth curve in a linear system on a smooth projective surface $S$. Using Exercises \ref{exc:Verlinde+SymC}, \ref{exc:listmult}, and Corollary \ref{cor:univ}, prove that $\mathsf{Z}_C(q) = \mathsf{Z}_{\PP^1}(q)^{-L(L+K_S) / 2}$. \emph{Hint: Take} 
$$
(S,L) = (\PP^2,\O(1)), (\PP^2,\O(2)), (\PP^1 \times \PP^1, \O(1,0)), (\mathrm{Bl}_p(K3),\O(E)),
$$
\emph{where $\mathrm{Bl}_p(K3)$ denotes the blow-up of a K3 surface in a point $p$ and $E$ denotes the exceptional divisor --- note that each of the linear systems $|L|$ contains a $\PP^1$.}
\item[(b)] It is a well-known fact that $\Sym^n(\PP^1) \cong \PP^n$ (again, essentially by the elementary theorem on symmetric polynomials). Use this to show that $\mathsf{Z}_{\PP^1}(q) = (1-q)^{-2}$. Deduce MacDonald's formula $\mathsf{Z}_C(q) = (1-q)^{-e(C)}$ for all smooth curves in linear systems on smooth projective surfaces. As noted in the footnote to Exercise \ref{exc:Verlinde+SymC}, there is a much faster way to derive this formula (for any curve $C$).
\end{enumerate}
\end{exercise}

\subsubsection{Hilbert schemes of divisors.} \label{sec:Hilbdiv}

So far we have studied the Hilbert scheme of points on a smooth quasi-projective variety (mostly in the surface case). There are many other Hilbert schemes parametrizing higher dimensional closed subschemes $Z \subset X$. We discuss one such case. For a $d$-dimensional smooth projective variety $X$, fix an effective class $\delta \in H_{2d-2}(X)$. Then one can consider the Hilbert scheme
\[
\Hilb_{\delta}(X) = \{ D \subset X \, : \, D \textrm{ \ effective divisor \ such that } [D] = \delta \}
\]
parametrizing effective divisors $D \subset X$ with class $\delta$. There exists a proper morphism
\[
\mathsf{aj} \colon \Hilb_\delta(X) \to \Pic^{\delta}(X), \quad D \mapsto [\O_X(D)],
\]
where $\Pic^{\delta}(X) \leq \Pic(X)$ denotes the subgroup of line bundles $L$ with $c_1(L)$ Poincar\'e dual to $\delta$. This morphism is known as the Abel--Jacobi map, and its fibre over $[L] \in \Pic^\delta(X)$ is the linear system $|L|$. In the special case $b_1(X) = 0$, $\Hilb_\delta(X)$ is itself a linear system, thus smooth, and we sometimes write
\[
\Hilb_\delta(X) = |\delta|.
\]

\begin{exercise} 
For $X = C$, a smooth projective curve of genus $g$, consider the Abel-Jacobi map $\mathsf{aj} \colon \Sym^{\delta}(C) \to \Pic^\delta(C)$. Assume that $\delta > 2g-2$. Deduce that all fibres of $\mathsf{aj}$ are projective spaces of the same dimension.\footnote{Indeed, in this case $\Sym^\delta(C)$ is a projective bundle over $\Pic^\delta(C)$.} 
\end{exercise}


Similar to the case of Hilbert schemes of points, the Zariski tangent space to $D \subset \Hilb_\delta(X)$ is given by $\Hom_X(I_{D/X},\O_D)$, which (as expected) simplifies to
\[
\Hom_X(\O_X(-D),\O_D) \cong H^0(X,\O_D(D)) \cong H^0(D,N_{D/X}),
\]
where $N_{D/X}$ denotes the normal bundle to $D \subset X$.

\subsection{Moduli of framed sheaves} \label{sec:moduliframed}

We end this lecture by giving an algebro-geometric description of the Uhlenbeck compactification in Y.~Tanaka's lecture in the case of $S^4$ and by giving a (semi-small) resolution of its singularities known as the moduli space of framed sheaves on $\PP^2$. These moduli spaces can be seen as generalisations of $(\C^2)^{[n]}$. The canonical reference, with many more details, on these moduli spaces is Nakajima's book \cite{Nak} (see also \cite{NY}).

Using homogeneous coordinates $[z_0:z_1:z_2]$ on the (complex) projective plane $\PP^2$, we denote by $\ell_{\infty} = Z(z_0) \subset \PP^2$ the line at infinity. We consider the moduli space
\[
M(r,n) = \Big\{ (E,\Phi) \, : \, \textrm{ rank } r \textrm{ torsion free sheaf } E \textrm{ on } \PP^2, c_2(E)=n, \Phi \colon E|_{\ell_{\infty}} \stackrel{\cong}{\rightarrow} \O_{\ell_\infty}^{\oplus r}    \Big\},
\]
where two pairs $(E_i,\Phi_i)$ are considered isomorphic when there exists an isomorphism $\psi \colon E_1 \to E_2$ such that $\Phi_1 =  \Phi_2 \circ \psi|_{\ell_\infty}$. This is a \emph{smooth} quasi-projective variety of dimension $2rn$ representing a corresponding moduli functor (in particular, it has a universal family). As a consequence of the framing isomorphism $\Phi$, we have $c_1(E) = 0$ and $E$ is locally free of rank $r$ in a Zariski open neighbourhood of $\ell_{\infty}$. 

\begin{exercise} 
 Prove the statements in the last sentence.\footnote{MK thanks W.~Lim for posing this exercise.} 
 \end{exercise}

\begin{exercise} \label{exc:rk1frmoduli}
Prove at the level of $\C$-points that we have a bijection $$(\C^2)^{[n]} \to M(1,n), \quad Z \mapsto (I_{Z/\PP^2}, \Phi)$$ by embedding $Z \subset \C^2 = \PP^2 \setminus \ell_{\infty}$ into $\PP^2$ and taking $\Phi$ to be the restriction to $\ell_\infty$ of the inclusion $I_{Z / \PP^2} \subset \O_{\PP^2}$.\footnote{This map indeed corresponds to an isomorphism of varieties.} 
\end{exercise}

We denote by $M_0^{\mathrm{reg}}(r,n) \subset M(r,n)$ the open subset parametrizing pairs $(E,\Phi)$ with $E$ locally free. At the level of sets, we can now define
\[
M_0(r,n) := \bigsqcup_{k=0}^n M_0^{\mathrm{reg}}(r,n-k) \times \Sym^{k}(\C^2).
\]
It turns out that that $M_0(r,n)$ can be given the structure of an affine algebraic variety and there exists a projective (affinization) morphism
\[
\mathsf{aff} \colon M(r,n) \to M_0(r,n), \quad [(E, \Phi)] \mapsto ([(E^{**},\Phi)],\mathrm{Supp}(E^{**} / E)),
\]
where $\mathrm{Supp(-)}$ denotes the set-theoretical support. The affine space $M_0(r,n)$ coincides with the Uhlenbeck partial compactification for $S^4$ discussed in Y.~Tanaka's lectures. 

\begin{exercise} 
Show that for $r=1$, and under the identification of Exercise \ref{exc:rk1frmoduli}, $\mathsf{aff}$ corresponds to the Hilbert--Chow morphism. 
\end{exercise}

\begin{exercise} 
Prove that the affinization morphism $\mathsf{aff}$ is semi-small. A proper morphism of varieties $f \colon X \to Y$ is called semi-small when $X$ is smooth and 
\[
\mathrm{codim}(\{ y \in Y \, : \, \dim f^{-1}(\{y\}) \geq r \}) \geq 2r, \quad \textrm{for all } r>0.
\]
\end{exercise}

\section{Lecture 2: Atiyah--Bott localization for moduli of sheaves}

\subsection{Atiyah--Bott localization} \label{sec:ABintro}

n this section, we give a brief introduction to the Atiyah--Bott localization formula, which is a powerful tool for calculating intersection numbers on smooth varieties with a torus action. A good reference for Atiyah--Bott localization in algebro-geometric language is \cite{EG} (though we will work in equivariant (co)homology instead of equivariant Chow groups). 

Consider the algebraic group $T = (\C^*)^r$, for some $r>0$, which we refer to as the algebraic torus of rank $r$. Suppose $T$ acts on a smooth variety $X$ of dimension $d$. 

\begin{definition}
A $T$-equivariant vector bundle on $X$ is a pair $(E,\Phi)$ on $X$, where $E$ is a vector bundle on $X$ and $\Phi \colon T \times E \to E$ is an action such that
\begin{displaymath}
\xymatrix
{
T \times E \ar^{\Phi}[r] \ar[d] & E \ar[d] \\
T \times X \ar[r] & X
}
\end{displaymath}
commutes and, for any $p \in X$, the induced map on fibres $\Phi_p \colon E|_p \to E|_{t \cdot p}$ is $\C$-linear.

Denote the action of $T$ on $X$ by $\sigma \colon T \times X \to X$ and the projection by $p_2 \colon T \times X \to X$. Furthermore, let $p_{23} \colon T \times T \times X \to T \times X$ denote projection on the last two coordinates and let $\mu \colon T \times T \to T$ be the multiplication map. A $T$-equivariant coherent sheaf on $X$ is a pair $(E,\Phi)$ where $E$ is a coherent sheaf on $X$ and $\Phi \colon \sigma^* E \to p_2^* E$ is an isomorphism satisfying the cocycle identity
\[
p_{23}^* \Phi \circ (\mathrm{id}_T \times \sigma)^* \Phi = (\mu \times \mathrm{id}_X)^* \Phi.
\]
We refer to $\Phi$ as a $T$-equivariant structure on $E$. We often suppress $\Phi$ from the notation and simply say that $E$ is a $T$-equivariant coherent sheaf on $X$.

There are obvious notions of morphisms between $T$-equivariant vector bundles and sheaves.
\end{definition}

\begin{exercise}
\hfill
\begin{enumerate}
\item[(a)] Let $(E,\Phi)$ be a $T$-equivariant coherent sheaf on $X$. For any $t \in T$ define $\lambda_t := i_{t}^* \Phi \colon t^* E \to E$, where $i_t \colon X \hookrightarrow T \times X$ with $i_t(p) = (t,p)$ denotes the inclusion and we denote the map $X \to X, p \mapsto \sigma(t,p)$ by $t$. Show that $\lambda_{st} = \lambda_t \circ t^* \lambda_s$ for all $s,t \in T$.
\item[(b)] To a rank $r$ vector bundle $E$ on $X$, one can associate the rank $r$ locally free sheaf $\mathcal{S}^*_{E/X}$ on $X$, where $\mathcal{S}_{E/X}$ is the sheaf of sections of $E$ and where $\mathcal{S}^*_{E/X}$ denotes its dual. This describes an equivalence of categories between the category of rank $r$ vector bundles on $X$ and rank $r$ locally free sheaves on $X$ \cite[Exercise II.5.5.18]{Har}. Show that this assignment extends to an equivalence of categories between the category of $T$-equivariant rank $r$ vector bundles on $X$ and $T$-equivariant rank $r$ locally free sheaves on $X$.
\end{enumerate}
\end{exercise}

We now discuss the special important case when $T$ acts \emph{trivially} on $X$. The character group $X(T)$ of $T$ is the abelian group of regular group homomorphisms $T \to \C^*$. It is isomorphic to $\Z^r$ as follows: to $w=(w_1, \ldots, w_r) \in \Z^r$ we associate the character
\[
\mathfrak{t}^w = \mathfrak{t}_1^{w_1} \cdots \mathfrak{t}_r^{w_r} \colon T \to \C^*, \quad (t_1, \ldots, t_r) \mapsto t_1^{w_1} \cdots t_r^{w_r}.
\]
For any vector bundle $E$ on $X$, one can consider the $T$-equivariant vector bundle $E \otimes \mathfrak{t}^w$, where the $T$-equivariant structure is defined by fibre-wise multiplication by the character $\mathfrak{t}^w$. Then any $T$-equivariant vector bundle $E$ on $X$ decomposes
\begin{equation} \label{eqn:eigendeco}
E = \bigoplus_{w \in X(T)} E_w \otimes \mathfrak{t}^w,
\end{equation}
where only finitely many many ``eigen vector bundles'' $E_w$ are non-zero. Moreover, we define
\[
E^f := E_0, \quad E^m := \bigoplus_{w \neq 0} E_w \otimes \mathfrak{t}^w
\]
as the fixed and moving parts of $E$ respectively.

We denote by 
\[
H_T^*(X,\Q), \quad H_*^T(X,\Q)
\]
the $T$-equivariant (co)homology groups of $X$ with rational coefficients. We also have a $T$-equivariant Poincar\'e duality isomorphism $H_T^{i}(X,\Q) \cong H^T_{2d-i}(X,\Q)$.\footnote{In the non-compact case, we use $T$-equivariant Borel--Moore homology.} To a $T$-equivariant coherent sheaf $E$ on $X$, one can associate its $T$-equivariant Chern classes\footnote{For Chern classes, we can use integer coefficients too.}
\[
c_i^T(E) \in H_T^{2i}(X,\Q).
\]
If $E$ is a $T$-equivariant rank $r$ locally free sheaf, we have $c_i^T(E) = 0$ for all $i>r$ as in the non-equivariant case. Furthermore, we define the Euler class of $E$ as 
\[
e(E) := c_r^T(E).
\]

An important r\^ole  is played by $X = \mathrm{pt}$ (a reduced point), in which case we have
\[
H_T^*(\pt,\Q) \cong \Q[\epsilon_1, \ldots, \epsilon_r],
\]
where $\epsilon_i = c_1^{T}(\C \otimes \mathfrak{t}_i)$. Here $\epsilon_i$ has cohomological degree 2. Note in particular that $H^{i}_T(\pt,\Q)$ does not vanish for $i > \dim_{\R}(\pt) =0$.

The first fact we would like to mention is that for \emph{trivial} $T$-action on $X$, we have
\[
H^T_*(X,\Q) \cong H_*(X,\Q) \otimes_{\Q} H_T^*(\pt,\Q) \cong H_*(X,\Q)[\epsilon_1, \ldots, \epsilon_r].
\]
We note that for $E$ as in \eqref{eqn:eigendeco} we have
\[
e(E) = \prod_w e(E_w \otimes \mathfrak{t}^w) = \prod_{w=(w_1, \ldots, w_r)} \Big( c_{r_w}(E_w) + c_{r_w-1}(E_w)\Big(\sum_i w_i \epsilon_i \Big) + \cdots  + \Big(\sum_i w_i \epsilon_i\Big)^{r_w} \Big),
\]
where we take the product over all $w$ for which $E_w$ has positive rank $r_w>0$.

The second fact we would like to mention is the following version of Atiyah--Bott localization: the following map is an isomorphism\footnote{It suffices to tensorize by $-\otimes_{\Q[\epsilon_1, \ldots, \epsilon_r]} \Q[\epsilon_1^{\pm 1}, \ldots, \epsilon_r^{\pm 1}]$ but we prefer working in a slightly bigger ring.}
\[
\iota_* \colon H_*(X^T,\Q) \otimes_{\Q} \Q(\epsilon_1, \ldots, \epsilon_r) \stackrel{\cong}{\rightarrow} H_*^T(X,\Q) \otimes_{\Q[\epsilon_1, \ldots, \epsilon_r]} \Q(\epsilon_1, \ldots\epsilon_r) =: H_*^T(X,\Q)_{\mathrm{loc}}.
\]
By a result of Iversen \cite{Ive}, the fact that $X$ is smooth implies that the fixed point locus $$X^T = \{p \in X \, : \, t\cdot p = p \textrm{ for all } t \in T\}$$ is also smooth. We denote its connected components by $\iota \colon X_i \hookrightarrow X$. Then \cite{AB, EG}
\[
\sum_i \iota_* \frac{[X_i]}{e(N_{X_i/X})} = [X],
\]
where $N_{X_i / X}$ denotes the normal bundle of $X_i$ in $X$. Here, it is important to note that $T_X$ has a natural $T$-equivariant structure and 
\[
T_X|_{X_i}^f \cong T_{X_i}, \quad T_X|_{X_i}^m \cong N_{X_i / X}
\]
so that $e(N_{X_i/X})$ is an invertible element. 

In particular, suppose one is interested in a (non-equivariant) intersection number $\int_X \omega$ for some $\omega \in H^*(X,\Q)$. Then one can apply the Atiyah--Bott localization formula by \emph{choosing} a lift of $\omega$ to $T$-equivariant cohomology $H^*_T(X,\Q)$ and then using
\begin{equation} \label{eqn:ABloc}
\int_X \omega = \sum_i \int_{X_i} \frac{\omega|_{X_i}}{e(N_{X_i/X})} \Big|_{\epsilon_1 = \cdots = \epsilon_r=0}.
\end{equation}
It is crucial to observe that the individual summands on the right hand side generally have poles in the equivariant parameters $\epsilon_1, \ldots, \epsilon_r$, but these poles cancel after summing all terms so that the specialization $\epsilon_1 = \cdots  = \epsilon_r = 0$ is well-defined. In applications of interest to us, $\omega$ is always a polynomial in Chern classes of vector bundles $E$ on $X$. Then equivariant lifts of $\omega$ are conveniently obtained by choosing $T$-equivariant structures on the $E$ and replacing Chern classes by $T$-equivariant Chern classes. 

Sometimes, we are interested in settings where $X$ is not compact. Then, for any $T$-equivariant cohomology class $\omega$ on $X$, we \emph{define} $\int_X \omega$ by the right hand side of \eqref{eqn:ABloc}, but without setting $\epsilon_1 = \cdots = \epsilon_r = 0$. 

\begin{exercise} \label{eqn:eXeXT}
Let $X$ be a $d$-dimensional smooth projective variety with an action by an algebraic torus $T$.
\begin{enumerate}
\item[(a)] By the Hopf index theorem, we have $e(X) = \int_X c_d(T_X)$. Use the Atiyah--Bott localization formula to prove that $e(X) = e(X^T)$.
\item[(b)] Suppose $X$ has finitely many $T$-orbits. Prove that $e(X) = e(X^T)$ by stratifying $X$ into orbits. 
\end{enumerate}
\end{exercise}

One can gain experience with applications of the Atiyah--Bott localization formula by carrying out calculations on $X = \PP^2$ on which we have an action by $T = (\C^*)^2$ as follows
\begin{equation} \label{eqn:actiononpts}
(t_1,t_2) \cdot [p_0:p_1:p_2] = [p_0:t_1^{-1} p_1 : t_2^{-1} p_2]. 
\end{equation}
Then
\[
(\PP^2)^T = \{ [1:0:0], [0:1:0], [0:0:1] \}
\]
thus the right hand side of \eqref{eqn:ABloc} is just a sum over three fixed points.

Denote by $\C^2 \cong U_\alpha = D(z_\alpha) \subset \PP^2$ the standard coordinate charts, which are $T$-invariant. Using coordinates $x = z_1 / z_0$, $y = z_2 / z_0$ on $U_0$, the action on the coordinate functions $x,y$ on $U_0$ is\footnote{For the action \eqref{eqn:actiononpts} on points of $\PP^2$, the action on the corresponding coordinate \emph{functions} $z_1/z_0, z_2/z_0$ has an extra inverse, i.e., $(t_1,t_2)\cdot(z_1/z_0,z_2/z_0) = (t_1 z_1/z_0,t_2 z_2/z_0)$.}
\begin{equation} \label{eqn:chartU0}
(t_1,t_2) \cdot x = t_1x, \quad (t_1,t_2) \cdot y = t_2y.
\end{equation}
On the other hand, using coordinates $x = z_2 / z_1$, $y = z_0 / z_1$ on $U_1$, the action on the coordinate functions $x,y$ on $U_1$ is
\begin{equation} \label{eqn:chartU1}
(t_1,t_2) \cdot x = t_1^{-1}t_2 x, \quad (t_1,t_2) \cdot y = t_1^{-1} y.
\end{equation}
Finally, using coordinates $x = z_0 / z_2$, $y = z_1 / z_2$ on $U_2$, the action on the coordinate functions $x,y$ on $U_2$ is
\begin{equation} \label{eqn:chartU2}
(t_1,t_2) \cdot x = t_2^{-1} x, \quad (t_1,t_2) \cdot y = t_1 t_2^{-1} y.
\end{equation}
Using the fact that on each chart $U_\alpha$, we have a unique fixed point $p_\alpha = (0,0)$, this gives us the $T$-representation of 
\[
T_{\PP^2}|_{p_\alpha} = \C \cdot \frac{\partial}{\partial x}\Big|_{(0,0)} \oplus \C \cdot \frac{\partial}{\partial y}\Big|_{(0,0)}.
\]
E.g.~for $\alpha=1$, we have $T_{\PP^2}|_{p_1} = \C \cdot \mathfrak{t}_1 \mathfrak{t}_2^{-1} \oplus \C \cdot \mathfrak{t}_1$. Expressions such as these can be viewed as elements of $K_0^T(\pt) \cong \Z[\t_1^{\pm 1},\t_2^{\pm 1}]$ --- the $T$-equivariant Grothendieck group of coherent sheaves on the point, and we simply write
\[
T_{\PP^2}|_{p_1} = \mathfrak{t}_1 \mathfrak{t}_2^{-1} + \mathfrak{t}_1.
\]
\begin{exercise}
\hfill
\begin{enumerate}
\item[(a)] Use Atiyah--Bott localization to determine the Chern numbers $\int_X c_2(T_X)$, $\int_X c_1(T_X)^2$ for $X = \PP^2$.
\item[(b)] Use Atiyah--Bott localization to determine the Chern numbers $\int_X c_3(T_X)$, $\int_X c_1(T_X)c_2(T_X)$, $\int_X c_1(T_X)^3$ for $X = \PP^3$.
\item[(c)] Check your calculations by determining the same Chern numbers using the Euler sequence $0 \to \O_{\PP^n} \to \O_{\PP^n}(1)^{\oplus (n+1)} \to T_{\PP^n} \to 0$.
\end{enumerate}
\end{exercise}

Finally, when $X$ is a smooth variety with $T = (\C^*)^r$-action, the structure sheaf $\O_X$ is naturally $T$-equivariant. Moreover, for $Y \subset X$ a $T$-invariant subscheme, the ideal sheaf $I_{Y/X} \subset \O_X$ is then also naturally $T$-equivariant. 

For example, consider the line bundle $\O_{\PP^2}(-d)$ for some $d \geq 1$. Then it has a natural $T$-equivariant structure when viewed as the ideal sheaf of $Y = Z(z_2^d) \subset \PP^2$. Concretely, $\O_{\PP^2}(-1)|_{U_\alpha} = \O_{U_\alpha}$ for $\alpha=2$. However, for $\alpha=0,1$, $\O_{\PP^2}(-1)|_{U_\alpha}$ corresponds to (respectively) the ideal
\[
(y^d) \subset \C[x,y], \quad (x^d) \subset \C[x,y].
\]
Therefore 
\[
\O_{\PP^2}(-d)|_{p_0} = \C \cdot \mathfrak{t}_2^d, \quad \O_{\PP^2}(-d)|_{p_1} = \C \cdot \mathfrak{t}_1^{-d} \t_2^d, \quad  \O_{\PP^2}(-d)|_{p_2} = \C. 
\]
Furthermore, we can turn $\O_{\PP^2}(d)$ into a $T$-equivariant line bundle by taking the dual $T$-equivariant structure. In this way, we have put a (non-unique!) $T$-equivariant structure on all line bundles on $\PP^2$. 

\begin{exercise}
\hfill
\begin{enumerate}
\item[(a)] Show that $\int_{\PP^2} c_1(\O(d))^2 = d^2$ by using Atiyah--Bott localization. 
\item[(b)] We can also put another $T$-equivariant structure on $\O_{\PP^2}(-2)$ by viewing it as the ideal sheaf of $Y = Z(z_1z_2) \subset \PP^2$. Repeat (a) for $d=2$ by using this alternative equivariant structure. 
\end{enumerate}
\end{exercise}

\subsection{Localization on Hilbert schemes} \label{sec:locHilb}

We now turn to our main topic of interest: applying Atiyah--Bott localization to computing intersection numbers on $S^{[n]}$ for smooth surfaces $S$ with a torus action. Although our focus is on $S = \PP^2$ and $S = \PP^1 \times \PP^1$, the methods we present work for any smooth projective toric surface $S$. As a consequence, we will obtain an effective algorithm for determining the universal functions in Corollary \ref{cor:univ} (at least in principle). 

We start with the case $S = \C^2$ with standard action
\[
(t_1,t_2) \cdot (p_1,p_2) = (t_1^{-1} p_1,t_2^{-1} p_2), \quad \textrm{for all } (t_1,t_2) \in T = (\C^*)^2.
\]
This action lifts to $(\C^2)^{[n]}$ by sending an ideal $I \subset \C[x,y]$ to
\[
(t_1,t_2) \cdot I := \{f(t_1x,t_2y) \, : \, f \in I\}, \quad \textrm{for all } (t_1,t_2) \in T.
\]
The $T$-fixed ideals are therefore precisely the monomial ideals $I \subset \C[x,y]$ such that $\dim_{\C}(\C[x,y] / I) = n$ (the colength of $I$). For example, for $n=4$, the fixed points are
\[
(x^4,y), \quad (x^3,xy,y^2), \quad (x^2,y^2), \quad (y^3,xy,x^2), \quad (y^4,x).
\]
To a monomial ideal $I \subset \C[x,y]$ we can associate a subset $\lambda \subset \Z_{\geq 0}^2$ as follows: for all $(a,b) \in \Z_{\geq 0}^2$ we have
\[
(a,b) \in \lambda \leftrightarrow x^ay^b \notin I.  
\]
The  subset $\lambda$ satisfies the following property: if $(a,b) \in \lambda$, then $(a',b') \in \lambda$ for all integers $0 \leq a' \leq a$ and $0 \leq b' \leq b$ and $|\lambda| = n$. Such subsets of $\Z_{\geq 0}^2$ are called integer partitions of $n$. It is easy to see that the monomial ideals of $\C[x,y]$ of colength $n$ are in bijective correspondence with the integer partitions of $n$ (viewed as subsets of $\Z_{\geq 0}^2$ as described above). In conclusion, we have
\[
\Big( (\C^2)^{[n]} \Big)^T = \big\{ \lambda \subset \Z_{\geq 0}^2 \, : \,\textrm{integer partition satisfying } |\lambda| = n \big\},
\]
which is a \emph{reduced} 0-dimensional scheme since $(\C^2)^{[n]}$ is smooth (Section \ref{sec:hilb}) and therefore $((\C^2)^{[n]})^T$ is smooth (by Iversen's result). 

In the following exercise, and later in these lecture notes, we will often encounter the ($q$-expansion of the) Dedekind eta function and its normalized version
\[
\eta(q) = q^{\frac{1}{24}} \prod_{n=1}^{\infty} (1-q^n), \quad \overline{\eta}(q) = \prod_{n=1}^{\infty} (1-q^n). 
\]

\begin{exercise} \label{exc:Euler}
\hfill
\begin{enumerate}
\item[(a)] Show that the number of integer partitions of $n=1$,$2$,$3$,$4$,$5$,$6$,$7$ is $1$,$2$,$3$,$5$,$7$,$11$,$15$ respectively.
\item[(b)] Prove \emph{Euler's formula} $\sum_{\lambda} q^{|\lambda|} = \overline{\eta}(q)^{-1}$. \emph{Hint: Expand each factor $(1-q^n)^{-1}$ using the geometric series.}
\item[(c)] For any variety $X$ with action by an algebraic torus $T$, we have $e(X) = e(X^T)$ (see Exercise \eqref{eqn:eXeXT}).\footnote{Using cohomology with compact support, this equality also holds in the non-compact case.} Show that $\sum_{n=0}^{\infty} e((\C^2)^{[n]}) q^n = \overline{\eta}(q)^{-1}$.
\end{enumerate}
\end{exercise}

The description of the fixed point locus in terms of integer partitions readily extends to $S^{[n]}$ for $S = \PP^2$ and $\PP^1 \times \PP^1$ with their standard torus actions (and in fact to any smooth toric surface).  The data of a $T$-invariant ideal sheaf $I \subset \O_{\PP^2}$ of finite colength are equivalent to the data of three $T$-invariant ideal sheaves $I|_{U_\alpha} \subset \C[x,y]$ of finite colength for $\alpha=0,1,2$ (where we use coordinates $x,y$ on each chart $U_\alpha \cong \C^2$ as in Section \ref{sec:ABintro}). In this case there are no gluing conditions since the sheaves $I_{U_\alpha \setminus \{(0,0)\}} \cong \O_{U_\alpha}$ automatically glue. It follows that $((\PP^2)^{[n]})^T$ is 0-dimensional and reduced, and
\[
\Big( (\PP^2)^{[n]} \Big)^T = \big\{ (\lambda_\alpha)_{\alpha=0}^{2} \subset \Z_{\geq 0}^2 \, : \,\textrm{integer partition with } |\lambda_0| + |\lambda_1| + |\lambda_2| = n \big\}.
\]

\begin{exercise}
\hfill
\begin{enumerate}
\item[(a)] Using Exercises \ref{eqn:eXeXT} and \ref{exc:Euler}, prove the following two formulae
\[
\sum_{n=0}^{\infty} e((\PP^2)^{[n]}) q^n = \overline{\eta}(q)^{-3}, \quad \sum_{n=0}^{\infty} e((\PP^1 \times \PP^1)^{[n]}) q^n = \overline{\eta}(q)^{-4},
\]
where $e(\PP^2) = 3$ and $e(\PP^1 \times \PP^1) = 4$.
\item[(b)] Recall the Hopf index theorem $e(X) =  \int_X c_d(T_X)$ for any smooth projective variety $X$. Using part (a), Exercise \ref{exc:listmult}, and Corollary \ref{cor:univ}, you can now prove G\"ottsche's formula \cite{Got}, namely, for \emph{any} smooth projective surface $S$, we have
\[
\sum_{n=0}^{\infty} e(S^{[n]}) q^n = \overline{\eta}(q)^{-e(S)}.
\]
\end{enumerate}
\end{exercise}

In order to calculate more general Chern numbers of $S^{[n]}$, for $S$ a smooth surface with a $T$-action, we give a formula for the tangent representation of $T_{S^{[n]}}$ at any fixed point. A proof can be found, e.g., in \cite[Proposition 4.1]{GK1}.
\begin{proposition}
Let $S = \C^2$ and let $Z \subset \C^2$ be a $T$-invariant 0-dimensional closed subscheme corresponding to an integer partition $\lambda \subset \Z_{\geq 0}^2$. Define $Z := \sum_{(a,b) \in \lambda} \t_1^a \t_2^b \in K_0^T(\pt) \cong \Z[\t_1^{\pm 1}, \t_2^{\pm 1}]$, which is the $T$-representation of $H^0(\C^2,\O_Z)$. Then
\[
T_{(\C^2)^{[n]}}|_Z = Z + \frac{Z^*}{\t_1\t_2} - \frac{(1-\t_1)(1-\t_2)}{\t_1\t_2} Z Z^*,
\]
where $(-)^*$ is the dual representation, which amounts to the substitutions $\t_i \mapsto \t_i^{-1}$.
\end{proposition}

From this result, it is also not hard to find the $T$-representations of $T_{S^{[n]}}$ at fixed points for $S = \PP^2$, $\PP^1 \times \PP^1$ (or any smooth toric surface). Indeed, for $S = \PP^2$, let $Z \subset \PP^2$ be a $T$-invariant 0-dimensional closed subscheme and let $Z_\alpha := Z|_{U_\alpha}$ corresponding to integer partition $\lambda_\alpha$. Referring to \eqref{eqn:chartU0}--\eqref{eqn:chartU2}, we define
\[
(\t_1^{(0)},\t_2^{(0)}) := (\t_1,\t_2), \quad (\t_1^{(1)},\t_2^{(1)}) := (\t_1^{-1}\t_2,\t_1^{-1}), \quad (\t_1^{(2)},\t_2^{(2)}) := (\t_2^{-1},\t_1 \t_2^{-1}).
\]
Define furthermore the $T$-representation of $H^0(U_\alpha,\O_{Z_\alpha})$ by
\[
Z_\alpha := \sum_{(a,b) \in \lambda_\alpha} (\t_1^{(\alpha)})^a (\t_2^{(\alpha)})^b.
\]
The tangent space at $Z$ is given by $\Hom_{\PP^2}(I_{Z/\PP^2},\O_Z)$ (Section \ref{sec:hilb}). Since $\O_Z = \O_{Z_0} \oplus \O_{Z_1} \oplus \O_{Z_2}$, it is not hard to see that
\[
T_{(\PP^2)^{[n]}}|_Z = \sum_{\alpha=0}^2 \Big( Z_\alpha + \frac{Z_\alpha^*}{\t_1^{(\alpha)}\t_2^{(\alpha)}} - \frac{(1-\t_1^{(\alpha)})(1-\t_2^{(\alpha)})}{\t_1^{(\alpha)}\t_2^{(\alpha)}} Z_\alpha Z_\alpha^*  \Big).
\]

\begin{exercise}
For $S = \PP^2$, determine all Chern numbers $\int_{S^{[2]}} c_4(T_{S^{[2]}})$, $\int_{S^{[2]}} c_3(T_{S^{[2]}}) c_1(T_{S^{[2]}})$, $\int_{S^{[2]}} c_2(T_{S^{[2]}})^2$, $\int_{S^{[2]}} c_2(T_{S^{[2]}}) c_1(T_{S^{[2]}})^2$, $\int_{S^{[2]}} c_1(T_{S^{[2]}})^4$. 
\end{exercise}

For any $T$-equivariant line bundle $L$ on $\PP^2$, the restriction $L|_{U_\alpha}$ is trivial. Thus, there exists a character $w^{(\alpha)} = (w^{(\alpha)}_1, w^{(\alpha)}_2) \in \Z^2$ such that $L|_{U_\alpha} \cong \O \otimes \mathfrak{t}^{w^{(\alpha)}}$. Therefore, for $Z \subset \PP^2$ as above, we have
\[
L^{[n]}|_Z = \sum_{\alpha=0}^{2} Z_\alpha \cdot \t_1^{w^{(\alpha)}_1} \t_2^{w^{(\alpha)}_2}.
\]
In principle, for $S = \PP^2$, $\PP^1 \times \PP^1$ (or any smooth projective toric surface) we now have the tools to express any intersection number $\int_{S^{[n]}} P$, where $P$ is any polynomial in the Chern classes of $T_{S^{[n]}}$, $L_1^{[n]}$, $\ldots$, $L_\ell^{[n]}$ with $[L_1], \ldots, [L_\ell] \in \Pic(S)$, as a sum over all integer partitions of $n$ of certain rational functions in the equivariant parameters. These type of calculations are easy to implement into a computer program.

In summary, consider Corollary \ref{cor:univ} for $E_i = L_i$ line bundles. This corollary reduces multiplicative generating series of intersection numbers on Hilbert schemes of points on an \emph{arbitrary} smooth projective surface $S$ to a finite number of universal power series. Suppose that we have tuples $(S,L_1, \ldots, L_\ell)$ such that \eqref{eqn:Chernvector} forms a $\Q$-basis and $S$ is a smooth projective toric surface (often $\PP^2$ and $\PP^1 \times \PP^1$ suffice) and $L_1, \ldots, L_\ell$ are $T$-equivariant line bundles on $S$. Then the universal series $A(q)$, $B(q)$, $\ldots$ are completely determined by the Atiyah--Bott localization formula as described in this section. 

\begin{exercise} (More work.)
\hfill
\begin{enumerate}
\item[(a)] It is proved in \cite{EGL} that for any smooth projective surface $S$ and line bundle $L$ on $S$, we have the following generating series of Verlinde numbers
\[
\sum_{n=0}^{\infty} \chi(S^{[n]},\det(L^{[n]})) q^n = (1+q)^{\chi(L)}.
\]
Using the method presented in this section, verify this formula for all $(S,L)$ modulo $q^5$.
\item[(b)] Lehn's conjecture states that for any smooth projective surface $S$ and line bundle $L$ on $S$, we have
\[
\sum_{n=0}^{\infty} q^n \int_{S^{[n]}} s_{2n}(L^{[n]}) = \frac{(1-w)^a(1-2w)^b}{(1-6w+6w^2)^c},
\]
where $a = L K_S - 2K_S^2$, $b = (L-K_S)^2 + 3 \chi(\O_S)$, $c = \frac{1}{2} L(L-K_S) + \chi(\O_S)$ and 
\begin{equation} \label{eqn:changevar}
q = \frac{(1-w)(1-2w)^4}{(1-6w+6w^2)^3} \cdot w. 
\end{equation}
Here we view \eqref{eqn:changevar} as a formal change of variables expressing $w$ as a formal power series in $q$ by inverting the relation \eqref{eqn:changevar}. Lehn's conjecture was proved by Marian--Oprea--Pandharipande \cite{MOP}. Using the methods presented in this section, verify this formula for all $(S,L)$ modulo $q^5$.
\end{enumerate}
\end{exercise}

\subsection{Nekrasov partition functions} \label{sec:Nekpart}

The moduli space of framed sheaves $M(r,n)$ (Section \ref{sec:moduliframed}) has a $T$-action as well, which can be used to define the Nekrasov partition function \cite{Nekr03, Moore:1997dj, Moore:1998et, Lossev:1997bz}. For more details on this section, we refer to \cite{NY}. Let $$T := T_1 \times T_2 \times T_3 = (\C^*)^{2} \times (\C^*)^{r} \times \C^*.$$ 
The first factor $T_1$ acts on points of $\PP^2$ by
$$
(t_1,t_2) \cdot [p_0:p_1:p_2] \mapsto [p_0:t_1^{-1}p_1:t_2^{-1}p_2].
$$ 
We denote the induced automorphism by $F_{(t_1,t_2)} \colon \PP^2 \to \PP^2$ for each $(t_1,t_2) \in (\C^*)^2$. The induced action of $T_1 \times T_2$ on $M(r,n)$ is defined by
$$
(t_1,t_2,e_0, \ldots, e_{r-1}) \cdot (E, \Phi) = ((F^{-1}_{t_1,t_2})^* E, \Phi_{(t_1,t_2,e_0, \ldots, e_{r-1})}),
$$
where $\Phi_{(t_1,t_2,e_0, \ldots, e_{r-1})}$ denotes the composition
$$
(F^{-1}_{t_1,t_2})^* E|_{\ell_{\infty}} \stackrel{(F^{-1}_{t_1,t_2})^* \Phi}{\longrightarrow} (F^{-1}_{t_1,t_2})^* \O_{\ell_{\infty}}^{\oplus r} \stackrel{\cong}{\longrightarrow} \O_{\ell_\infty}^{\oplus r} \stackrel{(e_0, \ldots, e_{r-1})}{\longrightarrow} \O_{\ell_\infty}^{\oplus r},
$$
where the second arrow is given by the natural $T_1$-equivariant structure on $\O_{\ell_{\infty}}$ and the third arrow is matrix multiplication by the diagonal matrix $\mathrm{diag}(e_0,\ldots,e_{r-1})$. Finally, we define the action of the third factor $T_3 = \C^*$ on $M(r,n)$ to be trivial. We denote the primitive characters of $T_1$ by $\mathfrak{t}_1$, $\mathfrak{t}_2$. Furthermore we write $\mathfrak{e}_0, \ldots, \mathfrak{e}_{r-1}$ for the primitive characters of $T_2$ and $\mathfrak{m}$ for the primitive character of $T_3$.

The fixed point locus $M(r,n)^{T_2}$ (using only the second torus factor) consists of framed sheaves
$$
E = \bigoplus_{i=0}^{r-1}I_{Z_i/\PP^2} \otimes \mathfrak{e}_i, \quad \Phi =( \iota_{Z_0} \oplus \cdots \oplus \iota_{Z_{r-1}})|_{\ell_\infty}, \quad \sum_{i=0}^{r-1} |Z_i| = n,
$$
where $Z_i \subset \PP^2$ are 0-dimensional subschemes supported at $\PP^2 \setminus \ell_{\infty} \cong \C^2$ with ideal sheaves $\iota_{Z_i} \colon I_{Z_i/\PP^2} \subset \O_{\PP^2}$. This description provides an isomorphism of schemes
\begin{equation} \label{eqn:Tfrfix}
M(r,n)^{T_2} \cong \bigsqcup_{n = n_0 + \cdots + n_{r-1}}  (\C^2)^{[n_0]} \times \cdots \times  (\C^2)^{[n_{r-1}]}.
\end{equation}
In particular, the fixed locus $M(r,n)^T$ (now using the full torus) is 0-dimensional and reduced, and can be indexed by $r$-tuples of $T_1$-fixed 0-dimensional closed subschemes $(Z_i)_{i=0}^{r-1}$ supported at $[1:0:0]$ satisfying
$\sum_i |Z_i| = n$.
In turn, by the previous section, the $T_1$-fixed 0-dimensional subschemes $(Z_i)_{i=0}^{r-1}$ are in bijection to a collection of integer partitions $(\lambda_i)_{i=0}^{r-1}$ satisfying $\sum_i |\lambda_i| = n$. Recall that each $Z_i$ also has a corresponding $T$-representation given by $Z_i = \sum_{(a,b) \in \lambda_i} \t_1^a \t_2^b$. 

We consider the $T$-representation of the tangent bundle at a fixed point $(E,\Phi)$
$$
T_{M(r,n)}|_{(E,\Phi)}  = \Ext_{\PP^2}^1(E,E(-\ell_\infty)) = R\Hom_{\PP^2}(E,E(-\ell_\infty))[1] \in K_0^{T}(\pt),
$$
where we used that $\Ext_{\PP^2}^{i}(E,E(-\ell_\infty)) = 0$ for $i \neq 1$. 
\begin{proposition} \label{lem:Tframed}
At any $T$-fixed point $(E,\Phi)$ of $M(r,n)$ corresponding to 0-dimensional subschemes $(Z_i)_{i=0}^{r-1}$ as above, we have the following equality in $K_0^{T}(\pt)$
\begin{align*}
T_{M(r,n)}|_{(E,\Phi)} = \sum_{i,j=0}^{r-1} (R\Gamma(\C^2,\O_{\C^2}) -R\Hom_{\mathbb{C}^2}(I_{Z_i/\C^2},I_{Z_j/\C^2})) \otimes \mathfrak{e}_i^{-1}\mathfrak{e}_j \\
=\sum_{i,j=0}^{r-1} \mathfrak{e}_i^{-1} \mathfrak{e}_j \Big( Z_j + \frac{Z_i^*}{\t_1\t_2} - \frac{(1-\t_1)(1-\t_2)}{\t_1\t_2} Z_i^* Z_j\Big).
\end{align*}
\end{proposition}

Let $\epsilon_1 = c_1^T(\t_1)$, $\epsilon_2 = c_1^T(\t_2)$, and define $w_i = c_1^T(\mathfrak{e}_i)$ and $m = c_1^T(\mathfrak{m})$. The Nekrasov partition function (without insertions) is defined by
\[
\mathsf{Z}_r(\epsilon_1,\epsilon_2,(w_i)_i;q) = \sum_{n=0}^{\infty} q^n \int_{M(r,n)} 1 \in \Q(\epsilon_1,\epsilon_2,(w_i)_i)[[q]],
\]
where the right hand side is defined by the Atiyah--Bott localization formula (recall that $M(r,n)$ is non-compact!). Similarly, the Nekrasov partition function with one adjoint matter is defined by
\[
\mathsf{Z}_r^{\mathsf{adj}}(\epsilon_1,\epsilon_2,(w_i)_i,m;q) = \sum_{n=0}^{\infty} q^n \int_{M(r,n)} e(T_{M(r,n)} \otimes \mathfrak{m}) \in \Q(\epsilon_1,\epsilon_2,(w_i)_i,m)[[q]],
\]
where again the right hand side is defined by the Atiyah--Bott localization formula and, in the $n$th summand, $e(-) = c_{2rn}^T(-)$ is the $T$-equivariant Euler class. Note that $T$ includes the trivial third factor $T_3 = \C^*$ with equivariant parameter $\mathfrak{m}$ with ``mass'' $m:=c_1^T(\mathfrak{m})$.

\begin{remark} \label{rem:Nekdiff}
By Proposition \ref{lem:Tframed}, $\mathsf{Z}_r(\epsilon_1,\epsilon_2,(w_i)_i;q)$ and $\mathsf{Z}^{\mathrm{adj}}_r(\epsilon_1,\epsilon_2,(w_i)_i,m;q)$, only depend on the differences $w_i - w_j$. Thus we also denote them by $\mathsf{Z}_r(\epsilon_1,\epsilon_2,(w_{i,i+1})_i;q)$ and $\mathsf{Z}^{\mathrm{adj}}_r(\epsilon_1,\epsilon_2,(w_{i,i+1})_i,m;q)$ respectively, where $w_{i,i+1} := w_i - w_{i+1}$.
\end{remark}

\subsection{Klyachko's formula (extra)} \label{sec:Kly}

So far, we have focused our attention on Hilbert schemes of points $S^{[n]}$ on a smooth projective surface $S$, which can be viewed as Gieseker--Maruyama--Simpson moduli spaces $M_S^H(1,\O_S,n)$. For moduli of sheaves of rank $r>1$, we cannot apply the universality result Corollary \ref{cor:univ}. Suppose $S$ has an action by a torus $T$, then this action lifts to $M:=M_S^H(r,L,c_2)$ by
\[
t \cdot [E] := [(t^{-1})^* E], \quad \textrm{for all } t \in T.
\]
However, unlike the $r=1$ case, $M^T$ may not be 0-dimensional, which limits the usefulness of Atiyah--Bott localization in calculations for $r>1$. Nonetheless, there is a somewhat explicit description of $M_S^H(r,L,c_2)^T$ pioneered by Klyachko \cite{Kly1, Kly2} (and further explored in \cite{Koo1, Koo2}) as we will now discuss. 

Suppose $\gcd(r,c_1(L) H)=1$. Consider the open locus
\[
V_S^H(r,L,c_2) := \{ [E] \in M_S^H(r,L,c_2) \, : \, E \textrm{ is locally free} \} \subset M_S^H(r,L,c_2).
\]
Note that $M_S^H(r,L,c_2)$ may have multiple irreducible components and, in general, $V_S^H(r,L,c_2)$ is by no means dense (in fact, it may be empty). Nonetheless, we consider the map
\begin{equation} \label{def:Psi}
\Psi \colon \bigsqcup_{c_2} M_S^H(r,L,c_2) \to \bigsqcup_{c_2} V_S^H(r,L,c_2), \quad [E] \mapsto [E^{**}].  
\end{equation}

We note that on a smooth variety of dimension 2, reflexive sheaves and locally free sheaves are the same, so the reflexive sheaf $E^{**}$, called the reflexive hull of $E$, is indeed locally free. This map is bad in the sense that it is only a \emph{constructible} morphism, i.e., at the level of $\C$-valued points the domain of each $\Psi|_{M_S^H(r,L,c_2)}$ is a finite disjoint union of (separated, finite type) locally closed subschemes $\bigsqcup_i D_i$ such that $\Psi|_{D_i}$ is a morphism. Still, this is good enough for calculations involving Euler characteristics as we will presently see.

The fibre over $[E] \in V_S^H(r,L,c_2)$ (at least at the level of reduced schemes) is $\bigcup_n \Quot_S(E,n)$ where $\Quot_S(E,n)$ denotes the Quot scheme parametrizing quotients $[E \twoheadrightarrow Q]$ with $Q$ a 0-dimensional sheaf on $S$ of length $n$.\footnote{In the Quot scheme, two quotients $\phi_i \colon E \twoheadrightarrow Q_i$, $i=1,2$, are identified when there exists an isomorphism $\psi \colon Q_1 \to Q_2$ such that $\phi_2 = \psi \circ \phi_1$. Since, presently, we are only interested in Euler characteristics, we only need to know that this description of the fibres of $\Psi$ holds at the level of reduced schemes. }

\begin{remark} \label{rem:Euler}
Before we proceed, we require some facts on the topological Euler characteristic. There exists a notion of topological Euler characteristic $e(X)$ of any separated finite type $\mathbb{C}$-scheme $X$. It satisfies the following properties. For any separated finite type $\C$-schemes $X,Y$ we have
\begin{itemize}
\item  $e(X) = e(X_{\mathrm{red}}) = e(X_{\mathrm{red}}(\mathbb{C}))$; 
\item $e(X \times Y) = e(X)e(Y)$;
\item if $X$ is a smooth projective variety (with its complex analytic topology), we have $e(X) = \sum_i (-1)^i b_i(X)$;
\item $e(X \setminus Z) = e(X) - e(Z)$ for any locally closed subscheme $Z \hookrightarrow X$;
\item if $T$ is an algebraic torus acting on $X$, we have $e(X) = e(X^T)$.
\end{itemize}
\end{remark}

We define the following generating series
\begin{align*}
\mathsf{Z}_{S,H,r,L}(q) = \sum_{c_2 \in \Z} e(M_S^H(r,L,c_2)) q^{c_2},\quad \mathsf{Z}^{\mathrm{lf}}_{S,H,r,L}(q) = \sum_{c_2 \in \Z} e(V_S^H(r,L,c_2)) q^{c_2}. 
\end{align*}

\begin{exercise}
Use the Bogomolov inequality \cite[Theorem 3.4.1]{HL} to show that these are Laurent series in $q$. 
\end{exercise}

For a locally free sheaf $E$ of rank $r$ on $S$, using stratification arguments, it is not hard to show that 
\begin{equation} \label{eqn:EulerQuot}
\sum_{n=0}^{\infty} e(\Quot_S(E,n)) q^n = \overline{\eta}(q)^{-re(S)}.
\end{equation}
In turn, one can the use the constructible map $\Psi$ to deduce
\begin{equation} \label{eqn:EulerMN}
\mathsf{Z}_{S,H,r,L}(q) = \overline{\eta}(q)^{-re(S)} \cdot \mathsf{Z}^{\mathrm{lf}}_{S,H,r,L}(q).
\end{equation}

\begin{exercise}
Prove \eqref{eqn:EulerQuot} and \eqref{eqn:EulerMN} by using \cite[Appendix A.2]{BK}. 
\end{exercise}

This still leaves the main task: calculating $\mathsf{Z}^{\mathrm{lf}}_{S,H,r,L}(q)$. When $S$ is a smooth projective surface, it suffices to determine $e(V_S^H(r,L,c_2)^T)$ by Remark \ref{rem:Euler}. Although the elements of $V_S^H(r,L,c_2)^T$ admit a nice description in terms of combinatorics and linear algebra data, due to Klyachko, this is generally still a rather complicated task. In particular, $V_S^H(r,L,c_2)$ is generally not 0-dimensional. However, for $S = \PP^2$ and $r=2$ we will see that $V_S^H(r,L,c_2)^T$ is 0-dimensional and we will describe its elements.

\subsubsection{Equivariant sheaves on $\C^2$}

For proofs of some of the claims made in this subsection, we refer to \cite{Per}. It is well-known \cite[Corollary II.5.5]{Har} that for an affine scheme $X = \Spec(R)$, where $R$ is a noetherian commutative ring, the global sections functor $H^0(X,-)$ gives an equivalence of abelian categories
\[
\mathrm{Coh}(X) \stackrel{\sim}{\longrightarrow} R\textrm{-Mod}_{f.g.}, \quad E \mapsto H^0(X,E),
\]
where $\mathrm{Coh}(X)$ denotes the category of coherent sheaves on $X$ and $R\textrm{-Mod}_{f.g.}$ is the category of finitely generated $R$-modules. We are interested in the case $X = \C^2 = \Spec \C[x,y]$. Firstly, considering $\C^2$ with the standard $T=(\C^*)^2$-action as before, $\C[x,y]$ is an $X(T) = \Z^2$-graded ring by
\[
\C[x,y] = \bigoplus_{m=(m_1,m_2) \in \Z_{\geq 0}^2} \C \cdot x^{m_1} y^{m_2} \otimes \t^{m}, 
\]
where $\t^m := \t_1^{m_1} \t_2^{m_2}$. When $E$ is a $T$-equivariant coherent sheaf on $\C^2$, the (generally infinite-dimensional) vector space $H^0(\C^2,E)$ has a $T$-action and decomposes into eigenspaces
\[
H^0(\C^2,E) = \bigoplus_{m \in \Z^2} H^0(\C^2,E)_m \otimes \t^{m}.
\]
Thus, it may not come as a surprise that there exists an equivalence of categories 
\[
T\textrm{-Coh}(\C^2) \stackrel{\sim}{\longrightarrow} \C[x,y]\textrm{-Mod}_{f.g.}^{\Z^2}, \quad E \mapsto H^0(\C^2,E) = \bigoplus_{m \in \Z^2} H^0(\C^2,E)_m \otimes \t^m,
\]
where $T\textrm{-Coh}(\C^2)$ denotes the category of $T$-equivariant coherent sheaves on $\C^2$ and $\C[x,y]\textrm{-Mod}_{f.g.}^{\Z^2}$ denotes the category of finitely generated $\Z^2$-graded $\C[x,y]$-modules. In turn, objects of the latter category can be described in terms of linear algebra data: namely, a collection of finite-dimensional $\mathbb{C}$-vector spaces $(E_m)_{m \in \mathbb{Z}^2}$ and $\mathbb{C}$-linear maps $(x_m \colon E_m \to E_{m+(1,0)})_{m \in \Z^2}$, $(y_m \colon E_m \to E_{m+(0,1)})_{m \in \Z^2}$ such that
\begin{itemize}
\item $y_{m+(1,0)} \circ x_{m} = x_{m+(0,1)} \circ y_{m}$ for all $m \in \Z^2$;
\item $\bigoplus_{m \in \mathbb{Z}^2} E_m$ is finitely generated as a $\mathbb{C}[x,y]$-module.
\end{itemize}
Moreover, one can show that $E$ is torsion free if and only if $x_m, y_m$ are injective for all $m \in \mathbb{Z}^2$. This describes the category of $T$-equivariant torsion free sheaves on $\C^2$ in terms of linear algebra data. Moreover, up to an equivalence of categories, we may take all maps $x_m, y_m$ as inclusions so that $T$-equivariant rank $r$ torsion free sheaves on $\C^2$ are just described by double filtrations $(E_m)_{m \in \Z^2}$ of $\C^r$
\[
E_{(m_1,m_2)} \subset E_{(m_1+1,m_2)}, \quad E_{(m_1,m_2)} \subset E_{(m_1,m_2+1)}, \quad \textrm{for all } (m_1,m_2) \in \Z^2
\]
such that the corresponding module is finitely generated. 

\begin{exercise}
In the above setup, prove that if $E$ is torsion free, then the maps $x_m, y_m$ are injective by using \cite[Exercise II.5.6(b)]{Har}. 
\end{exercise}

\begin{exercise} \label{exc:infnotation}
Let $E$ be a $T$-equivariant torsion free sheaf on $\C^2$. Prove the following statements:
\begin{itemize}
\item there exist $A_1,A_2 \in \mathbb{Z}$ such that $E_m = 0$ if $m_1 < A_1$ or $m_2 < A_2$;
\item for all $m_1, m_2 \gg 0$, we have $E_{(\infty,\infty)} := E_m \cong \mathbb{C}^{\mathrm{rk}(E)}$;
\item for $i \gg 0$, $(E_{(\infty,m)} := E_{(i,m)})_{m \in \mathbb{Z}}$ stabilizes to a flag of $\mathbb{C}^{\mathrm{rk}(E)}$;
\item for $j \gg 0$, $(E_{(m,\infty)} := E_{(m,j)})_{m \in \mathbb{Z}}$ stabilizes to a flag of $\mathbb{C}^{\mathrm{rk}(E)}$. 
\end{itemize}
\end{exercise}

One way to construct modules corresponding to $T$-equivariant rank $r$ torsion free sheaves on $\C^2$ is as follows. Given two flags $( F_m )_{m \in \mathbb{Z}}$, $( G_m )_{m \in \mathbb{Z}}$ of $\C^{r}$ such that (1) $F_m = G_m = 0$ for $m \ll 0$ and (2) $F_m = G_m = \mathbb{C}^r$ for all $m \gg 0$. Then $$(E_{(m_1,m_2)} := F_{m_1} \cap G_{m_2} )_{(m_1,m_2) \in \mathbb{Z}^2},$$ 
together with its natural inclusions, corresponds to a $T$-equivariant rank $r$ torsion free sheaf $E$ on $\mathbb{C}^2$. 

\begin{proposition} \label{prop:lfonC2}
The category of $T$-equivariant rank $r$ locally free sheaves on $\C^2$ is equivalent to the category of pairs of flags $( F_m )_{m \in \mathbb{Z}}$, $(G_m)_{m \in \mathbb{Z}}$ of $\C^r$ such that\footnote{There is an obvious notion of morphisms between such data.} 
\begin{itemize}
\item $F_m = G_m = 0$ for $m \ll 0$;
\item $F_m = G_m = \mathbb{C}^r$ for $m \gg 0$. 
\end{itemize}
\end{proposition}

The idea of the proof of this proposition is that a reflexive (i.e.~locally free) sheaf $E$ on $\C^2$ is in fact determined by the restrictions $E|_{\C \times \C^*}$, $E|_{\C^* \times \C}$, and when $E$ is $T$-equivariant, these restrictions give rise to the two flags. 

\begin{exercise}
\hfill
\begin{enumerate}
    \item[(a)] Show that a $T$-equivariant rank 2 locally free sheaf on $\mathbb{C}^2$ splits ($T$-equivariantly) as a direct sum of two $T$-equivariant line bundles.
    \item[(b)] Does the same hold for higher rank $T$-equivariant locally free sheaves on $\mathbb{C}^2$? 
\end{enumerate}
\end{exercise}

Continuing the notation above, suppose a $T$-equivariant torsion free sheaf $E$ on $\C^2$ is described by $(E_m)_{m \in \Z^2}$ (with $x_m, y_m$ all inclusions). Then, using the notation of Exercise \ref{exc:infnotation}, its reflexive hull $E^{**}$ corresponds to $$(E_{(m_1,\infty)} \cap E_{(\infty,m_2)} \subset \mathbb{C}^r)_{(m_1,m_2) \in \mathbb{Z}^2},$$ and the natural map $E \hookrightarrow E^{**}$ corresponds to the inclusions
$$
E_{(m_1,m_2)} \subset E_{(m_1,\infty)} \cap E_{(\infty,m_2)}, \quad (m_1,m_2) \in \Z^2.
$$ 

\begin{example}
By Proposition \ref{prop:lfonC2}, specifying a $T$-equivariant line bundle is equivalent to specifying a pair of integers $(A_1,A_2) \in \Z^2$ (i.e., the data of two flags jumping from $0$ to $\C$ at $m = A_i$). For instance, $\O_{\C^2}$ corresponds to $(0,0)$.
\end{example}

\begin{exercise}
Consider the ideal $I_{Z/\C^2} \subset \mathcal{O}_{\mathbb{C}^2}$ of a $T$-invariant closed subscheme $Z \subset \C^2$.
\begin{enumerate}
\item[(a)] Show that the description of rank 1 torsion free sheaves of this section is compatible with the description of $I_{Z/\C^2}$ in terms monomial ideals from the Section \ref{sec:locHilb}. 
\item[(b)] Let $Z = Z(y)$. Show that $I_{Z/\C^2}$ corresponds to a $T$-equivariant line bundle and give the weight decomposition of $I_{Z/\C^2}$. Answer the same question for $Z = Z(xy)$.
\end{enumerate}
\end{exercise}

\begin{example}
From Proposition \ref{prop:lfonC2}, we deduce that a $T$-equivariant rank 2 locally free sheaf on $\C^2$ corresponds to the following data: (1) $A_i \in \mathbb{Z}$, $\Delta_i \in \mathbb{Z}_{\geq 0}$ for $i=1,2$, (2) a 1-dimensional subspace $V_i \subset \mathbb{C}^2$ if $\Delta_i > 0$ for $i=1,2$. This is the data of two flags jumping from 0 to $V_i$ at $m=A_i$ and from $V_i$ to $\C^2$ at $A_i+\Delta_i$ (and jumping directly from 0 to $\C^2$ when $\Delta_i = 0$).
\end{example}

\begin{exercise} 
In the setting of the previous exercise, let $\Delta_1, \Delta_2 > 0$ and $V_1 \neq V_2$. Consider $\lim_{V_1 \to V_2} E$ while keeping $\dim_{\mathbb{C}}(E_m)$ fixed for all $m$, and conclude that the limit does \emph{not} correspond to a locally free sheaf. Conclude that $\Psi$ in \eqref{def:Psi} is not a morphism.
\end{exercise}

\subsubsection{Equivariant sheaves on $\PP^2$}

We are now in a position to describe $T$-equivariant torsion free sheaves $E$ on $\PP^2$. Indeed, using the standard $T$-invariant cover $(U_\alpha  \cong \C^2)_{\alpha=0}^2$ and the coordinates from Section \ref{sec:ABintro}, we obtain $T$-equivariant torsion free sheaves $E^{(\alpha)} := E|_{U_\alpha}$ on $U_\alpha \cong \C^2$ and each of these are described by double filtrations $(E_m^{(\alpha)})_{m \in \Z^2}$ as above. Conversely, given double filtrations $(E_m^{(\alpha)})_{m \in \Z^2}$ for $\alpha=0,1,2$ as above, we obtain $T$-equivariant torsion free sheaves $E^{(\alpha)}$ on $U_\alpha \cong \C^2$ and the question is when they glue to a sheaf on $\PP^2$. It turns out that the gluing conditions precisely translate into the following equalities
$$
E^{(0)}_{(\infty,m)} = E^{(1)}_{(m,\infty)}, \quad E^{(1)}_{(\infty,m)} = E^{(2)}_{(m,\infty)}, \quad E^{(2)}_{(\infty,m)} = E^{(0)}_{(m,\infty)}, \quad \textrm{for all } m \in \Z^2.
$$
Part of the proof involves Exercise \ref{exc:infnotation}. 

As in the previous subsection, the description of $T$-equivariant rank $r$ locally free sheaves on $\PP^2$ is particularly easy.
\begin{proposition}[Klyachko] \label{prop:lfonP2}
The category of $T$-equivariant locally free sheaves of rank $r$ on $\PP^2$ is equivalent to the category of triples of flags $(F_m^{(\alpha)})_{m \in \Z^2, \alpha =0,1,2}$ satisfying
\begin{itemize}
\item $F_m^{(\alpha)} = 0$ for $m \ll 0$;
\item $F_m^{(\alpha)} = \C^r$ for $m \gg 0$.
\end{itemize}
\end{proposition}

For flags as in this proposition, one defines corresponding double filtrations
\[
F_{(m_1,m_2)}^{(\alpha)} := F^{(\alpha)}_{m_1} \cap F_{m_2}^{(\alpha + 1)}
\]
for all $m = (m_1,m_2) \in \Z^2$ and $\alpha \in \Z/3 \Z=\{0,1,2\}$, which then correspond to a $T$-equivariant rank $r$ torsion free sheaf on $\PP^2$ as described above. 

\begin{example} \label{ex:lfrk1}
From Proposition \ref{prop:lfonP2}, we deduce that a $T$-equivariant line bundle on $\PP^2$ corresponds to a triple of integers $(A,B,C) \in \Z^3$ indicating where the flags jump from $0$ to $\C$. Suppose $L$ is the $T$-equivariant line bundle on $\PP^2$ corresponding to $(A,B,C) \in \Z^3$ and $E$ is a $T$-equivariant torsion free sheaf on $\PP^2$ corresponding to double filtrations $(E_m^{(\alpha)} )_{m \in \Z^2, \alpha=0,1,2}$. Then the double filtrations corresponding to $E \otimes L$ are given by
\begin{align*}
(E \otimes L)_{(m_1,m_2)}^{(0)} &= E_{m_1-A,m_2-B}^{(0)}, \quad (E \otimes L)_{(m_1,m_2)}^{(1)} = E_{m_1-B,m_2-C}^{(1)}, \\(E \otimes L)_{(m_1,m_2)}^{(2)} &= E_{m_1-C,m_2-A}^{(2)},
\end{align*}
for all $(m_1,m_2) \in \Z^2$.
\end{example}

\begin{example} \label{ex:lfrk2}
By Proposition \ref{prop:lfonP2}, the data of a $T$-equivariant rank 2 locally free sheaf on $\PP^2$ correspond to the following data in $\mathbb{Z}^3 \times \mathbb{Z}_{\geq 0}^3 \times (\mathbb{P}^1)^3$
\begin{itemize}
\item $(A,B,C) \in \mathbb{Z}^3$;
\item $(\Delta_1,\Delta_2,\Delta_3) \in \mathbb{Z}^3_{\geq 0}$; 
\item $(V_1,V_2,V_3) \in (\mathbb{P}^1)^3$, where we omit $V_i$ whenever $\Delta_i = 0$. 
\end{itemize}
\end{example}

\begin{exercise}
In the setting of Example \ref{ex:lfrk2}, show that $E$ is not $T$-equivariantly decomposable as a direct sum of $T$-equivariant line bundles if and only if $\Delta_i > 0$ for all $i$ and $V_i$ are mutually distinct. 
\end{exercise}

We would like to specify which of the locally free sheaves in Example \ref{ex:lfrk2} are stable. In turn, this requires us to determine their Chern characters. The first useful fact is that for a $T$-equivariant line bundle $L$ as in Example \ref{ex:lfrk1} corresponding to a triple of integers $(A,B,C) \in \Z^3$, we have
\[
c_1(L) = -(A+B+C)\ell,
\]
where $\ell \in H^2(\PP^2,\Z)$ is the Poincar\'e dual of the class of a line on $\ell \subset \PP^2$. In particular, we see that $L \cong \O_{\PP^2}$ (non-equivariantly!) if and only if $A+B+C = 0$. This reflects the fact that we have a short exact sequence
$$
0 \to \underbrace{X(T)}_{\cong \mathbb{Z}^2} \to \underbrace{\mathrm{Pic}^T(\mathbb{P}^2)}_{\cong \mathbb{Z}^3} \stackrel{c_1}{\to} \underbrace{\mathrm{Pic}(\mathbb{P}^2)}_{\cong \mathbb{Z}} \to 0,
$$
where the middle term if the $T$-equivariant Picard group.

Now suppose we have a $T$-equivariant rank 2 locally free sheaf $E$ on $\PP^2$ corresponding to $(A,B,C) \in \Z^3$, $(\Delta_1,\Delta_2,\Delta_3) \in \Z_{\geq 0}^3$, and $(V_1,V_2,V_3) \in (\mathbb{P}^1)^3$ as in Example \ref{ex:lfrk2}. Then its Chern character is given by
\begin{align*}
&\mathrm{ch}(E) = 2 - (2(A+B+C)+\Delta_1+\Delta_2+\Delta_3)\ell + \frac{1}{2}(A+B+C)^2 \ell^2 \\
&+ \frac{1}{2}(A+B+C+\Delta_1+\Delta_2+\Delta_3)^2 \ell^2 - \sum_{\alpha \in \Z/3 \Z} \Delta_\alpha\Delta_{\alpha+1}(1-\dim_{\mathbb{C}}(V_\alpha \cap V_{\alpha+1}))\ell^2, 
\end{align*}
where $\ell^2 \in H^4(\PP^2,\Z)$ is the Poincar\'e dual of the point class. 

\begin{exercise} 
Prove the formula for $\ch(E)$ using the following two facts:
\begin{itemize}
\item $\mathrm{ch}(E)$ only depends on $(\dim_{\mathbb{C}}(E_m))_{m \in \mathbb{Z}^2}$, not on the inclusions;
\item given a $T$-equivariant 0-dimensional sheaf $G$ on $\PP^2$, i.e.~$G_m^{(\alpha)} \neq 0$ for finitely many $\alpha,m$, then we have $\mathrm{ch}(G) = \sum_{\alpha,m} \dim_{\mathbb{C}}(G_m) \ell^2$. 
\end{itemize}
\end{exercise}

Continuing our study of $T$-equivariant rank 2 locally free free sheaves $E$ on $\PP^2$, it is easy to write down a formula for the slope $\mu_E$ and the slope $\mu_L$ of any $T$-equivariant line bundle $L \subset E$. We claim that $E$ is $\mu$-stable if and only if $\Delta_i >  0$ for all $i$, $V_i$ are mutually distinct, and 
\begin{align*}
\Delta_1 < \Delta_2 + \Delta_3, \quad \Delta_2 < \Delta_1 + \Delta_3, \quad \Delta_3 < \Delta_1 + \Delta_2.
\end{align*}
\begin{exercise}
\hfill
\begin{enumerate}
\item[(a)] Prove one implication of the above claim (assume $E$ is $\mu$-stable, then...).
\item[(b)] Prove the other implication assuming the following fact: $E$ is $\mu$-stable if and only if $\mu_F < \mu_E$ for all $T$-equivariant subsheaves $F \subset E$ satisfying $0 < \rk(F) < \rk(E)$ \cite[Proposition 4.13]{Koo1}.
\end{enumerate}
\end{exercise}

We require one more fact before we are able to determine $\mathsf{Z}_{\PP^2,H,r,L}^{\mathrm{lf}}(q)$: the relationship between $T$-equivariant and $T$-invariant sheaves. Namely, any element $[E] \in M_{\PP^2}^H(r,L,c_2)$ admits a $T$-equivariant structure and this $T$-equivariant structure is unique up to tensorizing by a character $\t^m \in \Z^2$. See \cite[Propositions 4.4, 4.5]{Koo1} for a general statement and proof.

Since we consider $\PP^2$, stability does not depend on the choice of ample divisor $H$. We take $H = \ell$ and suppress it from the notation. Furthermore, we take $r=2$ and $L = \O(1)$. Note that the condition $\det(E) \cong \O(1)$ is equivalent to $c_1(E) = \ell$ (because $b_1(\PP^2)=0$). Consider the fixed locus $V_{\PP^2}(2,\O(1),c_2)^T$. Any element $E$ of this fixed locus admits a $T$-equivariant structure and it is therefore described by data as above. Since the $T$-equivariant structure is unique up to tensorizing by a character, which does not change the isomorphism class of the underlying sheaf, it becomes unique when we set $A = B = 0$. Furthermore, $E$ is only changed by a $T$-equivariant isomorphism if we act on the flags $(F_m^{(\alpha)})_{m \in \Z, \alpha=0,1,2}$ in Proposition \ref{prop:lfonP2} via the standard action of $\PGL(2,\C)$ on $\PP^1$. Therefore, we may assume  $V_1 = [1:0]$, $V_2 = [0:1]$, $V_3 = [1:1]$ ($V_1,V_2,V_3$ are mutually distinct by simplicity of $E$). We deduce that  $N_{\PP^2}(2,\O(1),c_2)^T$ is 0-dimensional and thus reduced (because it is smooth). 

In summary: in order to calculate $\mathsf{Z}_{\PP^2,\ell,2,\O(1)}^{\mathrm{lf}}(q)$, we  sum over all $A,B,C \in \mathbb{Z}$, $\Delta_1, \Delta_2, \Delta_3 \in \mathbb{Z}_{>0}$, $V_1,V_2,V_3 \in \mathbb{P}^1$ such that:
\begin{itemize}
\item $V_1 = [1:0]$, $V_2 = [0:1]$, $V_3 = [1:1]$ (``up to isomorphism''); 
\item $A = B = 0$ (``equivariant structure unique up to tensorizing by character''); 
\item $2C +\Delta_1+\Delta_2+\Delta_3 = -1$ (``$c_1 = \ell$''); 
\item $\Delta_1 < \Delta_2 + \Delta_3$, $\Delta_2 < \Delta_1 + \Delta_3$, $\Delta_3 < \Delta_1 + \Delta_2$ (``stability''); 
\end{itemize}
and we take the sum over all such of
\begin{itemize}
\item $q^{c_2} = q^{\frac{1}{4} + \frac{1}{2} \sum_{i<j} \Delta_i \Delta_j - \frac{1}{4} \sum_i \Delta_i^2}$ .
\end{itemize}
In conclusion, we have
$$
\mathsf{Z}_{\PP^2,\ell,2,\O(1)}^{\mathrm{lf}}(q) = \sum_{{\scriptsize{\begin{array}{c} \Delta_1,\Delta_2,\Delta_3 > 0 \\ 2 \, \Big| \, 1 + \Delta_1 + \Delta_2 + \Delta_3 \\ \Delta_1 < \Delta_2+\Delta_3 \\  \Delta_2 < \Delta_1+\Delta_3 \\  \Delta_3 < \Delta_1+\Delta_2 \end{array}}}} q^{\frac{1}{4} + \frac{1}{2} \sum_{i<j} \Delta_i \Delta_j - \frac{1}{4} \sum_i \Delta_i^2}.
$$

\begin{exercise}
Using this formula, deduce the theorem below.
\end{exercise}

\begin{theorem}[Klyachko] We have
$$
\mathsf{Z}_{\mathbb{P}^2,\ell,2,\O(1)}(q) = \overline{\eta}(q)^{-6} \sum_{m=1}^{\infty} \sum_{n=1}^{\infty} \frac{q^{mn}}{1-q^{m+n-1}}.
$$
\end{theorem}

\begin{remark}
The series $\mathsf{Z}_{\mathbb{P}^2,\ell,2,\O(1)}^{\mathrm{lf}}(q)$ has remarkable properties. It is the generating series of Hurwitz class numbers \cite{Kly1}, which was shown by Zagier to be a \emph{mock modular form} of weight 3/2 \cite{Zag}. The completion of the generating series $\mathsf{Z}_{\mathbb{P}^2,\ell,2,\O(1)}(q)$ was discussed in Manschot's lecture \eqref{eqn:completedZagier}. It was used as an important test of the $S$-duality conjecture for topologically twisted $\mathcal{N} = 4$ supersymmetric Yang--Mills theory on $\PP^2$ in \cite{VW}. Finally, we note that there are calculations of $\mathsf{Z}_{\mathbb{P}^2,\ell,2,\O(1)}(q)$ using completely different methods, for example, Yoshioka's approach via the Weil conjectures \cite{Yos1} and Manschot's approach via wall-crossing \cite{Man1}. We return to $S$-duality in Section \ref{sec:Sdual}.
\end{remark}

\begin{exercise} 
(More work.)
Can you generalize the results of this section to rank $r=3$ on $\PP^2$? And to rank $r=2$ on $\PP^1 \times \PP^1$? \emph{Hint: Consult \cite{Koo2} if you get stuck. The result for $r=3$ on $\PP^2$ turns out to be an example of a mock modular form of depth 2 \cite{Man2}, which confirms the $S$-duality conjecture.}
\end{exercise}

\section{Lecture 3: Mochizuki's formula and virtual Euler characteristics} \label{sec:Moc}

\subsection{Introduction} \label{sec:lec3intro}

Let $S$ be a smooth projective surface with ample divisor $H$. Suppose $L$ is a line bundle on $S$ and $r>0$. We suppose there are no rank $r$ Gieseker strictly $H$-semistable torsion free sheaves $E$ on $S$ with $\det(E) \cong L$ (e.g.~take $\gcd(r,c_1(L)H) = 1$).\footnote{Strictly semistable means semistable but not stable.} The expected or virtual dimension of $M_S^H(r,L,c_2)$ is
\begin{align}
\begin{split} \label{eqn:vd}
\vd(r,L,c_2) &:= \dim_{\C}(\Ext^1_S(E,E)_0) - \dim_{\C}(\Ext^2_S(E,E)_0) \\
&= 2rc_2 - (r-1) c_1(L)^2 - (r^2-1) \chi(\O_S),
\end{split}
\end{align}
where $[E]$ is any point of $M_S^H(r,L,c_2)$.

\begin{exercise}
Prove \eqref{eqn:vd} by using the Hirzebruch--Riemann--Roch theorem $\chi(\O_S) - \chi(E,E) = \int_S (1- \ch(E)^{\vee} \ch(E)) \td(T_S)$, where $(-)^\vee$ sends $\ch_i$ to $(-1)^i \ch_i$. 
\end{exercise}

Motivated by the previous lectures, we are interested in the following generating series 
\begin{equation} \label{eqn:egenfun}
\mathsf{Z}_{S,H,r,L}(q) := q^{-\frac{\chi(\O_S)}{2r} + \frac{r K_S^2}{24}} \sum_{c_2 \in \Z} e(M_S^H(r,L,c_2)) q^{\frac{\vd(r,L,c_2)}{2r}}.
\end{equation}
Note the ``mysterious'' factor $q^{\bullet}$ in front --- this factor turns out to be necessary for ``good modular behaviour'' later on. For instance, as discussed below, it turns $\overline{\eta}(q)$ into $\eta(q)$. Note also that for $S = \PP^2$, this definition differs from the one of the previous subsection by a factor $q^{\bullet}$.

Very roughly speaking, $S$-duality predicts that this generating series has ``modular properties'' or ``is expressible in terms of modular forms''. We will give a mathematically more precise formulation of $S$-duality for surfaces $S$ satisfying $h^{2,0}(S) > 0$ (based on Vafa--Witten's work) in Section \ref{sec:Sdual}. Before we proceed, we require some modular objects.

We now introduce the definition of the lattices $A_{r}$ and $A_r^\vee$. Denoting the standard basis of $\Z^r$ by $(e_1, \ldots, e_r)$, the symmetric bilinear form of $A_{r}$ is determined by
$$
\langle e_i, e_j \rangle := \left\{ \begin{array}{cc} 2 & \textrm{if  } i=j \\ -1 & \textrm{if } |i-j|=1 \\ 0 & \textrm{otherwise.} \end{array}\right.
$$
Denote the matrix corresponding to this bilinear form by $M_{A_r}$. In terms of the standard basis of $\Z^r$, the symmetric bilinear form $\langle \cdot , \cdot \rangle^{\vee}$ of $A_{r}^\vee$ is given by the matrix $M_{A_r^\vee} := M_{A_r}^{-1}$. We define
\begin{align*}
\Theta_{A_{r},\ell}(q) &:= \sum_{v \in \mathbb{Z}^r} q^{\frac{1}{2} \langle v - \ell \lambda, v - \ell \lambda \rangle}, \quad  \lambda := \frac{1}{r+1}(r,r-1, \ldots, 1), \\
\Theta_{A_{r}^\vee,\ell}(q) &:= \sum_{v \in \mathbb{Z}^r} q^{\frac{1}{2} \langle v, v \rangle^\vee} e^{2 \pi \sqrt{-1} \langle v, \ell \cdot (1,0, \ldots, 0) \rangle^\vee}, \\
\Delta(q) &:= \eta(q)^{24}.
\end{align*}

We now provide a list of cases for which $\mathsf{Z}_{S,H,r,L}(q)$ was determined in the (late) 20th century.
\begin{itemize}
\item (G\"ottsche's formula) 
\begin{equation} \label{eqn:Gotnorm}
\mathsf{Z}_{S,H,1,\O}(q) = \eta(q)^{-e(S)}.
\end{equation}
\item (Klyachko's formula) 
\begin{equation} \label{eqn:Klynorm}
\mathsf{Z}_{\PP^2,\ell,2,\O(1)}(q) = \eta(q)^{-6} \sum_{m=1}^{\infty} \sum_{n=1}^{\infty} \frac{q^{mn-1/4}}{1-q^{m+n-1}}.
\end{equation}
\item ($K3$ surfaces) Let $S$ be a $K3$ surface and $r$ a prime number, then 
\begin{equation} \label{eqn:K3euler}
\mathsf{Z}_{S,H,r,L}(q) = \frac{1}{r} \sum_{j=0}^{r-1} e^{-\frac{\pi \sqrt{-1} j}{r} L^2} \Delta(e^{\frac{2 \pi \sqrt{-1} j}{r}} q^{\frac{1}{r}})^{-1}.
\end{equation}
\item (Li--Qin's blow-up formula) Suppose $\gcd(r,c_1(L)H)=1$. Denote by $\pi \colon \widehat{S} \to S$ the blow-up of $S$ in a point with exceptional divisor $C$, $\widehat{L}  := \pi^*L (\ell C)$ with $\ell \in \Z$, and $\widehat{H} := \pi^* H - \epsilon C$ with $\epsilon > 0$. Then the following formula holds up to arbitrarily high order in $q$ for a sufficiently small $\epsilon>0$ 
\begin{equation} \label{eqn:LQ}
\frac{\mathsf{Z}_{\widehat{S},\widehat{H},r,\widehat{L}}(q)}{\mathsf{Z}_{S,H,r,L}(q)} =  \frac{\Theta_{A_{r-1},\ell}(q)}{\eta(q)^r}.
\end{equation}
\end{itemize}

The proofs of \eqref{eqn:Gotnorm} and \eqref{eqn:Klynorm} were discussed in Sections \ref{sec:locHilb} and \ref{sec:Kly}. The proof of \eqref{eqn:K3euler} follows from a key fact of O'Grady \cite{OG} (and its extensions by \cite{Huy, Yos2}) as discussed in the next exercise. The blow-up formula \eqref{eqn:LQ} was derived by Li--Qin \cite{LQ1, LQ2}. For some modern developments on the blow-up formula and relations to geometric representation theory, see \cite{JLZ} and references therein. 

\begin{exercise} \label{exc:K3euler}
\hfill
\begin{enumerate}
\item[(a)]  Let $f(x) = \sum_{n=0}^{\infty} a_n x^n \in \C[[x]]$ be a formal power series. Prove that for any prime number $r$ and $k \in \Z$, we have
$$
\sum_{n \equiv k \mod r} a_n x^n = \frac{1}{r} \sum_{j=0}^{r-1} e^{-\frac{2 \pi \sqrt{-1} jk}{r}} \psi(e^{\frac{2 \pi \sqrt{-1} j}{r}} x),
$$
where the sum is over all $n \in \Z_{\geq 0}$ satisfying $n \equiv k \mod r$.
\item[(b)] Prove \eqref{eqn:K3euler} using the following fact. For $S$ a $K3$ surface, $r>0$, $H$, $L$ such that there are no Gieseker strictly $H$-semistable rank $r$ torsion free sheaves on $S$ with determinant $L$, the moduli space $M_S^H(r,L,c_2)$ is deformation equivalent to $\widetilde{S}^{[\vd(r,L,c_2) / 2]}$ for some (possibly different) $K3$ surface $\widetilde{S}$, i.e., the Hilbert scheme of points of the same dimension on a (possibly different) $K3$ surface $\widetilde{S}$. Then use that the topological Euler characteristic of these two spaces is the same.  
\end{enumerate}
\end{exercise}

One common feature of the examples above is that the moduli space is always smooth (by Exercises \ref{exc:modulionK3}, \ref{exc:moduliondelPezzo}, \ref{exc:rank1tf}). However, in general, the moduli space can be singular. For examples of singular moduli spaces when $S \subset \PP^3$ is a (very general) quintic surface, we refer to \cite{MS}. 

For singular moduli spaces (or moduli spaces of unexpected dimension), the topological Euler characteristic $e(-)$ appears to be the incorrect notion. We want to replace $e(-)$ by the virtual Euler characteristic $e^{\vir}(-)$. The goal of this lecture is to define virtual Euler characteristics  and discuss a method for calculating them.

\subsection{Virtual cycles} \label{sec:vircyc}

In order to proceed, we require the notion of virtual cycles. The upshot is that even when a moduli space $M$ is highly singular with several irreducible components (possibly of unknown dimension), it may still carry a so-called perfect obstruction theory giving rise to a virtual cycle $[M]^{\vir} \in H_{2\vd}(M,\Z)$, where $\vd$ is the expected or virtual dimension. In this context, we say that $M$ is quasi-smooth. Virtual cycles are central objects in Gromov--Witten and Donaldson--Thomas theory (e.g., see \cite{FP, Tho1, MNOP, Ric}). In these lectures, we encounter virtual cycles in the context of Seiberg--Witten, Donaldson, and Vafa--Witten theory. The foundations for virtual cycles were established by Behrend--Fantechi \cite{BF} and Li--Tian \cite{LT} (see also \cite{BCM} for an excellent introduction to virtual cycles). In this lecture, we will only cover the bare minimum. 

\begin{definition}
Let $M$ be a $\C$-scheme of finite type and $\mathbb{E} = [E^{-1} \to E^0]$ a 2-term complex of locally free sheaves on $M$ viewed as an object of $D^b(M)$ --- the bounded derived category of coherent sheaves on $M$.\footnote{Equivalently, the derived category of bounded complexes of $\O_M$-modules with coherent cohomology.} A perfect obstruction theory is a morphism $\phi \colon \mathbb{E} \to \tau^{\geq -1} \mathbb{L}_M$ in $D^b(M)$, where $\mathbb{L}_M$ denotes Illusie's cotangent complex and $\tau^{\geq -1} \mathbb{L}_M$ its truncation, such that $h^{0}(\phi)$ is an isomorphism and $h^{-1}(\phi)$ is a surjection.\footnote{Here $h^i(-)$ denotes the $i$th cohomology sheaf of the complex.}
\end{definition}

In this definition, the cotangent complex $\mathbb{L}_M$ can be seen as a replacement (or upgrade), in the case of singular schemes, of the cotangent sheaf $\Omega_M$. In particular, it has the property that $h^0(\mathbb{L}_M) \cong \Omega_M$. In turn, the cotangent complex $\mathbb{L}_M$ contains too much information and when we have a perfect obstruction theory, we can replace it by the much more economical 2-term complex of locally free sheaves $\EE$, which contains all the information about deformations and obstructions of points of $M$.

Given a perfect obstruction theory, there exists a ``natural'' class
\[
[M]^{\vir} \in H_{2\vd}(M,\Z), \quad \vd := \vd(M) := \rk(\mathbb{E}).
\]
We often refer to $\EE$ and $\EE^\vee$ as the virtual cotangent and virtual tangent bundle of $M$ respectively, writing $\Omega_M^{\vir} = \EE$ and $T_{M}^{\vir} = \EE^\vee$. Note that a perfect obstruction theory is extra data on $M$ and generally $M$ may admit more than one perfect obstruction theory. In applications, $M$ will be a moduli space of certain geometric objects. By a result of Siebert \cite{Sie}, when $M$ is quasi-projective, $[M]^{\vir}$ does not depend on the map $\phi$ but only on the $K$-theory class $[\EE] \in K_0(M)$, where $K_0(M)$ denotes that Grothendieck $K$-group of coherent sheaves on $M$ (see \cite[Exercise II.6.6.10]{Har}). 

Following Fantechi--G\"ottsche \cite{FG}, inspired by the Hopf index theorem we define the virtual Euler characteristic of $M$ as 
\[
e^{\vir}(M) := \int_{[M]^{\vir}} e(T_M^{\vir}),
\]
where, writing $T_M^{\vir} = [E_0 \to E_1]$ with $E_i$ locally free, we define the total Chern class $c(T_M^{\vir}) := c(E_0) c(E_1)^{-1}$ and $e(T_M^{\vir})$ is its degree $2 \vd$ part in $H^\mathrm{even}(M,\Z)$.\footnote{In fact, for degree reasons, we may also write $e^{\vir}(M)=\int_{[M]^{\vir}}c(T_M^{\vir})$.} We will use the same notation throughout whenever we take the Euler class of a complex. Warning: since $T_M^{\vir}$ is a complex, it could a priori happen that $c(T_M^{\vir})$ has nonzero components in degree $> 2\vd$.

Above, we considered schemes $M$ over $\C$, however, one can also consider the relative setup namely schemes $M$ over a base $\C$-scheme $B$ of finite type. Then there exists a notion of relative perfect obstruction theory which, in applications, usually leads to some version of deformation invariance. This will not play a major r\^ole in these lectures. 

\begin{exercise}
\hfill
\begin{enumerate}
\item[(a)] Let $M$ be a $\C$-scheme of finite type with perfect obstruction theory $\phi \colon \EE \to \tau^{\geq -1} \LL_M$. Then the obstruction sheaf is defined by $\mathrm{Ob}_M := h^1(\EE^\vee)$.  Show that $h^{-1}(\EE) \cong \mathrm{Ob}_M^\vee := \hom_{\O_M}(\mathrm{Ob}_M, \O_M)$.
\item[(b)] Suppose $M$ is smooth. Show that $\mathrm{Ob}_M$ is locally free. 
\end{enumerate}
\end{exercise}

In the situation of the previous exercise, $[M]^{\vir} = e(\mathrm{Ob}_M) \cap [M]$. If additionally $\dim(M) = \vd(M)$, we find $[M]^{\vir} = [M]$ is just the fundamental class. In particular, in this case $e^{\vir}(M) = e(M)$ (by the Hopf index theorem).

\begin{example} \label{exam:Zs}
Suppose $E \to A$ is a rank $r$ vector bundle over a smooth connected variety of dimension $d$. Let $s \colon A \to E$ be a section and set $M := Z(s)$. Note that already in this case $M$ can be pretty complicated. E.g.~for $A = \C^3$ and $s = (xy^2,y^2z)$, $M$ has two irreducible components --- one of dimension 1 and one of dimension 2 which is non-reduced. Examples like these are explore in \cite{BCM}. Denoting the ideal sheaf of $M \subset A$ by $I:= I_{M/A}$, the dual section $s^* \colon E^* \to \O_A$ surjects onto $I$ and we can form the commutative diagram
\begin{displaymath}
\xymatrix
{
E^*|_M \ar[r] \ar^{s^*}[d] & \Omega_A|_M \ar@{=}[d] \\
I/I^2|_M \ar^{d}[r] & \Omega_A|_M,
}
\end{displaymath} 
where the bottom arrow is the K\"ahler differential and the top arrow is the composition $d \circ s^*$. As an object of $D^b(M)$, the bottom complex is isomorphic to $\tau^{\geq -1} \LL_M$ and it is therefore is independent of the choice of embedding $M \subset A$. It is easy to see that this defines a perfect obstruction theory (exercise below). Furthermore, denoting the inclusion by $\iota \colon M \hookrightarrow A$, we have $\iota_* [M]^{\vir} = c_r(E) \cap [M]$. This generalizes a fact from intersection theory: when all irreducible components of $M$ have expected codimension $d-r$, we have $\iota_* [M] = c_r(E) \cap [M]$ \cite[Proposition 14.1]{Ful}.
\end{example}

\begin{exercise}
Check that the above diagram indeed defines a perfect obstruction theory on $M$.
\end{exercise}

\subsection{Seiberg--Witten invariants} \label{sec:SW}

Let $S$ be a smooth projective surface and let $\beta \in H_2(S,\Z)$ an effective curve class. Consider the Hilbert scheme $\Hilb_{\beta}(S)$ of effective divisors $C \subset S$ in class $\beta$ (Section \ref{sec:Hilbdiv}). In this section, we will denote the Poincar\'e dual to $\beta$ by the same symbol, i.e., $\beta \in H^2(S,\Z)$. Denote by $\mathcal{C} \subset \Hilb_\beta(S) \times S$ the universal subscheme, which tautologically satisfies $\mathcal{C}|_{\{C\} \times S} = C$. Denote by $\pi \colon \Hilb_\beta(S) \times S \to \Hilb_\beta(S)$ the projection and consider the complex
\[
\EE_{\SW} := (R \pi_* \O_{\cC}(\cC))^\vee,
\]
where $(-)^\vee = R\hom(-,\O_{\Hilb_\beta}(S))$ is the derived dual. Applying derived pull-back to an element $C \in \Hilb_\beta(S)$ and taking cohomology, we obtain $$h^i(\EE_{\SW}^\vee|_C) \cong H^i(C,\O_C(C)) \cong H^i(C,N_{C/S}),$$ where we used $\O_C(C) \cong N_{C/S}$ and which can only be non-zero for $i=0,1$. For $i=0$, this recovers the space of first order deformations of $C \subset S$ and for $i=1$ the space of obstructions (for the exact meaning of these terms, the reader can consult \cite{Har2}). Writing $M:=\Hilb_\beta(S)$, it was shown by D\"urr--Kabanov--Okonek \cite{DKO} that there exists a perfect obstruction theory
\[
\EE_{\SW} \to \tau^{\geq -1} \LL_M, \quad \vd = \frac{1}{2}\beta(\beta - K_S).
\]

\begin{exercise}
Derive this virtual dimension formula.
\end{exercise}

This perfect obstruction theory is already interesting and non-trivial in the case where $b_1(S) = 0$. Then $\mathsf{aj}$ maps to a point $[\O_S(\beta)]$ and $\Hilb_\beta(S)$ equals the linear system $|\O_S(\beta)| = \PP(H^0(S,\O_S(\beta)))$ which we also denote $|\beta|$. Although $|\beta|$ is smooth, its dimension $\dim_{\C}(H^0(\O_S(\beta))) - 1$ may not equal the expected dimension $\chi(L)$. The following result can be found in \cite[Section 6.3]{Moc}.

\begin{proposition}
Let $S$ be a smooth projective surface satisfying $b_1(S) = 0$ and $h^{2,0}(S) > 0$. Let $\beta \in H_2(S,\Z)$ be an effective curve class. Then $[|\beta|]^{\vir} = 0$ unless $\vd(|\beta|) = 0$, i.e., $\beta^2 = \beta K_S$.
\end{proposition}

\begin{definition}
In the setting of the previous proposition, we call $\beta \in H^2(S,\Z)$ for which $|\beta|^{\vir} :=[|\beta|]^{\vir} \neq 0$ a Seiberg--Witten basic class for $S$. For a Seiberg--Witten basic class $\beta$ for $S$, we define the (algebro-geometric) Seiberg--Witten invariant of $S$ associated to $\beta$ by $\SW(\beta) := \deg |\beta|^{\vir} = \int_{|\beta|^{\vir}} 1$.
\end{definition}

We record the following fact \cite[Proposition 6.3.4]{Moc}.
\begin{proposition} \label{prop:SWdual}
Let $S$ be a smooth projective surface satisfying $b_1(S) = 0$ and $h^{2,0}(S) > 0$. Then there are finitely many Seiberg--Witten basic classes for $S$. Furthermore, for a Seiberg--Witten basic class $\beta \in H^2(S,\Z)$, we have that $K_S - \beta \in H^2(S,\Z)$ is also a Seiberg--Witten basic class and $\SW(K_S - \beta) = (-1)^{\chi(\O_S)} \SW(\beta)$.
\end{proposition}

For a Seiberg--Witten basic class, define $\widetilde{\mathrm{SW}}(\beta) := \SW((\beta + K_S)/2)$. It is a non-trivial fact that algebro-geometric Seiberg--Witten invariants $\widetilde{\SW}(\beta)$ coincide with the original Seiberg--Witten invariants of differential geometry \cite{DKO, CK}. Seiberg--Witten invariants in the algebro-geometric context are often rather simple as illustrated by the following example. 

\begin{example} \label{exam:mingentype}
Suppose $S$ is a $K3$ surface, then its only Seiberg--Witten basic class is $0$ and $\SW(0) = 1$. Let $S$ be a minimal surface of general type satisfying $b_1(S) = 0$ and $h^{2,0}(S) > 0$. Then its only Seiberg--Witten basic classes are $0, K_S$ and $\SW(0) = 1$, $\SW(K_S) = (-1)^{\chi(\O_S)}$ (by Proposition \ref{prop:SWdual}). A proof can be found in \cite[Theorem 7.4.1]{Mor}.
\end{example}

\subsection{Donaldson invariants} \label{sec:Don}

Let $S$ be a smooth projective surface, $H$ an ample divisor on $S$, and $[L] \in \Pic(S)$. Let $M := M_S^H(r,L,c_2)$ and suppose there exists a universal sheaf $\cE$ on $M \times S$.\footnote{This assumption can eventually be dropped by Remark \ref{univexists}.} We denote by $\pi_M \colon M \times S \to M$ the projection and we consider the complex
\[
\EE_{\mathrm{Don}} := (R\hom_{\pi_M}(\cE,\cE)_0[1])^\vee,
\]
where $R\hom_{\pi_M}(-,-)$ denotes the composition of derived functors given by $R\pi_{M*} \circ R\hom_{\O_{M \times S}} (-,-)$. It was shown by T.~Mochizuki \cite{Moc} that there exists a perfect obstruction theory
\begin{equation} \label{eqn:potDon}
\EE_{\mathrm{Don}} \to \tau^{\geq -1} \LL_M, \quad \vd:=\vd(r,L,c_2),
\end{equation}
where $\vd(r,L,c_2)$ was defined in \eqref{eqn:vd}. Taking derived pull-back to a point, we obtain
\[
h^i(\EE_{\mathrm{Don}}^\vee|_{[E]}) \cong \Ext^{i+1}_S(E,E)_0,
\]
which can only be non-zero for $i=0,1$ by simplicity of $E$ (Section \ref{sec:modulisheaves}). Thus, for $i=0$, we obtain the Zariski tangent space $\Ext_S^1(E,E)_0$ to $M$ at $[E]$ and for $i=1$ the obstruction space $\Ext_S^2(E,E)_0$. Furthermore, we obtain a virtual cycle $[M]^{\vir}$ and virtual Euler characteristic $e^{\vir}(M)$. 

\begin{exercise}
In this exercise, we show how the map \eqref{eqn:potDon} is constructed (but we do not show it is a perfect obstruction theory). The universal sheaf $\cE$ has a (truncated) Atiyah class $\mathrm{At}(\cE) \in \Hom(\cE,\cE \otimes^L \tau^{\geq -1} \LL_{M \times S}[1])$ \cite{HT}. 
\begin{enumerate}
\item[(a)] Using the fact that $\LL_{M \times S} = \pi_M^* \LL_M \oplus \pi_S^* \LL_S$, show that $\mathrm{At}(\cE)$ gives rise to an element of $\Hom(R\hom(\cE,\cE)_0, \pi_M^* \tau^{\geq -1} \LL_M[1])$. 
\item[(b)] Using the adjunction $R\pi_{M*}(-) \dashv \pi_M^{!}(-) = \pi_M^*(-) \otimes \pi_S^* K_S[2]$, construct the map \eqref{eqn:potDon}.
\end{enumerate}
\end{exercise}

Denoting by $\pi_S \colon M \times S \to S$ the projection onto $S$, for any class $\sigma \in H^{\mathrm{even}}(S,\Q)$ and $\alpha \geq 0$, we define 
\[
\tau_{\alpha}(\sigma) := \pi_{M*} \big( \ch_{2+\alpha}(\cE) \cap \pi_{S}^{*} \, \sigma \big) \in H^*(M,\Q).
\]
In analogy with Donaldson--Thomas theory, we refer to $\tau_\alpha(\sigma)$ as a descendent insertion and we call it primary when $\alpha = 0$. We take  $P(\cE)$ an arbitrary polynomial expression with $\Q$-coefficients in descendent insertions, which is invariant under replacing $\cE$ by $\cE \otimes \mathcal{L}$ for any line bundle $\mathcal{L}$ on $M \times S$.\footnote{The invariance is necessary in order to have independence of the choice of universal sheaf. This can be circumvented by working on the Deligne--Mumford stack parametrizing pairs $(E,\theta)$ where $\theta \colon \det(E) \cong L$ (which is in fact what Mochizuki does \cite{Moc}).} We refer to
\[
\int_{[M]^{\vir}} P(\cE)
\]
as a (descendent) Donaldson invariant of $S$. In traditional Donaldson theory, one usually takes $r=2$ and all insertions primary.

For surfaces with $h^{2,0}(S) > 0$ and $b_1(S) = 0$, there exists a powerful formula due to Mochizuki \cite[Theorem 7.5.2]{Moc}, which expresses \emph{any} (descendent) Donaldson invariant of $S$ in terms of Seiberg--Witten invariants of $S$ and intersection numbers on products of Hilbert schemes of points on $S$ (which we essentially know how to compute by Sections \ref{sec:hilb} and \ref{sec:locHilb}). The condition $h^{2,0}(S)>0$ is essential in what follows. We will now discuss Mochizuki's formula for $r=2$ \cite[Theorem 1.4.6]{Moc}.

We consider $S^{[n_1]} \times S^{[n_2]}$ to be endowed with a trivial $\C^*$-action, we denote the generator of the character group of this $\C^*$ by $\s$, and we write $s:=c_1^{\C^*}(\s)$. 
Following Mochizuki's notation, for any divisor classes $a_1, a_2 \in H^2(S,\Z)$ and $n_1, n_2 \in \Z_{\geq 0}$, let
\begin{equation} \label{Psi}
\Psi(a_1,a_2,n_1,n_2) := \mathrm{Coeff}_{s=0} \Bigg( \frac{P(\I_1(a_1) \otimes \s^{-1} \oplus \I_2(a_2) \otimes \s)}{Q(\I_1(a_1) \otimes \s^{-1}, \I_2(a_2) \otimes \s)} \frac{e(\O(a_1)^{[n_1]}) \, e(\O(a_2)^{[n_2]} \otimes \s^2)}{(2s)^{n_1+n_2 - \chi(\O_S)}} \Bigg).
\end{equation}
Let us explain the notation. Here $e(-)$ denotes the $\C^*$-equivariant Euler class and $\mathrm{Coeff}_{s=0}(-)$ extracts the coefficient of $s^{0}$ (which should be thought of as the residue). The notation $\I_i(a_i)$ is short-hand for $\I_i \otimes \pi_S^* \O(a_i)$. Furthermore, for any $\C^*$-equivariant sheaves $\cE_1$, $\cE_2$ on $S^{[n_1]} \times S^{[n_2]} \times S$, flat over $S^{[n_1]} \times S^{[n_2]}$, we set
$$
Q(\cE_1,\cE_2) :=e(- R\hom_{\pi}(\cE_1,\cE_2) - R\hom_{\pi}(\cE_2,\cE_1)), 
$$
where $\pi \colon S^{[n_1]} \times S^{[n_2]} \times S \rightarrow S^{[n_1]} \times S^{[n_2]}$ denotes the projection. The Euler class $e(-)$ in the definition of $Q(\cE_1,\cE_2)$ should be interpreted as follows: write the class $(-)$ as a difference of $\C^*$-equivariant vector bundles $E_0-E_1$ and define $e(-)=e(E_0)/e(E_1)$, which is well-defined because all terms are $\C^*$-moving.
Finally, $P(-)$ is the expression obtained from $P(\cE)$ by formally replacing $\cE$ by $(-)$. We define 
$$
\widetilde{\Psi}(a_1,a_2,n_1,n_2,s)
$$
by expression \eqref{Psi} but before applying $\Res_{s=0}(-)$. 

Next, let $c_1,c_2$ be a choice of Chern classes and let $\ch = (2,c_1,\frac{1}{2}c_1^2 - c_2)$ denote the corresponding Chern character. For any decomposition $c_1 = a_1 + a_2$ into $(1,1)$-classes, we set
\begin{equation} \label{cA}
\cA(a_1,a_2,c_2) := \sum_{n_1 + n_2 = c_2 - a_1 a_2} \int_{S^{[n_1]} \times S^{[n_2]}} \Psi(a_1,a_2,n_1,n_2).
\end{equation}
We denote by $\widetilde{\cA}(a_1,a_2,c_2,s)$ the same expression with $\Psi$ replaced by $\widetilde{\Psi}$.

\begin{theorem}[Mochizuki] \label{mocthm}
Let $S$ be a smooth projective surface satisfying $b_1(S) = 0$ and $h^{2,0}(S) >0$. Let $H, L,c_2$ be chosen such that there exist no rank 2 strictly Gieseker $H$-semistable sheaves on $S$ with determinant $L$ and second Chern class $c_2$. Suppose a universal sheaf $\cE$ exists on $M_{S}^{H}(2,L,c_2) \times S$ and suppose the following hold:
\begin{enumerate}
\item[$\mathrm{(a)}$] $\chi(\ch) > 0$, where $\chi(\ch) := \int_S \ch \cdot \td(T_S)$ and $\ch = (2,L,\frac{1}{2}L^2 - c_2)$.
\item[$\mathrm{(b)}$] $p_{\ch}(m) > p_{K_S}(m)$ for all $m \gg 0$, where $p_{\ch}$ and $p_{K_S}$ are the reduced Hilbert polynomials associated to $\ch$ and $e^{K_S}$.
\item[$\mathrm{(c)}$] For all Seiberg--Witten basic classes $a_1$ satisfying $a_1 H \leq (L -a_1) H$ the inequality is strict. 
\end{enumerate}
Let $P(\cE)$ be any polynomial with $\Q$-coefficients in descendent insertions invariant under replacing $\cE$ by $\cE \otimes \mathcal{L}$ for any line bundle $\mathcal{L}$ on $M_S^H(2,L,c_2) \times S$. Then we have
$$
\int_{[M_{S}^{H}(2,L,c_2)]^{\vir}} P(\cE) = -2^{1-\chi(\ch)} \sum_{{\scriptsize{\begin{array}{c} c_1(L) = a_1 + a_2 \\ a_1 H < a_2 H \end{array}}}} \SW(a_1) \, \cA(a_1,a_2,c_2).
$$
\end{theorem}

\begin{remark} \label{univexists}
The assumption that $\cE$ exists on $M_{S}^{H}(2,L,c_2) \times S$, is unnecessary. Although $\cE$ may only exist \'etale locally, $R\hom_{\pi}(\cE,\cE)_0$ exists globally \cite[Section 10.2]{HL}. 
\end{remark}

\begin{remark} \label{assumpmocthm}
It is conjectured in \cite{GNY3} that assumptions (b) and (c) can be dropped from Theorem \ref{mocthm} and the sum can be replaced by a sum over \emph{all} Seiberg--Witten basic classes (not necessarily satisfying the inequality). However, assumption (a) is necessary. 
\end{remark}

\begin{remark} \label{rem:upper}
Assumption (a) is essential and puts a bound on $c_2$, namely 
$$
c_2 < \tfrac{1}{2} L (L-K_S) + 2\chi(\O_S).
$$ 
However, in principle this bound is not a severe restriction for the following reason. Applying $- \otimes \O_S(\ell H)$ induces an isomorphism of moduli spaces (Exercise \ref{exc:tensorlinebundle})
$$
M_{S}^{H}(2,L,c_2)\cong M_S^H(2,L+ 2 \ell H,c_2+ \ell H L+\ell^2 H^2)
$$ 
and their virtual Euler characteristics are the same. Under the above isomorphism, assumption (a) becomes 
$$
c_2 < \tfrac{1}{2} L(L-K_S) + 2\chi(\O_S) + (L -  K_S) \ell H +  \ell^2 H^2,
$$ 
so the upper bound on $c_2$ can be made \emph{arbitrarily large} by taking $\ell \gg 0$ ($H$ is ample).
\end{remark}

We now apply Mochizuki's formula as follows. In the setting of Theorem \ref{mocthm}, one considers
\begin{equation} \label{eqn:HilbxHilb}
\sum_{n_1,n_2 \geq 0} q^{n_1+n_2} \int_{S^{[n_1]} \times S^{[n_2]}} \widetilde{\Psi}(a_1,a_2,n_1,n_2).
\end{equation}
This expression makes sense for \emph{any} (possibly disconnected) smooth projective surface $S$ and $(1,1)$-classes $a_1, a_2 \in H^2(S,\Z)$ not necessarily satisfying the assumptions of the theorem. The idea is to determine \eqref{eqn:HilbxHilb} using the methods of Sections \ref{sec:hilb} and \ref{sec:locHilb}, i.e., universality (Corollary \ref{cor:univ}) and Atiyah--Bott localization. This usually involves normalizing \eqref{eqn:HilbxHilb} by it leading term. For a general choice of $P(\cE)$ this strategy may not work because multiplicativity \eqref{eqn:multprop} could fail. Moreover, generally it is hard to obtain a closed expression for the generating series in this way, but it can be very effective for computing \eqref{eqn:HilbxHilb} up to some order in $q$. Finally, strictly speaking, in Sections \ref{sec:hilb}, \ref{sec:locHilb} we did not discuss intersection numbers on \emph{products} of Hilbert schemes, but the methods extend rather straightforwardly (see \cite{GNY1, GK1} for more details). 

In summary, for $P(\cE)$ suitably satisfying multiplicativity, universality combined with Atiyah--Bott localization combined with Mochizuki's theorem provides an effective algorithm for computing (descendent) Donaldson invariants of smooth projective surfaces $S$ satisfying $b_1(S) = 0$ and $h^{2,0}(S)>0$.

Finally, we would like to point out that there also exists a version of Mochizuki's formula for any rank $r>1$ \cite[Theorem 7.5.2]{Moc}. This version is very similar except that it involves products of Seiberg--Witten invariants and intersection numbers on products of Hilbert schemes
\[
\prod_{i=1}^{r-1} \SW(\beta_i), \quad \int_{S^{[n_1]} \times \cdots \times S^{[n_r]}}(...),
\]
where now there are multiple auxiliary equivariant parameters $\s_1, \ldots, \s_{r-1}$. Although, in principle, the methods of this section allow one to determine a large class of descendent Donaldson invariants, carrying it out in practice is often complicated. The two major obstacles are taking the (iterated) residues and Remark \ref{rem:upper}.

\subsection{Virtual Euler characteristics} \label{sec:evir}

Continuing the notation of the previous section, it is perhaps not obvious that Mochizuki's formula can be used to calculate virtual Euler characteristics. However, $e^{\vir}(M)$ can indeed be written as a polynomial expression $P(\cE)$ in descendent insertions \cite[Proposition 2.1]{GK1}. This leads to the following. Let the setup be as in Theorem \ref{mocthm}. Then 
$$
e(M_{S}^{H}(2,L,c_2)) = -2^{1-\chi(\ch)}  \sum_{{\scriptsize{\begin{array}{c} c_1(L) = a_1 + a_2 \\ a_1 H < a_2 H \end{array}}}} \SW(a_1) \, \cA(a_1,a_2,c_2),
$$
where $P(\cE)$ is determined by 
$$
P(\cE) := c(R\hom_{\pi}(\cE,\cE)_0[1]),
$$ 
where $c(-)$ denotes the total Chern class and $\cE = \I_1(a_1) \otimes \s^{-1} \oplus \I_2(a_2) \otimes \s$. We note that the rank of 
$$
-R\hom_{\pi}(\I_1(a_1) \otimes \s^{-1} \oplus \I_2(a_2) \otimes \s,\I_1(a_1) \otimes \s^{-1} \oplus \I_2(a_2) \otimes \s)_0
$$ 
is given by 
\begin{align*}
&\chi(\O_S)+(2n_1+2n_2 - 2\chi(\O_S)) + (n_1+n_2-\chi(\O_S(a_2-a_1))) \\
&\qquad \qquad \qquad \qquad \qquad \qquad \quad \ \ \! + (n_1+n_2-\chi(\O_S(a_1-a_2))) \\
&= 4c_2-L^2-3\chi(\O_S) = \vd(2,L,c_2).
\end{align*}

We have the following conjectural ``universal structure'' for the virtual Euler characteristics of $M$ \cite{GKL}.

\begin{conjecture}[G\"ottsche--Kool--Laarakker]  \label{conjhor}
For any $r>1$, there exist formal power series $$A_0(q), (A_{ij}(q))_{1 \leq i \leq j \leq r-1} \in \C(\!(q^{\frac{1}{2r}})\!),$$ 
only depending on $r$, with the following property.\footnote{We usually suppress the dependence of $A_0(q)$, $A_{ij}(q)$ on $r$ from the notation.} For any smooth projective surface $S$ satisfying $b_1(S) = 0$, $h^{2,0}(S)>0$, any ample divisor $H$ on $S$, $[L] \in \Pic(S)$, and $c_2 \in H^4(S,\Z)$ such that there are no rank $r$ strictly Gieseker $H$-semistable sheaves on $S$ with determinant $L$ and second Chern class $c_2$, we have that $e^{\mathrm{vir}}(M_S^H(r,L,c_2))$ equals the coefficient of $q^{\frac{\vd(r,L,c_2)}{2r} - \frac{\chi(\O_S)}{2r} +  \frac{r K_S^2}{24}}$ of the following expression
\begin{align*}
r^{2+K_S^2 - \chi(\O_S)} \Delta(q^{\frac{1}{r}})^{-\frac{\chi(\O_S)}{2}} A_0(q)^{K_S^2} \sum_{\boldsymbol{\beta} \in H^2(S,\Z)^{r-1}}   \prod_{i} \epsilon_r^{i \beta_i L} \, \SW(\beta_i) \prod_{i \leq j} A_{ij}(q)^{\beta_i \beta_j},
\end{align*}
where $\epsilon_r := e^{2 \pi \sqrt{-1}/r}$ and the sum is over all $(r-1)$-tuples $\boldsymbol{\beta} = (\beta_1, \ldots, \beta_{r-1})$ in $H^2(S,\Z)$.
\end{conjecture}

We observe that the sum in this conjecture is finite since $S$ has only finitely many Seiberg--Witten basic classes (Section \ref{sec:SW}).

Furthermore, there are conjectural expressions for $A_0(q)$, $A_{ij}(q)$  for $r=2,3,4,5$, which we will discuss in Section \ref{sec:cosmic}. It is interesting to note that for $r>2$ the universal series generally have coefficients in Galois extensions of $\Q$ even though the final answer is a virtual Euler characteristic and thus an integer. It turns out that the universal series are related by actions of the Galois group \cite{GKL}.

This conjecture has been checked in examples for $r=2,3$, i.e.~for specific values of $S,H,L,c_2$, using Mochizuki's formula. However, the evidence for $r=4,5$ comes from $S$-duality as discussed in the next section. A proof of this conjecture (and similar conjectures for other virtual invariants) has been announced by D.~Joyce using a new vertex algebra wall-crossing formalism for virtual cycles \cite{Joy}. 

\begin{remark}
Based on Conjecture \ref{conjhor}, it was conjectured in \cite{GKL} that the blow-up formula \eqref{eqn:LQ} holds for smooth projective surfaces $S$ satisfying $b_1(S) = 0$ and $h^{2,0}(S) > 0$. This conjecture was essentially proved by Kuhn--Leigh--Tanaka \cite{KLT}.
\end{remark}

\begin{exercise}
Show that the specialization of Conjecture \ref{conjhor} to $S = K3$ is consistent with \eqref{eqn:K3euler}. 
\end{exercise}

\begin{exercise} (More work.) Write a computer program that calculates virtual Euler characteristics for $r=2$. Check your answers by comparing to the case $S=K3$ given by \eqref{eqn:K3euler}.
\end{exercise}

\section{Lecture 4: Virtual degeneracy loci and $S$-duality} \label{sec:VW}

\subsection{Vafa--Witten invariants} \label{sec:TT}

Let $S$ be a smooth projective surface with ample divisor $H$. We also fix a line bundle $L$ on $S$. In Section \ref{sec:Moc}, we argued that it is natural to study the following generating series 
\begin{equation} \label{eqn:evirgenfun}
\mathsf{Z}_{S,H,r,L}(q) := q^{-\frac{\chi(\O_S)}{2r} + \frac{r K_S^2}{24}} \sum_{c_2 \in \Z} e^{\vir}(M_S^H(r,L,c_2)) q^{\frac{\vd(r,L,c_2)}{2r}}.
\end{equation}
However, this definition does not reproduce the full Vafa--Witten generating series for surfaces satisfying $h^{2,0}(S) > 0$. Concretely, for $r=2$, it reproduces (at least conjecturally and based on verifications in examples \cite{GK1}) lines two and three of \cite[Equation (5.38)]{VW}. In order to explain line one of \cite[Equation (5.38)]{VW}, we need to enlarge the moduli space following the work of Tanaka--Thomas \cite{TT1} (for an approach via reduced Donaldson--Thomas invariants of the total space of $K_S$, see \cite{GSY2}).

A Higgs pair $(E,\phi)$ on $S$ (for gauge group $\SU(r)$) consists of a torsion free sheaf $E$ on $S$ and a morphism $\phi \colon E \to E \otimes K_S$ satisfying $\tr(\phi) = 0$. A Higgs pair $(E,\phi)$ is called $H$-stable if for all $\phi$-invariant subsheaves $0 \neq F \subsetneq E$ we have
\[
p_F(m) < p_E(m), \quad \textrm{for all } m \gg 0.
\]
In particular, if $(E,\phi)$ is a Higgs pair with $E$ Gieseker $H$-stable, then $(E,\phi)$ is $H$-stable but the converse need not be true. For $c_2 \in H^4(S,\Z) \cong \Z$, denote by
$$N:=N_S^H(r,L,c_2)$$
the moduli space of rank $r$ $H$-stable Higgs pairs $(E,\phi)$ on $S$ with $\det(E) \cong L$ and $c_2(E) = c_2$. There exists a $\C^*$-action on $N$ scaling the Higgs field
\[
t \cdot [(E,\phi)] \mapsto [(E,t \phi)]
\]
and the moduli space $N$ is non-compact. We denote by $\mathfrak{t}$ a primitive character for the $\C^*$-action. 

\begin{theorem}[Tanaka--Thomas]
\label{thm:TT}
In the setup above, there exists a $\C^*$-equivariant perfect obstruction theory
\[
\mathbb{E}_{\mathrm{VW}} \to \tau^{\geq -1} \LL_{N}, \quad \vd = 0.
\]
Moreover, there exists a $\C^*$-equivariant isomorphism $\theta \colon \EE^\vee[1] \cong \EE \otimes \mathfrak{t}$ satisfying $\theta^\vee[1] = \theta \otimes \mathfrak{t}^{-1}$, i.e., it is $\C^*$-equivariantly symmetric.  
\end{theorem}

\begin{exercise} 
Show that symmetry forces $\vd = 0$. 
\end{exercise}

From now on, we assume there are no rank $r$ strictly $H$-semistable Higgs pairs $(E,\phi)$ on $S$ with $\det(E) \cong L$ (e.g.~take $\gcd(r,HL)=1$). Then $N^{\C^*}$ is proper and can be decomposed
into ``open and closed'' components\footnote{We use the term component somewhat loosely: we do not claim that $N_\mu^{\C^*}$ is connected, but merely that it is a union of connected components.} 
$$
N^{\C^*} = \bigsqcup_{\mu} N^{\C^*}_{\mu},
$$
where $\mu$ runs over all sequences of positive integers $\mu = (\mu_0, \ldots, \mu_{\ell})$ satisfying $\mu_0+\cdots+\mu_\ell=r$. The component $N^{\C^*}_\mu$ contains the $\C^*$-fixed Higgs pairs $(E,\phi)$ with weight decomposition
\begin{equation} \label{eqn:eigendeco2}
E = \bigoplus_{i=0}^{\ell} E_i \otimes \mathfrak{t}^{-i}, \quad \rk(E_i) = \mu_i.
\end{equation}

Consider the sequences $\mu=(r)$ and $\mu = (1,\ldots, 1)=:(1^r)$. The component $N^{\C^*}_{(r)}$ is called the instanton or horizontal component. The component $N^{\C^*}_{(1^r)}$ is called the vertical or monopole component . Clearly, we have 
$$
N^{\C^*}_{(r)} \cong M_S^H(r,L,c_2).
$$
Let $N_i$ be a connected component of $N^{\C^*}$. We define the virtual normal bundle 
\[
\nu_{N_i}^{\vir} := \EE_{\mathrm{VW}}^\vee |_{N_i}^m,
\]
where $(-)^m$ denotes the moving part with respect to the $\C^*$-action.\footnote{Here we use that a $\C^*$-equivariant complex of vector bundles $\mathbb{E} = [\cdots \to E_i \to E_{i+1} \to \cdots ]$ on a scheme $X$ with trivial $\C^*$-action decomposes term-wise into characters as in Section \ref{sec:ABintro}. In particular $\mathbb{E} = \mathbb{E}^f \oplus \mathbb{E}^m$.} Furthermore, we can consider the composition
\[
\EE_{\VW}|_{N_i}^f \to (\tau^{\geq -1} \LL_{N})^f \to \tau^{\geq -1} \LL_{N_i},
\]
which is itself a perfect obstruction theory on $N_i$ by a general result of Graber--Pandharipande \cite{GP}.\footnote{In fact, for any $\C^*$-equivariant perfect obstruction theory, loc.~cit.~provides a virtual version of the Atiyah--Bott localization formula. It is one of the central tools for calculations with virtual cycles. However, in this lecture, we simply \emph{define} invariants by their virtual localization formula.} Focusing on the instanton branch $M:=M_S^H(r,L,c_2)$, we have \cite{TT1}
\[
\EE_{\VW}|_M^f \cong \mathbb{E}_{\mathrm{Don}}.
\]

\begin{definition} \label{def:SUrVW}
In the above setup, we define the $\SU(r)$ Vafa--Witten partition function by
$$
\sfZ_{S,H,L}^{\SU(r)}(q):= r^{b_1(S)-1} q^{-\frac{\chi(\O_S)}{2r} + \frac{r K_S^2}{24}} \sum_{c_2 \in \Z} (-1)^{\vd(r,L,c_2)} q^{\frac{\vd(r,L,c_2)}{2r}} \int_{[N_S^H(r,L,c_2)^{\C^*}]^{\vir}} \frac{1}{e(\nu^{\vir})}.
$$
Here, denoting the connected components of $N_S^H(r,L,c_2)^{\C^*}$ by $N_i$, we have
$$
\int_{[N_S^H(r,L,c_2)^{\C^*}]^{\vir}} \frac{1}{e(\nu^{\vir})} := \sum_i \int_{[N_i]^{\vir}} \frac{1}{e(\nu_{N_i}^{\vir})}.
$$
Furthermore, writing $\nu_{N_i}$ as a difference of $\C^*$-equivariant vector bundles $\nu_{N_i} = E_{0,i} - E_{1,i}$, we define $e(\nu_{N_i}) = e(E_{0,i}) / e(E_{1,i})$, which is well-defined (and invertible) since $\nu_{N_i}$ is $\C^*$-moving.
\end{definition}

Firstly, a priori, the integrals in Definition \ref{def:SUrVW} involves the equivariant parameter $t := c_1^{\C^*}(\mathfrak{t})$. However, using the symmetry in Theorem \ref{thm:TT}, one can show that the generating series does not depend on $t$.

Secondly up to a factor $r^{b_1(S) - 1}$, the contribution of the instanton branches to $\sfZ_{S,H,L}^{\SU(r)}(q)$ precisely equals \eqref{eqn:evirgenfun} studied in the previous lecture. 

\begin{exercise}
Prove the previous two statements. For the first statement, consult the proof of \cite[Proposition 2.22]{Tho2} when you get stuck. For the second statement, you need the following fact: $\nu_M^{\vir} \cong \mathbb{E}_{\mathrm{Don}} \otimes \mathfrak{t}[-1]$ \cite{TT1}.  
\end{exercise}

Still assuming there are no strictly semistables as above, it is shown in \cite[Proposition 7.4]{TT1}, that for $H K_S < 0$ or $K_S \cong \O_S$, only the instanton branches contribute and they are smooth of expected dimension so the virtual Euler characteristics coincide with the (topological) Euler characteristics. 

\begin{exercise}
Suppose $\gcd(r,HL)=1$. Establish the facts from the previous paragraph. For the first statement, use stability and the fact that a $\C^*$-fixed Higgs field $\phi$ satisfies $\det(\phi) = 0$.
\end{exercise}

When $h^{2,0}(S) > 0$, typically both the instanton and monopole branches contribute to the Vafa--Witten invariants and the horizontal branches tend to be singular (Section \ref{sec:lec3intro}). One can decompose the partition function according to contributions of the components $N_\mu^{\C^*}$ as follows
\begin{equation} \label{def:genfundeco}
\sfZ_{S,H,L}^{\SU(r)}(q) = r^{b_1(S)-1} \sum_\mu \sfZ_{S,H,L}^{\mu}(q).
\end{equation}
Using cosection localization \cite{KL1} with respect to a non-zero holomorphic 2-form on $S$, Thomas proves the following strong result \cite[Theorem 5.23]{Tho2}.
\begin{theorem}[Thomas] \label{thm:Thovanish}
In the above setup, assume $h^{2,0}(S) > 0$ and $r$ is prime. Then we have
$$
\sfZ_{S,H,L}^{\SU(r)}(q) = r^{b_1(S)-1} \big(\sfZ_{S,H,L}^{(r)}(q) + \sfZ_{S,H,L}^{(1^r)}(q)\big).
$$
\end{theorem}

\begin{remark}
Suppose $b_1(S) = 0$ and $h^{2,0}(S)>0$, which will be our main case of interest. The assumption in this section that ``stable equals semistable'' can be removed by working with Joyce--Song Higgs pairs \cite{TT2} (the above vanishing result then also holds). The original definition in \cite{TT2} depended on a conjecture which was established by H.~Liu \cite{Liu}. It was also shown by Kuhn--Liu--Thimm that the invariants are independence of $H$ \cite{KLTh} (this crucially depends on $h^{2,0}(S)>0$). 
\end{remark}

\subsection{$S$-duality} \label{sec:Sdual}

Let $S$ be a smooth projective surface satisfying $h^{2,0}(S) > 0$ and $H_1(S,\Z) = 0$. Since $H_1(S,\Z) = 0$, fixing the determinant or the first Chern class of a Higgs pair is equivalent. In this context it is more natural to do the latter. Thus, let $r>0$ and $c_1 \in H^2(S,\Z)$ algebraic. As before, we assume there exist no strictly $H$-semistable Higgs pairs (for gauge group $\SU(r)$) on $S$ with first Chern class equal to $c_1$.

In order to mathematically state Vafa--Witten's $S$-duality conjecture, one requires the Langlands dual generating series
\[
\sfZ_{S,H,c_1}^{^{L}\mathrm{SU}(r)}(q),
\]
where $^{L} \SU(r) = \mathrm{PSU}(r) = \SU(r) / \mu_r$, where $\mu_r \leq \SU(r)$ is the cyclic subgroup generated by $\mathrm{diag}(\zeta, \ldots, \zeta)$, where $\zeta$ is a primitive $r$th root of unity. We will not introduce it here. For $r$ prime, this generating series is defined via virtual Euler characteristics of moduli spaces of \emph{twisted} sheaves in \cite{Jia,JK} (see also \cite{BGJK}). We then have the following $S$-duality conjecture.\footnote{Interestingly, when $H_1(S,\Z)$ has torsion, the $S$-duality conjecture requires modification as discussed in \cite{Wit}.}

\begin{conjecture}[Vafa--Witten] \label{Sdualconj}
Let $S$ be a smooth projective surface satisfying $H_1(S,\Z) = 0$ and $h^{2,0}(S)>0$, and let $H$ be an ample divisor on $S$. Let $r$ be prime and $c_1 \in H^2(S,\Z)$ algebraic. Then $\sfZ_{S,H,c_1}^{\SU(r)}(q)$ and $\sfZ_{S,H,c_1}^{^{L}\SU(r)}(q)$ are Fourier expansions in $q = \exp(2 \pi \sqrt{-1} \tau)$ of holomorphic functions $\sfZ_{S,H,c_1}^{\SU(r)}(\tau)$ and $\sfZ_{S,H,c_1}^{^{L}\SU(r)}(\tau)$ on the upper half plane $\mathbb{H}$ satisfying
\begin{equation*} 
\mathsf{Z}^{\SU(r)}_{S,H,c_1}(-1/\tau) = (-1)^{(r-1)\chi(\O_S)} \Bigg( \frac{ r \tau}{\sqrt{-1}} \Bigg)^{-\frac{e(S)}{2}} \mathsf{Z}^{^{L}\SU(r)}_{S,H,c_1}(\tau).
\end{equation*}
\end{conjecture}

In order to verify $S$-duality in examples, we need a bit of lattice theory based on \cite{VW, LL} (see, e.g., \cite[Section 4.2]{GK3} for proofs). Recall that we assume $H_1(S,\Z) = 0$, so $H^*(S,\Z)$ has no torsion (Poincar\'e duality combined with the universal coefficient theorem). The lattice $H^2(S,\Z)$ with its Poincar\'e pairing, denoted by $\cdot$, is torsion free and unimodular. For any $a,b \in H^2(S,\Z)$, define
\begin{align} \label{def:delta}
\delta_{a,b} := \left\{\begin{array}{cc} 1 & \mathrm{if \, } a-b \in rH^2(S,\mathbb{Z}) \\ 0 & \mathrm{otherwise.} \end{array} \right.
\end{align}
For any prime $r$, we have
$$
H^2(S,\Z) / r H^2(S,\Z) \cong H^2(S,\Z) \otimes \Z/r\Z \cong H^2(S,\Z/ r\Z). 
$$
Denoting the Betti numbers of $S$ by $b_i(S)$ and its signature by $\sigma(S)$, we have $b_2(S) = b_2^+(S) + b_2^-(S)$ and $\sigma(S) = b_2^+(S) - b_2^-(S)$. Then for any $c_1 \in H^2(S,\Z)$ (not necessarily algebraic), we have
\begin{align}
\begin{split} \label{eqn:fluxrk2}
\sum_{[x] \in H^2(S,\Z/2\Z)} (-1)^{c_1 \cdot x} &= 2^{b_2(S)} \delta_{c_1,0}, \\
\sum_{[x] \in H^2(S,\Z / 2 \Z)} (-1)^{c_1 \cdot x} (\sqrt{-1})^{x^2} &= 2^{\frac{b_2(S)}{2}} (\sqrt{-1})^{\frac{\sigma(S)}{2} - c_1^2}.
\end{split}
\end{align}
For $r>2$ prime and $m =1, \ldots, r-1$, we have
\begin{align}
\begin{split} \label{eqn:fluxrkr}
\sum_{[x] \in H^2(S,\Z / r\Z)} e^{\frac{2 \pi \sqrt{-1}}{r} (c_1 \cdot x)} &= r^{b_2(S)} \delta_{c_1,0}, \\
\sum_{[x] \in H^2(S,\Z / r\Z)} e^{\frac{2 \pi \sqrt{-1}}{r} (c_1 \cdot x)} e^{\frac{\pi \sqrt{-1} (r-1)}{r} m x^2} &= \epsilon(m)^{b_2(S)} r^{\frac{b_2(S)}{2}} e^{-\frac{\pi \sqrt{-1}}{8} (r-1)^2 \sigma(S)} e^{\frac{\pi \sqrt{-1} (r-1)}{r} n c_1^2},
\end{split}
\end{align}
where $mn \equiv -1 \mod r$,
$$
\epsilon(m) := \left\{ \begin{array}{cc} \Big( \frac{m/2}{r} \Big) & \textrm{if} \ m \ \textrm{is even} \\ \Big( \frac{(m+r)/2}{r} \Big) & \textrm{if} \ m \ \textrm{is odd} \end{array} \right.
$$
and $\big( \frac{a}{r} \big)$ denotes the Legendre symbol.

\begin{exercise}
Prove the first two equations of \eqref{eqn:fluxrk2} and \eqref{eqn:fluxrkr}. 
\end{exercise}

\begin{exercise}
In the setting of Conjecture \ref{Sdualconj}, let $S$ be a $K3$ surface and consider $\sfZ_{S,H,c_1}^{\SU(r)}(q) = r^{-1} \big(\sfZ_{S,H,c_1}^{(r)}(q) + \sfZ_{S,H,c_1}^{(1^r)}(q)\big)$ (Theorem \ref{thm:Thovanish}). Then
\begin{align*}
\sfZ_{S,H,c_1}^{(1)^r}(q) = \frac{\delta_{c_1,0}}{r^2} \Delta(q^r)^{-1}, \quad \sfZ_{S,H,c_1}^{(r)}(q) = \frac{1}{r} \sum_{j=0}^{r-1} e^{-\frac{\pi \sqrt{-1} j}{r} c_1^2} \Delta(e^{\frac{2 \pi \sqrt{-1} j}{r}} q^{\frac{1}{r}})^{-1}.
\end{align*}
The second equality follows from Exercise \ref{exc:K3euler} and the first was calculated in \cite{TT2, Laa1}. Using these expressions,\footnote{It turns out that the expressions in this exercise, when appropriately defined \cite{TT2, Laa2, JK}, also hold in the strictly semistable case, i.e., we can use them for any algebraic $c_1 \in H^2(S,\Z)$.} \footnote{It is important to note that not all classes in $H^2(S,\Z/r\Z)$ can be represented by algebraic classes (i.e., classes of Hodge type $(1,1)$). This is the reason why the geometric definition of $\mathsf{Z}_{S,H,c_1}^{^{L}\SU(r)}(q)$ involves twisted sheaves. The formula \eqref{eqn:LSUrK3} is established in \cite{JK}.}
\begin{equation} \label{eqn:LSUrK3}
\mathsf{Z}_{S,H,c_1}^{^{L}\SU(r)}(q) = \sum_{w \in H^2(S,\Z / r \Z)} e^{\frac{2 \pi \sqrt{-1}}{r} c_1 \cdot w} \big(\sfZ_{S,H,c_1}^{(r)}(q) + \sfZ_{S,H,c_1}^{(1^r)}(q)\big).
\end{equation}
For any $m = 1, \ldots, r-1$, we define $n \in \{1, \ldots, r-1\}$ by the equation $mn \equiv -1 \mod r$. We record the following modular transformations under $\tau \mapsto -1/\tau$
\begin{align*}
\Delta(r \tau) \mapsto \Big( \frac{\tau}{r} \Big)^{12} \Delta\Big( \frac{\tau}{r} \Big), \quad \Delta\Big( \frac{\tau+m}{r} \Big) \mapsto \tau^{12} \Delta\Big( \frac{\tau+n}{r} \Big).
\end{align*}
Using \eqref{eqn:fluxrk2}, \eqref{eqn:fluxrkr} and the equations given in this exercise, verify that the $S$-duality conjecture holds for $K3$ surfaces and prime rank.
\end{exercise}

\subsection{Virtual degeneracy loci}

Let $S$ be a smooth projective surface satisfying $h^{2,0}(S)>0$, $b_1(S) = 0$, and with ample divisor $H$. As before, let $r>1$ and let $L$ be a line bundle on $S$ such that there are no rank $r$ strictly $H$-semistable Higgs pairs $(E,\phi)$ on $S$ (for gauge group $\SU(r)$) with second Chern class $c_2$. In this section, we focus on the monopole branch $N^{\C^*}_{(1^r)}$. A Higgs pair in this branch decomposes into rank 1 torsion free sheaves as in \eqref{eqn:eigendeco}. A rank 1 torsion free sheaf on $S$ is of the form $I_{Z/S} \otimes L$, where $Z \subset S$ is a 0-dimensional subscheme with ideal sheaf $I_{Z/S} \subset \O_S$ and $L$ is a line bundle on $S$. This observation allows us to express the monopole branch in terms of nested Hilbert schemes of points and curves on $S$. For simplicity, we focus on the case $r=2$, but the general case is similar.

For any $(n_0,n_1) \in \Z_{\geq 0}^2$ and an effective class $\beta \in H^2(S,\Z)$, consider the smooth variety $S^{[n_0]} \times S^{[n_1]} \times |\beta|$ (Section \ref{sec:hilb}). Inside of this space, we have the following incidence locus studied by  Gholampour--Thomas \cite{GT1, GT2} (see also \cite{GSY1})
\begin{align} \label{def:incloc}
S_\beta^{[n_0,n_1]} = \Big\{(Z_0,Z_1,C) \, : \, I_{Z_{0}/S}(-C) \subset I_{Z_1/S}  \Big\} \stackrel{\iota}{\subset} S^{[n_0]} \times S^{[n_1]} \times |\beta|.
\end{align}
Then we have an open and closed embedding
\begin{equation} \label{eqn:Higgsinnested}
N_S^H(2,L,c_2)_{(1,1)}^{\C^*} \hookrightarrow \bigcup_{\beta,n_0,n_1} S_\beta^{[n_0,n_1]},
\end{equation}
where the union is over all $\beta, n_0, n_1$ satisfying
\begin{align}
\begin{split} \label{Cherneq}
&c_1(L) \equiv K_S - \beta \mod 2 H^2(S,\Z), \\
&c_2 = n_0+n_1 + \frac{1}{4} c_1^2  -\frac{1}{4}(K_S - \beta)^2.
\end{split}
\end{align}

\begin{exercise}
Given a Higgs pair $(E,\phi) \in N_{(1,1)}^{\C^*}$, use the decomposition \eqref{eqn:eigendeco2} in order to construct the map \eqref{eqn:Higgsinnested} (at the level of $\C$-points) and derive \eqref{Cherneq}.
\end{exercise}

More precisely, for each $S_\beta^{[n_0,n_1]}$, its elements all correspond to rank 2 Higgs pairs with determinant $L$ and second Chern class $c_2$, but they may not be $H$-stable. Either all elements of $S_\beta^{[n_0,n_1]}$ are $H$-stable or they are all $H$-unstable, and $N_S^H(2,L,c_2)_{(1,1)}^{\C^*}$ is the union of the $S_\beta^{[n_0,n_1]}$ containing only $H$-stable Higss pairs.

It is tempting to think that \eqref{def:incloc} fits in the setting of Example \ref{exam:Zs}, i.e., there exists a vector bundle $E$ on $A = S^{[n_0]} \times S^{[n_1]} \times |\beta|$ cutting out $S_\beta^{[n_0,n_1]}$. Although this can indeed sometimes be realized (see Exercise \ref{exc:IHES} below), it generally gives the wrong virtual cycle on $S_\beta^{[n_0,n_1]}$. Instead, roughly speaking, Gholampour--Thomas define a virtual cycle by looking at the degeneracy locus of a certain map $d \colon E_0 \to E_1$ of vector bundles and removing a trivial piece $H^2(S,\O_S)$ from the obstruction bundle. In this way, they obtain a virtual cycle $[S_\beta^{[n_0,n_1]}]^{\vir}$ which coincides with the one coming from ($\C^*$-localized) Tanaka--Thomas's theory $[N_{(1,1)}^{\C^*}]^{\vir}$ as described in the Section \ref{sec:TT}. 

The advantage of the incidence description, is that it yields a virtual cycle which is suitable for calculations. Let $\I_i$ be the pull-back to $S^{[n_0]} \times S^{[n_1]} \times |\beta| \times S$ of the universal ideal sheaf on $S^{[n_i]} \times S$. Let $\pt = \mathrm{Spec}(\C)$ and consider the following projection 
$$
\pi \colon S^{[n_0]} \times S^{[n_1]} \times |\beta| \times S \to S^{[n_0]} \times S^{[n_1]} \times |\beta|.
$$   

\begin{theorem}[Gholampour--Thomas] \label{thm:GT}
The push-forward 
$$
\iota_* [S_\beta^{[n_0,n_1]}]^{\vir} \in H_{2n_0+2n_{1}}(S^{[n_0]} \times S^{[n_1]} \times |\beta|,\Z)
$$ 
is given by \footnote{As in Section \ref{sec:vircyc}, for the rank $n_0+n_1$ complex $\mathbb{V} := R \pi_* \O(\beta)  - R \hom_{\pi}(\I_{0},\I_1(\beta))$, we define $e(\mathbb{V})$ as the degree $2(n_0+n_1)$ part in $H^{\mathrm{even}}(S^{[n_0]} \times S^{[n_1]} \times |\beta|,\Z)$ of $c(\mathbb{V})$. In fact, it is known that $c(\mathbb{V})$ has no components in degree $>2(n_0+n_1)$ \cite[Theorem 8.3]{GT1}. This fact is known as generalized Carlsson--Okounkov vanishing \cite{GT1, CO}.}
\begin{align*}
\iota_* [S_\beta^{[n_0,n_1]}]^{\vir} = \SW(\beta) \cdot e\Big( R \pi_* \O(\beta)  - R \hom_{\pi}(\I_{0},\I_1(\beta)) \Big) \cap [S^{[n_0]} \times S^{[n_1]} \times \pt],
\end{align*}
where $\pt \in |\beta|$.
\end{theorem}

\begin{exercise} \label{exc:IHES}
Consider $S^{[n_0,n_1]} := S_\beta^{[n_0,n_1]}$ for $\beta = 0$.\footnote{This exercise is extracted from a lecture of R.~P.~Thomas at an IH\'ES Summer School in 2021.}
\begin{enumerate}
\item[(a)] Show that $Z_1 \subset Z_0$ if and only if $\Hom_S(I_{Z_0/S},I_{Z_1/S}) \neq 0$. Thus, at the level of $\C$-valued points, $S^{[n_0,n_1]}$ is defined as the locus of $(Z_0,Z_1) \in S^{[n_0]} \times S^{[n_1]}$ at which $\Hom_S(I_{Z_0/S},I_{Z_1/S}) \neq 0$.
\item[(b)] Suppose $H^1(S,\O_S) = H^2(S,\O_S) = 0$. Show that $\Ext^2_S(I_{Z_0/S},I_{Z_1/S}) = 0$ for all $(Z_0,Z_1) \in S^{[n_0]} \times S^{[n_1]}$. \emph{Hint: Use Serre duality and compare to Exercise \ref{exc:rank1tf}.}
\end{enumerate}
Assuming $H^1(S,\O_S) = H^2(S,\O_S) = 0$, one can show that $R \hom_\pi(\I_{0},\I_1)$ is quasi-isomorphic to a 2-term complex $[E_0 \stackrel{d}{\to} E_1]$ of vector bundles of ranks $e_0, e_1$ respectively. Thus at the level of $\C$-valued points, $S^{[n_0,n_1]}$ is defined as the locus where $d$ drops rank from $e_0$ to $e_0-1$.
\begin{enumerate}
\item[(c)] Show that $d$ gives a section of $(\Lambda^{e_0} E_0^*) \wedge \Lambda^{e_0} E_1$ which precisely cuts out $S^{[n_0,n_1]}$. Deduce that, via Example \ref{exam:Zs}, this gives a virtual cycle which (generally) has virtual dimension $\neq n_0+n_1$.
\end{enumerate}
\end{exercise}

Recall that we are only interested in those components $S_\beta^{[n_0,n_1]}$ which contain $H$-stable Higgs pairs. Importantly, Laarakker observed that if $S_\beta^{[n_0,n_1]}$ contains $H$-unstable Higgs pairs, we have $\iota_*[S_\beta^{[n_0,n_1]}]^{\vir} = 0$ allowing us to work with \emph{all} $n_0,n_1,\beta$ satisfying \eqref{Cherneq} \cite{Laa1}. Thus, we can push the entire calculation of the monopole part of the Vafa--Witten partition function to the smooth ambient spaces $S^{[n_0]} \times S^{[n_1]}$.

Fix $\beta,n_0,n_1$ satisfying \eqref{Cherneq}. We fix two line bundles $[L_0], [L_1] \in \Pic(S)$ determined by the following equations (recall that we assume $b_1(S) = 0$) 
\begin{align*}
c_1(L_0 \otimes L_1^*) = K_S - \beta, \quad c_1(L_0 \otimes L_1) = c_1(L). 
\end{align*}
We consider the following class in $K_0(S^{[n_0]} \times S^{[n_1]})$
\begin{equation*} \label{def:T[n]}
T^{[n_0,n_1]} := R \hom_{\pi}( \E^{[n_0,n_1]},\E^{[n_0,n_1]} \otimes K_S \otimes \mathfrak{t})_0 - R \hom_{\pi}( \E^{[n_0,n_1]},\E^{[n_0,n_1]})_0, 
\end{equation*}
where
\begin{equation*} \label{def:E[n]}
\E^{[n_0,n_1]} :=  \I_0 \otimes L_0 \oplus \I_1 \otimes L_1 \otimes \mathfrak{t}^{-1}.
\end{equation*}
This time, $\pi$ denotes the projection $\pi \colon S^{[n_0]} \times S^{[n_1]}  \times S \to S^{[n_0]} \times S^{[n_1]}$ (we already got rid of the linear systems $|\beta|$ by Theorem \ref{thm:GT}). Also recall that $(-)_0$ denotes the trace-free part.\footnote{In Section \ref{sec:modulisheaves}, we already encountered the trace maps $\tr^i \colon \Ext_X^i(E,E) \to H^i(X,\O_X)$. For any line bundle $L$ on $X$, they generalize to trace maps $\tr^i \colon \Ext_X^i(E,E \otimes L) \to H^i(X,L)$.} Furthermore, we define
\begin{equation*} \label{def:ai}
a := K_S - \beta.
\end{equation*}
Using the normalized complex
\begin{equation*} \label{def:T0[n]}
T^{[n_0,n_1]}_0 := T^{[n_0,n_1]} - T^{[0,0]}, 
\end{equation*}
we define
\[
\Upsilon(a,n_0,n_1):= \frac{e\big(R \pi_* \O(K_S - a) - R \hom_{\pi}(\I_{0}, \I_1(K_S - a))\big)}{e((T_0^{[n_0,n_1]})^m)},
\]
where $(-)^m$ denotes the $\C^*$-moving part.

Using the relation $\SW(a) = (-1)^{\chi(\O_S)} \SW(\beta)$ (Proposition \ref{prop:SWdual}), one can derive the following expression (see \cite[Section 4]{Laa1} for the derivation)
\begin{align}
\begin{split} \label{eqn1}
\sfZ_{S,H,L}^{(1,1)}(q) = &\Big( - \frac{1}{2} \Big)^{\chi(\O_S)} q^{-\chi(\O_S) + \frac{K_S^2}{12}} \times \\
&\sum_{a \in H^2(S,\Z)} (2^{-a^2} q^{-\frac{a^2}{4}}) \, \delta_{L,a}   \,   \SW(a) \sum_{n_0,n_1 \geq 0} q^{n_0+n_1}  \int_{S^{[n_0]} \times S^{[n_1]}} \Upsilon(a,n_0,n_1). 
\end{split}
\end{align}

Note that this formula has some structural similarities with Mochizuki's formula (applied to virtual Euler characteristics) in Sections \ref{sec:Don}, \ref{sec:evir}. However, it is much better behaved. Most notably, it does not involve taking residues and we do not have the restriction on second Chern classes from Remark \ref{rem:upper}. The strategy for its calculation is the same as for \eqref{eqn:HilbxHilb}: one isolates the expression  
$$
\sum_{n_0,n_1 \geq 0} q^{n_0+n_1}  \Upsilon(a,n_0,n_1),
$$
and studies it for any (possibly disconnected) smooth projective surface $S$ and divisor class $a \in H^2(S,\Z)$ using the methods described in Sections \ref{sec:hilb} and \ref{sec:locHilb} (i.e., universality and Atiyah--Bott localization). Since this formula (and its higher rank analogue) is better behaved, the analogue of Conjecture \ref{conjhor} on the monopole branch is actually a theorem:
\begin{theorem}[Laarakker] \label{thm:Laarakker}
For any $r>1$, there exist universal power series 
$$
B_0(q), (B_{ij}(q))_{1 \leq i \leq j \leq r-1} \in \Q(\!(q^{\frac{1}{2r}})\!)
$$
only depending on $r$, with the following property.\footnote{We usually suppress the dependence of $B_0(q)$, $B_{ij}(q)$ on $r$ from the notation.} For any smooth projective surface $S$ satisfying $H_1(S,\Z) = 0$, and $h^{2,0}(S)>0$, any ample divisor $H$ on $S$ and $[L] \in \Pic(S)$, we have
\begin{align*}
\sfZ_{S,H,L}^{(1^r)}(q)   = \Bigg( \frac{(-1)^{r-1}}{r \Delta(q^r)^{\frac{1}{2}}} \Bigg)^{\chi(\O_S)} B_0(q)^{K_S^2} \sum_{\boldsymbol{a} \in H^2(S,\Z)^{r-1}} \, \delta_{L,\sum_i i a_i} \, \prod_{i} \SW(a_i) \prod_{i \leq j} B_{ij}(q)^{a_i a_j},
\end{align*}
where the sum is over all $(r-1)$-tuples $\boldsymbol{a} = (a_1, \ldots, a_{r-1})$ in $H^2(S,\Z)$.
\end{theorem}

\begin{remark}
This theorem is also valid without the assumption $H_1(S,\Z) = 0$, in which case the right hand side should be divided by the order of the group of $r$-torsion elements $\mathrm{Pic}(S)[r]$ \cite{Laa2}. In fact, it also holds in the presence of strictly $H$-semistable objects \cite{Laa2}. The assumption $h^{2,0}(S) > 0$ is crucial (since this theorem relies on Theorem \ref{thm:GT}).
\end{remark}

One notable difference compared to the universality on the instanton branch (Conjecture \ref{conjhor}) is that the universal functions have rational coefficients instead of coefficients in a Galois extension of $\Q$. In fact, there is a close conjectural relationship between the universal functions in Conjecture \ref{conjhor} and Theorem \ref{thm:Laarakker}. In order to formulate it, we normalize the two universal series $A_0(q)$, $B_0(q)$ as follows
\begin{align} \label{eqn:normalizeAB}
A_0'(q) = \frac{\Theta_{A_{r-1}^{\vee},0}(q)}{\eta(q)^r} A_0(q), \quad B_0'(q) = \frac{\Theta_{A_{r-1},0}(q)}{\eta(q)^r} B_0(q).
\end{align}

The following conjecture is consistent with (and partially implies) the $S$-duality conjecture as explained in \cite[Section 4.3]{GKL}.\footnote{In \cite{GK5}, further such modular transformations at the level of universal functions are conjectured which (on a dense open $\HH^{\circ} \subset \HH$) imply the $S$-duality conjecture.} Roughly speaking, it states that the universal functions of the instanton branch and monopole branch form the components of a vector-valued modular function.
\begin{conjecture}[G\"ottsche--Kool--Laarakker] \label{conj:AB}
For any $r>1$, $A'_0(q)$, $A_{ij}(q)$, $B'_0(q)$, $B_{ij}(q)$ are Fourier expansions in $q = \exp(2 \pi \sqrt{-1} \tau)$ of meromorphic functions $A'_0(\tau)$, $A_{ij}(\tau)$, $B'_0(\tau)$, $B_{ij}(\tau)$ on a dense open subset $\HH^{\circ} \subset \mathbb{H}$ satisfying
$$
A'_0(\tau) = B'_0(-1/\tau), \quad A_{ij}(\tau) = B_{ij}(-1/\tau).
$$
\end{conjecture}

Note that Conjectures \ref{conjhor}, \ref{conj:AB}, and Theorem \ref{thm:Laarakker} do not require $r$ to be necessarily prime.

As we will see, the universal functions of this conjecture satisfy a rich modular structure related to the classical modular curve associated to the Hecke congruence subgroup $\Gamma_0(r) \leq \SL(2,\Z)$.

\subsection{Quantum vacua of cosmic strings} \label{sec:cosmic}

The expression \eqref{eqn1} for the monopole contribution to the Vafa--Witten partition function (and its higher rank analogues) has more consequences than Theorem \ref{thm:Laarakker}. Since it provides an explicit expression in terms of integrals on products of Hilbert schemes of points, it can be used to determine the universal functions $B_0(q), B_{ij}(q)$ for a given rank up to some order in $q$ (depending on computer power) by combining universality and Atiyah--Bott localization (Sections \ref{sec:hilb}, \ref{sec:locHilb}). This leads to the conjectural formulae for the universal series presented at the end of this section.

Following this path, the universal functions are ultimately expressed in terms of integrals 
\[
\int_{(\C^2)^{[n_0]} \times \cdots \times (\C^2)^{[n_{r-1}]}} (\cdots).
\]
We have also encountered $(\C^2)^{[n_0]} \times \cdots \times (\C^2)^{[n_{r-1}]}$ as connected components of fixed loci of moduli of framed sheaves under the framing torus action \eqref{eqn:Tfrfix}. This observation allows one to rewrite the universal functions in terms of the Nekrasov partition function with one adjoint matter (defined in Section \ref{sec:Nekpart}). Remarkably, the outcome is very elegant as we will now describe \cite{AKL}. 

We present the result in the case $r=2$, though an analogous result holds for any rank. For $S = \PP^2$ or $\PP^1 \times \PP^1$ (or any smooth projective toric surface), denote by $(U_\alpha)_{\alpha}$ the maximal $T$-invariant affine open cover for the standard action of $T = (\C^*)^2$. As in Section \ref{sec:locHilb}, we denote the primitive characters of the torus action on $U_\alpha \cong \C^2$ by $(\t_1^{(\alpha)},\t_2^{(\alpha)})$ (as monomials in $\t_1, \t_2$) and we set
\[
\epsilon_1^{(\alpha)} = c_1^{T}(\t_1^{(\alpha)}), \quad \epsilon_2^{(\alpha)} = c_1^{T}(\t_2^{(\alpha)}).
\]
Furthermore, for any $T$-equivariant divisor class $a$ on $S$, consider $\O_S(a)|_{U_\alpha}$ and define
$$
a^{(\alpha)} = c_1^{T}(\O_S(a)|_{U_\alpha}).
$$
Recall from Section \ref{sec:TT} that $t := c_1^{\C^*}(\mathfrak{t})$ denotes the equivariant parameter corresponding to the $\C^*$-scaling action on the Higgs field, which will act as the mass parameter in the Nekrasov partition function with one adjoint matter $\mathsf{Z}^{\mathrm{adj}}_2(\epsilon_1,\epsilon_2,w_{1,2},m;q)$. The following is shown in \cite{AKL}. 
\begin{proposition} \label{prop:main}
For any smooth projective toric surface $S$ and $T$-equivariant algebraic class $a \in H^2_T(S,\Z)$ satisfying $a^2 = a K_S$, we have
\begin{align*}
\Delta(q^2)^{-\frac{\chi(\O_S)}{2}}  B_0(q)^{K_S^2} B_{11}(q)^{a^2} = 2^{-a^2} q^{- \chi(\O_S) + \frac{K_S^2}{12} - \frac{a^2}{4}} \prod_{\alpha=1}^{e(S)} \mathsf{Z}_2^{\mathsf{adj}}(\epsilon_1^{(\alpha)},\epsilon_2^{(\alpha)},a^{(\alpha)}+t,-t;q),
\end{align*}
where on the right hand side we put $\epsilon_1 = \epsilon_2 = 0$.\footnote{Although each factor of the product over $\alpha$ generally has poles in $\epsilon_1, \epsilon_2$, the product does not. Note that we use the notation of Remark \ref{rem:Nekdiff} for the Nekrasov partition function with one adjoint matter.}
\end{proposition}

It is also possible to express the contribution of the instanton branch in terms of a Nekrasov partition function. However, the answer is much less nice compared to the monopole branch because it involves adjoint and fundamental matter due to the presence of $e(\O(a_1)^{[n_1]})  e(\O(a_2)^{[n_2]} \otimes \s^2)$ in \eqref{Psi} \cite[Proposition B.1]{GK1}.

This proposition allows one to use properties of the Nekrasov partition function to deduce relations between the universal functions $B_0(q)$, $B_{ij}(q)$. In particular, using a blow-up formula for the Nekrasov partition function due to Kuhn--Leigh--Tanaka \cite{KLT} and a wallcrossing result of Ohkawa \cite{Ohk} (proved differently and generalized to the $K$-theoretic setting in \cite{AKL}) we obtain the following relations. For any subset $I \subset [r-1]:=\{1, \ldots, r-1\}$, define
\begin{align} \label{def:CI} 
B_{I}(q) := B_0(q) \prod_{i \leq j \in I} B_{ij}(q).
\end{align} 
In particular, $B_{\varnothing}(q) = B_0(q)$. Note that that the universal functions $B_I(q)$ determine $B_0(q), B_{ij}(q)$ and vice versa. Furthermore, we define
$$
|I| := \sum_{i \in I} 1, \quad |\!|I|\!| := \sum_{i \in I} i, \quad \epsilon_r := e^{2 \pi \sqrt{-1}/r}.
$$
The following relations were conjectured in \cite[Conjecture 3.6]{GKL}.
\begin{theorem}[Arbesfeld--Kool--Laarakker] \label{thm:AKL}
For any $r, \ell$, and $i \leq j$, the following equations hold  
\begin{align*}
B_{ij}(q) = B_{r-j,r-i}(q), \quad \sum_{I \subset [r-1]} \epsilon_r^{\ell |\!|I|\!|} B_I(q)^{-1} = \frac{\Theta_{A_{r-1}^{\vee},\ell}(q)}{\eta(q)^r}.
\end{align*}
\end{theorem}

For $r=2$, these relations are enough to determine all universal functions $B_0(q)$, $B_{11}(q)$. 
For rank $r>2$, more relations are required. However, using explicit calculations on Hilbert schemes of points (as explained in Sections \ref{sec:hilb}, \ref{sec:locHilb}, and the beginning of this section), the universal functions $B_0(q)$, $B_{ij}(q)$ have been determined up to some order in $q$ for various $r>2$.\footnote{In order to find the precise orders to which they are known, we refer to \cite{GKL, GK5}.}
Based on this data, there is a complete set of conjectural closed formulae for $B_0(q)$, $B_{ij}(q)$ for $r=3,4,5$ (recently extended to all $r \leq 13$ in \cite{GK5}).

\begin{definition}
Let $S$ be a smooth projective surface satisfying $H_1(S,\Z) = 0$, $h^{2,0}(S)>0$, and let $c_1 \in H^2(S,\Z)$ (we fix $c_1$ instead of $L$). For any $r>1$, we define
\begin{align*}
\Phi_{r,S,c_1}(q) &= \overline{B}_0(q)^{K_S^2} \sum_{\boldsymbol{a} \in H^2(S,\Z)^{r-1}}  \delta_{c_1,\sum_i i a_i} \prod_{i} \SW(a_i) \prod_{i \leq j} B_{ij}(q)^{a_i a_j}, \\
\Psi_{r,S,c_1}(q) &= \overline{A}_0(q)^{K_S^2} \sum_{\boldsymbol{a} \in H^2(S,\Z)^{r-1}}  \prod_{i} \epsilon_r^{i a_i c_1} \SW(a_i) \prod_{i \leq j} A_{ij}(q)^{a_i a_j},
\end{align*}
where $\overline{B}_0(q)$, $B_{ij}(q)$ are the universal functions in Theorem \ref{thm:Laarakker} with $\overline{B}_0(q)$ defined in \eqref{eqn:normalizeAB}, and $\overline{A}_0(q)$, $A_{ij}(q)$ are the universal functions in Conjecture \ref{conjhor} with $\overline{A}_0(q)$ defined in \eqref{eqn:normalizeAB}.
\end{definition}

Instead of presenting the expressions for $B_0(q)$, $B_{ij}(q)$, it is slightly easier to give formulae for $\Phi_{r,S,c_1}(q)$ under the additional assumption that $S$ is minimal of general type (so that Example \ref{exam:mingentype} applies). The general case is discussed in \cite{GKL, GK5}. We define
\begin{align*} 
t_{A_{r},\ell}(q) := \frac{\Theta_{A_{r},0}(q)}{\Theta_{A_{r},\ell}(q)}.
\end{align*}
We turn to $\Psi_{r,S,c_1}(q)$ at the end of this section.

\medskip

\subsubsection{Rank 2}

The following formula was conjectured by Vafa--Witten \cite[(5.38), line 1]{VW} for surfaces with smooth canonical curve $C \in |K_S|$ (see also \cite{DPS} in a more general context). It is implied by Theorems \ref{thm:Laarakker} and \ref{thm:AKL} that we have
$$
\Phi_{2,S,c_1}(q) = \delta_{c_1,0} + \delta_{c_1,K_S} (-1)^{\chi(\O_S)} t_{A_1,1}(q)^{K_S^2}.
$$

This formula was derived from complete different, purely physical, considerations involving quantum vacua and mass gaps for cosmic strings in 1994 by Vafa--Witten \cite{VW}. Remarkably, more than three decades later, using purely algebro-geometric methods, we arrive at exactly the same result. We now move on to ``quantum vacua of cosmic string theory'' in higher rank.\footnote{There exists a formula for $\SU(r)$ Vafa--Witten invariants for surfaces with $h^{2,0}(S) >0$ and any prime $r>2$ in the physics literature \cite{LL}. It turns out to be incorrect.} 

\medskip

\subsubsection{Rank 3}

The following equality was conjectured in \cite{GK3}
$$
\Phi_{3,S,c_1}(q) = \delta_{c_1,0} \, t_{A_2,1}(q)^{K_S^2}  \, \big(X_{+}(q)^{K_S^2} + X_{-}(q)^{K_S^2} \big) + \big( \delta_{c_1,K_S} +\delta_{c_1,-K_S} \big) \, (-1)^{\chi(\O_S)} \, t_{A_2,1}(q)^{K_S^2}, 
$$
where $X_{\pm}(q)$ are the two roots of 
$$
X^2 - 4  t_{A_2,1}(q)^{2} \, X + 4 t_{A_2,1}(q) = 0.
$$

\subsubsection{Rank 4}

Consider Ramanujan's octic continued fraction \cite{Duk}
\begin{equation} \label{defu}
u(q)=\frac{\sqrt{2} q^{\frac{1}{8}}}{1+\frac{q}{1+q+ \frac{q^2}{1+q^2 + \cdots}}} = \frac{\sqrt{2} \, \eta(q) \eta(q^4)^2 }{\eta(q^2)^3}.
\end{equation}
Then we conjecturally have \cite{GKL}
\begin{align*}
\Phi_{4,S,c_1} = &\ \delta_{c_1,0} \Big\{ \Big( \frac{Z - Z^{-1}}{t_{A_3,2}^{-1} u(q^2)^{-4} - Z^{-1}} \Big)^{K_S^2} + \Big( \frac{Z^{-1} - Z}{t_{A_3,2}^{-1} u(q^2)^{-4} - Z} \Big)^{K_S^2} \Big\} \\
&+ \delta_{c_1,2K_S}(-1)^{\chi(\O_S)}  \Big\{ \Big( \frac{Z - Z^{-1}}{t_{A_3,2}^{-1} Z - u(q^2)^{4}} \Big)^{K_S^2} + \Big( \frac{Z^{-1} - Z}{t_{A_3,2}^{-1} Z^{-1} - u(q^2)^{4}} \Big)^{K_S^2} \Big\} \\
&+(\delta_{c_1,K_S} + \delta_{c_1,-K_S}) \, t_{A_3,1}^{K_S^2} \Big\{\big( 1 + u(q^2)^{-4} \big)^{K_S^2} + (-1)^{\chi(\O_S)}\big( u(q^2)^4+1 \big)^{K_S^2} \Big\},
\end{align*}
where $Z$ is a root of
\begin{align*}
Z -3( u(q^2)^{-2} + u(q^2)^{2}) + Z^{-1} = 0.
\end{align*}

\subsubsection{Rank 5}

For rank 5, we encounter the Rogers--Ramanujan continued fraction appearing in Ramanujan's famous letter to Hardy \cite{Ram}
$$
r(q) = \frac{q^{\frac{1}{5}}}{1+\frac{q}{1+\frac{q^2}{1+\cdots}}}.
$$
We define the following two functions 
\begin{align*}
\beta_1 =  \frac{1}{25} t_{A_4,1} \big( 3 r^{-5} + 2 -8 r^5 \big), \quad \beta_2 = \frac{1}{25} t_{A_4,2} \big( 8 r^{-5}  + 2  -3 r^{5} \big). 
\end{align*}
Then we conjecturally have \cite{GKL}
\begin{align*}
\Phi_{5,S,c_1} = \, &\delta_{c_1,0} \Bigg\{ \Bigg( \frac{Z X_+^2Y_+^2}{\beta_1\beta_2}   \Bigg)^{\frac{K_S^2}{2}} + \Bigg( \frac{X_+^2Y_-^2}{Z\beta_1\beta_2}   \Bigg)^{\frac{K_S^2}{2}} + \Bigg( \frac{X_-^2Y_+^2}{Z\beta_1\beta_2}   \Bigg)^{\frac{K_S^2}{2}} + \Bigg(  \frac{Z  X_-^2Y_-^2}{\beta_1\beta_2}   \Bigg)^{\frac{K_S^2}{2}} \Bigg\} \\
&+\big(\delta_{c_1,K_S} + \delta_{c_1,-K_S} \big) \Big\{ \beta_1^{K_S^2} + (-1)^{\chi(\O_S)}\big(X_+^{K_S^2} + X_-^{K_S^2}\big) \Big\}  \\
&+ \big(\delta_{c_1,2K_S} + \delta_{c_1,-2K_S} \big) \Big\{ \beta_2^{K_S^2} + (-1)^{\chi(\O_S)}\big(Y_+^{K_S^2} + Y_-^{K_S^2}\big) \Big\}, 
\end{align*}
where $X_{\pm}$, $Y_{\pm}$, $Z$ are the roots of
\begin{align*}
&X^2 - \frac{4}{5} \beta_1 \big(\beta_1 t_{A_4,1}^{-1}  - 1 \big) \big(3r^{-5}+1\big) X + \frac{4}{5}\beta_1^2 \big(3r^{-5} +1\big) = 0, \\
&Y^2 - \frac{4}{5} \beta_2  \big(\beta_2  t_{A_4,2}^{-1}  - 1 \big) \big(1-3r^{5} \big) Y + \frac{4}{5}\beta_2^2 \big(1-3r^{5} \big) = 0,  \\
&Z - \frac{6}{25} \big( 8r^{-5} -13 - 8r^5 \big) + Z^{-1} = 0. 
\end{align*}

The functions $t_{A_1,1}(q), t_{A_2,1}(q)$ are related to continued fractions as follows 
\begin{align*}
u(q)^{-2} =  t_{A_1,1}(q), \quad c(q)^{-1} + 4c(q)^2 = 3  t_{A_2,1}(q),
\end{align*}
where $c(q)$ is Ramanujan's cubic continued fraction
$$
c(q)=\frac{q^{\frac{1}{3}}}{1+\frac{q+q^2}{1+ \frac{q^2+q^4}{1+ \cdots}}}.
$$
Denote by $\Gamma_0(r) \leq \mathrm{SL}(2,\Z)$ the Hecke congruence subgroup of level $r$. Let $\mathbb{H} \subset \C$ denote the upper half plane. The quotient $\Gamma_0(r) \setminus (\mathbb{H} \cup \{\textrm{cusps}\})$ is called the classical modular curve and is denoted by $X_0(r)$. It has genus zero precisely when $r=1,2,\ldots, 10,12,13,16,18,25$. In these cases, elements of the function field $\C(X_0(r))$ can be expressed as rational functions in terms of a single modular function called a Hauptmodul for $\Gamma_0(r)$. The following are Hauptmoduln for $\Gamma_0(r)$ for $r=3,4,5$ respectively
$$
j_3(q) = \frac{\eta(q)^{12}}{\eta(q^3)^{12}}, \quad j_4(q) = \frac{\eta(q)^8}{\eta(q^4)^8}, \quad j_5(q) = \frac{\eta(q)^6}{\eta(q^5)^6}, \quad q = e^{2 \pi \sqrt{-1} \tau}.
$$
The continued fractions $c(q),u(q),r(q)$ are related to $j_3(q),j_4(q),j_5(q)$ as follows \cite{Duk}
\begin{align*} \label{contdvsHaupt}
c(q)^{-3} -15 + 48 c(q)^3 + 64 c(q)^6 &= j_3(q), \quad 16 u(q)^{-8} - 16 = j_4(q), \\
r(q)^{-5} - 11 - r(q)^5 &= j_5(q).
\end{align*}

Finally, combining Conjectures \ref{conjhor}, \ref{conj:AB}, Theorem \ref{thm:Laarakker} and the formulae for $\Phi_{r,S,c_1}(q)$ above, gives conjectural formulae for $\Psi_{r,S,c_1}(q)$ for $r=2,3,4,5$. The ones for $r=2,3$ were verified up to some order in a list of examples using Mochizuki's formula in \cite{GK1, GK3}. The formula for $\Psi_{r,S,c_1}(q)$ is given by the formula for $\Phi_{r,S,c_1}(q)$ after making the following replacements:
\begin{enumerate}
\item Replace $\delta_{c_1,\ell K_S}$ by $\epsilon_r^{\ell K_S c_1}$ for all $\ell$.
\item Replace $t_{A_{r-1},\ell}(q)$ by $t_{A_{r-1}^\vee,\ell}(q)$ for all $\ell$.
\item For $r=4$, replace $u(q^2)$ by
$$
\frac{\eta(q^{\frac{1}{2}}) \eta(q^\frac{1}{8})^2 }{\eta(q^\frac{1}{4})^3}.
$$
\item For $r=5$, replace $r(q)$ by
$$
\frac{1 - \varphi \, r(q)}{\varphi + r(q)},
$$
where $\varphi = \frac{1+ \sqrt{5}}{2}$ denotes the golden ratio.
\end{enumerate}

By Thomas's vanishing result (Theorem \ref{thm:Thovanish}), we have obtained complete conjectural expressions for the $\SU(r)$ Vafa--Witten partition function for $r=2,3,5$. One can prove that the resulting expressions satisfy the $S$-duality conjecture (Conjecture \ref{Sdualconj}). This was done for $r=2$ in \cite{VW, DPS, GK1} and for $r=3,5$ in \cite{GK3, GKL}. We refer to \cite{GK5} for some new developments for $r>5$.

In this chapter, our focus was on the (unrefined) Vafa--Witten invariants. Besides considering gauge groups of higher rank, the methods of this chapter can be used for studying several other virtual invariants associated to moduli of sheaves on complex smooth projective surfaces, which leads to a rich interaction between physics, differential, and algebraic geometry, for example: 
\begin{itemize}
\item $K$-theoretic refinements\footnote{For a smooth projective variety with only $(p,p)$-cohomology, its $\chi_y$-genus and Poincar\'e polynomial coincide. Moduli of stable sheaves on del Pezzo surfaces often have this property \cite{Bea}, so the earlier literature on their Poincar\'e polynomials can also be seen as providing $K$-theoretic calculations, e.g., \cite{Yos1}.} \cite{AMP, GK1, GK3, GKL, Gottsche:2019vbi, GoNaYo2, KPT, Kim:2025fpz, Tho2}; 
\item virtual Segre and Verlinde invariants \cite{Aspman:2023ate, EGL,  Furrer:2026byd, Got2, GK4, GM, MOP, Obe}; 
\item twisted Vafa--Witten invariants \cite{GW, Manschot:2021qqe, TT1}.
\end{itemize}
More discussion on the physical perspectives can be found for $K$-theoretic invariants in Section \ref{sec:K-theoretic}, for virtual Segre invariants in Section \ref{sec:SQCD}, and for twisted Vafa--Witten invariants in Section \ref{sec:N=2*}.

We also note an interesting relation between Vafa--Witten invariants and quasimap invariants arising in the case $S = E \times C$ where $E$ is an elliptic curve and $C$ a smooth projective curve \cite{Nes}, and a definition of $\mathrm{SU}(2)$ Vafa--Witten invariants via a new form of localization with respect to an involution \cite{Schirr}. Another novel direction is the study of modularity properties of generating functions of Donaldson--Thomas invariants arising from moduli spaces of stable 2-dimensional sheaves on a compact Calabi--Yau threefold via wall-crossing in \cite{FT1, FT2}.


\begin{thebibliography}{DMVV}
\bibitem{ASY}
O.~Aharony, N.~Seiberg and Y.~Tachikawa,
\textit{Reading between the lines of four-dimensional gauge theories},
JHEP \textbf{08}, 115 (2013)
doi:10.1007/JHEP08(2013)115
[arXiv:1305.0318 [hep-th]].

\bibitem{AMP} S.~Alexandrov, J.~Manschot, and B.~Pioline, \emph{$S$-duality and refined BPS indices}, Comm.~Math.~Phys.~380 (2020) 755--810.

\bibitem{AlGa03}L. \'{A}lvarez-C\'{o}nsul and O. Garc\'{i}a-Prada, 
{\it Hitchin--Kobayashi correspondence, quivers and vortices}, 
Commun. Math. Phys. {\bf 23} (2003), 1--33. 

\bibitem{AGF}
L.~Alvarez-Gaume and D.~Z.~Freedman,
\textit{Geometrical Structure and Ultraviolet Finiteness in the Supersymmetric Sigma Model},
Commun. Math. Phys. \textbf{80}, 443 (1981).

\bibitem{Apruzzi:2021nmk}
F.~Apruzzi, F.~Bonetti, I.~Garc\'\i{}a Etxebarria, S.~S.~Hosseini and S.~Schafer-Nameki,
\textit{Symmetry TFTs from String Theory},
Commun. Math. Phys. \textbf{402} (2023) no.1, 895-949
doi:10.1007/s00220-023-04737-2
[arXiv:2112.02092 [hep-th]].

\bibitem{AKL} N.~Arbesfeld, M.~Kool, and T.~Laarakker, \emph{Vafa--Witten invariants from wall-crossing for framed sheaves}, arXiv:2602.07898.

\bibitem{Aspman:2021vhs}
J.~Aspman, E.~Furrer and J.~Manschot,
\textit{Cutting and gluing with running couplings in N=2 QCD},
Phys. Rev. D \textbf{105} (2022) no.2, 025021
doi:10.1103/PhysRevD.105.025021
[arXiv:2107.04600 [hep-th]].

\bibitem{Aspman:2022sfj}
J.~Aspman, E.~Furrer and J.~Manschot,
\textit{Topological twists of massive SQCD, Part I},
Lett. Math. Phys. \textbf{114} (2024) no.3, 62
doi:10.1007/s11005-024-01803-1
[arXiv:2206.08943 [hep-th]].

\bibitem{Aspman:2023ate}
J.~Aspman, E.~Furrer and J.~Manschot,
\textit{Topological twists of massive SQCD, Part II},
Lett. Math. Phys. \textbf{114} (2024) no.4, 98
doi:10.1007/s11005-024-01829-5
[arXiv:2312.11616 [hep-th]].

\bibitem{Asuke}
T. Asuke, \emph{Godbillon-{V}ey class of transversely holomorphic
	foliations}, MSJ Memoirs, vol.~24, Mathematical Society of Japan, Tokyo,
2010. MR 2663329

\bibitem{Atiy} M. F. Atiyah, {\it Geometry of Yang--Mills fields}, Scuola Normale Superiore Pisa, 1979.

\bibitem{AtBo} M. F. Atiyah and R. Bott, {\it The Yang--Mills equations over Riemann surfaces},
Philos. Trans. Roy. Soc. London Ser. A 308 (1983), 523--615.

\bibitem{AB} M.~F.~Atiyah and R.~Bott, \emph{The moment map and equivariant cohomology}, Topol.~23 (1984) 1--28.

\bibitem{AHS}
M. F. Atiyah, N. J. Hitchin, and I. M. Singer, 
{ \it Self-duality in four-dimensional {R}iemannian geometry}, 
{Proc. Roy. Soc. London Ser. A} 362 (1978), 425--461.

\bibitem{BG}
P.~H.~Balduf and D.~Gaiotto,
\textit{Combinatorial proof of a non-renormalization theorem},
JHEP \textbf{05}, 120 (2025).

\bibitem{BKP} Y.~Bae, M.~Kool, and H.~Park, \emph{Counting surfaces on Calabi-Yau 4-folds I: Foundations}, Geom.~Topol.~30 (2026) 2431--2564.

\bibitem{BCM} L.~Battistella, F.~Carocci, and C.~Manolache, \emph{Virtual classes for the working mathematician}, SIGMA 16 (2020), 026.

\bibitem{BaFu} S. Bauer and M. Furuta, {\it A stable cohomotopy refinement of Seiberg--Witten invariants. I.} Invent. Math. 155 (2004), 1--19. \href{https://arxiv.org/abs/math/0204340}{arXiv:math/0204340}

\bibitem{Baulieu:1997nj}
L.~Baulieu, A.~Losev, and N.~Nekrasov, \textit{Chern-Simons and twisted supersymmetry in various dimensions}, {\em Nucl. Phys. B}, vol.~522,
  pp.~82--104, 1998, hep-th/9707174.

\bibitem{Baulieu:1988xs}
L.~Baulieu and I.~M.~Singer,
``Topological Yang-Mills Symmetry,''
Nucl. Phys. B Proc. Suppl. \textbf{5}, 12-19 (1988)
doi:10.1016/0920-5632(88)90366-0

\bibitem{Bea} A.~Beauville, \emph{Sur la cohomologie de certains espaces de modules de fibr\'es vectoriels}, In:
Geometry and analysis (Bombay, 1992), pages 37--40. Tata Inst.~Fund.~Res., Bombay,
1995. 

\bibitem{BBZBDN}
C.~Beem, D.~Ben-Zvi, M.~Bullimore, T.~Dimofte and A.~Neitzke, \textit{Secondary products in supersymmetric field theory},
Annales Henri Poincare \textbf{21}, no.4, 1235-1310 (2020).

\bibitem{BLLPRvR}
C.~Beem, M.~Lemos, P.~Liendo, W.~Peelaers, L.~Rastelli and B.~C.~van Rees,
\textit{Infinite Chiral Symmetry in Four Dimensions},
Commun. Math. Phys. \textbf{336}, no.3, 1359-1433 (2015).

\bibitem{BPR}
C.~Beem, W.~Peelaers and L.~Rastelli,
\textit{Deformation quantization and superconformal symmetry in three dimensions},
Commun. Math. Phys. \textbf{354}, no.1, 345-392 (2017).

\bibitem{BRvR}
C.~Beem, L.~Rastelli, and B.~C. van Rees, 
\textit{$\mathcal{W}$ symmetry in six dimensions}, 
JHEP \textbf{05} (2015), 017.

\bibitem{BF} K.~Behrend and B.~Fantechi, \textit{The intrinsic normal cone}, Invent.~Math.~128 (1997) 45--88.

\bibitem{BenettiGenolini:2020doj}
P.~Benetti Genolini and L.~Tizzano,
\textit{Instantons, symmetries and anomalies in five dimensions},
JHEP \textbf{04} (2021), 188
doi:10.1007/JHEP04(2021)188
[arXiv:2009.07873 [hep-th]].

\bibitem{BV82}
N. Berline and M. Vergne,
\newblock Classes caract{\'e}ristiques {\'e}quivariantes. Formule de localisation en cohomologie {\'e}quivariante,
\newblock {\em C. R. Acad. Sci. Paris S{\'e}r. I Math.}, 295(9):539--541, 1982.

\bibitem{BoWu}
P.~Bomans and J.~Wu,
\textit{Unravelling the Holomorphic Twist: Central Charges},
Commun. Math. Phys. \textbf{405}, no.12, 290 (2024).

\bibitem{BGWW}
P.~Bomans, N.~Garner, B.~R.~Williams and J.~Wu,
\textit{Unravelling the Holomorphic Twist II: Anomalies and Extended Supersymmetry},
arXiv:2509.16737 [hep-th].

\bibitem{Bonelli:2019boe}
G.~Bonelli, F.~Del Monte, P.~Gavrylenko and A.~Tanzini,
\textit{${\mathcal {N}}$ = $2^*$ Gauge Theory, Free Fermions on the Torus and Painlev{\'e} VI},
Commun. Math. Phys. \textbf{377}, no.2, 1381-1419 (2020)
doi:10.1007/s00220-020-03743-y
[arXiv:1901.10497 [hep-th]].

\bibitem{Bonelli:2020xps}
G.~Bonelli, F.~Fucito, J.~F. Morales, M.~Ronzani, E.~Sysoeva and A.~Tanzini,
  \emph{{Gauge theories on compact toric manifolds}},
  \href{http://dx.doi.org/10.1007/s11005-021-01419-9}{\emph{Lett. Math. Phys.}
  {\bf 111} (2021) 77}, [\href{https://arxiv.org/abs/2007.15468}{{\tt
  2007.15468}}].

\bibitem{BJ} D.~Borisov and D.~Joyce, \textit{Virtual fundamental classes for moduli spaces of sheaves on Calabi-Yau four-folds}, Geom.~Topol.~21 (2017) 3231–3311.

\bibitem{BL1} L.~A.~Borisov and A.~Libgober, \emph{Elliptic genera of singular varieties}, Duke Math.~Jour.~116 (2003) 319--351.

\bibitem{BL2} L.~A.~Borisov and A.~Libgober, \emph{McKay correspondence for elliptic genera}, Annals of Math.~161 (2005) 1521--1569.

\bibitem{BottTu}
R.~Bott, L.~W.~Tu
\textit{Differential Forms in Algebraic Topology},
Springer New York, Graduate Texts in Mathematics, 1982.

\bibitem{Bradlow:1996alg}
S.~{Bradlow} and O.~{Garcia-Prada}, \emph{{Non-abelian monopoles and
  vortices}}, {\emph{arXiv e-prints} (Feb., 1996) alg--geom/9602010},
  [\href{https://arxiv.org/abs/alg-geom/9602010}{{\tt alg-geom/9602010}}].

\bibitem{BrEt} A. Braverman and P. Etingof, {\it Instanton counting via affine Lie algebras II: from Whittaker vectors to the Seiberg-Witten prepotential}, Studies in Lie theory, 61--78, Progr. Math., 243, Birkh\"auser, Boston, 2006. \href{https://arxiv.org/abs/math/0409441}{arXiv:math/0409441}.

\bibitem{Larry}
K.~Bringmann, A.~Folsom, K.~Ono, and L.~Rolen, {\em Harmonic {M}aass forms and
  mock modular forms: theory and applications}, vol.~64 of {\em American
  Mathematical Society Colloquium Publications}.
\newblock American Mathematical Society, Providence, RI, 2017.

\bibitem{BK} J.~Bryan and M.~Kool, \emph{Donaldson--Thomas invariants of local elliptic surfaces via the topological vertex}, Forum of Math.~Sigma 7 (2019) 1--45.

\bibitem{BGKWWY}
K.~Budzik, D.~Gaiotto, J.~Kulp, B.~R.~Williams, J.~Wu and M.~Yu,
\textit{Semi-chiral operators in 4d $ \mathcal{N} $ = 1 gauge theories},
JHEP \textbf{05}, 245 (2024).

\bibitem{CO} E.~Carlsson and A.~Okounkov, \textit{Exts and vertex operators}, Duke Math.~J.~161 (2012) 1797--1815.

\bibitem{CK} H.-l.~Chang and Y.-H.~Kiem, \textit{Poincar\'e invariants are Seiberg--Witten invariants}, Geom.~Topol.~17 (2013) 1149--1163. 

\bibitem{Chen} X. Chen, {\it On Vafa--Witten equations over K\"ahler manifolds}, J. Reine Angew. Math. 814 (2024), 135--163. 
\href{https://arxiv.org/abs/2307.02964}{arXiv:2307.02964}.

\bibitem{CiKa}
			I. Ciocan-Fontanine and M. Kapranov, {\it Virtual fundamental classes via dg-manifolds}, Geom. Topol. 13 (2009), 1779--1804. \href{https://arxiv.org/abs/math/0703214}{arXiv0703214}. 

\bibitem{Closset:2021lhd}
C.~Closset and H.~Magureanu, \textit{The $U$-plane of rank-one 4d $\mathcal{N}=2$ KK
  theories}, {\em SciPost Phys.}, vol.~12, no.~2, p.~065, 2022, 2107.03509.

\bibitem{Closset:2022vjj}
C.~Closset and H.~Magureanu, \textit{Partition functions and fibering operators on
  the Coulomb branch of 5d SCFTs}, {\em JHEP}, vol.~01, p.~035, 2023,
  2209.13564.

\bibitem{Cordes:1994fc}
S.~Cordes, G.~W.~Moore and S.~Ramgoolam,
\textit{Lectures on 2-d Yang-Mills theory, equivariant cohomology and topological field theories},
Nucl. Phys. B Proc. Suppl. \textbf{41} (1995), 184-244
doi:10.1016/0920-5632(95)00434-B
[arXiv:hep-th/9411210 [hep-th]].

\bibitem{Costello1}
K.~J.~Costello,
\textit{Notes on supersymmetric and holomorphic field theories in dimensions 2 and 4},
Pure Appl. Math. Quart. \textbf{09}, no.1, 73-165 (2013).

\bibitem{Costello2}
K.~Costello,
\textit{Supersymmetric gauge theory and the Yangian},
arXiv:1303.2632.

\bibitem{Dabholkar:2020fde}
A.~Dabholkar, P.~Putrov and E.~Witten,
\textit{Duality and Mock Modularity},
SciPost Phys. \textbf{9} (2020) no.5, 072
doi:10.21468/SciPostPhys.9.5.072
[arXiv:2004.14387 [hep-th]].

\bibitem{DaGu}
B. Dai and R. Guan, {\it Transversality for the full rank part of Vafa--Witten moduli spaces}, Comm. Math. Phys. 389 (2022), 1047--1060. 

\bibitem{Dedushenko:2017tdw}
M.~Dedushenko, S.~Gukov and P.~Putrov, \emph{{Vertex algebras and 4-manifold
  invariants}},
  \href{http://dx.doi.org/10.1093/oso/9780198802013.003.0011}{\emph{{Nigel
  Hitchin's 70th Birthday Conference}} {\bf 1} (5, 2017) 249--318},
  [\href{https://arxiv.org/abs/1705.01645}{{\tt 1705.01645}}].


\bibitem{Dennemann:2026idg}
M.~Dennemann and J.~Manschot,
\textit{Equivariant Localization for N=2 Theories with Hypermultiplets},
[arXiv:2608.25000 [hep-th]].

\bibitem{DMVV} R.~Dijkgraaf, G.~Moore, E.~Verlinde, and H.~Verlinde, \emph{Elliptic genera of symmetric products and second quantized strings}, Commun.~Math.~Phys.~185 (1997) 197--209.

\bibitem{DPS} R.~Dijkgraaf, J.-S.~Park, and B.~J.~Schroers, \textit{$N=4$ supersymemtric Yang--Mills theory on a K\"ahler surface}, hep-th/9801066 ITFA-97-09.

\bibitem{Dona83} S. K. Donaldson, {\it An application of gauge theory to four-dimensional topology}, J. Diff. Geom. 18 (1983), 279--315.

\bibitem{Dona85}
S. K. Donaldson, 
{\it Anti self-dual Yang--Mills connections over complex algebraic
	   surfaces and stable vector bundles}, 
Proc. London Math. Soc. 50 (1985), 1--26. 
 
\bibitem{Dona86}S. K. Donaldson, {\it Connections, cohomology, and the intersection forms of 4-manifolds}, 
J. Differential Geom. 24 (1986), 275--341.

\bibitem{Donaldson:1987}
S. K. Donaldson, 
{\it Infinite determinants, stable bundles and curvature},  
Duke Math. J.  54 (1987),  231--247.

\bibitem{Dona87} S. K. Donaldson, {\it The orientation of Yang--Mills moduli spaces and\/ $4$-manifold topology}, J. Diff. Geom. 26 (1987), 397--428. 

\bibitem{Dona90}S. K. Donaldson, {\it Polynomial invariants for smooth four-manifolds}, Topology, {\bf 25}, (1990), 257--315. 

\bibitem{DoKr} S. K. Donaldson and P. B. Kronheimer, {\it The geometry of four-manifolds}, Oxford University Press, 1990.

\bibitem{Douglas:1996xp}
M.~R.~Douglas, S.~H.~Katz and C.~Vafa,
\textit{Small instantons, Del Pezzo surfaces and type I-prime theory},
Nucl. Phys. B \textbf{497} (1997), 155-172
doi:10.1016/S0550-3213(97)00281-2
[arXiv:hep-th/9609071 [hep-th]].

\bibitem{DK}
T.~Duchamp and M.~Kalka, \emph{Deformation theory for holomorphic foliations},
J. Differential Geometry \textbf{14} (1979), no.~3, 317--337. MR 594704

\bibitem{Duk} W.~Duke, \textit{Continued fractions and modular functions}, Bull.~AMS 42 (2005) 137--162. 

\bibitem{DKO} M.~D\"urr, A.~Kabanov, and C. Okonek, \textit{Poincar\'e invariants}, Topol.~46 (2007) 225--294.

\bibitem{EG} D.~Edidin and W.~Graham, \emph{Localization in equivariant intersection theory and the Bott residue formula}, Amer.~Jour.~Math.~120 (1998) 619--636.

\bibitem{ellingsrud1995wall}
G.~Ellingsrud and L.~G{\"o}ttsche, \textit{Wall-crossing formulas, Bott residue
  formula and the Donaldson invariants of rational surfaces}, {\em Quart. J.
  Math. Oxford Ser.}, vol.~49, no.~2, pp.~307--329, 1998, alg-geom/9506019.

\bibitem{ElGo} G. Ellingsrud and L. G\"ottsche, {\it Variation of moduli spaces and Donaldson invariants under change of polarization}, J. Reine Angew. Math. 467 (1995), 1--49. \href{https://arxiv.org/abs/alg-geom/9410005}{alg-geom/9410005}.
  
\bibitem{EGL} G.~Ellingsrud, L.~G\"ottsche, and M.~Lehn, \textit{On the cobordism class of the Hilbert scheme of a surface}, Jour.~Alg.~Geom.~10 (2001) 81--100.

\bibitem{ElGwLo}
C.~Elliott, O.~Gwilliam and M.~Lotito,
\textit{Twists of superconformal algebras},
J. Geom. Phys. \textbf{222}, 105759 (2026).

\bibitem{ES} C.~Elliott and P.~Safronov, \textit{Topological twists of supersymmetric algebras of observables}, Commun. Math. Phys. \textbf{371}, no.2, 727-786 (2019).

\bibitem{EaSaWa} R.~Eager, I.~Saberi and J.~Walcher, \textit{Nilpotence varieties}, Annales Henri Poincare \textbf{22}, no.4, 1319-1376 (2021).

\bibitem{ElSaWi}
C.~Elliott, P.~Safronov and B.~R.~Williams,
\textit{A taxonomy of twists of supersymmetric Yang{\textendash}Mills theory},
Selecta Math. \textbf{28}, no.4, 73 (2022).

\bibitem{EY} C.~Elliott and P.~Yoo, \textit{Geometric Langlands twists of $N=4$ gauge theory from derived algebraic geometry}, Adv. Theor. Math. Phys. \textbf{22}, 615-708 (2018).

\bibitem{FG} B.~Fantechi and L.~G\"ottsche, \textit{Riemann-Roch theorems and elliptic genus for virtually smooth schemes}, Geom.~Topol.~14 (2010) 83--115.

\bibitem{FHK}
G.~Faonte and B.~Hennion and M.~Kapranov,
\textit{Higher {{Kac-Moody}} Algebras and Moduli Spaces of {{G-bundles}}},
arXiv:1701.01368.

\bibitem{Feehan:1997gj}
P.~M.~N. Feehan and T.~G. Leness, \emph{{$\rm PU(2)$ monopoles. I. Regularity,
  Uhlenbeck compactness, and transversality}}, {\emph{J. Diff. Geom.} {\bf 49}
  (1998) 265--410}, [\href{https://arxiv.org/abs/dg-ga/9710032}{{\tt
  dg-ga/9710032}}].

\bibitem{FeLe1}
P. M. N. Feehan and G. Leness, 
{\it Superconformal simple type and Witten's conjecture}, 
Adv. Math. 356 (2019), 106821, 35 pp.
\href{https://arxiv.org/abs/1408.5085}{arXiv:1408.5085}

\bibitem{FeLe2}
P. M. N. Feehan and G. Leness, 
{\it The SO(3) monopole cobordism and superconformal simple type}, 
Adv. Math. 356 (2019), 106817, 38 pp.
\href{arxiv:1408.5307}{arXiv:1408.5307}

\bibitem{FeGu} B. Feigin and S. Gukov, {\it VOA[M4]}, J. Math. Phys. 61 (2020), 012302. \href{https://arxiv.org/abs/1806.02470}{arXiv:1806.02470}.

\bibitem{FGY}
L.~O.~Felder, Z.~Gui and C.~A.~S.~Young,
\textit{Higher Chiral Algebras in a Polysimplicial Model},
[arXiv:2506.09728 [math.QA]].

\bibitem{FT1} S.~Feyzbakhsh and R.~P.~Thomas, \emph{Curve counting and $S$-duality}, Épijournal de Géom.~Alg.~7 (2023).

\bibitem{FT2} S.~Feyzbakhsh and R.~P.~Thomas, \emph{Rank $r$ DT theory from rank 0}, Duke Math.~Jour.~173 (2024) 2063--2116.

\bibitem{Freed:toappear} D.~S. Freed, M.~J. Hopkins, G.~W. Moore, in preparation.

\bibitem{FrUh} D. S. Freed and K. K. Uhlenbeck, {\it Instantons and four-manifolds}, Second edition, Mathematical Sciences Research  Institute Publications, 1991.

\bibitem{FrMo} R. Friedman and J.W. Morgan, {\it Smooth four-manifolds and complex surfaces}, Springer-Verlag, Berlin, 1994. 

\bibitem{FrQi}R. Friedman and Z. Qin, {\it Flips of moduli spaces and transition formulas for Donaldson polynomial invariants of rational surfaces}, Comm. Anal. Geom. 3 (1995), 11--83. \href{https://arxiv.org/abs/alg-geom/9410007}{alg-geom/9410007}. 

\bibitem{Ful} W.~Fulton, \textit{Intersection theory}, Springer-Verlag (1998).

\bibitem{FP} W.~Fulton and R.~Pandharipande, \textit{Notes on stable maps and quantum cohomology}, Algebraic geometry Santa Cruz 1995, AMS, 45--96.

\bibitem{Furrer:2026byd}
E.~Furrer and J.~Manschot,
\textit{Universal Functions for Topological Correlators}, Annales Henri Poincare (2026),
[arXiv:2602.20279 [hep-th]].

\bibitem{Furrer:2026aea}
E.~Furrer and R.~K.~Singh,
\textit{SU($N$) QCD, four-manifolds and Segre numbers},
[arXiv:2610.03874 [hep-th]].

\bibitem{FMM:future}
E.~Furrer, J.~Manschot and G.~W. Moore, in preparation.

\bibitem{Furu} M. Furuta, {\it Monopole equation and the 118-conjecture}, Math. Res. Lett. 8 (2001), 279--291. 

\bibitem{GKW}
D.~Gaiotto, J.~Kulp and J.~Wu,
\textit{Higher operations in perturbation theory},
JHEP \textbf{05}, 230 (2025).

\bibitem{GER}
I.~Garc{\'i}a-Etxebarria and D.~Regalado,
\textit{$ \mathcal{N}=3 $ four dimensional field theories},
JHEP \textbf{03}, 083 (2016).

\bibitem{GaPa}
N.~Garner and N.~M.~Paquette,
\textit{Mathematics of String Dualities},
PoS \textbf{TASI2021}, 007 (2023).

\bibitem{GaWi}
N.~Garner and B.~R.~Williams,
\textit{Raviolo vertex algebras},
Selecta Math. \textbf{31}, no.3, 46 (2025).

\bibitem{GIKOS}
A.~Galperin, E.~Ivanov, S.~Kalitzin, V.~Ogievetsky and E.~Sokatchev,
\textit{Unconstrained Off-Shell N=3 Supersymmetric Yang-Mills Theory},
Class. Quant. Grav. \textbf{2}, 155 (1985).

\bibitem{GSY1} A.~Gholampour, A.~Sheshmani, and S.-T.~Yau, \textit{Nested Hilbert schemes on surfaces: virtual fundamental class}, Adv.~Math.~365 (2020) 107046.

\bibitem{GSY2} A.~Gholampour, A.~Sheshmani, and S.-T.~Yau, \textit{Localized Donaldson--Thomas theory of surfaces}, Amer.~Jour.~Math.~142 (2020) 405--442.

\bibitem{GT1} A.~Gholampour and R.~P.~Thomas, \textit{Degeneracy loci, virtual cycles and nested Hilbert schemes I}, Tunisian Jour.~Math.~2 (2020) 633--665.

\bibitem{GT2} A.~Gholampour and R.~P.~Thomas, \textit{Degeneracy loci, virtual cycles and nested Hilbert schemes II}, Compos.~Math.~156 (2020) 1623--1663.

\bibitem{Gottsche:1996}
L.~G\"ottsche, \textit{Modular Forms and Donaldson Invariants for 4-Manifolds with
  $b_2^+= 1$}, {\em Journal of the American Mathematical Society}, vol.~9,
  no.~3, pp.~827--843, 1996, alg-geom/9506018.

\bibitem{Gott3} L. G\"ottsche, {\it Theta functions and Hodge numbers of moduli spaces of sheaves on rational
surfaces}, Comm. Math. Phys. 206 (1999), 105--136.

\bibitem{Got} L.~G\"ottsche, \emph{The Betti numbers of the Hilbert scheme of points on a smooth projective surface}, Math.~Ann.~286 (1990) 193--207.

\bibitem{Got2} L.~G\"ottsche, \emph{Blowup formulas for Segre and Verlinde numbers of surfaces and higher rank Donaldson invariants}, In: ``Frontiers in Geometry and Topology'', Eds.: P.~M.~N.~Feehan, L.~L.~Ng, P.~S.~Ozsváth, Proc.~Symp.~Pure Math.~109 (2024) AMS. 

\bibitem{GoKo3} L. G\"{o}ttsche and M. Kool, 
			{\it Sheaves on surfaces and virtual invariants}, \href{https://arxiv.org/abs/2007.12730}{arXiv:2007.12730}. 

\bibitem{GK1} L.~G\"ottsche and M.~Kool, \textit{Virtual refinements of the Vafa--Witten formula}, Comm.~Math.~Phys.~376 (2020) 1--49.

\bibitem{GKDMVV} L.~G\"ottsche and M.~Kool, \emph{A rank 2 Dijkgraaf--Moore--Verlinde--Verlinde formula}, Comm.~Numb.~Theory and Phys.~13 (2019) 165--201. 

\bibitem{GK3} L.~G\"ottsche and M.~Kool, \textit{Refined $\mathrm{SU}(3)$ Vafa--Witten invariants and modularity}, Pure Appl.~Math.~Quart.~14 (2018) 467--513.

\bibitem{GK4} L.~G\"ottsche and M.~Kool, \emph{Virtual Segre and Verlinde numbers of projective surfaces}, Jour.~LMS 106 (2022) 2562--2608.

\bibitem{GK5} L.~G\"ottsche and M.~Kool, \emph{$S$-duality, Siegel functions, and modular curves}, in preparation.

\bibitem{GKL} L.~G\"ottsche, M.~Kool, and T.~Laarakker, \textit{$\mathrm{SU}(r)$ Vafa--Witten invariants, Ramanujan's continued fractions, and cosmic strings}, Mich.~Math.~Jour.~75 (2025) 3--63. 

\bibitem{Gottsche:2019vbi}
L.~G{\"o}ttsche, M.~Kool, and R.~A. Williams, ``{Verlinde formulae on complex
  surfaces: K-theoretic invariants},'' {\em Forum of Mathematics, Sigma},
  vol.~9, p.~e5, 2021.

\bibitem{GM} L.~G\"ottsche and A.~Mellit, \emph{Refined Verlinde and Segre formula for Hilbert schemes}, arXiv:2210.01059.
 
 
\bibitem{GNY1} L.~G\"ottsche, H.~Nakajima, and K.~Yoshioka, \textit{Instanton counting and Donaldson invariants}, J.~Diff.~Geom.~80 (2008) 343--390.


\bibitem{GoNaYo2} L. G\"{o}ttsche, H. Nakajima and K. Yoshioka, {\it K-theoretic Donaldson invariants via instanton counting}, Pure Appl. Math. Q. 5 (2009), 1029--1111. \href{https://arxiv.org/abs/math/0611945}{arXiv:0611945}.


\bibitem{GNY3} L.~G\"ottsche, H.~Nakajima, and K.~Yoshioka, \textit{Donaldson = Seiberg--Witten from Mochizuki's formula and instanton counting}, Publ.~Res.~Inst.~Math.~Sci.~47 (2011) 307--359.

\bibitem{GW}
L.~G\"ottsche and W.~Wang, \emph{Twisted Vafa--Witten invariants}, in preparation.

\bibitem{GoZa} L. G\"{o}ttsche and D. Zagier, {\it Jacobi forms and the structure of Donaldson invariants for\/ $4$-manifolds with $b^{+} =1$}, Selecta Math. 4 (1998), 69--115. \href{https://arxiv.org/abs/alg-geom/9612020}{alg-geom/9612020}.

\bibitem{GP} T.~Graber and R.~Pandharipande, \textit{Localization of virtual classes}, Invent.~Math.~135 (1999) 487--518. 

\bibitem{Groj} I. Grojnowski, {\it Instantons and affine algebras. I. The Hilbert scheme and vertex operators},
Math. Res. Lett. 3 (1996), 275--291.

\bibitem{GJT} J. Gross, D. Joyce, and Y. Tanaka, {\it Universal structures in $\mathbb{C}$-linear enumerative invariant theories}, SIGMA 18 (2022), 068, 61pp, Special Issue on Enumerative and Gauge-Theoretic Invariants in honor of Lothar G\"ottsche on the occasion of his 60th birthday. 
\href{https://arxiv.org/abs/2005.05637}{arXiv:2005.05637}.


\bibitem{Gro} A.~Grothendieck, \textit{Techniques de construction et th\'eor\`emes d'existence en g\'eom\'etrie alg\'ebrique. IV. Les sch\'emas de Hilbert}, S\'eminaire Bourbaki (1960/61), Vol.~6, Exp.~No.~221, 249--276.

\bibitem{Guan} R. Guan, {\it On the general part of the perturbed Vafa--Witten moduli spaces on 4-manifolds} J. Funct. Anal. 284 (2023), Paper No. 109780, 23 pp. 
\href{https://arxiv.org/abs/2207.03701}{arXiv:2207.03701}.

\bibitem{Guan2} 
R. Guan, {\it Transversality of the perturbed reduced Vafa--Witten moduli spaces on 4-manifolds}, Differential Geom. Appl. 94 (2024), Paper No. 102139, 14 pp. 
\href{https://arxiv.org/abs/2212.01586}{arXiv:2212.01586}.

\bibitem{GuWaWi}
Z.~Gui, M.~Wang and B.~R.~Williams,
\textit{Higher-dimensional Chiral Algebras in the Jouanolou Model and free-field realization},
[arXiv:2510.26608 [math.QA]].

\bibitem{GwWi}
O.~Gwilliam and B.~R.~Williams,
\textit{Higher Kac{\textendash}Moody algebras and symmetries of holomorphic field theories},
Adv. Theor. Math. Phys. \textbf{25}, no.1, 129-239 (2021).

\bibitem{HaNa} G. Harder and M. S. Narasimhan,
{\it On the cohomology groups of moduli spaces of vector bundles on curves},
Math. Ann. 212 (1974/75), 215--248.

\bibitem{Har} R.~Hartshorne, \textit{Algebraic geometry}, Springer-Verlag, 1977.

\bibitem{Har2} R.~Hartshorne, \textit{Deformation theory}, Springer-Verlag, 2009.

\bibitem{Hitc}
N. Hitchin, 
{\it The self-duality equations on a Riemann surface}, 
Proc. London Math. Soc. 55 (1987), 91--114. 

\bibitem{Huang:2011qx}
M.~X.~Huang, A.~K.~Kashani-Poor and A.~Klemm,
\textit{The $\Omega$ deformed B-model for rigid $\mathcal{N}=2$ theories},
Annales Henri Poincare \textbf{14}, 425-497 (2013)
doi:10.1007/s00023-012-0192-x
[arXiv:1109.5728 [hep-th]].

\bibitem{Huy} D.~Huybrechts, \emph{Compact hyper-K\"ahler manifolds: basic results}, Invent.~Math.~135 (1999) 63--113.

\bibitem{HL} D.~Huybrechts and M.~Lehn, \textit{The geometry of moduli spaces of sheaves}, Cambridge University Press (2010).

\bibitem{HT} D.~Huybrechts and R.~P.~Thomas, \emph{Deformation-obstruction theory for complexes via Atiyah and Kodaira--Spencer classes}, Math.~Ann.~346 (2010) 545--569.

\bibitem{Hyun:1995mb}
S.~Hyun, J.~Park, and J.-S. Park, \textit{Spin-c Topological QCD}, {\em Nucl. Phys. B},
  vol.~453, pp.~199--224, 1995, hep-th/9503201.

\bibitem{Hyun:1995hz}
S.~Hyun, J.~Park, and J.-S. Park, \textit{$N=2$ supersymmetric QCD and four manifolds:
  1. The Donaldson and Seiberg-Witten invariants}, 1995, hep-th/9508162.

\bibitem{Intriligator:1997pq}
K.~A. Intriligator, D.~R. Morrison, and N.~Seiberg, \textit{Five-dimensional
  supersymmetric gauge theories and degenerations of Calabi-Yau spaces}, {\em
  Nucl. Phys. B}, vol.~497, pp.~56--100, 1997, hep-th/9702198.

\bibitem{Ive} B.~Iversen, \emph{A fixed point formula for actions of tori on algebraic varieties}, Invent.~Math.~16 (1972) 229--236.

\bibitem{Jel} J.~Jelisiejew, \textit{Pathologies on the Hilbert scheme of points}, Invent.~Math.~220(2) (2020) 581--610.

\bibitem{Jeong}
S.~Jeong, \textit{SCFT/VOA correspondence via $\Omega$-deformation},
JHEP \textbf{10} (2019), 171.

\bibitem{Jia} Y.~Jiang, \emph{Counting twisted sheaves and S-duality}, Adv.~Math 400 (2022) 108332.

\bibitem{JK} Y.~Jiang and M.~Kool, \emph{Twisted sheaves and $\SU(r)/\mathbb{Z}_r$ Vafa--Witten theory}, Math.~Ann.~382 (2022) 719--743.

\bibitem{JLZ} Q.~Jiang, W.-P.~Li, and Y.~Zhao, \emph{Instantons on the blown-up surface and the affine vertex algebra}, arXiv:2511.18959.

\bibitem{JiTh} Y. Jiang and R. P. Thomas, {\it Virtual signed Euler characteristics}, Jour. Alg. Geom. 26 (2017), 379–397. 
\href{https://arxiv.org/abs/arXiv:1408.2541}{arXiv:1408.2541}. 

\bibitem{Johansen}
A.~Johansen,
\textit{Twisting of $N=1$ SUSY gauge theories and heterotic topological theories},
Int. J. Mod. Phys. A \textbf{10}, 4325-4358 (1995).

\bibitem{Joyc1} D. Joyce, {\it Ringel–Hall style vertex algebra and Lie algebra
structures on the homology of moduli spaces}, preprint. 2018. 
\href{https://people.maths.ox.ac.uk/joyce/hall.pdf}{available in his website}. 

\bibitem{Joy} D.~Joyce, \emph{Enumerative invariants and wall-crossing formulae in abelian categories}, arXiv:2111.04694.

\bibitem{JoSo} D. Joyce and Y. Song, {\it A theory of generalized Donaldson--Thomas invariants}, Memoirs of the AMS 217, no. 1020 (2012). \href{https://arxiv.org/abs/arXiv:0810.5645}{ arXiv:0810.5645}.

\bibitem{JTU} D. Joyce, Y. Tanaka, and M. Upmeier, {\it On orientations for gauge-theoretic moduli spaces}, Adv. Math. 362 (2020), 106957. 
\href{https://arxiv.org/abs/1811.01096}{arXiv:1811.01096}. 

\bibitem{Kanno:1998qj}
H.~Kanno and S.~K.~Yang,
\textit{Donaldson-Witten functions of massless N=2 supersymmetric QCD},
Nucl. Phys. B \textbf{535} (1998), 512-530
doi:10.1016/S0550-3213(98)00560-4
[arXiv:hep-th/9806015 [hep-th]].

\bibitem{Kapustin:2006hi} A.~Kapustin, \textit{Holomorphic reduction of N=2 gauge theories, Wilson-'t Hooft operators, and S-duality}, [arXiv:hep-th/0612119 [hep-th]].

\bibitem{Kapustin:2006pk}
A.~Kapustin and E.~Witten,
\textit{Electric-Magnetic Duality And The Geometric Langlands Program},
Commun. Num. Theor. Phys. \textbf{1}, 1-236 (2007)
doi:10.4310/CNTP.2007.v1.n1.a1
[arXiv:hep-th/0604151 [hep-th]].

\bibitem{KaWi} A. Kapustin and E. Witten, {\it Electric-magnetic duality and the geometric Langlands
program}, Commun. Number Theory Phys. 1 (2007), 1--236. \href{https://arxiv.org/abs/hep-th/0604151}{hep-th/0604151}. 

\bibitem{KPT} O.~Khlaif, B.~Pioline, A.~Tanzini, \emph{Refined Vafa--Witten invariants for toric surfaces from supersymmetric localization in 5D gauge theory}, arXiv:2607.18410.

\bibitem{KL1} Y.-H.~Kiem and J.~Li, \textit{Localizing virtual cycles by cosections}, JAMS 26 (2013) 1025--1050.

\bibitem{Kim:2025fpz}
H.~Kim, J.~Manschot, G.~W.~Moore, R.~Tao and X.~Zhang,
\textit{Path Integral Derivations Of K-Theoretic Donaldson Invariants},
[arXiv:2509.23042 [hep-th]].

\bibitem{Kirw} F. C. Kirwan,
{\it Cohomology of quotients in symplectic and algebraic geometry},
Mathematical Notes, 31. Princeton University Press, Princeton, NJ, 1984. 

\bibitem{Klya} A. A. Klyachko, {\it Moduli of vector bundles and numbers of classes}  (Russian) Funktsional. Anal. i Prilozhen. 25 (1991), 81--83; translation in Funct. Anal. Appl. 25 (1991), 67--69.

\bibitem{Kly1} A.~A.~Klyachko, \emph{Vector bundles and torsion free sheaves on the projective plane}, preprint Max Planck Institut f\"ur Mathematik (1991).

\bibitem{Kly2} A.~A.~Klyachko, \emph{Equivariant bundles on toral varieties}, Math.~USSR Izvestiya 35 (1990) 337--375.

\bibitem{Kobayashi:1982}
S.~Kobayashi, \textit{Curvature and stability of vector bundles}, {\em
Proc. Japan Acad. Ser. A Math. Sci.} 58(4): 158-162 (1982)

\bibitem{Koba} S. Kobayashi, {\it Differential geometry of complex vector bundles}, Princeton University Press, 1987.

\bibitem{Kont} M.~Kontsevich, \textit{Operads and motives in deformation quantization}, Lett. Math. Phys. \textbf{48}, 35-72 (1999).

\bibitem{Koo1} M.~Kool, \emph{Fixed point loci of moduli spaces of sheaves on toric varieties}, Adv.~Math.~227 (2011) 1700--1755.

\bibitem{Koo2} M.~Kool, \emph{Euler characteristics of moduli spaces of torsion free sheaves on toric surfaces}, Geom.~Ded.~176 (2015) 241--269.

\bibitem{Korpas:2017qdo}
G.~Korpas and J.~Manschot, \textit{Donaldson-Witten theory and indefinite theta
  functions}, {\em JHEP}, vol.~11, p.~083, 2017, 1707.06235.

\bibitem{Korpas:2019cwg}
G.~Korpas, J.~Manschot, G.~W.~Moore and I.~Nidaiev,
\textit{Mocking the u-plane integral},
Res. Math. Sci. \textbf{8}, no.3, 43 (2021)
doi:10.1007/s40687-021-00280-5
[arXiv:1910.13410 [hep-th]].

\bibitem{Kots} D. Kotschick, {\it $SO(3)$-invariants for $4$-manifolds with $b^{+}_{2}=1$}, Proc. London Math. Soc. 63 (1991), 426--448. 

\bibitem{KoMo} D. Kotschick and J.W. Morgan, {\it $SO(3)$-invariants for 4-manifolds with $b^{2}_{+}=1$. II}, J. Diff. Geom. 39 (1994), 433--456. 

\bibitem{KrTaYa} J. Kryczka, Y. Tanaka, S.-T. Yau, Lagrangian correspondences of nonabelian Hodge type and shifted twistor structures, June 2026, 72 pages. \href{https://arxiv.org/abs/2606.30637}{arXiv:2606.30637}. 

\bibitem{Kron} P. B. Kronheimer, {\it Four-manifold invariants from higher rank bundles}, J. Diff. Geom. 70 (2005), 59--112. \href{https://arxiv.org/abs/math/0407518}{math.GT/0407518}.

\bibitem{KrMr} P. B. Kronheimer and T. S. Mrowka, {\it Embedded surfaces and the structure of Donaldson's polynomial invariants}, J. Diff. Geom. 41 (1995), 573--734.

\bibitem{KLT} N.~Kuhn, O.~Leigh and Y.~Tanaka, \textit{The blowup formula for the instanton part of Vafa-Witten invariants on projective surfaces}, to appear in Alg.~Geom., arXiv:2205.12953.

\bibitem{KLTh} N.~Kuhn, H.~Liu, and F.~Thimm, \emph{Wall-crossing for invariants of equivariant 3CY categories}, arXiv:2512.23012.

\bibitem{KuTa} N. Kuhn and Y. Tanaka, \textit{A blowup formula for virtual enumerative invariants on projective surfaces}, July 2021. 
\href{https://arxiv.org/abs/2107.08155}{arXiv:2107.08155}.

\bibitem{KPS} A.~Kuznetsov, Y.~Prokhorov, and C.~Shramov, \emph{Hilbert schemes of lines and conics and automorphism groups of Fano threefolds}, Japanese Jour.~Math.~13 (2018) 109--185.

\bibitem{Laa1} T.~Laarakker, \textit{Monopole contributions to refined Vafa--Witten invariants}, Geom.~Topol.~24 (2020) 2781--2828.

\bibitem{Laa2} T.~Laarakker, \textit{Vertical Vafa--Witten invariants}, Selecta Math.~27 (2021).

\bibitem{Labastida:1998sk}
J.~M.~F. Labastida and C.~Lozano, \textit{Duality in twisted N=4 supersymmetric
  gauge theories in four-dimensions}, {\em Nucl. Phys.}, vol.~B537,
  pp.~203--242, 1999, hep-th/9806032.

\bibitem{LL} J.~M.~F.~Labastida and C.~Lozano, \textit{The Vafa--Witten theory for gauge group $\SU(N)$}, Adv.~Theor.~Math.~Phys.~5 (1999) 1201--1225.

\bibitem{Labastida:1995zj}
J.~M.~F. Labastida and M.~Marino, \textit{NonAbelian monopoles on four manifolds},
  {\em Nucl. Phys.}, vol.~B448, pp.~373--398, 1995, hep-th/9504010.

\bibitem{Labastida:1995gp}
J.~M.~F. Labastida and M.~Marino, \emph{{Polynomial invariants for SU(2)
  monopoles}},
  \href{http://dx.doi.org/10.1016/0550-3213(95)00536-0}{\emph{Nucl. Phys. B}
  {\bf 456} (1995) 633--668}, [\href{https://arxiv.org/abs/hep-th/9507140}{{\tt
  hep-th/9507140}}].

\bibitem{Labastida:1996tz}
J.~M.~F.~Labastida and M.~Marino,
\textit{Twisted N=2 supersymmetry with central charge and equivariant cohomology},
Commun. Math. Phys. \textbf{185} (1997), 37-71
doi:10.1007/s002200050081
[arXiv:hep-th/9603169 [hep-th]].

\bibitem{Labastida:2005zz}
J.~Labastida and M.~Marino,
\textit{Topological quantum field theory and four manifolds},
Springer, 2005,
doi:10.1007/1-4020-3177-7

\bibitem{LV} P.~Lambrechts and  I.~Voli\'{c}, \textit{Formality of the little N-disks operad}, Mem. of the AMS; no. 1079, 2014

\bibitem{Leh1} M.~Lehn, \emph{Lectures on Hilbert schemes, algebraic structures and moduli spaces}, CRM proceedings and lecture notes v.~38 AMS Providence, R.I., 2004.

\bibitem{Leh2} M.~Lehn, \emph{Chern classes of tautological sheaves on Hilbert schemes of points on surfaces}, Invent.~Math.~136 (1999) 157--207.

\bibitem{Liju93}J. Li, {\it Algebraic geometric interpretation of Donaldson's polynomial invariants}, J. Differential Geom.  {\bf 37}  (1993), 417--466. 

\bibitem{LQ1} W.-P.~Li and Z.~Qin, \emph{On blowup formulae for the $S$-duality conjecture of Vafa and Witten}, Invent.~Math.~136 (1999) 451--482.

\bibitem{LQ2} W.-P.~Li and Z.~Qin, \emph{On blowup formulae for the $S$-duality conjecture of Vafa and Witten II: the universal functions}, Math.~Res.~Lett.~5 (1998) 439--453.

\bibitem{LT} J.~Li and G.~Tian, \textit{Virtual moduli cycles and Gromov--Witten invariants of algebraic varieties}, JAMS 11 (1998) 119--174. 
\bibitem{Liu} H.~Liu, \emph{Semistable refined Vafa-Witten invariants}, to appear in Alg.~Geom., arXiv:2309.03673.

\bibitem{LZ} B.~H.~Lian and G.~J.~Zuckerman, \textit{New perspectives on the BRST algebraic structure of string theory}, Commun. Math. Phys. \textbf{154}, 613-646 (1993).

\bibitem{LiRaTa}
C.-C. Liu, S. Rayan, and Y. Tanaka, {\it The Kapustin--Witten equations and nonabelian Hodge theory}, Eur. J. Math. 8 (2022), 23--41. 
\href{https://arxiv.org/abs/2012.06175}{arXiv:2012.06175}. 

\bibitem{LMNS1}
A.~Losev, G.~W.~Moore, N.~Nekrasov and S.~Shatashvili,
\textit{Four-Dimensional Avatars of Two-Dimensional RCFT},
Nucl. Phys. B Proc. Suppl. \textbf{46}, 130-145 (1996).

\bibitem{LMNS2}
A.~Losev, G.~W.~Moore, N.~Nekrasov and S.~Shatashvili,
\textit{Chiral Lagrangians, anomalies, supersymmetry, and holomorphy,}
Nucl. Phys. B \textbf{484}, 196-222 (1997).

\bibitem{LoNeSha}
A.~Losev, N.~Nekrasov, and S.~L. Shatashvili, \textit{Issues in topological gauge
  theory}, {\em Nucl. Phys.}, vol.~B534, pp.~549--611, 1998, hep-th/9711108.

\bibitem{Lossev:1997bz}
A.~Losev, N.~Nekrasov and S.~L.~Shatashvili,
\textit{Testing Seiberg-Witten solution},
NATO Sci. Ser. C \textbf{520}, 359-372 (1999)
[arXiv:hep-th/9801061 [hep-th]].

\bibitem{LuTe} M. L\"{u}bke and A. Teleman, {\it The Kobayashi-Hitchin correspondence}, World Scientific, 1995.

\bibitem{Malmendier:2010ss}
A.~Malmendier, \textit{Donaldson invariants of $\mathbb{P}^1 \times \mathbb{P}^1$
  and Mock Theta Functions}, {\em Commun. Num. Theor. Phys.}, vol.~5,
  pp.~203--229, 2011, 1008.0175.

\bibitem{Malmendier:2008db}
A.~Malmendier and K.~Ono,
\textit{SO(3)-Donaldson invariants of $\mathbb{CP}^2$ and Mock Theta Functions},
Geom. Topol. \textbf{16} (2012), 1767-1833
doi:10.2140/gt.2012.16.1767
[arXiv:0808.1442 [math.DG]].

\bibitem{Manschot:2011dj}
J.~Manschot,
\textit{BPS invariants of $\mathcal{N}=4$ gauge theory on Hirzebruch surfaces},
Commun. Num. Theor. Phys. \textbf{6} (2012), 497-516
doi:10.4310/CNTP.2012.v6.n2.a4
[arXiv:1103.0012 [math-ph]].

\bibitem{Man1} J.~Manschot, \emph{Sheaves on $\mathbb{P}^2$ and generalized Appell functions}, Adv.~Theor.~Math.~Phys.~21 (2017) 655--681.

\bibitem{Man2} J.~Manschot, \emph{Vafa-Witten theory and iterated integrals of modular forms}, Comm.~Math.~Phys.~371 (2019) 787--831. 

\bibitem{Manschot:2019pog}
J.~Manschot, G.~W.~Moore and X.~Zhang,
\textit{Effective gravitational couplings of four-dimensional $ \mathcal{N} $ = 2 supersymmetric gauge theories},
JHEP \textbf{06}, 150 (2020)
doi:10.1007/JHEP06(2020)150
[arXiv:1912.04091 [hep-th]].

\bibitem{Manschot:2021qqe}
J.~Manschot and G.~W.~Moore,
\textit{Topological correlators of $\mathrm{SU}(2)$, $\mathcal{N}=2^*$ SYM on four-manifolds},
Adv. Theor. Math. Phys. \textbf{28} (2024) no.2, 407-597
doi:10.4310/atmp.240914021307
[arXiv:2104.06492 [hep-th]].

\bibitem{MOP} A.~Marian, D.~Oprea, and R.~Pandharipande, \emph{The combinatorics of Lehn’s conjecture}, J.~Math.~Soc.~Japan 71 (2019) 299--308.

\bibitem{Marino:1998bm}
M.~Marino and G.~W.~Moore,
\textit{The Donaldson-Witten function for gauge groups of rank larger than one},
Commun. Math. Phys. \textbf{199}, 25-69 (1998)
doi:10.1007/s002200050494
[arXiv:hep-th/9802185 [hep-th]].

\bibitem{Marino:1998rg}
M.~Marino and G.~W.~Moore,
\textit{Donaldson invariants for nonsimply connected manifolds},
Commun. Math. Phys. \textbf{203} (1999), 249
doi:10.1007/s002200050611
[arXiv:hep-th/9804104 [hep-th]].

\bibitem{MaWe}
K. Matsuki and R. Wentworth, {\it Mumford--Thaddeus principle on the moduli space of vector bundles on an algebraic surface},  
Internat. J. Math. 8 (1997), 97--148. \href{https://arxiv.org/abs/alg-geom/9410016}{alg-geom/9410016}. 

\bibitem{MNOP} D.~Maulik, N.~Nekrasov, A.~Okounkov, and R.~Pandharipande, \textit{Gromov--Witten theory and Donaldson--Thomas theory I}, Compositio Math.~142 (2006) 1263--1285.

\bibitem{MaTh} D. Maulik and R. P. Thomas, {\it Sheaf counting on local K3 surfaces},
Pure Appl. Math. Q. 14 (2018), 419--441. 
\href{https://arxiv.org/abs/1806.02657}{arXiv:1806.02657}. 

\bibitem{MS} N.~Mestrano and C.~Simpson, \emph{Obstructed bundles of rank 2 on a quintic surface}, Inter.~Jour.~Math.~22 (6) (2011) 789--836.

\bibitem{Moc} T.~Mochizuki, \textit{Donaldson type invariants for algebraic surfaces}, Lecture Notes in Math.~1972, Springer-Verlag, Berlin (2009). 

\bibitem{Moore:1997dj}
G.~W.~Moore, N.~Nekrasov and S.~Shatashvili,
\textit{Integrating over Higgs branches},
Commun. Math. Phys. \textbf{209}, 97-121 (2000)
doi:10.1007/PL00005525
[arXiv:hep-th/9712241 [hep-th]].

\bibitem{Moore:1998et}
G.~W.~Moore, N.~Nekrasov and S.~Shatashvili,
\textit{D particle bound states and generalized instantons},
Commun. Math. Phys. \textbf{209}, 77-95 (2000)
doi:10.1007/s002200050016
[arXiv:hep-th/9803265 [hep-th]].

\bibitem{Moore:2024vsd}
G.~W.~Moore, V.~Saxena and R.~K.~Singh,
\textit{Topological Twisting of 4d $\mathcal{N}=2$ Supersymmetric Field Theories},
[arXiv:2411.14396 [hep-th]].

\bibitem{MoWi97}G. Moore and E. Witten, {\it Integration over the $u$-plane in Donaldson theory}, 
Adv. Theor. Math. Phys. {\bf 1} (1997), 298--387. 
\href{https://arxiv.org/abs/hep-th/9709193}{arXiv:hep-th/9709193}.

\bibitem{Morg93}
J. W. Morgan, {\it Comparison of the Donaldson polynomial invariants with their algebro-geometric analogues}, Topology  {\bf 32}  (1993), 449--488. 

\bibitem{Mor} J.~W.~Morgan, \emph{The Seiberg--Witten equations and applications to the topology of
smooth four-manifolds}, Math.~Notes 44, Princeton Univ.~Press, Princeton NJ (1996).

\bibitem{Nakahara:2003nw}
M.~Nakahara, 
\textit{Geometry, Topology and Physics}, 
2nd~ed., Graduate Student Series in Physics, 
CRC Press, Boca Raton, FL, 2003.

\bibitem{Naka1} H. Nakajima, {\it Instantons on ALE spaces, quiver varieties, and Kac--Moody algebras},
Duke Math. J. 76 (1994), 365--416.

\bibitem{Naka2} H. Nakajima, {\it Instantons and affine Lie algebras}, Nucl. Phys. B (Proc. Suppl.) 46 (1996), 154--161. \href{https://arxiv.org/abs/alg-geom/9510003}{alg-geom/9510003}.

\bibitem{Naka3} H. Nakajima, {\it Heisenberg algebra and Hilbert schemes of points on projective surfaces}, Ann. of Math. 145 (1997), 379--388. \href{https://arxiv.org/abs/alg-geom/9507012}{alg-geom/9507012}.

\bibitem{Nak} H.~Nakajima, \emph{Lectures on Hilbert schemes of points on surfaces}, Univ.~Lect.~Ser.~18, AMS, 1999.

\bibitem{Naka4} H. Nakajima, {\it Towards a mathematical definition of Coulomb branches of 3-dimensional $\mathcal{N}=4$  gauge theories, I.} Adv. Theor. Math. Phys.  20  (2016), 595--669. \href{https://arxiv.org/abs/1503.03676}{arXiv:1503.03676}.

\bibitem{NaYo05} H. Nakajima and K. Yoshioka, {\it Instanton counting on blowup. I. 4-dimensional pure gauge theory}, 
Invent. Math. 162 (2005), 313--355. 
\href{https://arxiv.org/abs/math/0306198}{arXiv:math/0306198}.

\bibitem{NaYo1} H. Nakajima and K. Yoshioka, {\it Perverse coherent sheaves on blow-up. I. A quiver description}, Adv. Stud. Pure Math., 61 (2011), 349--386. \href{https://arxiv.org/abs/0802.3120}{arXiv:0802.3120}. 
			
			\bibitem{NaYo2} H. Nakajima and K. Yoshioka,
			{\it Perverse coherent sheaves on blow-up. II. Wall-crossing and Betti numbers formula.} 
			J. Algebraic Geom., 20 (2011), no. 1, 47--100. \href{https://arxiv.org/abs/0806.0463}{arXiv:0806.0463}. 
            
\bibitem{NaYo3} H. Nakajima and K. Yoshioka, {\it Perverse coherent sheaves on blowup, III: Blow-up formula from wall-crossing}, Kyoto J. Math.  51  (2011), 263--335. \href{https://arxiv.org/abs/0911.1773}{arXiv:0911.1773}. 

\bibitem{NY} H.~Nakajima and K.~Yoshioka, \textit{Lectures on instanton counting}, Algebraic structures and moduli spaces, CRM Proc.~Lecture Notes vol.~38, AMS Providence RI, 31--101 (2004).

\bibitem{Nekrasov:1996cz}
N.~Nekrasov, \textit{Five dimensional gauge theories and relativistic integrable
  systems}, {\em Nucl. Phys. B}, vol.~531, pp.~323--344, 1998,
  hep-th/9609219.

\bibitem{Nekr03} N. Nekrasov, {\it Seiberg--Witten prepotential from instanton counting}, Adv. Theor. Math. {\bf 7} (2003), 831--864. 
\href{https://arxiv.org/abs/hep-th/0206161}{arXiv:hep-th/0206161}.

\bibitem{Nekrasov:2003vi}
N.~Nekrasov, \emph{{Localizing gauge theories}},  in \emph{{14th International
  Congress on Mathematical Physics}}, pp.~645--654, 7, 2003.

\bibitem{NeOk} N. Nekrasov and A. Okounkov, {\it Seiberg-Witten prepotential and random partitions}, The unity of mathematics, 525--596, Progr. Math., 244, Birkh\"auser Boston, Boston, MA, 2006. \href{https://arxiv.org/abs/hep-th/0306238}{arXiv:hep-th/0306238}.


\bibitem{NS} N.~A.~Nekrasov and S.~L.~Shatashvili, \textit{Quantization of Integrable Systems and Four Dimensional Gauge Theories},
ICMP09, pp.~265--289 (2009).

\bibitem{Nes} D.~Nesterov, \emph{Enumerative mirror symmetry for moduli spaces of Higgs bundles and $S$-duality}, arXiv:2302.08379.

\bibitem{Obe} G.~Oberdieck, \emph{Universality of descendent integrals over moduli spaces of stable sheaves on K3 surfaces}, SIGMA 18 (2022) 076 15 pages.

\bibitem{OG} K.~O'Grady, \emph{The weight-two Hodge structure of moduli space of sheaves on a K3 surface}, J.~Alg.~Geom.~6 (1999) 599--644.

\bibitem{OT} J.~Oh and R.~P.~Thomas, \textit{Counting sheaves on Calabi-Yau 4-folds, I}, Duke Math.~J.~172 (2023) no. 7 1333--1409.

\bibitem{OY1}
J. Oh and J. Yagi,
\textit{Poisson vertex algebras in supersymmetric field theories},
Lett. Math. Phys. \textbf{110} (2020), no.~8, 2245--2275.

\bibitem{OY2}
J.~Oh and J.~Yagi,
\textit{Chiral algebras from $\Omega$-deformation}, JHEP
\textbf{08} (2019), 143.

\bibitem{Ohk} R.~Ohkawa, \textit{Wall-crossing between stable and co-stable ADHM data}, Lett.~Math.~Phys.~108 (2018), no. 6, 1485--1523.

\bibitem{okonek1996recent}
C.~Okonek and A.~Teleman, \emph{Recent developments in seiberg-witten theory
  and complex geometry}, {\emph{arXiv preprint alg-geom/9612015} (1996) }.

\bibitem{PP} R.~Pandharipande and R.~P.~Thomas, \emph{Curve counting via stable pairs in the derived category}, Invent.~Math.~178 (2009) 407--447.

\bibitem{PS} M.~Penkava and A.~Schwarz, \textit{On Some Algebraic Structures Arising in String Theory}, 1992. 
\href{https://arxiv.org/abs/hep-th/9212072}{arxiv:hep-th/9212072}.

\bibitem{Per} M.~Perling, \emph{Graded rings and equivariant sheaves on toric varieties}, Math.~Nachr.~263--264 (2004) 181--197.

\bibitem{Peskin:1995}
M.~E. Peskin and D.~V. Schroeder,
\emph{An Introduction to Quantum Field Theory},
Westview Press, Boulder, CO, 1995.

\bibitem{Pidstrigach:1995}
V.~Pidstrigach and A.~Tyurin, \textit{Localisation of the Donaldson's invariants along Seiberg-Witten classes}, (1995) arXiv:dg-ga/9507004

\bibitem{Ram} S.~Ramanujan, \textit{Collected Papers}, Chelsea, New York (1962).

\bibitem{Rawnsley}
J.~H. Rawnsley,
\emph{Flat partial connections and holomorphic structures in {$C^{\infty }$} vector bundles},
Proc. Amer. Math. Soc. \textbf{73} (1979), no.~3, 391--397. MR 518527

\bibitem{Ric} A.~T.~Ricolfi, \emph{An invitation to modern enumerative geometry}, Springer, 2022.

\bibitem{SaWi1}
I.~Saberi and B.~R.~Williams,
\textit{Twisted characters and holomorphic symmetries},
Lett. Math. Phys. \textbf{110}, no.10, 2779-2853 (2020).

\bibitem{SaWi2}
I.~Saberi and B.~R.~Williams,
\textit{Superconformal Algebras and Holomorphic Field Theories},
Annales Henri Poincare \textbf{24}, no.2, 541-604 (2023).

\bibitem{Sedl} S. Sedlacek, {\it A direct method for minimizing the Yang--Mills functional over $4$-manifolds}, 
Comm. Math. Phys. 86 (1982), 515--527.

\bibitem{Schirr} S.~Schirren, \emph{A perfect obstruction theory for $\mathrm{SU}(2)$-Higgs pairs}, arXiv:2403.16625.

\bibitem{Seiberg:1996bd}
N.~Seiberg, \textit{Five-dimensional SUSY field theories, nontrivial fixed points
  and string dynamics}, {\em Phys. Lett. B}, vol.~388, pp.~753--760, 1996,
  hep-th/9608111.

\bibitem{Seiberg:1994aj}
N.~Seiberg and E.~Witten, \textit{Monopoles, duality and chiral symmetry breaking in
  N=2 supersymmetric QCD}, {\em Nucl. Phys.}, vol.~B431, pp.~484--550, 1994,
  hep-th/9408099.

\bibitem{SeWi} N. Seiberg and E. Witten, {\it Electric-magnetic duality, monopole condensation, and confinement in $N=2$ supersymmetric Yang-Mills theory}, 
Nuclear Phys. B 426 (1994), 19--52.

\bibitem{Shapere:2008zf}
A.~D.~Shapere and Y.~Tachikawa,
\textit{Central charges of N=2 superconformal field theories in four dimensions},
JHEP \textbf{09}, 109 (2008)
doi:10.1088/1126-6708/2008/09/109
[arXiv:0804.1957 [hep-th]].

\bibitem{Sie} B.~Siebert, \textit{Virtual fundamental classes, global normal cones and Fulton's canonical classes}, in: Frobenius manifolds, ed.~K.~Hertling and M.~Marcolli, Aspects Math.~36, 341--358, Vieweg (2004). 

\bibitem{Tachikawa:2004ur}
Y.~Tachikawa, \textit{Five-dimensional Chern-Simons terms and Nekrasov's instanton
  counting}, {\em JHEP}, vol.~02, p.~050, 2004, hep-th/0401184.

\bibitem{Tachikawa:2013kta}
Y.~Tachikawa,
\textit{N=2 supersymmetric dynamics for pedestrians},
Lect. Notes Phys. \textbf{890}, 2014
doi:10.1007/978-3-319-08822-8
[arXiv:1312.2684 [hep-th]].

\bibitem{Tam} D.~Tamarkin, \textit{Another proof of M. Kontsevich formality theorem}, 1998. 
\href{https://arxiv.org/abs/math/9803025}{arxiv:math/9803025}.

\bibitem{Tana} Y. Tanaka, {\it Stable sheaves with twisted sections and the Vafa-Witten equations on smooth projective surfaces}, Manuscripta Math. 146 (2015), 351-363. \href{https://arxiv.org/abs/1312.2673}{arXiv:1312.2673}.

\bibitem{Tana2}
Y. Tanaka, 
{\it Some boundedness properties of solutions to 
the Vafa--Witten equations on
	closed 4-manifolds},  
 Q. J. Math. {68}  (2017), 1203--1225. 
\href{https://arxiv.org/abs/1308.0862}{arXiv:1308.0862}. 

\bibitem{Tana3}
Y. Tanaka, 
{\it A perturbation and generic smoothness of the Vafa--Witten moduli spaces on closed symplectic four-manifolds}, 
Glasg. Math. J.  {61}  (2019), 471--486. 
\href{https://arxiv.org/abs/1410.1691}{arXiv:1410.1691}. 

\bibitem{Tana4}
Y. Tanaka, {\it On the singular sets of solutions to the Kapustin--Witten equations and the Vafa--Witten ones on compact K\"ahler surfaces}, Geom. Dedicata {199}  (2019), 177--187. 
\href{https://arxiv.org/abs/1510.07739}{arXiv:1510.07739}. 

\bibitem{TT1} Y.~Tanaka and R.~P.~Thomas, \textit{Vafa--Witten invariants for projective surfaces I: stable case}, Jour.~Alg.~Geom.~29 (2020) 603--668. 

\bibitem{TT2} Y.~Tanaka and R.~P.~Thomas, \textit{Vafa--Witten invariants for projective surfaces II: semistable case}, Pure Appl.~Math.~Quart.~13 (2017) 517--562.

\bibitem{Taub1}C. H. Taubes, {\it The behavior of sequences of solutions to the Vafa--Witten equations}, preprint, 2017.  
\href{https://arxiv.org/abs/1702.04610}{arXiv:1702.04610}. 

\bibitem{Taub2}C. H. Taubes, {\it Spectral flow calculations for reducible solutions to the massive Vafa--Witten equations}, preprint, 2017.  
\href{https://arxiv.org/abs/2401.13419}{arXiv:2401.13419}. 

\bibitem{Taub3}C. H. Taubes, {\it Non-convergent sequences of solutions to the massive Vafa--Witten equations with 'interesting' $\mathbb{Z}/ 2 \mathbb{Z}$ self-dual harmonic 2-form limits}, preprint, 2017.  
\href{https://arxiv.org/abs/2409.14959}{arxiv:2409.14959}. 

\bibitem{Teleman:1996}
A.~{Teleman}, \emph{{Non-abelian Seiberg-Witten theory and projectively stable
  pairs}}, {\emph{arXiv e-prints} (Sept., 1996) alg--geom/9609020},
  [\href{https://arxiv.org/abs/alg-geom/9609020}{{\tt alg-geom/9609020}}].

\bibitem{Tho1} R.~P.~Thomas, \emph{A holomorphic Casson invariant for Calabi-Yau 3-folds, and bundles on K3 fibrations}, J.~Diff.~Geom.~54 (2000) no.~2 367--438.

\bibitem{Tho2} R.~P.~Thomas, \emph{Equivariant K-theory and refined Vafa--Witten invariants}, Comm.~Math.~Phys.~378 (2020) 1451--1500.
\bibitem{Toda} Y. Toda, {Stable pairs on local K3 surfaces}, Jour. Diff. Geom. 92 (2012), 285--370.
\href{https://arxiv.org/abs/1103.4230}{arxiv:1103.4230}. 

\bibitem{Uhle1}K .K .Uhlenbeck, 
{\it Removable singularities in Yang-Mills fields}, Comm. Math. Phys. 83 (1982), 11--29.

\bibitem{Uhle2}K. K. Uhlenbeck, {\it Connections with $L^p$ bounds on curvature}, Comm. Math. Phys. 83 (1982), 31--42. 

\bibitem{UhYa}
K. Uhlenbeck and S-T. Yau, 
{\it On the existence of 
Hermitian-Yang--Mills connections in stable vector bundles}, 
Comm. Pure Appl. Math. 39 (1986), suppl. S257--S293.

\bibitem{VW} C.~Vafa and E.~Witten, \textit{A strong coupling test of $S$-duality}, Nucl.~Phys.~B 431 (1994) 3--77.

\bibitem{Vak} R.~Vakil, \textit{Murphy’s law in algebraic geometry: badly-behaved deformation spaces}, Invent.~Math.~164(3) (2006) 569--590.

\bibitem{BGJK} D.~van Bree, A.~Gholampour, Y.~Jiang, and M.~Kool, \emph{A virtual $\mathrm{PGL}_r\textrm{-}\mathrm{SL}_r$ correspondence for projective surfaces}, Moduli 2 e5 (2025) 1--41. 

\bibitem{WaWi}
M.~Wang and B.~R.~Williams,
\textit{On the renormalization and quantization of topological-holomorphic field theories},
Adv. Math. \textbf{496} (2026).

\bibitem{Weinberg:2000cr}
S.~Weinberg,
\textit{The quantum theory of fields. Vol. 1: Foundations, Vol. 2: Modern Applications, Vol. 3: Supersymmetry}, Cambridge University Press, 1995.

\bibitem{Wess:1992cp}
J.~Wess and J.~Bagger,
\emph{Supersymmetry and Supergravity},
2nd ed. (Princeton University Press, Princeton, NJ, 1992).

\bibitem{Witten:1982fp}
E.~Witten,
\textit{An SU(2) Anomaly},
Phys. Lett. B \textbf{117} (1982), 324-328
doi:10.1016/0370-2693(82)90728-6

\bibitem{WittenMorse}
E.~Witten,
\textit{Supersymmetry and Morse theory},
J. Diff. Geom. \textbf{17}, no.4, 661-692 (1982).

\bibitem{Witt88}E. Witten, {\it Topological quantum field theory}
Comm. Math. Phys. 117 (1988), 353--386.

\bibitem{Witt94}
E. Witten, 
{\it Monopoles and four-manifolds}, 
Math. Res. Lett.  {\bf 1}  (1994), 769--796. 
\href{https://arxiv.org/abs/hep-th/9411102}{arXiv:hep-th/9411102}. 

\bibitem{Witten:1995gf}
E.~Witten,
\textit{On $S$-duality in Abelian gauge theory},
Selecta Math. \textbf{1}, 383 (1995)
doi:10.1007/BF01671570
[arXiv:hep-th/9505186 [hep-th]].

\bibitem{Wit} E.~Witten, \textit{AdS/CFT correspondence and topological field theory}, J.~High Energy Phys.~9812 (1998) 012.

\bibitem{WZ} E.~Witten and B.~Zwiebach, \textit{Algebraic structures and differential geometry in 2-D string theory}, Nucl. Phys. B \textbf{377}, 55-112 (1992).
            
\bibitem{Yosh2} K. Yoshioka, {\it The Betti numbers of the moduli space of stable sheaves of rank 2 on a ruled surface}, Math. Ann.  302, no. 1 (1995), 519-540.

\bibitem{Yosh3} K. Yoshioka, {\it Euler characteristics of $SU(2)$ instanton moduli spaces on rational elliptic surfaces}, Comm. Math. Phys. 205 (1999), 501--517.


\bibitem{Yos1} K.~Yoshioka, \emph{The Betti numbers of the moduli space of stable sheaves of rank 2 on $\PP^2$}, J.~Reine Angew.~Math.~453 (1994) 193--220.

\bibitem{Yos2} K.~Yoshioka, \emph{Some examples of Mukai’s reflections on K3 surfaces}, J.~Reine Angew.~Math.~515 (1999) 97--123.

\bibitem{MR2605321}
D.~Zagier, \textit{Ramanujan's mock theta functions and their applications (after
  {Z}wegers and {O}no-{B}ringmann)}, {\em Ast\'erisque}, no.~326, pp.~Exp. No.
  986, vii--viii, 143--164 (2010), 2009.
\newblock S{\'e}minaire Bourbaki. Vol. 2007/2008.
  
\bibitem{Zag} D.~Zagier, \emph{Nombres de classes et formes modulaires de poids 3/2}, C.~R.~Acad.~Sc.~Paris 281A (1975)
883--886.\bibitem{ZwegersThesis}
S.~P. Zwegers, {\em Mock Theta Functions}.
\newblock PhD thesis, Utrecht University, 2008.

\end{thebibliography}
\end{document}